\documentclass{aa} 
\usepackage{graphicx}
\usepackage{amsmath}
\usepackage{txfonts}
\newcommand{\kms}{\,km\,s$^{-1}$~}	  
\begin{document}
\title{VLT multi-object spectroscopy of 33 eclipsing binaries
   in the Small Magellanic Cloud\thanks{Based on
observations made with the  FLAMES-GIRAFFE multi-object spectrograph mounted
on the Kuyen VLT telescope at ESO-Paranal Observatory (Swiss GTO programme 
072.A-0474A; PI: P. North)}}

\subtitle{New distance and depth of the SMC, and a record-breaking apsidal
motion}

\author{P. North\inst{1} \and R. Gauderon\inst{1}
          \and F. Barblan\inst{2}
          \and F. Royer\inst{3}
          }

\offprints{P. North}

\institute{Laboratoire d'Astrophysique, Ecole Polytechnique F\'{e}d\'{e}rale de Lausanne (EPFL),
		Observatoire, CH-1290 Versoix, Switzerland\\
              \email{pierre.north@epfl.ch}
\and
     Geneva Observatory, Geneva University, CH-1290 Sauverny, Switzerland\\
	 \and
             GEPI, UMR 8111 du CNRS, Observatoire de Paris-Meudon, F-92195 Meudon Cedex, France\\
	      \email{frederic.royer@obspm.fr}}
   
   \date{Received May ??, 2010; accepted May ??, 2010}

\abstract {}
{Our purpose is to provide reliable stellar parameters for a significant sample of
eclipsing binaries, which are representative of a whole dwarf and metal-poor galaxy.
We also aim at providing a new estimate of the mean
distance to the SMC and of its depth along the line of sight for the observed
field of view.}
{We use radial velocity curves obtained with the ESO FLAMES facility at the VLT
and light curves from the OGLE-II photometric survey. The radial velocities were
obtained by least-squares fits of the observed spectra to synthetic ones,
excluding the hydrogen Balmer lines.}
{Our sample contains 23 detached, 9 semi-detached and 1 overcontact systems.
Most detached systems have properties consistent with stellar evolution
calculations from single-star models at the standard SMC metallicity $Z = 0.004$,
though they tend to be slightly overluminous.
The few exceptions are probably due to third light contribution or insufficient
signal-to-noise ratio. The mass ratios are consistent with a flat
distribution, both for detached and semi-detached/contact binaries.
A mass-luminosity relation valid from
$\sim$4 to $\sim$18 $\mathcal{M}_{\odot}$ is derived. The uncertainties are in the
$\pm$2 to $\pm$11 \% range for the masses, in the $\pm$2 to $\pm$5 \% range for
the radii and in the $\pm$1 to $\pm$6 \% range for the effective temperatures. The
average distance modulus is $19.11\pm0.03$ ($66.4\pm 0.9$ kpc). The moduli derived
from the $V$ and from the $I$ data are consistent within $0.01$~mag.
The $2\,\sigma$ depth of the SMC is, for our field,
of $0.25$~mag or $7.6$~kpc under the assumption of a gaussian distribution of
stars along the line of sight. Three systems show significant apsidal motion, one
of them with an apsidal period of 7.6 years, the shortest known to date for a
detached system with main sequence stars.}
{}


\titlerunning{VLT spectroscopy of eclipsing binaries in the SMC}
   \maketitle
%

\section{Introduction}

   Since the late 1990s, the usefulness of extragalactic eclipsing 
binaries has been emphasized in a number of papers. The reader can notably refer 
to the excellent reviews from Clausen (\cite{jC04}) and Guinan (\cite{eG04},
\cite{eG07}). The two major contributions of eclipsing binaries (hereafter
EBs) to astrophysics are to provide (1) fundamental mass and radius
measurements for the component stars, allowing to test stellar evolution models,
and (2) precise distance moduli (DMi)
derived from the luminosities calculated from the combination of the 
absolute radii with the effective temperatures. Until a purely geometrical
distance determination is feasible, 
Paczy\'{n}ski (\cite{bP01}) considers that detached EBs are the
most promising distance indicators to the Magellanic Clouds. Besides, Wyithe \&
Wilson (\cite{WW02}, hereafter WW02) remarked that semi-detached EBs are even
more promising, since their parameters are better constrained.

The renewal of interest in extragalactic EBs, especially EBs in the 
Magellanic Clouds, has been stimulated by the release of a huge number 
of light curves as a byproduct of automated microlensing surveys 
(EROS, MACHO, OGLE) with 1-m class telescopes. As photometry is only half 
of the story, high resolution spectrographs attached 
to 4-m class or larger telescopes had to be used to obtain reliable radial 
velocity (RV) curves. Four B-type EB systems belonging to the Large Magellanic 
Cloud (LMC) were accurately characterized in a series of papers 
from Guinan et al. (\cite{GFD98}), Ribas et al. (\cite{RGF00}, \cite{RFMGU02})
and Fitzpatrick et al. (\cite{FRG02}, \cite{FRGMC03}).
More recently, from high resolution, high $S/N$ spectra obtained with UVES at
the ESO VLT,
the analysis of eight more LMC systems was presented by Gonz\'{a}lez et al.
(\cite{GOMM05}). 
A corner stone in the study of EBs in the SMC was set up with the release 
of two papers from Harries et al. (\cite{HHH03}, hereinafter HHH03) and
Hilditch et al. (\cite{HHH05}, hereinafter HHH05) giving the 
fundamental parameters of a total of 50 EB systems of spectral types O and B. 
The spectroscopic data were obtained with the 2dF multi-object spectrograph on
the 3.9-m Anglo-Australian Telescope. To our knowledge this is the first use of 
multi-object spectroscopy in the field of extragalactic EBs. Recently, even the
distance to large spiral galaxies were measured on the basis of
two EBs in M31 (Ribas et al. \cite{RJVFHG05}, Vilardell et al. \cite{Vil10})
and one in M33 (Bonanos et al. \cite{BSK06}).    

The huge asymmetry, between the number of EB light curves 
published so far and the very small number of RV curves, is striking. 
If one considers the SMC, the new OGLE-II catalog of EBs in the SMC 
(Wyrzykowski et al. 2004) contains 1350 light curves and currently only 50 
of these systems have moderately reliable RV curves. This paper reduces a
little this imbalance by releasing the analysis of 28 more EB 
systems plus revised solutions for 5 systems previously described by HHH03 and HHH05.
The RV measurements were derived from muti-object spectroscopic observations 
made with the VLT FLAMES facility.

Another strong motivation for increasing the number of fully resolved binaries
is to settle the problem of the distribution of the mass ratio
of detached binaries with early B primaries. Recently, two papers were
published supporting two diametrically opposed conclusions:  
van Rensbergen et al. (\cite{VDJ06}), whose work is based on the 9th
Catalogue of Spectroscopic Binaries (Pourbaix et al. \cite{PTB04}), support
the view that the $q$-distribution (where $q$ is the mass ratio) follows a Salpeter-like decreasing power law;
however, from the examination of the homogeneous sample of the 21 detached systems
characterized by HHH03 and HHH05, Pinsonneault \& Stanek
(\cite{PS06}) draw the conclusion that the proportion of close detached systems
with mass ratio $q > 0.87$ far outnumbers what can be expected from either
a Salpeter or a flat $q$-distribution (the ``twins" hypothesis).
Finally, let us mention that the $q$-distribution of semi-detached (i.e. evolved)
systems is no more settled, the statistics strongly depending
on the method used to find the mass ratios, i.e. from the light curve solution
or from SB2 spectra (van Rensbergen et al. \cite{VDJ06}).

Although the controversy about the characteristic distance to the SMC
seems to be solved in favour of a mid position between the ``short"
and the ``long" scales, distance data and line of sight depth remain vital for
comparison with theoretical models concerning the three-dimensional structure
and the kinematics of the SMC (Stanimirovi\'{c} et al. \cite{SSJ04}).

Our contribution provides both qualitative and quantitative improvements over previous
studies. Thanks to the VLT GIRAFFE facility, spectra were obtained with a resolution
three times that in HHH03/05's study. Another strong point is the treatment of nebular
emission. The SMC is known to be rich in \ion{H}{ii} regions (Fitzpatrick \cite{elF85},
Torres \& Carranza \cite{TC87}). Consequently, strong Balmer lines in emission are
very often present in the spectra of the binary systems under study.
Therefore, it appeared rapidly that it was essential to find a consistent way to deal
with this ``third component" polluting most double-lined (SB2) spectra.

We present the observations in Section~2. The reduction of the spectroscopic data and the
interpretation of both photometric and spectroscopic data are described in Section~3, where the
errors are also discussed in detail. The individual binary systems are described in Section~4,
while the sample as a whole is discussed in relation with the SMC properties in Section~5.
 
\begin{figure*}
\centering
\includegraphics[width=12cm]{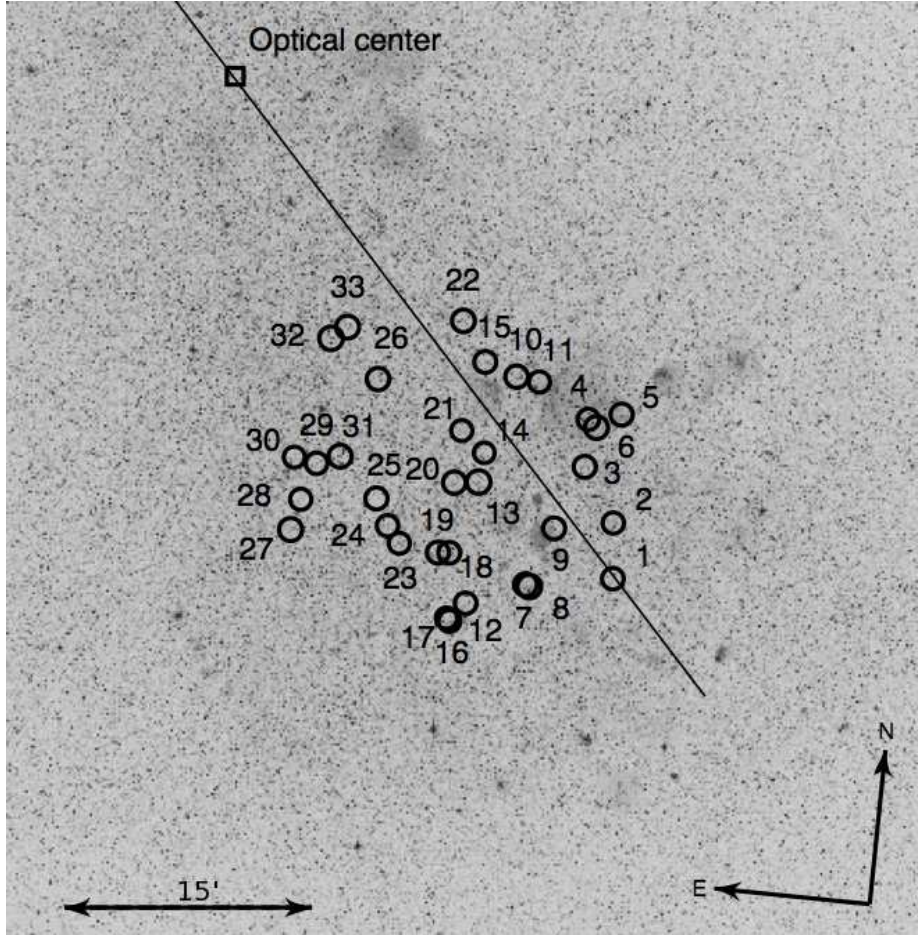}
\caption{Positions of the 33 binaries in the 1$\degr$ $\times$ 1$\degr$ surrounding field centered on (J2000)
   $00^\mathrm{h} 49^\mathrm{m} 20^\mathrm{s}$,
   $-73\degr 13\arcmin 37.3\arcsec$. They lie alongside the southwestern part of the
   SMC bar (oblique line) at $\sim$0.5$\degr$ of the SMC optical
   center.
   Image retrieved from the STScI Digitized Sky Survey (POSS-II/UK Schmidt Telescope,
   Red) (copyright \copyright 1993-5 Anglo-Australian
   Observatory.)}
\label{photodss}%
\end{figure*}

\section{Observations}
\label{obs}

The targets, astrometry included, were selected from the first OGLE photometric
catalog. The GIRAFFE field of view (FoV) constrained 
to choose systems inside a 25\arcmin-diameter circle. Other constraints were
$I\leq18$ mag, at least 15 well-behaved detached light curves (for the SMC
field) and finally seven bump cepheids in the FoV (for another program). The
positions in the sky of 33 objects studied in this paper are shown in Fig. 
\ref{photodss}. The epoch, exposure time, air mass, seeing and age of moon for
each of the 16 CCD frames are gathered in Table \ref{spectro}.

\begin{figure}[!ht]
\includegraphics[trim = 5mm 180mm 80mm 10mm, clip, width=9.5cm]{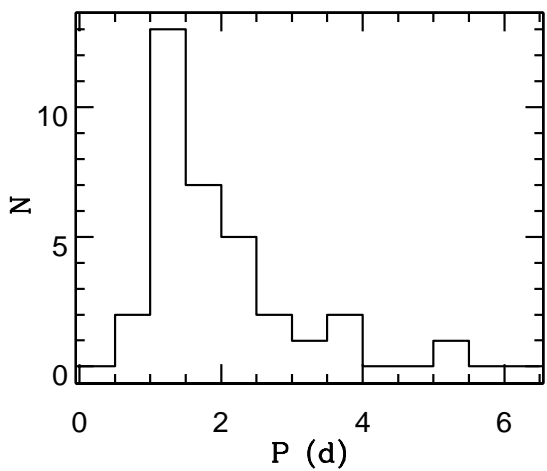}
\caption{ Histogram of periods of our sample of 33 eclipsing binaries in 0.5
day bins.}
\label{periods}
\end{figure}

The relation between our own 1--33 labeling and the OGLE names can be found in
Table \ref{basicparam}, which lists the basic parameters of the systems. The
coordinates are from Wyrzykowski et al. (\cite{WUK04}). The orbital periods and
epochs of the primary minimum are close to those listed by Wyrzykowski et al.
(\cite{WUK04}), but the periods were improved
as far as possible using the radial velocity curves determined in this work.
That was worth the effort, since spectroscopic observations were performed more
than three years after the last photometric ones. Since the times of the
photometric minima are quite sharply defined, the uncertainty on the period
(mentioned between parentheses in Table \ref{basicparam}) is based on the
uncertainty of the spectroscopically defined epoch of the primary minimum. The
latter is quite precise for circular orbits; for eccentric orbits, it is less
accurate, because the Kepler equation had to be solved and the solution is
affected by the uncertainty on the eccentricity. We have decided here to adopt
the {\sl dynamic} definition of the primary and secondary components, rather
than the photometric one. In other words, {\sl the primary component is always
the more massive one, and the primary minimum always corresponds to the eclipse
of the primary by the secondary component}. As a consequence, it may happen that
the so-called primary minimum is {\sl not} the deeper one.
Figure \ref{periods} shows the histogram of the periods. The strong
observational bias in favour of short periods is conspicuous.
  
For all but two binaries, the light curves come from the new version of the
OGLE-II catalog of eclipsing binaries detected in the SMC (Wyrzykowski et al.
\cite{WUK04}). This catalog is based on the Difference Image Analysis (DIA) 
catalog of variable stars in the SMC (see
\verb+http://sirius.astrouw.edu.pl/+$\sim$\verb+ogle/ogle2/+ \verb+smc_ecl/index.html+).
The data were collected from 1997 to 2000. The systems 4~121084 and 5~100731 were
selected from the first version of the catalog (standard PSF photometry) but for an
unknow reason they do no appear anymore in the new version. Nevertheless, they were
retained as there is no objective reason to exclude them.

The DIA photometry is based on $I$-band observations (between 202 and 312 points
per curve). $B$ and $V$ light curves were also used in spite of a much poorer sampling
(between 22--28 points/curve and 28--46 points/curve in $B$ and $V$ respectively).
To give an idea of the accuracy of the OGLE photometry, the objects studied in this
paper have an average $I$ magnitude and scatter in the range
$15.083\pm0.009$ to $18.159\pm 0.047$. These values were calculated from the
best-fitting synthetic light curves. For the two
other bands, we get $14.701\pm0.011$ to $18.090\pm0.025$ for $B$ and
$14.966\pm0.009$ to $18.173\pm0.023$ for $V$.
The quality of an observed light curve can be better expressed by comparing the
depth of the primary minimum $\Delta I_{\mathrm{min I}}$ to the average RMS
scatter $\sigma_I$. These ratios are shown in Table \ref{LCscatter}.
%
%
This permits to classify the light curves in five categories: low
($\Delta I_{\mathrm{min I}}/\sigma_I < 10$), low-to-medium
($10 \leq \Delta I_{\mathrm{min I}}/\sigma_I < 20$), medium
($20 \leq \Delta I_{\mathrm{min I}}/\sigma_I < 30$), medium-to-high
($30 \leq \Delta I_{\mathrm{min I}}/\sigma_I < 40$) and high
($\Delta I_{\mathrm{min I}}/\sigma_I \geq 40$) quality. According to this
scheme, most $I-$band light curves (58\%) belong to the low-to-medium and medium
quality categories, one-third (33\%) in the medium-to-high and high quality
categories, and the remaining 9\% in the poor quality category. This
classification scheme is not useful for the other bands, the
low sampling being the limiting factor.
 
VLT FLAMES/GIRAFFE spectroscopy was obtained by one of us (FR) during eight
consecutive nights from November 16 to 23, 2003. The spectrograph was used in
the low resolution (LR2) Medusa mode: resolving power $R=6400$, bandwidth
$\Delta\lambda=603$~\AA\ centered on $4272$~\AA . 
The most prominent absorption lines in the blue part of early-B stars spectra
are: H$\epsilon$, \ion{He}{i} $\lambda4026$, H$\delta$,
\ion{He}{i} $\lambda4144$, H$\gamma$, \ion{He}{i} $\lambda4388$, and
\ion{He}{i} $\lambda4471$. For late-O stars, \ion{He}{ii} $\lambda4200$ and
\ion{He}{ii} $\lambda4542$ gain in importance. Two fields, one in the SMC and
one in the LMC, were alternatively observed at a rate of four exposures per night
with an integration time of 2595~s for all but one epoch. Therefore, 16 spectra
per target were obtained, with a total of 104 targets in the SMC and 44 in the
LMC. The LMC SB2 systems are being analyzed and will be presented in another
paper.

Beside the spectra of the objects, 21 sky spectra were obtained for each exposure
in the SMC. The parameters related to the spectroscopic observations are gathered
in Table \ref{spectro}. The observed signal-to-noise ratios ($S/N$) were
determined for each {\sl smoothed} spectrum (see Section \ref{spectro_data_red})
in the continuum between 4195 and 4240 \AA. For each
object, two values are presented in Table \ref{orbparam}: the highest and lowest 
$S/N$ values for an exposure of 2595~s. Not surprisingly, the short
exposure of 707~s (due to a technical problem) was useful for the brightest
objects only. For a given binary with $t_\mathrm{exp} = 2595$ s, the ratio of the
highest $S/N$ to the lowest $S/N$ is $\sim$2. 
 
\begin{table*}
\caption{Spectroscopic observations: epochs, heliocentric Julian dates,
exposure times and sky conditions. The value of the air mass is given at mid
exposure. The value of the seeing is an average between the start and end
values. The moon age is the number of days elapsed since the last new moon.}             
\label{spectro}      
\centering          
\begin{tabular}{c c c c c c}     
\hline\hline       
Date of observation (start) & HJD & $t_\mathrm{exp}$ & Air mass & Seeing & Age of Moon \\ 
	& ($-$2\,450\,000) & (s)	 &          &   ($\arcsec$)     & (d)      \\  
\hline                    
  2003-11-16T00:39:59.373 & 2959.5423 &2595 & 1.526 & 0.77 & 21.00\\
  2003-11-16T04:43:39.664 & 2959.7115 &2595 & 1.744 & 0.53 & 21.16\\ 
  2003-11-17T00:35:51.283 & 2960.5394 &2595 & 1.526 & 1.27 & 21.96\\ 
  2003-11-17T03:55:11.067 & 2960.6778 &2595 & 1.645 & 0.95 & 22.09\\ 
  2003-11-18T00:27:41.157 & 2961.5336 &2595 & 1.529 & 0.80 & 22.94\\ 
  2003-11-18T04:40:08.506 & 2961.7090 &2595 & 1.755 & 0.96 & 23.12\\ 
  2003-11-19T00:41:56.657 & 2962.5435 &2595 & 1.519 & 1.00 & 23.97\\ 
  2003-11-19T05:06:05.545 & 2962.7269 &2595 & 1.845 & 0.67 & 24.16\\ 
  2003-11-20T00:27:32.704 & 2963.5335 &2595 & 1.524 & 1.01 & 25.02\\
  2003-11-21T00:46:40.291 & 2964.5358 & 707 & 1.518 & 0.98 & 26.13\\
  2003-11-21T01:00:00.631 & 2964.5560 &2595 & 1.512 & 0.83 & 26.14\\
  2003-11-21T05:39:31.454 & 2964.7501 &2595 & 1.998 & 0.74 & 26.35\\
  2003-11-22T00:20:45.386 & 2965.5287 &2595 & 1.523 & 0.94 & 27.23\\ 
  2003-11-22T04:32:44.400 & 2965.7037 &2595 & 1.779 & 0.92 & 27.43\\ 
  2003-11-23T00:44:54.938 & 2966.5454 &2594 & 1.514 & 0.68 & 28.40\\ 
  2003-11-23T04:27:51.663 & 2966.7002 &2595 & 1.778 & 1.02 & 28.58\\ 
\hline                  
\end{tabular}
\end{table*}

\section{Data reductions and analysis}
\subsection{Spectroscopic data reduction}
\label{spectro_data_red}
The basic reduction and calibration steps including velocity correction to the
heliocentric reference frame for the spectra were performed with the GIRAFFE
Base Line Data Reduction Software (BLDRS) (see
\verb+http://girbldrs.sourceforge.net+).
Sky subtraction, a critical step for faint objects, was done as follows: for each
epoch an average sky spectrum was computed from the 21 sky spectra measured over
the whole FoV. For a given epoch the sky level was found to vary slightly across
the field, but interpolating between spectra  was not considered a valuable
alternative. Local sky variations with respect to the average spectrum are given
in Table \ref{skyemission}. The values $\Delta_\mathrm{sky}$ are read as follows:
for example, the sky position labeled S19 is on average (i.e. over all epochs)
$\sim$20\% brighter than the mean (i.e. over all sky positions) sky spectrum. The
variations were found to be between about $-11$ and $+35$\%. Normalization to the
continuum, cosmic-rays removal and Gaussian smoothing ($FWHM$ = 3.3 pix) were
performed with standard NOAO/PyRAF tasks.

The first 60 \AA \ of the spectra, i.e for wavelengths between 3940 and 4000 \AA,
were suppressed. The reason is that below $\sim$4000\AA \ the $S/N$
is getting very poor and therefore there is no reliable way to place the
continuum. Furthermore, the region around H$\epsilon$ was found to be strongly
contaminated by the interstellar \ion{Ca}{ii} H and K absorption lines. The last
few \,\AA \,(above 4565 \AA) were equally found unusable because of a strongly
corrupted signal.

\subsection{Analysis}
\label{analysis}
For historical reasons, the analysis has been made in essentially two steps.

First, RG did a complete analysis of all systems, using the KOREL code
(Hadrava \cite{pH95}, \cite{pH04}) to obtain both the radial velocity curves and
the disentangled spectra of the individual components. Then, the simultaneous
analysis of light and RV curves was made with the 2003 version of the
Wilson-Devinney (WD) Binary Star Observables Program (Wilson \& Devinney
\cite{WD71}; Wilson \cite{rW79}, \cite{rW90}) via the PHOEBE 
interface (Pr\v{s}a \& Zwitter \cite{PZ05}). However, simulations performed
following the referee's request regarding this early version of the work, showed
that the amplitude $K_\mathrm{P}$ of the RV curve and the mass ratio
$q\equiv \mathcal{M}_\mathrm{S}/\mathcal{M}_\mathrm{P}=K_\mathrm{P}/K_\mathrm{S}$
were not recovered with the expected robustness. More details about these simulations
are given below (Section \ref{disentangling}). Although, on average, the right
values are recovered, one particular solution may be off by as much as five percent
(one sigma) or ten percent (two sigma), which was deemed unsatisfactory\footnote{This
should not be interpreted as a criticism of the KOREL code, but only as a warning
that this code should be used in a very careful way. See also Subsection
\ref{simul-comparison} for a remark about the other method.}.

Thus, a second, almost independent analysis was made by PN, using a least-squares
fit of synthetic binary spectra to observed {\sl unsmoothed} spectra for the RV
determinations. The latter technique seemed much more robust, according to the same
simulation: the $K_\mathrm{P}$ and $K_\mathrm{S}$ amplitudes are recovered to better than one
percent -- at least for the particular binary that was simulated -- and
the dispersion of the values of the small eccentricity ($e\sim 0.03$) is no larger
than two percent. Small systematic errors may result from temperature or rotation
mismatch, but they remain smaller than the uncertainties of the previous
analysis.

We took the opportunity of this new analysis to define a more objective
determination of the effective temperatures of the components. Instead of a 
visual fit of a spectrum close to quadrature, we used a least-squares fit
of synthetic binary spectra to all observed spectra falling out of eclipses.
Then, the error on the effective temperatures could be naturally defined as
the RMS scatter of the results. More details are given in the next sections and
in the following discussion of individual binaries.

\subsection{Photometry: quality check}
\label{qualitycheck}
\begin{figure}[!ht]
\begin{center}
\includegraphics[width=9.0cm]{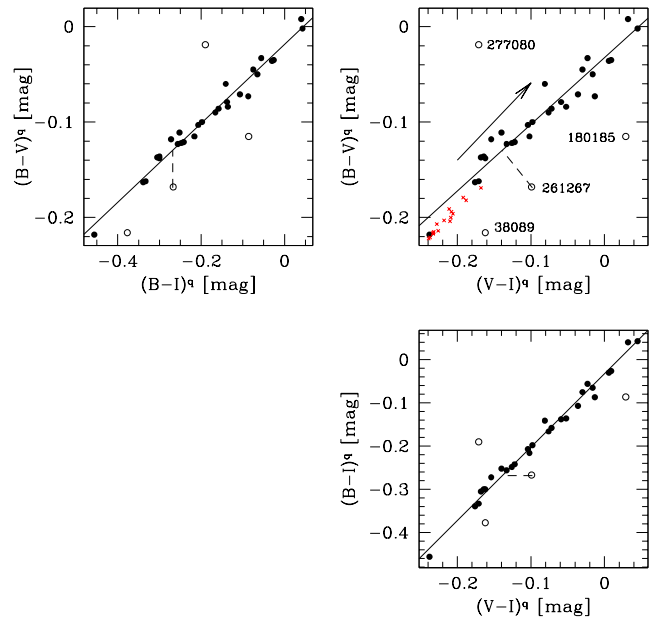}
\caption{Observed (reddened) colour-colour diagrams of our sample. These are the
brightest out-of-eclipse values from Table \ref{basicparam}. They were determined from
the best light-curve fits. The low scatter is due to a reddening line almost parallel
to the main sequence (see upper right panel). Note that some values (open symbol) are
doubtful as they lie off the linear trend. The dashed lines illustrate how to
recover correct $B-V$ and $V-I$ indices from doubtful ones, under the assumption
that the $B-I$ index is reliable, for the system 5 261267.}
\label{colour_colour.2}%
\end{center}
\end{figure}

The quality of the $I-$band light curves was discussed in Section \ref{obs}.
Despite the high range of RMS scatter, we can expect a very accurate determination of the
out-of-eclipse $I-$magnitude because of the large number of data points ($\sim$300).
This is not the case with the $B-$ and $V-$bands. A much poorer sampling can lead to
erroneous zero-level computation in the light-curve analysis step and result in wrong
colour-index determination. Therefore, it is necessary to perform a quality check
of the photometric data. This was done in the form of colour-colour diagrams of our
sample. Figure \ref{colour_colour.2} presents the three diagrams that can be obtained
from the three colour indices $(B-V)^\mathrm{q}$, $(V-I)^\mathrm{q}$ and $(B-I)^\mathrm{q}$.
These are the values at quadratures, i.e an average value characterizing a ``hybrid" star
of intermediate properties with respect to the two components of a particular system.
Not surprisingly, most binaries are found on a relatively narrow linear strip. For any
diagram, the scatter of the objects is low because the reddening line is almost parallel
to the sequence. For example, the ratio $E_{B-V} / E_{V-I} = 0.81$, determined from
Eq. \ref{excessratio} in Section \ref{distancemodulus}, is close to 0.69, the slope of the
sequence $(B-V)^\mathrm{q}$ vs. $(V-I)^\mathrm{q}$. Nevertheless, four outliers appear,
which are marked with open symbols. In principle, there is a possibility to restore a bad
colour index, as illustrated by the example of the system 5 261267:
Inspecting the three light curves, one can suspect that the cause of the discrepancy
lies in a poorly sampled $V-$band light curve in the out-of-eclipse domain. The other
two light curves ($B$ and $I$) seem more reliable. Therefore, only the $B-I$ colour-index
is reliable for this system. But the two other indices, $B-V$ and $V-I$, can safely
be interpolated from the $B-I$ value under the assumption that the system lies on
the linear strip. The method is illustrated by the dashed lines in the diagrams.
Of course, this reconstruction of two bad indices from a good one is not possible,
unless only one of the three light curves is unreliable (either $B$ or $V$). In this
particular example, the situation is not so clear-cut because the reconstructed indices
($B-V=-0.129$ and $V-I=-0.139$) would imply too low a $V^q$ value. Therefore, all four
outliers will be excluded in the final estimate of the distance modulus.

On the upper right diagram of Fig. \ref{colour_colour.2}, red crosses indicate
the intrinsic colours computed below (Section \ref{secsynthphot}). Their positions,
slightly below the regression line of the sequence, is entirely compatible with the
reddening arrow and the observed colours, which inspires confidence regarding the
colour excesses determined in Section \ref{secsynthphot}.

\subsection{Synthetic spectra}
Except for the first two steps of the analysis, i.e. (a) the simultaneous disentangling of the
composite spectra and retrieving of the RV curves through the KOREL code (Hadrava \cite{pH95},
\cite{pH04}) and (b) the simultaneous analysis of the light and RV curves through the WD/PHOEBE
code, the search for the parameters of a binary relies heavily on synthetic spectra.
Indeed, the systemic velocity, the projected rotational velocities, the ratio of
radii and the primary temperature are found by comparing the observed spectra (disentangled and
composite as well) with a library of synthetic spectra. Actually, two libraries were used. For
objects with effective temperature
$15\,000\ \mathrm{K} \leq T_\mathrm{eff}\leq30\,000\ \mathrm{K}$, we used the BSTAR2006 library
recently released by {Lanz \& Hubeny (\cite{LH07}). For the few objects above $30\,000\ \mathrm{K}$,
we took the OSTAR2002 library previously released by the same authors (\cite{LH03}).
Both libraries are available at the TLUSTY Web site
(verb+http://nova.astro.umd.edu/index.html+). 

The grids with a metallicity suitable for the SMC, $Z = 0.004$, i.e. one-fifth of the solar
metallicity, were chosen. Concerning the BSTAR2006 library, we took the grid with a microturbulent
velocity of 2 km s$^{-1}$. Both grids of spectra are based on NLTE line-blanketed model atmospheres.
The temperature step is 1000 K below $30\,000\ \mathrm{K}$ and 2500 K for early-O stars. We
restrained to surface gravities $3.25\ \mathrm{dex} \leq \log g\leq4.75\ \mathrm{dex}$
(0.25 dex step) . The spectra were convolved with the appropriate rotational profiles
($V_{\mathrm{rot}} \sin i$ = 30, 40, 50, 75, 100, 150, 200, 250 and
300 km s$^{-1}$) and with a Gaussian instrumental profile (resolution of 0.67 \AA) by mean of
the program ROTIN3 (http://nova.astro.umd.edu/Synspec43/synspec-frames-rotin.html). Beside a grid
of continuum normalized spectra, a grid of flux spectra was generated for colour indices
calculation through synthetic photometry.

Formally, a normalized synthetic composite spectrum is computed at a given orbital phase
$\phi$ from the normalized $j-$component spectra $F^{\mathrm{n}}_{j,c \ln \lambda}$,
the radial velocities of both components $V_{j} (\phi, \vec p)$ and the light dilution factors
$\ell_{j,c \ln \lambda}$:\\
\begin{equation}
F^{\mathrm{n}}_{c \ln \lambda} (\phi) = \sum_{j =
\mathrm{P}}^{\mathrm{S}} \ell_{j,c \ln \lambda} (\phi) F^{\mathrm{n}}_{j,c \ln \lambda}
\ast \delta(c \ln\lambda - V_{j} (\phi, \vec p)) 
\end{equation}
$\delta$ is the Dirac $\delta-$function and $\vec p$ is the set of orbital parameters. The
spectra are expressed in the logarithmic wavelength scale. The light dilution factors were
calculated from light-curve modeling (Iliji\'{c} et al. \cite{IHPF04}):\\
\begin{equation}
\ell_{\mathrm{P},\lambda} (\phi) = \frac{1}
{1  + \mathcal{L}_{\lambda} \, \frac{\epsilon_{\mathrm{S}}(\phi)}{\epsilon_{\mathrm{P}}(\phi)}}
\quad \mbox{and} \quad \ell_{\mathrm{S},\lambda} (\phi) = 1 - \ell_{\mathrm{P},\lambda} (\phi) 
\end{equation}
where $\epsilon_{j}(\phi)$, the flux of component $j$ in a given passband normalized by the value
at quadrature, accounts for possible out-of-eclipse variability due to
departure from sphericity and reflection effects (Fitzpatrick et al. \cite{FRGMC03});
$\mathcal{L}_{\lambda}$ is the ratio of the continuum monochromatic luminosities, i.e. the
ratio of the mean surface brightnesses $J_{\mathrm{j},\lambda}$ times the ratio of radii,\\ 
\begin{equation}
\epsilon_{\mathrm{j}}(\phi) \equiv
\frac{f_{\mathrm{j},\lambda}(\phi)}{f_{\mathrm{j},\lambda}^{\mathrm{q}}}  \quad \mbox{and} \quad
\mathcal{L}_{\lambda} \equiv \frac{L_{\mathrm{S},\lambda}}{L_{\mathrm{P},\lambda}} =
\frac{J_{\mathrm{S},\lambda}}{J_{\mathrm{P},\lambda}}
\left(\frac{R_{\mathrm{S}}}{R_{\mathrm{P}}} \right)^{2}
\end{equation}
We checked that $\mathcal{L}_{\lambda}$ can safely be considered constant
through the $4000-4580$~\AA\ spectral interval and can be safely identified 
with the luminosity ratio in the Cousins $B$-band of effective wavelength
$\lambda_{\mathrm{eff}} = 4360$~\AA.

\subsection{Radial velocities}
\label{RV}
As explained above (Section \ref{analysis}), the radial velocities were determined
using two independent methods. As a first step, a disentangling method was used, which
allowed to give a first estimate of the parameters of all systems. As a second
and final step, a least-squares method was used. Each step
is described in turn below.

\subsubsection{First step: spectrum disentangling}
\label{disentangling}
Simon \& Sturm (\cite{SS94}) were the first to propose a method allowing the
simultaneous recovery of the individual spectra of the components and of the
radial velocities. Another method aimed at the same results, but using Fourier
transforms to save computing time, was proposed almost simultaneously by Hadrava
(\cite{pH95}). The advantages of these methods are that they need no hypothesis
about the nature of the components of the binary system, except that their
individual spectra remain constant with time. Contrary to the correlation
techniques, no template is needed. In addition to getting at once the radial
velocities and orbital elements, one gets the individual spectra of the
components (``disentangling''), with a signal-to-noise ratio which significantly
exceeds that of the observed composite spectra. For instance, in the case of a
binary system with two components of equal brightness and observed 16 times with
$S/N=50$, the signal to noise ratio of each disentangled spectrum would be
$S/N=4\times 50\times 0.5=100$ (the factor $0.5$ being due to the fact that
there are two components). That is important, because the nature of the
components can then be determined safely. In this
work we use this advantage to determine the effective temperatures of some components,
but for brighter binaries observed at higher resolution and S/N, it would also be possible
to determine photospheric abundances. Other details about these techniques and their
applications (including abundance determinations) can be found in e.g. Hensberge et al.
(\cite{HPV00}), Pavlovski \& Hensberge (\cite{PH05}) and Hensberge \& Pavlovski
(\cite{HP07}).

The radial velocities were determined from the lines of \ion{He}{i} ($\lambda$4471,
$\lambda$4388, $\lambda$4144, $\lambda$4026) only. We preferred to avoid the H
Balmer lines (as did Fitzpatrick et al. \cite{FRG02}) because (1) their large width make
them more sensitive to systematics due e.g. to wrong placement of the continuum, 
and (2) because of moderate to strong nebular emission polluting most systems
(only 6/33 systems were found devoid of any emission). Consequently, four regions with a 
width of 80 \AA \ centered on the four \ion{He}{i} lines were cut from each spectrum
of the series.
Since the KOREL code makes use of the Fourier transform of the spectra, both edges of
each spectral region were fixed to 1 by hammering of the signal
to 1 with a cosine bell function (Hanning window). 
KOREL was run with out-of-eclipse spectra only, although the line-strength factors,
i.e. the contributions of each component to the system continuum, could be obtained in
principle as results of the KOREL analysis. However, most of our spectra
were found to have a too low $S/N$ to provide reliable results. Therefore, the selection
was performed using the out-of-eclipse phase ranges given by 
the light curves. The period $P$ was taken from Table \ref{basicparam}, which are slightly
improved values relative to those of Wirzykowski et al. (\cite{WUK04}), as explained in
Section \ref{obs}. A first estimate of the epoch of periastron passage $T_{0}$, of the
eccentricity $e$ and 
of the longitude of periastron $\omega_{0}$ were determined from the light curves. In the
case of eccentric  systems, a first solution was found neglecting apsidal motion.
The only orbital parameters allowed to converge were the primary semi-amplitude
$K_\mathrm{P}$ and the mass ratio $q$. For each system, KOREL was run with a grid of
values $(K_\mathrm{P}, q)$. The solution with 
the minimum sum of squared residuals as defined by Hadrava (\cite{pH04}) was retained 
as the best solution. For eccentric systems, a second run was performed letting
$K_\mathrm{P}$, $q$, $T_{0}$ and $\omega$ free to converge ($e$ being determined
by photometry).  
It is important to notice that the four spectral regions were analyzed simultaneously,
i.e in a single run of KOREL. 
Each region was weighted according to the $S/N$ of each \ion{He}{i} line
($\mathrm{weight} \propto (S/N)^{2}$). To circumvent the difficulty to measure $S/N$s
inside the lines, they were estimated from the values calculated with the GIRAFFE
Exposure Time Calculator of the ESO. The calculated values
were then normalized to the measured value between 4195 and 4240 \AA. The non-Keplerian
correction and Rossiter effect were calculated from the WD/PHOEBE solution.

Beside the simultaneous retrieving of RV curves, orbital parameters and disentangled
spectra, the KOREL code is able to disentangle spectra for a given orbital solution
($K_\mathrm{P}$, $q$, $T_{0}$ and $\omega$ fixed). A final run of KOREL with this mode
was then used to disentangle the regions around the Balmer and \ion{He}{ii} 4200 and
4542 lines. Indeed, \ion{He}{ii} lines and a number of \ion{Si}{iii-iv} lines are very
useful to set the temperature of hot components.

\subsubsection{Testing the robustness by simulations}
In order to test possible biases on the determination of $K_\mathrm{P}$ and
$q$ by KOREL, we have simulated ten sets of nine out-of-eclipse composite
spectra of the system 5 266131. We used the fitted radial velocity curves to
shift the synthetic spectra of each component and add them at the observed
phases, using the adopted luminosity ratio in the $B$ band. We used the
synthetic spectra with parameters closest to the observed ones, from the grid of
Munari et al. (\cite{MSCZ05}). A Gaussian noise was added to the composite
synthetic spectra, such that the signal-to-noise ratio varied from 37 to 71 (see
Table \ref{orbparam}) and assuming that the $S/N$ ratio is inversely
proportional to the seeing given in Table \ref{spectro}. The KOREL code was run
on each of these ten simulated datasets, and the averages of the resulting
$K_{\mathrm{P}}$ and $q$ values were computed. We found
$\bar{K}_{\mathrm{P}}=228~\mathrm{km\,s^{-1}}$ and $\bar{q}=0.877$ instead of
the input values $223.4\pm 4.8~\mathrm{km\,s^{-1}}$ and $0.889\pm 0.039$
respectively. The simulation then gives $\bar{K}_{\mathrm{S}}=K_{\mathrm{P}}/q=
260~\mathrm{km\,s^{-1}}$ instead of the input
$263.0\pm 3.9~\mathrm{km\,s^{-1}}$. Thus, the parameters obtained from the
simulated spectra agree with the input value to within $1\sigma$, so that no
significant systematic error is to be feared. However, the RMS scatter of
the $K_{\mathrm{P}}$ and $q$ parameters proved disappointingly large, about
$10.8~\mathrm{km\,s^{-1}}$ and $0.044$ respectively. This means that the
uncertainty on the RV amplitude reaches about 5\%, which translates into 15\%
for the masses.

\subsubsection{Second step: least-squares RV determination}
On the basis of the first analysis, we selected two synthetic spectra from the
OSTAR2002 and BSTAR2006 libraries for the two components of each system, with
the parameters closest to the estimations. A chi-square was computed as the
quadratic sum of the differences between the observed spectrum and the composite
synthetic one, for arbitrary radial velocities. However, we did not use the
complete spectra: the hydrogen Balmer lines were suppressed because of their
large width and because they are mixed with nebular emission in a number of
cases. A SuperMongo (Lupton \& Monger \cite{LM00}) procedure implementing the
{\it amoeba} minimization
algorithm was used, letting the two radial velocities and the blue intensity
ratio free to converge. The radial velocities are essentially constrained by
five He\,\textsc{i} lines ($\lambda 4009$, 4026, 4144, 4388 and 4471).
Convergence was generally fast and robust, in the sense that the results did
not depend on the initial guess values. Some iterations were necessary,
however, to clearly identify the primary and secondary components, so that the
right model be attributed to the right component.

A preliminary analysis of the radial velocities was then performed using an
interactive code (Lucke \& Mayor \cite{LM80}), which allowed to assess the quality of
the RV curves (especially the RMS scatter of the residuals) and obtain first
orbital elements.

\subsubsection{Simulations}
\label{simul-comparison}
The same ten sets of nine composite synthetic spectra, described above, was
used to test the least-squares method of RV determination. The results proved
very encouraging, since they follow distributions, the sigma of which amount
to only 0.8 and 0.6\% of the means, for the amplitudes $K_P$ and $K_S$
respectively. The sigma of the eccentricity distribution is 1.8\% of $e=0.036$,
and the argument of the periastron has $\sigma(\omega)=11^\circ$.

The effect of a mismatch was tested by using template spectra with effective
temperatures smaller by 3300~K, respectively 4080~K for the primary and
secondary components, compared to the temperatures used to build the artificial
"observed" spectra (the projected rotational velocities were also smaller by
about 40 \kms). The amplitudes changed by $+1.1$, resp. $+0.3$\% only for the
primary and secondary components. Increasing the temperatures by 2700, resp.
1860~K (and the $V_\mathrm{rot}\sin i$ by 40 \kms) lead to relative differences
$\Delta K_P/K_P=-1.1$\% and $\Delta K_S/K_S=-1.4$\%.
Thus, the mismatch that can be expected will not induce systematic errors much
larger than about one percent, which is in general smaller than the random
error. In view of its robustness, we have adopted the
least-squares technique for RV determination, rather than the results of the
KOREL code. However, we are aware that the above comparison between the two
techniques may not be quite fair, because the KOREL code recovers the individual
spectra from the data, while the least-squares fit uses external template
spectra. In that sense the advantage of the least-squares fit may prove somewhat
artificial.


\subsection{Apsidal motion}
\label{apsid}
The WD code allows to determine the time derivative $\dot{\omega}$ of
the argument of the periastron. That possibility was used for all eccentric
systems but one (4 175333), the latter having the smallest eccentricity of all.
The systems for which a significant apsidal motion was found in that way were
examined further by subdividing the photometric data into four consecutive time
series, and examined using an interactive version of the EBOP16 code (Etzel
\cite{E80}). The $\omega$ angle (if precise enough) and the $e\,\cos\omega$
quantities ajusted for each time series were then examined for systematic
variation. That allowed to better visualize the effect of apsidal motion on the
light curves, and to assess better its significance.

\subsection{Wilson-Devinney analysis}
\label{WDanalysis}
\subsubsection{First step} For each system, a preliminary photometric solution had
been found (before taking thre radial velocities into account) by the application of 
the method of multiple subsets (MMS) (Wilson \& Biermann \cite{WB76}). The groups of
subsets used were essentially the same as those advocated by Wyithe \& Wilson
(\cite{WW01}, \cite{WW02}) (Table \ref{MMS}). That allowed to provide fairly precise
values of $e$ and $\omega$ that were then introduced in the KOREL analysis. Then,
all three light curves and both RV curves provided by KOREL were analyzed
simultaneously using the WD code. That does not imply, however, that photometric
and spectroscopic data were analyzed in a really simultaneous way, since results
from the preliminary light curve analysis were used in the KOREL analysis; it is
rather an iterative analysis. The $I$ light curve is the most constraining one,
thanks to the large number of points, but the $B$ and $V$ light curves are very
important too, since they provide accurate out-of-eclipse $B$ and $V$ magnitudes.
The mass ratio $q$ was fixed to the value found by KOREL. The semi-major orbital axis
$a$, treated as a free parameter, allows to scale the masses and radii. In a first
run, the temperature of the primary was arbitrarily fixed to 26\,000 K. Second-order
parameters like albedos and gravity darkening exponents were fixed to 1.0.
Metallicities $\left[M/H\right]$ were set at $-0.5$. The limb-darkening
coefficients were automatically interpolated after each fit from the van Hamme
tables (van Hamme \cite{VH93}). 
The code needs an estimation of the standard deviations of the 
observed curves in order to assign a weight to each curve. These $\sigma$'s were
calculated from the sums of squares of residuals of the individual curves, as advocated
by Wilson and van Hamme (\cite{WV04}). These values were refined for subsequent runs. 

A fine tuning run was performed with the primary temperature found after analyzing the
observed spectra. The standard uncertainties on the whole set of parameters were
estimated in a final iteration by letting them free to converge.

The standard procedure described above is sufficient for symmetric light curves only. For
systems displaying a small depression before the primary minimum, as it is occasionally
the case with semi-detached systems, it is necessary to introduce a cool spot on the
primary component. Obviously, this step is performed after obtaining the symmetric
best-fit solution. The spot is characterized by four parameters, i.e. colatitude,
longitude, angular radius and temperature factor. As the observed feature can be described by
a large number of combinations of the four parameters related to the spot (high
degeneracy), the spot was arbitrarily put on the equator of the primary (i.e.
colatitude of $\pi/2$) and the three other parameters were optimized alternately
following the MMS. In case of high propensity to diverge, one of the three free
parameters was set to an arbitrary value, the MMS being performed on the two remaining
parameters.
       
In this first step, the WD analysis was performed using the photometric convention,
according to which the primary star is the one that is eclipsed near phase zero,
i.e. the star with the higher mean surface brightness in a given passband
($J_{\mathrm{P},\lambda} > J_{\mathrm{S},\lambda}$). It followed that in some cases $q$
may not necessarily be $\leq 1$. In this paper, we have finally adopted
the dynamic convention $q \leq 1$ in order to avoid confusion.

For the detached systems, the orbit was considered as circular when the
eccentricity given by the WD code was smaller than its estimated error.
 
\begin{table}[!ht]
\caption{Method of multiple subsets: groups of parameters allowed to converge for each model.
$P_\mathrm{orb}$, $t_{0}$, $T_\mathrm{P}$ and $q$ are fixed; $e$ is fixed to 0 for clearly
non-eccentric systems; $a$ may be included in any subset as it does not correlate with any
other parameters. See text for treatment of asymmetric light curves.}             
\label{MMS}      
\centering          
\begin{tabular}{c c c}     
\hline\hline 
Subset 1 & Subset 2 & Subset 3 \\ 
\hline
 & Detached systems &  \\
\hline
 ($e$) & $i$ & $T_\mathrm{S}$ \\
 $\Delta{\phi}$ & $\Omega_\mathrm{S}$ & $\Omega_\mathrm{P}$ \\
 $L_\mathrm{P}^{B, V, I}$ & ($\dot{\omega}$) & ($\omega_{0}$) \\ 
\hline
 & semi-detached and overcontact systems &  \\
\hline
 $\Delta{\phi}$ & $i$ & \\
 $L_\mathrm{P}^{B, V, I}$ & $\Omega_\mathrm{P}$ & \\ 
 $T_\mathrm{S}$ & & \\
\hline   

\end{tabular}
\end{table}

\begin{table*}
\caption{
Basic parameters of the observed eclipsing binaries: identifying number
(Fig. 1), OGLE identifying code, coordinates, orbital period, epoch of primary
minimum (see text), infrared and $(B-V)$ and $(V-I)$ colour indices. 
The difference between $I(\mathrm{DIA})$ and $I(\mathrm{DoPhot})$ photometry
is indicated as well. The ``q" superscript denotes values at a quadrature (the
brightest in case of asymmetric light curve). 
$\alpha$, $\delta$ are from Wyrzykowski et al.
(\cite{WUK04}), while $P_\mathrm{orb}$ and $t_{0}$ are updated values based on
both photometry and radial velocities (see text). The uncertainty on the period
is given between parentheses as the value of the last digit or, in a few cases,
of the last two digits.
$I^\mathrm{q}$, $V^\mathrm{q}$, $(B-V)^\mathrm{q}$ and $(V-I)^\mathrm{q}$ were
determined from the light-curve solutions. The colour indices between parentheses
are unreliable: they correspond to the outliers in Fig. \ref{colour_colour.2}.}  
            
\label{basicparam}      
\centering          
\begin{tabular}{r c c c l l c r c c} 
\hline\hline       
id & OGLE  & $\alpha$ (J2000) & $\delta$ (J2000) & $P_\mathrm{orb}$
& $t_{0}$ (HJD & $I_\mathrm{DIA}^\mathrm{q}$ & $\Delta I_\mathrm{DIA-DoP}^\mathrm{q}$ & $(B-V)^\mathrm{q}$ & $(V-I)^\mathrm{q}$ \\ 
	&object	& (h m s) & ($\degr$ $\arcmin$ $\arcsec$) & (d) & $-$2\,450\,000) &(mag) &(mag) & (mag) & (mag)\\  
\hline                    
 1 & 4 110409 & 00:47:00.19 & $-73:18:43.1$ & 2.973170(4)&619.49136& 15.840 & $-0.002$ & $-0.036\pm0.007$   &$ 0.006\pm0.004$	\\
 2 & 4 113853 & 00:47:03.76 & $-73:15:19.8$ & 1.320757(4)&620.90811& 17.340 & $-0.008$ & $-0.002\pm0.006$   &$ 0.045\pm0.003$	\\
 3 & 4 117831 & 00:47:31.74 & $-73:12:02.2$ & 1.164566(2)&621.36981& 17.799 & $ 0.003$ & $-0.115\pm0.004$   &$-0.102\pm0.003$	\\
 4 & 4 121084 & 00:47:32.05 & $-73:09:08.4$ & 0.823722(1)&624.39596& 16.959 & $-0.000$ & $-0.084\pm0.004$   &$-0.052\pm0.004$	\\
 5 & 4 121110 & 00:47:04.60 & $-73:08:40.1$ & 1.111991(1)&622.29034& 17.003 & $ 0.009$ & $-0.090\pm0.005$   &$-0.076\pm0.002$	\\
 6 & 4 121461 & 00:47:24.69 & $-73:09:35.5$ & 1.94670	 &624.3954 & 17.926 & $ 0.003$ & $-0.060\pm0.005$   &$-0.081\pm0.005$	\\
 7 & 4 159928 & 00:48:13.53 & $-73:19:30.8$ & 1.150460(2)&621.13880& 16.704 & $-0.000$ & $-0.079\pm0.002$   &$-0.059\pm0.002$	\\
 8 & 4 160094 & 00:48:10.17 & $-73:19:37.1$ & 1.699634(66)&620.04883&17.125 & $-0.000$ & $-0.163\pm0.002$   &$-0.176\pm0.002$	\\
 9 & 4 163552 & 00:47:53.24 & $-73:15:56.5$ & 1.545811(2)&620.73188& 15.771 & $ 0.008$ & $-0.045\pm0.003$   &$-0.030\pm0.002$	\\
10 & 4 175149 & 00:48:34.80 & $-73:06:52.6$ & 2.000375(3)&623.85898& 14.970 & $-0.003$ & $-0.218\pm0.007$   &$-0.238\pm0.004$	    \\
11 & 4 175333 & 00:48:15.38 & $-73:07:05.3$ & 1.251126(9)&622.86576& 17.732 & $-0.009$ & $-0.121\pm0.003$   &$-0.122\pm0.003$  \\
12 & 5 016658 & 00:49:02.93 & $-73:20:55.9$ & 1.246158(2)&466.70225& 17.446 & $-0.005$ & $-0.123\pm0.004$   &$-0.133\pm0.002$  \\
13 & 5 026631 & 00:48:59.84 & $-73:13:28.8$ & 1.411680(1)&465.98392& 16.242 & $-0.001$ & $-0.137\pm0.003$   &$-0.168\pm0.001$  \\
14 & 5 032412 & 00:48:56.86 & $-73:11:39.7$ & 3.607857(1)&464.67202& 16.318 & $ 0.005$ & $-0.033\pm0.006$   &$-0.023\pm0.002$  \\
15 & 5 038089 & 00:49:01.85 & $-73:06:06.9$ & 2.389426(2)&468.55092& 15.256 & $-0.002$ &($-0.216\pm0.011$) &($-0.162\pm0.007$)      \\
16 & 5 095337 & 00:49:15.34 & $-73:22:05.8$ & 0.904590(1)&466.18186& 17.090 & $ 0.004$ & $-0.118\pm0.052$   &$-0.154\pm0.030$  \\
17 & 5 095557 & 00:49:18.07 & $-73:21:55.3$ & 2.421185(21)&466.80139&17.440 & $-0.012$ & $-0.100\pm0.003$   &$-0.098\pm0.003$  \\
18 & 5 100485 & 00:49:19.86 & $-73:17:55.6$ & 1.519124(1)&467.15922& 17.150 & $-0.008$ & $-0.138\pm0.008$   &$-0.162\pm0.003$  \\
19 & 5 100731 & 00:49:29.33 & $-73:17:57.9$ & 1.133344(3)&467.82186& 17.378 & $ 0.000$ & $-0.111\pm0.003$   &$-0.140\pm0.002$  \\
20 & 5 106039 & 00:49:20.00 & $-73:13:37.3$ & 2.194069(5)&465.38253& 16.695 & $-0.005$ & $-0.050\pm0.033$   &$-0.016\pm0.019$  \\
21 & 5 111649 & 00:49:17.19 & $-73:10:24.5$ & 2.959578(3)&470.15054& 16.726 & $ 0.007$ & $ 0.008\pm0.009$   &$ 0.032\pm0.003$  \\
22 & 5 123390 & 00:49:22.66 & $-73:03:42.8$ & 2.172917(41)&464.12108&16.203 & $-0.001$ & $-0.162\pm0.015$   &$-0.171\pm0.008$  \\
23 & 5 180185 & 00:50:02.63 & $-73:17:34.4$ & 5.491165(95)&469.37759&17.321 & $ 0.034$ &($-0.115\pm0.011$) &($ 0.029\pm0.009$)      \\
24 & 5 180576 & 00:50:13.44 & $-73:16:33.1$ & 1.561124(2)&466.91033& 17.607 & $ 0.006$ & $-0.071\pm0.009$   &$-0.036\pm0.003$  \\
25 & 5 185408 & 00:50:24.52 & $-73:14:56.0$ & 1.454991(2)&466.28931& 17.524 & $-0.006$ & $-0.122\pm0.003$   &$-0.126\pm0.002$  \\
26 & 5 196565 & 00:50:30.17 & $-73:07:38.2$ & 3.942732(12)&468.26098&16.942 & $ 0.003$ &      (no data)     &	(no data)    \\
27 & 5 261267 & 00:51:35.04 & $-73:17:11.5$ & 1.276632(2)&464.97658& 16.833 & $-0.005$ &($-0.129\pm0.006$) &($-0.099\pm0.004$)   \\
28&5 265970&00:51:28.13&$-73:15:17.6$&3.495685(54)&$465.39125\ ^\mathrm{a}$&16.226&$0.005$&$-0.136\pm0.005$ &$-0.164\pm0.002$  \\
29 & 5 266015 & 00:51:16.82 & $-73:13:01.9$ & 1.808925(2)&465.10449& 15.964 & $ 0.002$ & $-0.086\pm0.004$ &$-0.072\pm0.003$  \\
30 & 5 266131 & 00:51:35.81 & $-73:12:44.8$ & 1.302945(22)&465.50898&17.119 & $ 0.001$ & $-0.103\pm0.016$ &$-0.104\pm0.004$   \\
31 & 5 266513 & 00:50:57.49 & $-73:12:30.3$ & 1.107510(2)&467.15823& 18.066 & $ 0.005$ & $-0.035\pm0.005$ &$ 0.009\pm0.003$  \\
32 & 5 277080 & 00:51:11.68 & $-73:05:20.3$ & 1.939346(4)&465.96082& 16.070 & $ 0.002$ &($-0.019\pm0.013$) &($-0.171\pm0.003$)       \\
33 & 5 283079 & 00:50:58.67 & $-73:04:35.8$ & 1.283583(1)&466.92376& 17.422 & $-0.002$ & $-0.073\pm0.004$ &$-0.013\pm0.002$  \\
\hline                  
\end{tabular}
\begin{list}{}{}
\item[$^{\mathrm{a}}$] Incomplete observations for one of the eclipse. 
\end{list}
\end{table*}

\subsubsection{Second step} Both photometric and RV curves were analyzed simultaneously,
fixing the effective temperature of the primary component to the spectroscopic value (see
below for the determination of the latter). For semi-detached and contact systems, there is
no need to fix any other parameter. For detached systems, however, the ratio of radii is
very poorly constrained by photometry alone when the eclipses are partial, which is the
case of all detached systems in our sample. Therefore, we adopted the brightness ratio
determined by spectroscopy, and fixed the potential of the primary, $\Omega_P$, to a value
such that the brightness ratio in the blue band matched the spectroscopic one
within the
uncertainties. The potential depends on both radius and mass, but the latter is constrained
by the RV curve, so that fixing a potential is equivalent to fixing a radius. In some
cases it was not possible to reproduce the spectroscopic brightness ratio without
degrading the photometric fit, so we gave priority to the latter. 

\subsection{Systemic velocity and projected rotational velocities}
\subsubsection{First step}
The component spectra of the four regions centred on the \ion{He}{i} lines were normalized
with the help of the KORNOR program (Hadrava \cite{pH04}). The systemic velocity
$V_{\gamma}$ was found from the disentangled spectra of the four regions centred on
the \ion{He}{i} lines. The observed spectra were cross-correlated via the IRAF
$\it{fxcor}$ task against synthetic spectra computed for the estimated $T_{\mathrm{eff}}$,
$\log g$ and $V_{\mathrm{rot}} \sin i$. The $V_{\gamma}$ values gathered in Table
\ref{orbparam} were obtained as the $S/N$-weighted averages of the individual
velocities calculated for each line.

The projected rotational velocities, $V_{\mathrm{rot}}^{\mathrm{P, S}} \sin i$, were
tentatively measured by calibrating the FWHMs of the \ion{He}{i} lines against a grid
of (FWHM, $V_{\mathrm{rot}} \sin i$) values obtained from synthetic spectra
(Hensberge et al. \cite{HPV00}). The FWHMs were computed from Gaussian or Voigt profiles
fitting via the IRAF $\it{splot}$ task. The $V_{\mathrm{rot}} \sin i$ values retrieved
by this method were often found unsatisfactory when comparing observed and synthetic
spectra retrospectively. The problem proved to lie in the high sensitivity of the FWHM
measurement to the continuum placement. Therefore, a synchronous rotational velocity
was assumed for most circular binaries unless profile fitting proved this hypothesis
wrong. Anyway, this assumption is certainly justified for short-period systems, i.e.
binaries with $(R/a) \geq 0.25$ (North \& Zahn \cite{NZ03}) where the ratio $(R/a)$
is the star radius divided by the separation. In the case of eccentric systems,
pseudo-synchronization was assumed (Mazeh \cite{tM08}, Eq. 5.1). For a given star,
its pseudo-synchronous rotational velocity is computed from its radius and the
pseudo-synchronization frequency of the binary. This equilibrium frequency, close to
the orbital periastron frequency, is given in Hut (\cite{pH81}).

\subsubsection{Second step}
Contrary to the first step, when the KOREL code was used, we do not need to define
the systemic velocity {\it a posteriori} here. The least-squares method directly
provides ``absolute'' radial velocities (i.e. not only relative ones), even though
mismatch might bias them by a few km\,s$^{-1}$. Thus the systemic velocity naturally flows
from the WD analysis, which includes the RV curves.

As in the first step, rotational velocities were derived from the assumption of
synchronous (for circular orbits) or pseudo-synchronous (for eccentric orbits) spin
motion. No clear departure from this assumption could be seen on the spectra.

\subsection{Spectroscopic luminosity and ratio of radii}
\label{spectroradiusratio}
\subsubsection{First step}
As mentioned above, and as emphasized repeatedly by Andersen (e.g. \cite{jA80})
and rediscovered by Wyithe \& Wilson (\cite{WW01}, hereafter WW01), the ratio of radii
$k \equiv R_{\mathrm{S}} / R_{\mathrm{P}}$ of an EB displaying partial eclipses
is poorly constrained by its light curve. The ratio of monochromatic luminosities
$\mathcal{L}_{\lambda}$ is equally not well recovered in fitting light curves of
simulated EBs (i.e. EBs with previously known parameters). On the contrary, the
surface brightness ratio and consequently the derived effective temperature ratio
$T_{\mathrm{eff}}^{\mathrm{S}} / T_{\mathrm{eff}}^{\mathrm{P}}$ is, in general,
reliably recovered. The sum of the radii $\Sigma R_{j} / a$ is also very well
constrained. The poor constraining of $k$ is very well illustrated by Fig. 3 in   
Gonz\'{a}lez et al. (\cite{GOMM05}). 

Since our sample comprises only systems with partial eclipses, $\mathcal{L}_{\lambda}$,
then $k$ must be determined in order to find reliable radii and surface gravities.
We followed the procedure described in Gonz\'{a}lez et al. (\cite{GOMM05}). The
ratio of the monochromatic luminosities can be expressed by Eq. \ref{eqnRratio}
(Hilditch \cite{rH01}):\\
\begin{eqnarray}
\mathcal{L}_{\lambda} & = & \left(\frac{W_{\lambda}^{\mathrm{S}}}
{W_{\lambda}^{\mathrm{P}}}\right)_{\mathrm{obs}} \left(\frac{W_{\lambda}^{\mathrm{P}}
(T_{\mathrm{eff}}^{\mathrm{P}},\log
g_{\mathrm{P}}, Z_{\mathrm{P}}, Y_{\mathrm{P}})}{W_{\lambda}^{\mathrm{S}}
(T_{\mathrm{eff}}^{\mathrm{S}},\log g_{\mathrm{S}}, Z_{\mathrm{S}},
Y_{\mathrm{S}})}\right)_{\mathrm{true}} \nonumber\\  
& = & \left(\frac{R_{\mathrm{S}}}{R_{\mathrm{P}}}\right)^{2}
\left(\frac{T_{\mathrm{eff}}^{\mathrm{S}}}{T_{\mathrm{eff}}^{\mathrm{P}}}\right)^{4}
10^{-0.4 \left(BC_{\lambda}^{\mathrm{P}} (T_{\mathrm{eff}}^{\mathrm{P}})
- BC_{\lambda}^{\mathrm{S}} (T_{\mathrm{eff}}^{\mathrm{S}})\right)}
\label{eqnRratio}
\end{eqnarray}
where $R_{j}$ is the radius, $T_{\mathrm{eff}}^{j}$ the effective temperature and
$BC_{\lambda}$ the bolometric correction in the $\lambda$-band of component $j$. 
In our case, $\lambda = B$ and the bolometric correction in the $B$-band is given
by $BC_{B} (T_{\mathrm{eff}}) = BC_{V} (T_{\mathrm{eff}}) - (B - V)_{0} (T_{\mathrm{eff}})$,
where the bolometric correction in the $V$-band is interpolated from values given
by Lanz \& Hubeny (\cite{LH03}, \cite{LH07}) and the intrinsic colour index is
computed from synthetic photometry. $W_{l}^{j}$ is the equivalent width (EW) of
line $l$ for component $j$. The `obs' index means that the EWs are measured in an
observed (out of eclipse) spectrum of the binary. These ``apparent'' EWs are then
normalized by the true values measured in synthetic component spectra. The dependence
of the true EWs on the effective temperature, surface gravity, metallicity and
helium abundance ($Y$) is emphasized.
 
In principle, from the analysis of a set of four lines, knowing the masses,
metallicities and helium abundances, it should be possible to derive a purely
spectroscopic solution for the two radii and the two effective temperatures. If we
further assume that the sum of the radii is known from the light-curves analysis as
well as the temperature ratio, the analysis of only two lines is in principle
sufficient to determine a mixed photometric-spectroscopic ratio of radii. However,
since EWs are sensitive to a possible continuum misplacement, we preferred to
fit the observed line profiles with a synthetic composite spectrum or to determine the
ratio of the EWs of two different lines in the same component. The latter methods are
more reliable than blind application of Eq. \ref{eqnRratio} to estimate the effective
temperatures, at least in case of spectra with moderate $S/N$. 

For a given chemical composition, true undiluted EWs depend on both the effective
temperature and the surface gravity of the stars. Moreover, this dependence is not
always monotonic even if we restrict to late O to early B stars, the \ion{He}{i}
lines having a maximum strength at $\sim$20\,000 K. Consequently, in order to avoid
the hassle of working with non-explicit equations, for a given line, Eq. \ref{eqnRratio}
was solved with the photometric temperature ratio and the true EWs values
corresponding to the photometric $\log g$ and a first guess of
$T_{\mathrm{eff}}^{\mathrm{P}}$. $\mathcal{L}_{B}$ is then used with
$T_{\mathrm{eff}}^{\mathrm{S}} /T_{\mathrm{eff}}^{\mathrm{P}}$ and a first guess of
$\Delta BC_{B}$ to compute the ratio of radii $k$. Combining $k$ and
$\Sigma R_{j} / a$, the new $R_{j}$ and $\log g_{j}$ values are obtained
straightforwardly. The small error introduced in the chain 
$\mathcal{L}_{B} \to k \to  R_{j} \to \log g_{j} \to T_{\mathrm{eff}}^{\mathrm{P}}
\to W_{\lambda}^{j} \to ...$ because of using approximate values
for the true EWs could be removed after iterating one more time. 

Nevertheless, this method is not very efficient when the observed EWs have large
uncertainties as in case of a composite spectrum of low $S/N$. In this case, a more
pragmatic approach consists in optimizing  both $\mathcal{L}_{B}$ and
$T_{\mathrm{eff}}^{\mathrm{P}}$ in a single step by looking for the best-fitting
synthetic composite spectrum for a given pair ($\mathcal{L}_{B}$,
$T_{\mathrm{eff}}^{\mathrm{P}}$) and the $\Sigma R_{j} / a$ and 
$T_{\mathrm{eff}}^{\mathrm{S}} /T_{\mathrm{eff}}^{\mathrm{P}}$ constraints. 

\subsubsection{Second step}
\label{lum_ratio_2}
Here the luminosity ratio in the blue is simply one of the three parameters determined
by the non-linear least-squares algorithm {\it amoeba}, the other two parameters being
the effective temperatures (see more details below, Section \ref{Teff1-2}). So the
luminosity ratio is determined in a very homogeneous way, and an error estimate naturally
arises through the RMS scatter of the resulting values. This does not guarantee,
however, that the results are free from any bias. In particular, one may suspect that,
in the temperature regime where the strength of all lines (H and He ones) vary in the
same way with temperature, some degeneracy may arise between the temperatures and the
luminosity ratio. Since that temperature regime spans roughly from $22000$ to $30000$,
it means that the luminosity ratio of the majority of systems may be fragile.
Nevertheless, {\it a posteriori} examination of the resulting HR diagrams does not
confirm this fear, even though a few systems fail to match the evolutionary tracks.

\subsection{Effective temperatures}
\label{Teff1-2}
Once reliable $V_{\mathrm{rot}} \sin i$ and $\log g$ values have been found, a way
for setting the temperature of the primary must be found (the temperature of the
secondary is a by-product via the photometric temperature ratio).
    
For late O and early B stars, the H$\gamma$ and H$\delta$ Balmer lines are far better
temperature indicators than \ion{He}{i} lines (Huang \& Gies\cite{HG06a}, \cite{HG06b}).
Therefore, the most direct way to determine the effective temperatures of both
components of a given system would consist in calibrating the equivalent widths
measured on the normalized disentangled spectra with those obtained with a library of
synthetic spectra. Unfortunately, this is not always possible because of the high
proportion of systems contaminated by H$\gamma$ and H$\delta$ nebular emission
lines. Thus, most spectra of individual components are not reliable around the Balmer
lines.

\subsubsection{First step}
A safer method consists in comparing an observed composite spectrum close to quadrature
with a synthetic composite spectrum computed at the same orbital phase.
The spectra retained for the temperature determination are those with
$0.21 < \phi < 0.29$ or $0.71 < \phi < 0.79$. This method is quite sensitive to the
continuum placement. A low $S/N$ and/or strong emission lines can hinder a reliable
profile fitting.

Another method is a variant of the traditional spectral type vs. temperature calibration.
In the traditional method, the line strengths ratio of two lines are measured and
compared to the values obtained from a series of reference spectra whose spectral
types are known. The effective temperature is then found via a spectral type -
$T_{\mathrm{eff}}$ calibration scale. As emphasized by HHH03/05, this technique is
efficient for O-B1 stars but far less straightforward for later types. Above this limit,
the relative strength of the \ion{He}{ii} 4542 and \ion{Si}{iii} 4553 lines is a
reliable tool, as is the relative strength of the \ion{Si}{iv} 4089-4116 and
\ion{He}{i} 4121 lines. For temperatures below $\sim$29-30\,000 K, the problem is the
lack of exploitable metallic lines. Unfortunately, the faint \ion{Si}{ii} lines are
totally undetectable. For later B stars, the only detectable metallic line is
\ion{Mg}{ii} 4481, but this line is often severely buried in the noise for most
components.

\subsubsection{Second step}
The method is the same, qualitatively speaking, as that of the first step, consisting
in fitting a composite synthetic spectrum to the observed one near quadrature. However,
the least-squares fit method allows in principle to use all out-of-eclipse spectra,
and provides a much more objective estimate of the temperatures. Since the radial
velocities are known, the only parameters which have to be fit are the effective
temperatures of both components and the blue luminosity ratio. As mentioned in
Section \ref{lum_ratio_2}, the fit is quite robust at both ends of the temperature
range of our sample: at the cool end, the He\,\textsc{i} lines increase in strength
with temperature, while the H\,\textsc{i} Balmer lines decrease; in addition, the
\ion{Mg}{ii} line decreases very fast. At the hot end, both H and He\,\textsc{i}
lines decrease with temperature, but the He\,\textsc{ii} lines begin to appear.
In the intermediate range, all lines vary more or less in parallel, which may lead
to degeneracy when the S/N ratio is poor.

For all systems showing a significant nebular emission in the core of the hydrogen
Balmer lines, we simply removed a $4$~\AA~ wavelength interval centered on the emission
line, in both observed and synthetic composite spectra. But, contrary to the synthetic
spectra used for RV determinations, here we include the H Balmer lines in the fit,
except for their very centres.

The RMS scatter of the fitted effective temperatures is typically of the order of
$1000$~K, and is even smaller than that for one third of the sample. Although formally,
the error bar on $T_\mathrm{eff}$ should be set to that scatter divided by the square
root of the number of spectra, we chose to put it equal to the scatter itself. Indeed,
visual examination of the observed and model spectra show that the temperature effect
is often very subtle, so we feel this choice is more realistic.

The fit proved to depend somewhat on the normalization of the spectra. The latter were
first normalized using an automatic procedure with fixed continuum regions. Then, another
automatic procedure was used, which corrected the first normalization with the help
of a pair of synthetic spectra with preliminary stellar parameters. That normalization
resulted in a slightly higher continuum and was found satisfactory in general, except
for the bluer end of the spectrum.
A final normalization was made by hand, which was adopted in most cases, but not in all
because the continuum proved sometimes too high. The temperature determination was
run on the automatically normalized spectra as well as on the manually normalized ones.
The results were found to depend little on the normalization, which was not unexpected
since the last two normalizations did not differ much from one another.

The average temperatures were computed on all out-of-eclipse spectra on the one hand,
and on a selection of those for which the RV difference is greater than 300~\kms
(250 or even 200~\kms for longer period systems) on the other hand. The selection
often resulted in a smaller scatter of the temperature, though not in every case.
The average temperature was weighted with the inverse of the chi-square provided by
the {\sl amoeba} procedure. 

\subsection{Synthetic photometry and reddening}
\label{secsynthphot}
Intrinsic $(B-V)$ colour indices are needed for two purposes: the computation of the
$E_{B-V}$ colour excess for a given system and the computation of the $B-$band
bolometric corrections $BC_B$ of the individual components. In the first case,
$(B-V)_0$ is phase-dependent and characterizes the binary as a whole, while in the
second case $(B-V)_0^j$ is the usual (constant) colour characteristic of a given
star. Both types of colour indices were computed from synthetic photometry, i.e. from
synthetic stellar spectra and the response functions of the filters. The general
formula for the phase-dependent $(B-V)_0$ of a binary is given by   
\begin{equation}
(B-V)_{0} (\phi) = -2.5 \log \frac{\int S_{B} F^{0}_{\lambda \oplus} (\phi) \,
\mathrm{d}\lambda \int S_{V} \, \mathrm{d}\lambda}
{\int S_{V} F^{0}_{\lambda \oplus}
(\phi) \, \mathrm{d}\lambda \int S_{B} \, \mathrm{d}\lambda} + C_{B-V}
\label{eqnBV}
\end{equation}
where $S_{X}$ is the response function of the $X-$band filter (Bessell \cite{mB90}),
$F^{0}_{\lambda \oplus}$ is the synthetic composite flux spectrum at a given epoch and
$C_{B-V} = 0.606$ mag is the zero point (Bessell et al. \cite{BCP98}). The synthetic
composite spectra (i.e. the theoretical unreddened flux received on Earth) were
calculated from the synthetic component spectra $F^{0}_{j,\lambda}$ (i.e. the surface fluxes), 
the radial velocities $V_{j} (\phi, \vec p)$ and the radii $R_{j}$:
\begin{eqnarray}
F^{0}_{\lambda \oplus} (\phi) & = & \frac{1}{d^{2}} \sum_{j =
\mathrm{P}}^{\mathrm{S}} R_{j}^{2} F^{0}_{j,\lambda} \ast \delta(c \ln
\lambda - V_{j} (\phi, \vec p)) \nonumber\\
 & = & \left( \frac{R_{\mathrm{P}}}{d} \right)^{2} \sum_{j =
\mathrm{P}}^{\mathrm{S}} w_{j} F^{0}_{j,\lambda} \ast \delta(c \ln
\lambda - V_{j} (\phi, \vec p))  
\label{eqnF}
\end{eqnarray}
where $w_{\mathrm{P}} = 1$ and $w_{\mathrm{S}} =
\left( R_{\mathrm{S}} / R_{\mathrm{P}} \right)^{2}$. Substituting $F^{0}_{\lambda \oplus}$
in Eq. (\ref{eqnBV}) by the expression given by Eq. (\ref{eqnF}), one sees that
Eq. (\ref{eqnBV}) does not depend on the distance $d$ to the system.   

A similar procedure was used to compute the $(V-I)_0$ colour indices, taking
$C_{V-I} = 1.268$ and the appropriate response functions $S_V$ and $S_I$. This index
is needed for the determination of the distance modulus from the $I-$band
photometric observations.

Alternatively, the intrinsic colour index $(B-V)_0$ could be estimated via a
colour-temperature relation from the literature. This kind of calibration being
established with stars from the solar neighborhood (e.g. Flower \cite{pF96}),
the coefficients of the fit are in turn representative of stars with a solar
metallicity. Consequently, synthetic photometry with $Z = 0.004$ spectra was
considered more reliable for our objects. Both $(B-V)_0$ and $(V-I)_0$ were computed
for a grid of temperatures and surface gravities. Because the fluxes are not given
beyond $7500$~\AA~ in the OSTAR2002 library (which contains fluxes for stars
hotter than 30\,000~K), we had to complement the flux distribution
of stars hotter than $30\,000$~K by the appropriate Kurucz fluxes\footnote{see
\texttt{http://kurucz.harvard.edu/grids.html}} in order to be able to compute the
$(V-I)_0$ index. We chose the grid with a metallicity of $-0.5$. Because of that
inhomogeneity, the $(V-I)_0$ indices of binary systems hosting components
with $T_\mathrm{eff} > 30\,000$~K may be slightly less reliable than those of the other
systems (the systems 4~110409, 4~121084, 4~175149, 5~32412, 5~38089 and 5~266015 are
in this case). For a given object, the colour indices were linearly interpolated from
these grids. 

To estimate the uncertainties, however, the following first-order approximations valid
for early-B stars of intermediate $\log g$ values were used (see Section
\ref{uncertainties}),
\begin{eqnarray}
\label{eqCI}
(B-V)_0 & \approx & 1.768 - 0.455 \log T_{\mathrm{eff}} ~~~\mathrm{for}~~~
15\,000 \leq T_{\mathrm{eff}} \leq 32\,000 \\
&\approx& 0.228 - 0.113 \log T_{\mathrm{eff}} ~~~\mathrm{for}~~~
 32\,000 \leq T_{\mathrm{eff}} \leq 37\,500 \nonumber \\
(V-I)_0 & \approx & 1.958 -  0.502 \log T_{\mathrm{eff}} ~~~\mathrm{for}~~~
15\,000 \leq T_{\mathrm{eff}} \leq 32\,000 \nonumber \\
&\approx& 0.502 - 0.178 \log T_{\mathrm{eff}} ~~~\mathrm{for}~~~
32\,000 \leq T_{\mathrm{eff}} \leq 37\,500 \nonumber 
\end{eqnarray}

\begin{table}
\caption{Intrinsic colour indices at quadrature computed by synthetic photometry. Because of the
too short wavelength range provided in the OSTAR2002, $(V-I)_0$ values of systems with a
component hotter than $30 \, 000$ K were computed using Kurucz red fluxes (see text).}             
\label{synthphot}      
\centering          
\begin{tabular}{c c c } 
\hline\hline  
Object & $(B-V)_0^\mathrm{q}$ & $(V-I)_0^\mathrm{q}$  \\
	& (mag) & (mag)  \\
\hline                    
 4 110409  & $-0.243\pm 0.006$& 	 $-0.252\pm 0.007$  \\
 4 113853  & $-0.196\pm 0.007$& 	 $-0.206\pm 0.008$  \\    
 4 117831  & $-0.179\pm 0.004$& 	 $-0.192\pm 0.004$ \\
 4 121084  & $-0.271\pm 0.004$& 	 $-0.294\pm 0.004$ \\
 4 121110  & $-0.248\pm 0.006$& 	 $-0.269\pm 0.006$  \\
 4 121461  & $-0.207\pm 0.007$& 	 $-0.228\pm 0.008$  \\
 4 159928  & $-0.247\pm 0.006$& 	 $-0.263\pm 0.007$ \\
 4 160094  & $-0.238\pm 0.009$& 	 $-0.260\pm 0.009$  \\
 4 163552  & $-0.232\pm 0.008$& 	 $-0.246\pm 0.009$ \\
 4 175149  & $-0.271\pm 0.002$& 	 $-0.288\pm 0.002$ \\
 4 175333  & $-0.191\pm 0.006$& 	 $-0.211\pm 0.007$ \\
 5 016658  & $-0.203\pm 0.005$& 	 $-0.218\pm 0.005$ \\
 5 026631  & $-0.242\pm 0.003$& 	 $-0.257\pm 0.003$ \\
 5 032412  & $-0.283\pm 0.001$& 	 $-0.303\pm 0.001$    \\
 5 038089  & $-0.273\pm 0.001$& 	 $-0.292\pm 0.001$ \\  
 5 095337  & $-0.242\pm 0.009$& 	 $-0.261\pm 0.010$ \\  
 5 095557  & $-0.216\pm 0.003$& 	 $-0.233\pm 0.003$ \\	
 5 100485  & $-0.215\pm 0.004$& 	 $-0.232\pm 0.005$  \\
 5 100731  & $-0.217\pm 0.004$& 	 $-0.233\pm 0.004$  \\ 
 5 106039  & $-0.204\pm 0.003$& 	 $-0.210\pm 0.003$  \\
 5 111649  & $-0.169\pm 0.002$& 	 $-0.168\pm 0.002$  \\
 5 123390  & $-0.254\pm 0.004$& 	 $-0.274\pm 0.004$   \\
 5 180185  & $-0.182\pm 0.004$& 	 $-0.188\pm 0.005$   \\
 5 180576  & $-0.221\pm 0.009$& 	 $-0.237\pm 0.010$  \\
 5 185408  & $-0.222\pm 0.004$& 	 $-0.239\pm 0.004$   \\
 5 196565  & $-0.200\pm 0.003$& 	 $-0.209\pm 0.004$  \\
 5 261267  & $-0.253\pm 0.006$& 	 $-0.266\pm 0.007$  \\
 5 265970  & $-0.214\pm 0.001$& 	 $-0.226\pm 0.001$  \\
 5 266015  & $-0.265\pm 0.004$& 	 $-0.280\pm 0.004$  \\
 5 266131  & $-0.239\pm 0.006$& 	 $-0.258\pm 0.007$  \\
 5 266513  & $-0.194\pm 0.011$& 	 $-0.208\pm 0.012$  \\
 5 277080  & $-0.244\pm 0.004$& 	 $-0.256\pm 0.004$  \\       
 5 283079  & $-0.234\pm 0.004$& 	 $-0.252\pm 0.004$  \\ 
\hline			  
\end{tabular}		  
\end{table}

Once the colour excess $E_{B-V} \equiv (B-V)-(B-V)_0$ is determined, one must
make an assumption about the value of the extinction parameter
$\mathcal{R}_{V} \equiv A_V / E_{B-V}$. This parameter is assumed equal
to the standard value $3.1\pm0.3$ for each system. It is worth mentioning that this
parameter suffers from a rather
large uncertainty. For the SMC bar, Gordon et al. (\cite{GCMLW03}) propose the
mean value $\mathcal{R}_{V} = 2.74\pm0.13$ from a sample of four stars with
$\mathcal{R}_{V}$ ranging from $2.4\pm0.3$
 to $3.3\pm0.38$. Nevertheless, because of the small size of Gordon et al.'s sample,
 the more conservative standard value was retained for this paper.
 Moreover, the contribution of $\mathcal{R}_{V}$ to the total error
 budget for the DMi is low. The difference $\Delta_{DM}$ between a DM
 calculated with Gordon et al. 's value and the
 standard value is given by $\Delta_{DM} = -0.36 \, E_{B-V}$. Thus, with
 $E_{B-V} \sim 0.1$, the DM is slightly increased by 0.036 mag if
 $\mathcal{R}_{V}=2.74$ is adopted.
 
 In order to determine the distance modulus from the $I-$band data, the
 ratio of the absorptions in the infrared and optical bands $A_I / A_V$ is
 calculated from the relationship,  
\begin{eqnarray} 
\frac{A_\lambda}{A_V} = a(x)  \, + \,  b(x) \, \frac{1}{\mathcal{R}_{V}}
\quad \mathrm{or}  \quad A_\lambda
= E_{B-V} \left( a(x) \mathcal{R}_{V}  \, + \,  b(x) \right) 
\end{eqnarray}
with $x \equiv 1/\lambda_{\mathrm{eff}}$. The coefficients $a(x)$ and $b(x)$
are computed from the polynomials given in Eq. 3a-b from Cardelli et al.
(\cite{CCM89}). An effective wavelength of 7980\AA\ is taken for the Cousins
$I-$band filter 
(Bessell \cite{mB90}). Hence, it follows that  $A_I / A_V \approx 0.600$. That
value contrasts with the value of 0.479 given by Cardelli et al. (\cite{CCM89})
in their Table~3 for the $I$ band, because they adopt another effective
wavelength.

\subsection{Distance modulus: $V-$ or $I-$band approach?}
\label{distancemodulus}
There are two ways of determining the distance modulus of a particular system,
depending on whether one takes visual or infrared data:
\begin{eqnarray} 
5 \log d - 5 & = &
V^\mathrm{q} \, - \, M_V^\mathrm{q} \, - \, A_V^\mathrm{q} \nonumber\\ 
   & = & I^\mathrm{q} \, - \, M_I^\mathrm{q} \, - \, A_I^\mathrm{q} \nonumber\\ 
   & \approx & I^\mathrm{q} \, - \, M_V^\mathrm{q} \,
   + \, (V-I)_0^\mathrm{q} \, - \, 0.600 \, A_V^\mathrm{q}    
\label{eqnmodule}
\end{eqnarray}
The safest choice is to take the expression which leads to the value with the
smallest uncertainty. The $I-$band magnitude at quadrature is far more
reliable than the $V-$band value because there is typically $\sim$10$\times$
more points in the $I-$band light curve. Moreover, the contribution of
the interstellar absorption is lower at infrared wavelengths. Indeed, it is easy
to demonstrate that the variance of the visual absorption
is always larger than the sum of the contributions due to the infrared
absorption and the intrinsic $(V-I)$ colour:
\begin{eqnarray}   
\sigma^{2}_{A_V^\mathrm{q}} > \sigma^{2}_{A_I^\mathrm{q}} \, 
+ \, \sigma^{2}_{(V-I)_0^\mathrm{q}} \quad \mbox{with} \quad 
\sigma^{2}_{(V-I)_0^\mathrm{q}} \approx \sigma^{2}_{(B-V)_0^\mathrm{q}}  
\label{eqnmodule2}
\end{eqnarray}
Consequently, the $I-$band calculation must be preferred, at least for system
for which there is a reliable way to obtain the $(V-I)_0$ index. We have seen
that this is the case of systems where both components have
$T_{\mathrm{eff}} \leq 30\,000$ K, while for hotter binaries, the synthetic
spectra of the OSTAR2002 library had to be complemented by Kurucz models beyond
7500 \AA. In spite of this slight inhomogeneity, we have adopted the $(V-I)_0$
indices computed in this way, rather than refrain from computing the $I$
distance modulus for these hot binaries.

In order to check the purely synthetic intrinsic $(V-I)$ index, both $V-$ and
$I-$band formulations of Eq. \ref{eqnmodule} can be used to derive the following
relation between the two colour excesses:
\begin{eqnarray}
\frac{E_{V-I}}{E_{B-V}} = 0.400 \, \mathcal{R}_{V}
\label{excessratio}
\end{eqnarray}
A simple modification of this equation gives,
\begin{eqnarray}
\label{reddeningline}
(V-I)_0 = (V-I) - 0.400  \, \mathcal{R}_{V}  \, E_{B-V}
\end{eqnarray}
This is a semi-empirical dereddening, as $E_{B-V}$ is calculated from a
synthetic intrinsic $(B-V)$ index. The $(V-I)_0$ indices so obtained agree
within $0.01-0.02$ mag for 4 of the 6 hot binaries; the differences reaches
$0.03$~mag for the system 5 32412, but this is less than the estimated error on
the distance modulus. For 5 38089, the difference is $0.06$~mag, but this system
has abnormal colours and so will be excluded from the statistics of the distance
moduli.

\subsection{Uncertainties}
\label{uncertainties}
\subsubsection{Distance modulus and related parameters}
The calculus of the uncertainties in a number of parameters relies on some
assumptions and simplifications. These points are discussed in this paragraph. 
The uncertainty in the distance modulus of a particular binary
$\sigma_{5 \log d - 5}$ is determined via the standard rules of the propagation
of errors for independent variables,
\begin{eqnarray}
\sigma^{2}_{5 \log d - 5} & \approx & \sigma^{2}_{M_{V}^{\mathrm{q}}} \,
+ \, \sigma^{2}_{(V-I)_0^\mathrm{q}} \, + \, \nonumber\\ 
& & \left( 0.600 \, \mathcal{R}_{V} \right)^{2} 
\left(\sigma^{2}_{(B-V)^{\mathrm{q}}}\,+\,\sigma^{2}_{(B-V)_{0}^{\mathrm{q}}}\right)
\label{sigmamodule}
\end{eqnarray}
This expression was easily derived from the infrared formulation of
Eq. \ref{eqnmodule}. The uncertainty in the $I$ magnitude
($\sigma_{I^{\mathrm{q}}} \leq 2$ mmag) and the $E_{B-V}^{2}  \,
\sigma^{2}_{\mathcal{R}_{V}}$ term are negligible and consequently omitted from
Eq. \ref{sigmamodule}. It is worth noting that formally the intrinsic colour
indices $(B-V)_{0}^{\mathrm{q}}$ and $(V-I)_{0}^{\mathrm{q}}$
and the visual absolute magnitude of the system $M_{V}^{\mathrm{q}}$ are not
independent variables, as they depend ultimately on the effective
temperatures and radii of the stars. Nevertheless, for the sake of simplicity
these variables were considered as independent in the following development.
The uncertainty on $M_{V}^{\mathrm{q}}$ can be expressed as a function of the
uncertainties on the absolute magnitudes of the components $M_{V}^{P}$ and
$M_{V}^{S}$ 
\begin{equation}
\sigma^{2}_{M_{V}^{\mathrm{q}}} =
\frac{1}{\left( 10^{-0.4 M_{V}^{\mathrm{P}}} + 
10^{-0.4 M_{V}^{\mathrm{S}}} \right)^{2}}
\sum_{j = \mathrm{P}}^{\mathrm{S}} \left( 10^{-0.4 M_{V}^{j}}
\sigma _{M_{V}^{j}}\right)^{2}
\end{equation}

The absolute visual magnitude of a component is found via the absolute
bolometric magnitudes $M_{\mathrm{bol}}$ and the visual bolometric correction
$BC_{V}$,
\begin{equation}
M_V(R, T_{\mathrm{eff}}, Z) =
M_\mathrm{bol}(R, T_{\mathrm{eff}}) - BC_V(T_{\mathrm{eff}}, Z, \log g)
\label{eqnMv}
\end{equation}

There is now an important point emphasized by Clausen (\cite{jC00}).
$M_{\mathrm{bol}}$ and $BC_{V}$ depend both on the effective temperature of the
star and thus are not independent variables. Consequently, from the bolometric
corrections calculated by Lanz and Hubeny (\cite{LH03}, \cite{LH07}), we find that
for late-O and early-B stars the bolometric correction can be given by
\begin{eqnarray} 
BC_{V} & \approx & \alpha - \beta \log T_{\mathrm{eff}} \quad \mbox{with} \nonumber\\
\alpha & = & 21.72 \pm 0.28 \, \mbox{mag} \quad \mbox{and} \quad  \beta =
5.51 \pm 0.07 \, \mbox{mag} 
\label{eqnBC}
\end{eqnarray}
with the associated uncertainty
\begin{equation}
\sigma_{BC^{j}_{V}} \approx \frac{\beta}{\ln 10}
\frac{\sigma_{T^{j}_{\mathrm{eff}}}}{T^{j}_{\mathrm{eff}}}
\end{equation}

Combining Eq. \ref{eqnMv} and \ref{eqnBC} and expressing $M_{\mathrm{bol}}$ as a
function of effective temperature $T_{\mathrm{eff}}$ and radius $R$, the visual
absolute magnitude $M_{V}$ can be written in turn as a function of $T_{\mathrm{eff}}$
and $R$ (see Clausen \cite{jC00} for more details). It follows that the uncertainty
in $M_{V}$ is given by  
\begin{equation}
\sigma^{2}_{M_{V}^{j}} \approx
\frac{1}{\left( \ln 10 \right)^{2}} \left[ \left( \frac{5  \, \sigma_{R_{j}}}{R_{j}}
\right)^{2} + \left( \frac{\left( 10 - \beta \right) \, 
\sigma_{T_{\mathrm{eff}}^{j}}}{T_{\mathrm{eff}}^{j}} \right)^{2} \right]
\end{equation}

Thus the uncertainty in $M_{\mathrm{bol}}$ (`10' factor) is partially cancelled by
the uncertainty in $BC_{V}$ (`$\beta$' factor). 

The uncertainties in the intrinsic colour indices at quadrature, $(B-V)_0$ and
$(V-I)_0$, are estimated via the approximation for early-B stars given in
Section \ref{secsynthphot}. Since the colour indices relate, not to a single star,
but to a binary system, one defines an ``equivalent'' effective temperature
$T_{\mathrm{eff,syst}}$ of the system (which is the effective temperature of a
single star with the same colour index as the system) by inverting Eq. \ref{eqCI}.
Then, a rough estimate of the error on the intrinsic colour index is obtained by
propagating the error on $T_{\mathrm{eff,syst}}$, which is identified with that on
$T_{\mathrm{eff}}^{\mathrm{P}}$, the temperature of the primary. In general,
the error on the absolute magnitude of the system will outweigh by far the other
terms on the right side of Eq. \ref{sigmamodule}.

\subsubsection{Masses and radii}
Masses and radii depend heavily on the radial velocity semi-amplitudes $K_{j}$.
The WD code does not provide directly the errors on the masses and radii, but only
those on the semi-major axis $a$, on the mass ratio $q$, the inclination $i$, the
effective temperature of the secondary and the potentials. It is in principle
possible to derive the errors on the masses from those on $a$ and $q$ using the
third Kepler law (which gives the total mass) and appropriate propagation formulae.
However, the resulting errors tend to appear underestimated, so we preferred to use
the errors on $\mathcal{M}_{P,S}\sin^3i$ given the {\sl bina} code that was used for
a preliminary interpretation of the RV curves alone. The errors on the masses
are then obtained through the formula
\begin{equation}
\sigma^{2}_{\mathcal{M}_{P,S}}=
\left(\frac{\sigma_{\mathcal{M}_{1,2}\sin^3i}}{\sin^3i}\right)^2+
\left(\frac{3\cos i}{\sin^4i}\right)^2\,\sigma^2_i
\end{equation}
and may be twice larger than those obtained from WD. Still, they have to
be considered rather as lower limits to the real uncertainties, because they are
based only on the scatter of the radial velocities around the fitted curve.
Systematic errors may arise, however, from the choice of the template spectra used to
obtain the RV values, and from the unavoidable fact that these synthetic spectra can
never perfectly match the real ones.

The relation between the uncertainties in the masses and pertaining variables can be
found in Hilditch (\cite{rH01}), for example. 
The absolute radius of component $j$ is obtained from
\begin{equation} 
R_{j} = a \, r_{j}\left(\Sigma_r,\, k\right)
\end{equation}
where $a$ is the semi-major axis and $r_j$ the relative radius of component $j$. The
relative radii are considered as functions of two independent variables,
the sum $\Sigma_r \equiv r_\mathrm{P} + r_\mathrm{S}$ and the ratio
$k \equiv r_\mathrm{S} / r_\mathrm{P}$ of the relative radii. The sum of the radii
is obtained reliably from the analysis of the light curves. The ratio of the radii
is obtained either from the spectroscopic $B$ luminosity ratio, in the case of
detached systems (unless the minima are so deep that the photometric estimate
proves accurate enough), or from the light curve in the case of semi-detached and
contact systems. The individual radii are thus given by
\begin{equation}
r_\mathrm{P} =
\frac{1}{1+k}\,\Sigma_r \quad \mathrm{and} \quad r_\mathrm{S}= \Sigma_r - r_\mathrm{P}
\end{equation}
with the associated uncertainties 
\begin{equation}
\sigma_{r_\mathrm{\mathrm{P, S}}}^2 =
\left(\frac{\alpha_{\mathrm{P, S}}}{1+k}\right)^2\, \sigma_{\Sigma_r}^2 \,
+ \, \left( \frac{\Sigma_r}{(1+k)^2} \right)^2 \, \sigma_{k}^2
\end{equation}
where $\alpha_\mathrm{P} = 1$ and $\alpha_\mathrm{S} = k$. $\Sigma_r$ is robustly
determined from the light curve. For the sake of simplicity, the variance in
$\Sigma_r$ was taken from the EBOP solution of the $I$ curve, by scaling the error on
$r_P$ obtained by fixing the ratio of radii. 

The spectroscopic ratio of radii $k$ is obtained by inverting Eq. \ref{eqnRratio}
\begin{equation}
k\equiv\frac{R_S}{R_P}=\mathcal{L}_B^{1/2}\,\left(\frac{T_S}{T_P}\right)^2\cdot
10^{0.2\,\Delta BC_B}
\end{equation}
where
\begin{eqnarray}
BC_{B} &=& BC_{V}-(B-V)_0\approx 21.72-5.51\,\log T_{\mathrm{eff}}\\
&&-(1.768-0.455\,\log T_{\mathrm{eff}}) \\
\Delta BC_{B}  & \equiv & BC_{B}^{\mathrm{P}} (T_{\mathrm{eff}}^{\mathrm{P}})
- BC_{B}^{\mathrm{S}} (T_{\mathrm{eff}}^{\mathrm{S}})\approx -5.055\, 
\log\left(\frac{T_{\mathrm{eff}}^{\mathrm{S}}}{T_{\mathrm{eff}}^{\mathrm{P}}}\right)
\end{eqnarray}
Therefore, the ratio of radii can be written
\begin{equation}
k\approx \mathcal{L}_B^{1/2}\,
\left(\frac{T_{\mathrm{eff}}^{\mathrm{S}}}{T_{\mathrm{eff}}^{\mathrm{P}}}\right)^{0.99}
\end{equation}
Neglecting the uncertainties on the empirical coefficients in the expression for the
bolometric correction, the error on $k$ is then:
\begin{eqnarray}
\left(\frac{\sigma_{k}}{k}\right)^{2} & \approx &  \frac{1}{4}
\left(\frac{\sigma_{\mathcal{L}_{B}}}{\mathcal{L}_{B}}\right)^2+0.99^2\,
\left[\left(\frac{\sigma_{T_{\mathrm{eff}}^{\mathrm{P}}}}{T_{\mathrm{eff}}^{\mathrm{P}}}\right)^2
+\left(\frac{\sigma_{T_{\mathrm{eff}}^{\mathrm{S}}}}{T_{\mathrm{eff}}^{\mathrm{S}}}\right)^2\right]
\end{eqnarray}

\section{The individual binaries}
\label{individual}
Each system is discussed thoroughly in this section. We give details concerning
the light-curve solution, the radial-velocity solution, the temperature and
luminosity-ratio determinations and the characteristics of the spectra. Also
discussed are the positions of the components in the mass-surface gravity plane
and the temperature-luminosity (HR) diagram. Review of the distances and
collective properties of the whole sample of 33 binaries follows in
Section \ref{label_discussion}. Except where otherwise stated, when referring to
the light curve of a specific system, it means the $I-$band light curve. 

The $I-$band light curves and the best-fit solutions are shown in
Figs \ref{allLC1}-\ref{allLC3_small}. The RV curves are shown in
Figs \ref{allRV1}-\ref{allRV3}. The mass-$\log g$ diagrams are shown 
in Figs \ref{allGM1}-\ref{allGM3_small}. The HR diagrams are shown in
Figs \ref{allHR1}-\ref{allHR3}.
The parameters found from the WD/PHOEBE analysis are given in Tables
\ref{orbparam} (orbital parameters) and \ref{WDana} (temperature
ratios, potentials and luminosity ratios).
The astrophysical parameters of the primary and secondary components are given in
Tables \ref{astroparam1} and \ref{astroparam2}, respectively.
Finally, the RMS scatters of the light curves and RV curves are summarized in
Tables \ref{LCscatter} and \ref{RVscatter}.
     
\subsection{4 110409}
With a difference of 0.04 mag
in the brightness level between phase 0.25 and phase 0.75, this semi-detached
system displays the most asymmetric light curve among all the systems studied in
this paper. The light curve is bright ($I^{\mathrm{q}} < 16$ mag) and of high
quality, with a low RMS scatter combined with a deep primary eclipse ($\Delta
I_{\mathrm{min I}}/\sigma_I$ $\sim$ 65).
This EB-type light curve shows a relatively strong depression occurring just
before the primary minimum. Actually, this is strong evidence for absorption by
a gas stream stemming from the (inner) L1 Lagrangian point and seen in
projection against the primary surface (HHH05).
As a consequence, the use of a ``simple'' symmetric model for the light-curve fit
is not satisfactory, resulting in a rather poor fit despite the intrinsic
quality of the observations. Therefore, this solution was subsequently improved
by adding a cool spot on the equator of the primary component (see Section
\ref{WDanalysis}). The parameters of the spot are: a colatitude of $\pi/2$~rad
(fixed), a longitude of 0.569 rad, an angular radius of 0.3 rad and a
temperature factor of 0.6, i.e. the effective temperature of the spot is
$0.6$ that of the rest of the stellar surface. Although this
new synthetic light curve gives a far more satisfactory fit, the
$O-C$ curve reveals that this system is certainly more complex than this ``one
circular cool spot" model. Actually, there are still some discrepancies at the
bottom of the eclipses and just after the secondary minimum. Nevertheless, this
model is certainly sufficient to set reliably the inclination, the brightness
ratio of the components and the maximum out-of-eclipse flux.   
On the finding chart, the image of this star is slightly elongated in the EW
direction, suggesting a blend with another, fainter star which would lie
$1\arcsec$ or slightly farther away to the West. Nevertheless, no clear sign
of a 3rd light is seen in the lightcurve.

The RV curves are well constrained with 11 out-of-eclipse spectra and notably
observations close to phase 0.75. This system was previously studied by HHH05.
There is significant differences between their RV parameters and ours. Our RV
semi-amplitudes are 135 and 259 km s$^{-1}$, to be compared to their values of
160 and 247 km s$^{-1}$. Beside having lower $S/N$ and resolving power than us,
in this particular case the discrepancy is certainly due to their admitted lack
of observations close to the quadratures. Consequently, our value for the mass
ratio, $q = 0.52$, is certainly more secure than theirs (0.65).   

We found a spectroscopic $B$ luminosity ratio of 1.45. This is higher than the
photometric value (1.29), perhaps because of the large distortion of the Roche lobe
filling companion. Interestingly, the brighter, i.e. primary, component has lower
monochromatic luminosities than the secondary component: even though the primary
has a higher bolometric luminosity, it emits mostly in the UV part of the
spectrum, so that its $I$ luminosity, for instance, is lower than for the
secondary. 

The most interesting parts of the disentangled spectra of both components are
presented in Fig. \ref{fig4_110409}. As a consequence of the low $B$
luminosity of the primary, the spectrum of the latter is the noisier of the
pair. Not surprisingly, in both spectra the most prominent features are the
\ion{H}{i} and \ion{He}{i} lines. It is tempting to identify a number of
features in the primary spectrum with the \ion{C}{ii} 4267,
\ion{O}{ii} 4276-4277, \ion{Si}{iii} 4553, \ion{Si}{iv} 4089 and 
\ion{Si}{iv} 4116 lines. Nevertheless, both the lack of positive identification
of the \ion{He}{ii} 4542 line for a $\sim$14 $\mathcal{M}_{\odot}$ star and the
noisy profile of the \ion{He}{i} lines mean that one must be careful in not
over-interpreting a spectrum of rather low quality.
The better secondary spectrum displays cleaner features. The \ion{He}{i} 4471
and \ion{Mg}{ii} 4481 lines allow to secure the temperature of the secondary.       

By fixing the photometric temperature and $B$ luminosity ratios, a least-squares
fit of the 11 out-of-eclipse spectra provided a primary temperature very close to
$32\,500$ K, that is to say 7000 K more than what was determined by HHH05. 

Both mass-$\log g$ and the HR diagrams are typical of a massive Algol-type
binary. The brighter and more massive component of the system appears to be
close to the zero-age main sequence (ZAMS), while the secondary component is
larger and far more luminous than a non-evolved star of the same mass.

\begin{figure*}[htb]
\begin{center}
\includegraphics[trim = 0mm 5mm 0mm 3mm, clip, width=6.5cm]{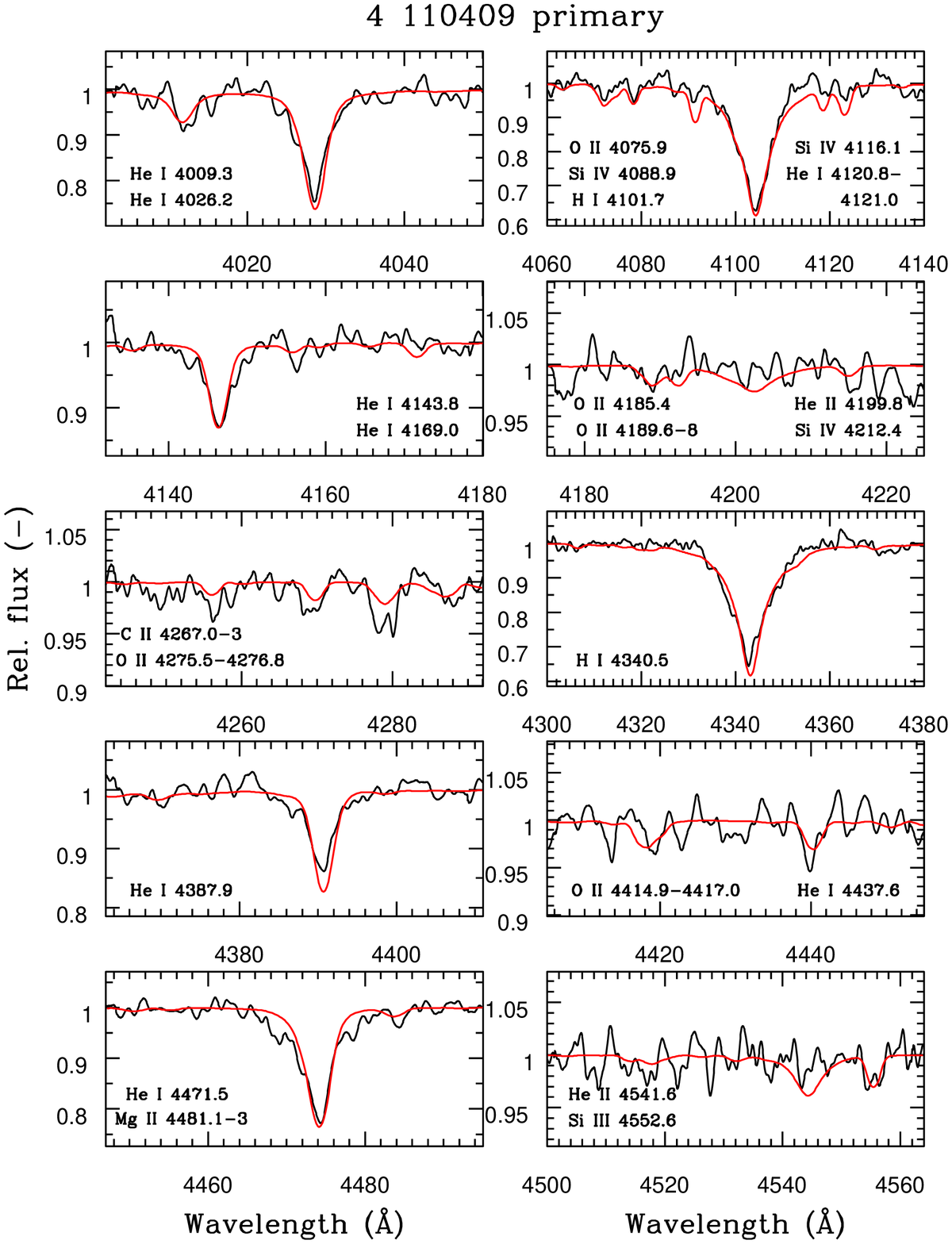}
\includegraphics[trim = 0mm 5mm 0mm 3mm, clip, width=6.5cm]{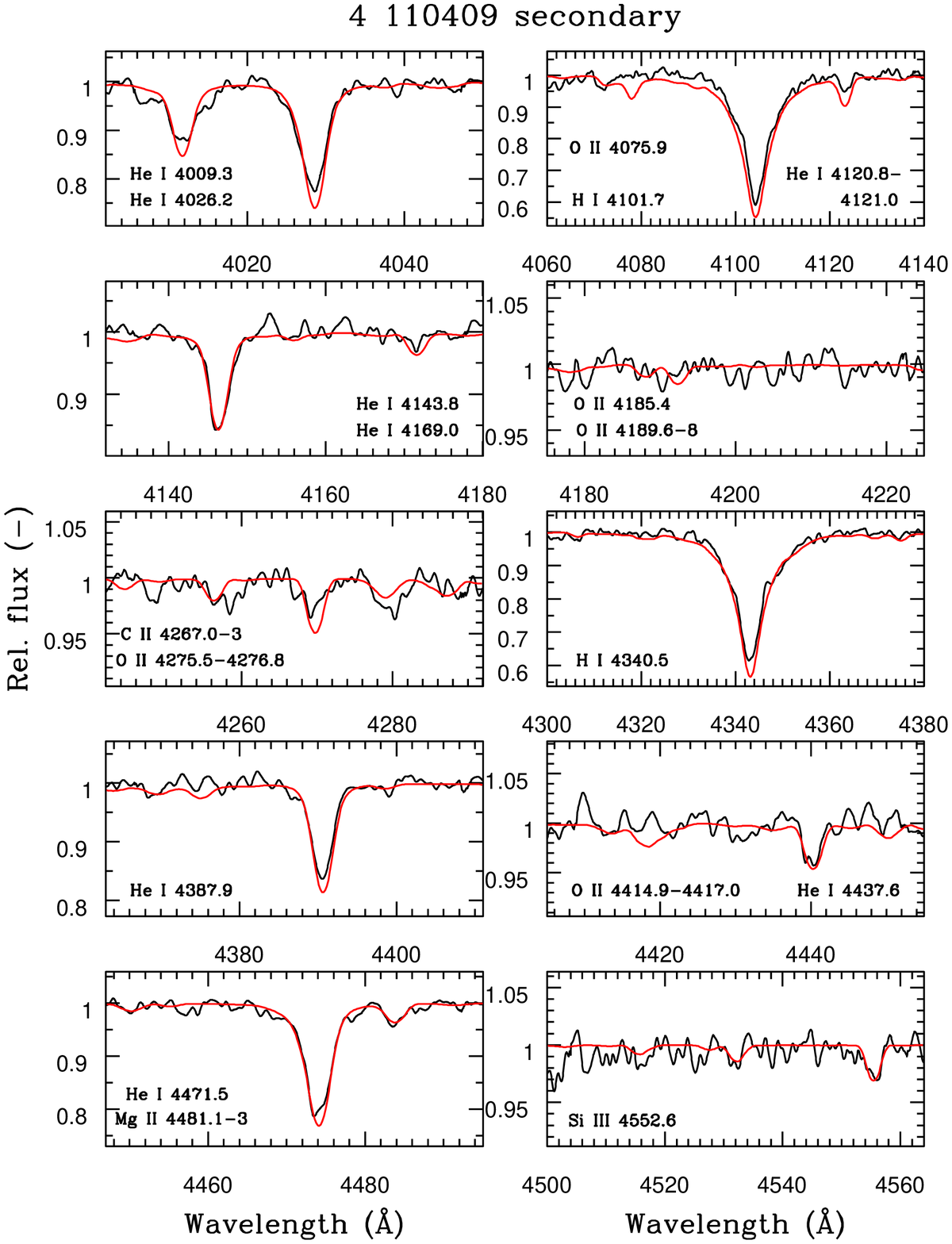}
\caption{Sections of the disentangled spectra of the primary and secondary
components of the binary 4 110409. The scale on the vertical axis is adapted to
the size of the features in each section. The red line represents the best fit
synthetic spectrum shifted to the systemic velocity. Beside prominent \ion{H}{i}
and \ion{He}{i} lines, \ion{Mg}{ii} and \ion{Si}{iii} lines are visible in the
spectrum of the secondary. The spectrum of the primary is far less convincing;
nevertheless \ion{Si}{iv} lines seem to be present next to H$\delta$.}
\label{fig4_110409}%
\end{center}
\end{figure*}

\subsection{4 113853}
The best fit was obtained
with a semi-detached model. Because of a moderate RMS scatter (0.017) combined
with shallow eclipses ($\sim$0.18 mag) if not purely ellipsoidal variations,
the light curve of this binary is one of the poorest of the whole sample. This
low amplitude is due to a low inclination ($\sim$60$\degr$). The $O-C$ curve
reveals that the profiles of the eclipses are not perfectly symmetrical. The
quality of the data is not sufficient, though, to trace the possible astrophysical
cause of this asymmetry. On the finding chart, the star seems fairly well isolated.
  
Despite only 7 out-of-eclipse spectra, the RV curves are rather well
constrained with observations close to both quadratures.   
 
The $S/N$ of the composite spectra are low ($25-65$) and there is a sizeable
nebular emission in the Balmer lines. Because of the lack of metallic lines and
the severe contamination of the Balmer lines by nebular emission which hinder
the disentangling procedure, the disentangled spectra of the components were
not used. The least-squares fit was performed, letting both temperatures and the
$B$ luminosity ratio free to converge. It provided a temperature ratio remarkably
close to the photometric one, and a spectroscopic $B$ luminosity ratio of
$0.68$, also in perfect agreement with the photometric ratio ($0.68$). Thus, fixing
the temperature and luminosity ratios to the photometric values was not needed to
determine a reliable temperature of the primary.

Both the mass-$\log g$ and  the HR diagrams show an evolved system with a primary
component seemingly half-way between the ZAMS and the terminal-age main sequence
(TAMS). On the HR diagram, the primary lies much higher than the evolutionary track
corresponding to its mass. Whether this is due to a temperature overestimate
(linked e.g. with an underestimated sky background) or to some evolutionary effect
remains to be examined. Besides, the distance modulus perfectly agrees with the
currently accepted value for the SMC.

\subsection{4 117831}
This faint system has a low-to-medium quality light curve of the EA type.
There is a slight ellipsoidal variation between the eclipses and 
the latter have a similar depth ($\sim$0.4 mag). This is a close detached
system with similar components. The finding chart suggests a possible slight
blend with a fainter star located some $1\arcsec$ to the East of the system.
No clear sign of a 3rd light, however, is seen in the lightcurves.

The RV curves are well constrained with 12 out-of-eclipse spectra and
observations close to phase 0.25 and phase 0.75. The mass ratio close to one
($q = 0.98$) is indicative of a binary with `twin' components.

The disentangled spectra are of too low quality to see any useful metallic line.
The \ion{Mg}{ii} 4481 and \ion{C}{ii} 4267 are barely visible. Disentangling of
the Balmer lines was hindered by the strong emission. 
A first least-squares fit provided both temperatures and a poorly constrained
spectroscopic $B$ luminosity ratio of $0.97\pm 0.11$. The WD code converged to a
higher luminosity ratio ($1.17\pm 0.05$), even when we tried to minimize it by
fixing the potential of the primary. The temperature of the primary was finally
set by a fit where the ratio of temperatures was fixed to the photometric value,
and the luminosity ratio assumed equal to one. The small number of photometric
data in the minima, especially the primary one, probably makes the photometric
luminosity ratio unreliable and explains why the radius of the primary component
appears slightly smaller than that of the secondary one.

According to the mass-$\log g$ diagram, the age of the system is about $50$ Myr,
assuming the standard SMC metallicity $Z=0.004$. The positions of both
components in the HR diagram agree to within the error bars with the
evolutionary tracks.

This system was studied by Wyithe et al. \cite{WW02} (see Table \ref{compWW}).
Their results not being constrained by spectroscopy, it is not surprising that
they found a very different solution. They considered this system as a
semi-detached binary with a photometric mass ratio of 0.157. Our spectroscopic
results completely rule out that model.

\subsection{4 121084}
This system displays deep eclipses ($> 0.6$ mag) of similar depth. A slight
ellipsoidal variation is visible. This is clearly a close detached system with
slightly distorted twin components. No clear sign of crowding is seen on the
finding chart, except possibly with very faint neighbour stars.

The RV curves are well constrained with 9 out-of-eclipse observations regularly
distributed around the quadratures.

The composite spectra are polluted by strong nebular emission in both H$\gamma$
and H$\delta$ lines. Nevertheless, the widely separated Balmer lines allow a
reliable temperature and luminosity ratio determination. The disentangled
spectra  are useful to confirm the rather high $V_{\mathrm{rot}} \sin i$ values
of the components. Not surprisingly, no metallic lines are visible because of
the moderate $S/N$ combined with fast rotational velocities. The potential of
the primary was fixed so that the luminosity ratio given by the WD code matches
the spectroscopic one. The temperature of the primary was obtained by fixing
the temperature ratio to the photometric one, 

Both stars lie on the ZAMS, both in the mass-$\log g$ and HR diagrams. On the
HR diagram,
however, they are clearly more luminous and hotter than their expected positions
for a metallicity $Z=0.004$. They would better agree with the ZAMS and
evolutionary tracks for $Z=0.001$, as many other systems do. Moving the
representative points to the their expected positions for $Z=0.004$ would
require a $2000$~K decrease in effective temperature; that seems large, but
the residuals between the observed and synthetic composite spectra show only
very subtle changes. Only a modest systematic effect might be responsible.

\subsection{4 121110}
The medium-to-high quality light curve shows a
deep ($\sim$0.5 mag) primary eclipse. A slight ellipsoidal variation is visible
between the eclipses. This is again a close detached system with slightly
distorted components. No star closer than $3\arcsec$ is seen on the finding chart,
except for a very faint one lying about $2\arcsec$ away to the SW.

The RV curves are well constrained with 11 out-of-eclipse spectra.

There is strong nebular emission in both Balmer lines. The spectroscopic $B$
luminosity ratio ($0.415\pm 0.047$) nicely agrees with the photometric one
(0.424), without any need for fixing the potential of the primary. The
temperature of the primary was fitted after fixing the temperature and
luminosity ratios to their photometric values, as ususal.  
The \ion{Si}{iii} 4553 line is clearly visible on the disentangled spectrum of
the primary. The lack of \ion{Mg}{ii} 4471 confirms the relatively high
temperature of the primary. The spectrum of the secondary is too noisy for the
identification of metallic lines. 

On the mass-$\log g$ diagram, the stars match an isochrone corresponding to
about $7-8$ Myr. In the HR diagram, the positions of both components are above
the $Z=0.004$ evolutionary tracks but are consistent with the lower metallicity
ones ($Z=0.001$). Increasing the helium content would also help to reconcile
their positions with the evolutionary tracks, unless a systematic effect raises
the apparent effective temperatures.

\subsection{4 121461}
This is an eccentric system with two (relatively) widely separated
components. Both eclipses are very similar in depth and width. With
$I^{\mathrm{q}} \sim 17.9$ mag, this is one of the faintest systems in our
sample. Nevertheless, the finding chart indicates no crowding problem
whatsoever. No significant apsidal motion was found on the basis of photometry.
An analysis with the EBOP
code shows that the $\omega_0$ value depends critically on the $e\sin\omega_0$
quantity, which is poorly constrained, while the more robust $e\cos\omega_0$
quantity is such that $\cos\omega_0\simeq 0.75$ and thus does not constrain
$\omega_0$ very tightly. Fig. \ref{apsid_121461} suggests a marginal decrease of
$e\cos\omega_0$ with time which, if real, could only be due to gravitational
perturbations from a third body, because $\dot{\omega}< 0$, while pure tidal
effects always result in $\dot{\omega}> 0$. We have assumed no apsidal motion.
\begin{figure}[htb]
\begin{center}
\includegraphics[width=9cm]{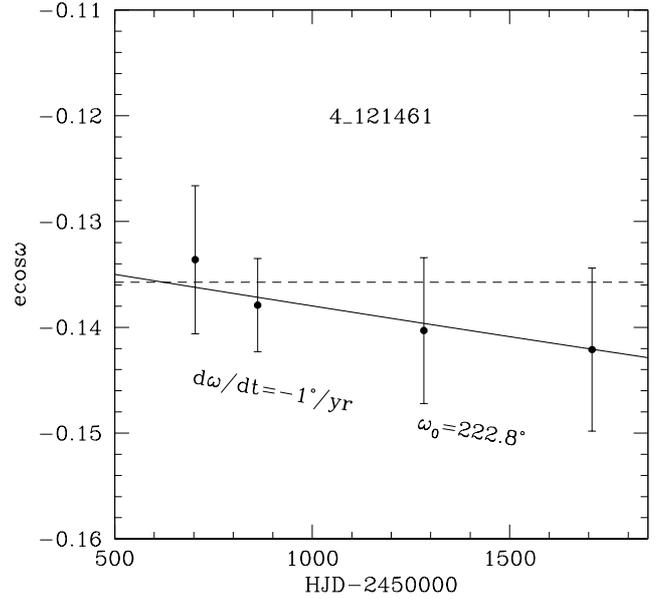}
\caption{Variation with time of the $e\cos\omega$ quantity of the system
4 121461, determined with the
EBOP code on the basis of the $I$ magnitudes grouped in four time intervals. The
solid line corresponds to a hardly significant negative apsidal motion,
while the dashed horizontal line indicates no apsidal motion, for $e=0.185$ and
$\omega_0=222.8^\circ$.}
\label{apsid_121461}
\end{center}
\end{figure}

This system is one of the two in our sample having 16 
out-of-eclipse spectra, of which 15 were used (the 8th one has too
poor SNR). There is a sufficient number of observations close to
the quadratures.

The composite spectra are very noisy. The disentangled
spectra are of very low quality, with no exploitable metallic lines. There is
strong nebular emission. Letting the temperature of both components free to
converge (together with the $B$ luminosity ratio) resulted in very uncertain
values, so we fixed the temperature ratio to the photometric one in order to
estimate the temperature of the primary. The rotational velocities were fixed
to the pseudosynchronized values. A luminosity ratio of $0.95\pm 0.06$ was found
on the basis of the spectra, which agrees well with the photometric one
($0.91\pm 0.03$) obtained without fixing the potential of the primary. The
photometric luminosity ratio was adopted, which results in almost identical
radii for the components. This results in a slightly lower surface gravity on
the secondary component than on the primary, because of the mass ratio, but this
difference is not significant.

On the mass-$\log g$ diagram, this system lies close to the ZAMS but might be up
to $15-20$~Myr old. The positions of the stars in the HR diagram agree, within
the error bars, with the evolutionary tracks, although they tend to lie too
high, as is the case of other systems.

\begin{figure*}[htb]
\begin{center}
\includegraphics[trim = 0mm 5mm 0mm 3mm, clip, width=6.5cm]{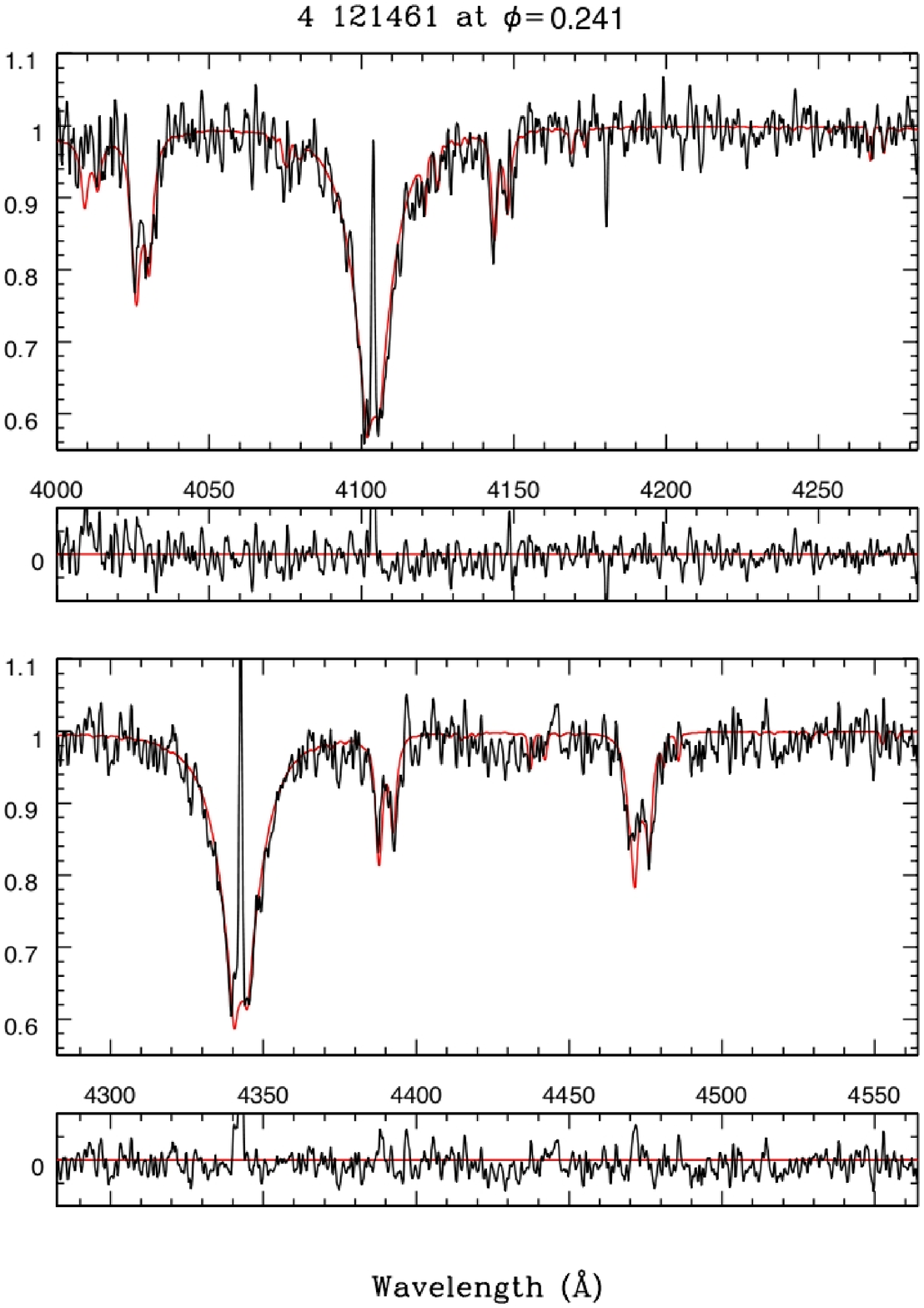}
\includegraphics[trim = 0mm 5mm 0mm 3mm, clip, width=6.5cm]{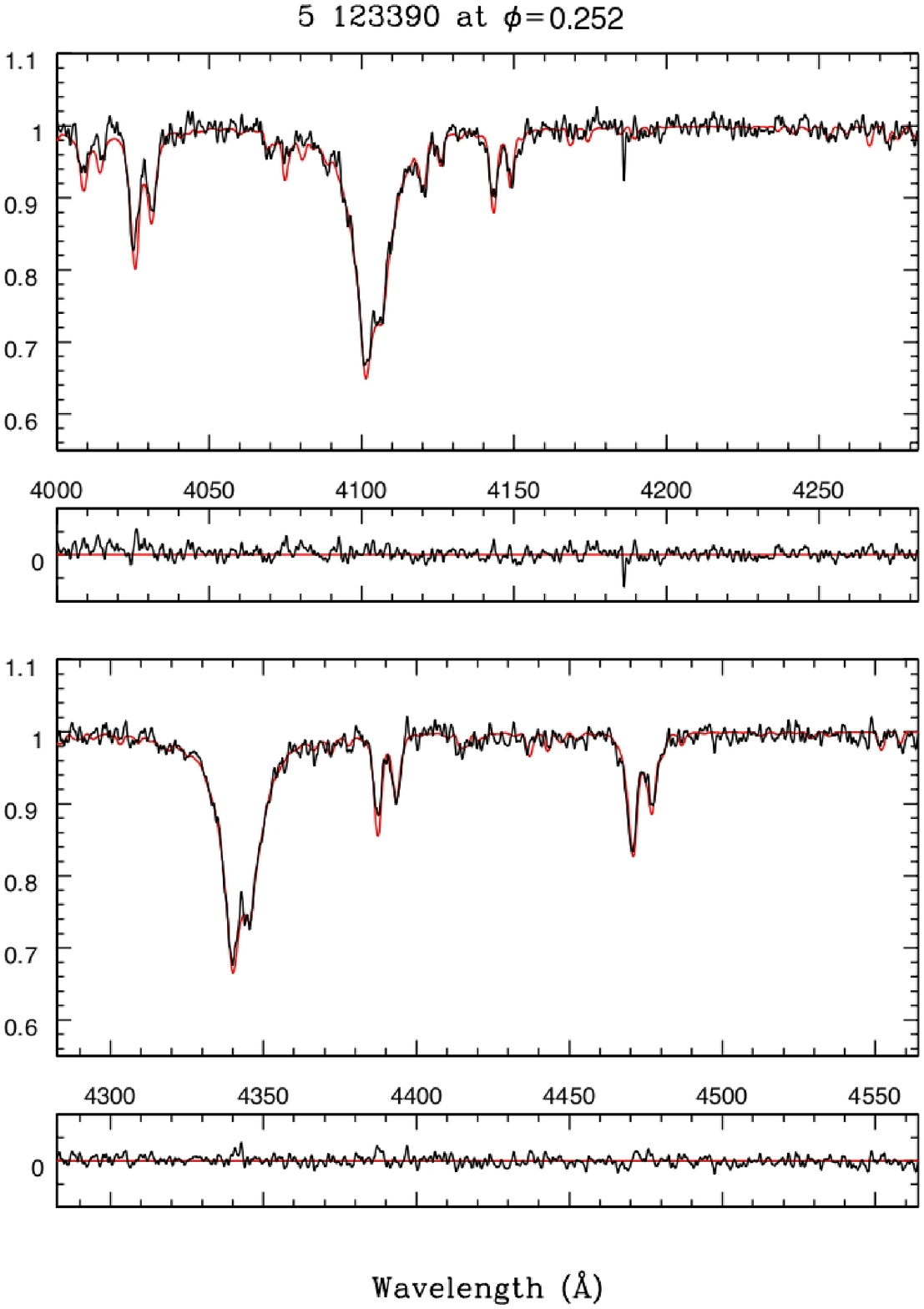}
\caption{Example of observed spectra close to a quadrature. Both spectra are at the
same scale. The red line represents the composite synthetic spectrum of the binary
system, i.e. the addition of the scaled and velocity-shifted synthetic spectra of the
two components for the corresponding orbital phase. Beside a low $S/N$, the spectrum
of 4 121461 shows strong nebular emission in the Balmer lines. From the best fit,
the $B$ luminosity ratio was found to be $\sim$0.95 with a primary temperature of
$\sim$22\,000 K. The spectrum of 5 123390 points to a $B$ luminosity ratio of 0.59
and a primary temperature of 27\,000 K.}
\label{sp_entier}%
\end{center}
\end{figure*}

\subsection{4 159928}
This system presents low-to-medium quality light curves of the EW type with
eclipses of unequal depth. The RMS scatter of the light curve is rather low,
but the minima are not very deep ($\Delta I_\mathrm{min I} \sim 0.25$ mag).
It can be inferred from this that the system comprises close, strongly distorted
components of unequal brightness. The best-fitting model corresponds to a
semi-detached binary with a low inclination, close to $60\degr$. The finding
chart shows a well isolated target, except for a quite faint neighbour at about
$2\arcsec$ to the NNE.

There are only 8 out-of-eclipse spectra, but they are close to the two
quadratures. 

A spectroscopic luminosity ratio of 0.45 was found in the $B$ band, which is
higher than the photometric value (0.36). The temperature of the primary
was determined by fixing the temperature and luminosity ratios to the
photometric values, after suppression of the very strong nebular emission lines.
The \ion{Si}{iii} 4553 line is the only metallic line clearly visible in the
disentangled spectrum of the primary. The spectrum of the secondary shows no
exploitable metallic line. The \ion{Mg}{ii} 4481 is barely visible. The
synchronized values for $V_\mathrm{rot} \, \sin i$ are close to
200 km s$^{-1}$ and therefore all but the strongest lines are buried in the noise. 

Both the mass-$\log g$ and HR diagrams show positions typical of  a
semi-detached system with an evolved secondary component. The primary is
slightly overluminous relative to the evolutionary track of a single star.

\subsection{4 160094}
This detached system of moderate eccentricity presents
low-to-medium quality light curves with rather shallow eclipses
($\Delta I_\mathrm{min I} \sim 0.20$ mag). Except for a few very faint
neighbours, the target seems free from crowding on the finding chart.

There are 11 out-of-eclipse spectra. Both quadratures are well covered by the
observations.

The $T_\mathrm{eff}$ of the primary was determined together with that of the
secondary and with the $B$ luminosity ratio. The temperature of the secondary
proved rather ill-defined, so the photometric temperature ratio was used to
define it, as usual. The potential of the primary was fixed to a value that
implies a luminosity ratio close to the spectroscopic one. No metallic lines
are visible in the very noisy disentangled spectra. The nebular emission is
strong in both \ion{H}{i} lines.

Despite the moderate quality of the photometric and spectroscopic data, the
positions of both stars fall right on the ZAMS in the mass-$\log g$ diagram. In
the HR diagram, their position agree well with the evolutionary tracks, though
they appear slightly overluminous.

A notable characteristic of this system is its fast apsidal motion
$\dot{\omega} = 9.8 \pm 1.9$ $\degr$ yr$^{-1}$.
Figure \ref{4_160_ecosw} shows the $e\cos\omega$ product as a function of time, as
obtained using the EBOP code. The solid line represents the WD solution,
which appears consistent with the EBOP results, even though the latter would be
compatible with a faster apsidal motion coupled with a slighly smaller
eccentricity. Further discussion of this result is deferred to Section 5.
\begin{figure}[htb]
\begin{center}
\includegraphics[width=9cm]{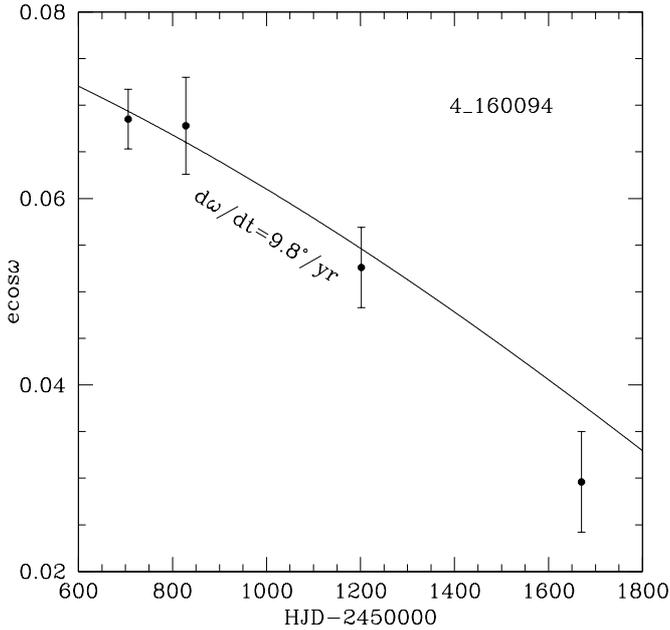}
\caption{Variation with time of the $e\cos\omega$ quantity of the 4 160094
system, determined with the
EBOP code on the basis of the $I$ magnitudes grouped in four time intervals. The
solid line corresponds to the parameters listed in Table~\ref{orbparam}.}
\label{4_160_ecosw}
\end{center}
\end{figure}

\subsection{4 163552}
This is one of the five systems with $I^{\mathrm{q}} < 16$ mag, displaying a
high quality light curve of the EB type. The eclipses have very similar depths
($\sim0.4$ mag), indicating that the temperature ratio is very close to unity.
Some faint neighbours are seen on the finding chart about $2\arcsec$ from
the target, and it is difficult to judge whether closer neighbours might lurk
within the relatively large spot left by this bright system. According to the
light curve, there is a substantial third-light contribution to this system: we
find $\ell_{3,I}\sim 0.12$, while Graczyk (\cite{dG03}, hereafter dG03) found
$\ell_{3,I}\sim 0.26$. A significant ellipsoidal variation is indicative of
tidally distorted components, though the system is still in a detached
configuration.  

There are only 9 out-of-eclipse spectra, but the RV curves are well constrained
near quadratures.

The observed spectra of this binary were roughly corrected for the presence of a
third component, by subtracting a constant from the normalized spectra. The value
of the constant was adjusted until the effective temperatures obtained from the
least-squares fit to the corrected spectra give a reasonable match to the evolutionary
tracks at the metallicity $Z=0.004$ in the HR diagram. A constant of 0.1 only proved
sufficient to this purpose, which is widely different from dG03's photometric estimate
of the third light, $\ell_{3,B}\simeq 0.26$. On the other hand, our WD analysis
resulted in $\ell_{3,B}\simeq 0.14$, which is more consistent with our rough
spectroscopic estimate.
Actually, a first attempt to analyze this system resulted in highly overluminous
components. This was interpreted as a clear sign that the observed spectra were
severely contaminated by the object responsible for the photometric third light.     

The disentangled spectrum of the primary shows a faint \ion{Mg}{ii} 4481 line and
a noisy \ion{Si}{iii} 4553 line. These lines were used to constrain the temperature
of the primary. The \ion{Si}{iii} 4553 line is visible in the spectrum of the
secondary too.

This system was studied by HHH05, dG03 (see his Fig. 6) and WW02. Only dG03 took the
third-light contribution into account. Not surprisingly,
our photometric solution is very close to theirs.

Since we adjusted the amount of third light so that the primary component have a
realistic luminosity, the HR diagram shows a primary close to the stellar
evolutionary track of a 9.6 $\mathcal{M}_{\odot}$ star, and a secondary close
to the evolutionary track of a 9.1 $\mathcal{M}_{\odot}$ star. Their positions
are clearly far from the zero-age main sequence.
From the mass-$\log g$ diagram, the age of the system is estimated to be
20 Myr. As the position of the primary was inferred from the expected
position in the HR diagram, the proposed solution is not entirely reliable
and thus is not used for the determination of the mean distance modulus of the
SMC (see Section \ref{distsection}).     

\subsection{4 175149}
The medium-to-high quality light curve of this system is of the EB type;
unfortunately, most of the right part of each eclipse is missing. That
is due to a period very close to $2$ days exactly. The minima are
fairly deep ($\Delta I_\mathrm{min I} \sim 0.50$ mag), well-defined and of
unequal depth. There is a strong ellipsoidal variation. The binary is a
semi-detached system, with distorted components of different brightnesses.
A slight depression occurring before the primary eclipse is indicative of a
gas stream. However, we did not venture into adding a spot on the primary in
order to mimick this effect, because we felt that the large gaps in the light
curve already limit the reliability of the proposed solution.

This system is close to the edge of the CCD used in the OGLE II survey, so it
was also listed under the designation 5~38079 in the OGLE database. Thanks to
that circumstance, there are 574 photometric $I$ magnitudes instead of less than
300. An examination of the OGLE finding chart reveals
that this binary is clearly blended.  

This is the second system with 16 out-of-eclipse spectroscopic observations.
Because of the 2-day period, all these observations took place before the
quadratures. Nevertheless, the RV curves seem to be sufficiently constrained.

A spectroscopic $B$ luminosity ratio of 1.39 was found, though with a large
scatter ($\sigma\sim 0.14$). This is higher than the photometric value (1.18).
The disentangled spectra are fairly good.
The spectrum of the hot primary component displays the \ion{He}{ii} 4200 and
4542 lines. The \ion{Si}{iii} 4553 and a faint \ion{Si}{iv} 4089 lines are
equally visible. These lines provide strong constraints for the temperature of
the primary. The best-fitting primary temperature was obtained after fixing the
temperature and luminosity ratios to the photometric values.

The mass-$\log g$ diagram shows the typical oblique orientation
of the segment connecting the components of an evolved binary. The HR diagram
shows a highly overluminous primary, relative to the evolutionary
track of a 11.8 $\mathcal{M}_{\odot}$ star, and an evolved secondary far more
luminous than a main-sequence star of 7.8 $\mathcal{M}_{\odot}$. One may wonder,
whether the strange position of the primary is due to an unrecognized
third light, both in photometry and spectroscopy, or to some evolutionary
effect. In any case, there is no obvious third light in the light curve.

\subsection{4 175333}
This slightly eccentric system presents low-to-medium quality light curves
of the EA type. This is one of the few systems with $I^{q} > 17.5$ mag.
Consequently, the $O-C$ curve shows a relatively high scatter. The minima are of
unequal depth. A slight ellipsoidal variation is visible. This is
clearly a detached system with components of unequal brightness. The target
appears perfectly isolated on the finding chart.

There are 14 out-of-eclipse spectra. The observations constrain well the RV
curves.

This system was studied by WW01. The fact that they consider the eclipses as
total (while we consider them as partial) and their lack of spectroscopic
constraints on their ratio of radii account for the differences between their
solution and ours. The evidence for total eclipses does not appear compelling,
so additional photometry would be needed to settle the issue. We found a
spectroscopic $B$ luminosity ratio of 0.55, slightly smaller
than that finally adopted taken photometry into account. The disentangled
spectrum of the primary shows a noisy \ion{Mg}{ii} 4481 line. The spectrum of
the secondary is too noisy to detect any metallic line. The temperature of the
primary was fitted with the temperature and luminosity ratios fixed to the
photometric values. The fit with both temperatures free to converge, together
with the luminosity ratio, gave a rather large scatter of about $1600$~K and a
secondary temperature about 900~K cooler. In spite of the partial eclipses,
there was no need to fix the potential of the primary in order to find a $B$
luminosity ratio that matches the spectroscopic value, so the photometric value
of the luminosity ratio was adopted.

On the mass-$\log g$ diagram, both stars fall right on the 20 Myr isochrone.
On the HR diagram, however, both stars appear significantly overluminous
relative to their evolutionary tracks, suggesting that the
effective temperatures may be overestimated by at least $2000$~K! Strangely
enough, the colour excess of this system appears to be small ($E(B-V)=0.07$) and
the distance modulus ($18.6$) clearly smaller than the accepted value for the
SMC ($\sim 18.9-19.0$). If the effective temperatures had indeed been
overestimated,
this would have implied both a too blue intrinsic colour and a too large
intrinsic luminosity (hence a more negative absolute magnitude), se one would
rather expect a large colour excess and a large distance modulus.

\subsection{5 016658}
This close detached system presents medium quality light curves with eclipses
of equal depths, and is composed of tidally distorted twin components. The
finding charts reveals no crowding problem.

There are 11 out-of-eclipse spectra. The RV curves are well constrained by the
observations around phase 0.25.

This system was studied by WW01. As for the previous binary, the differences
observed between their (photometric) solution and ours is due to their lack
of a spectroscopic constraint on the ratio of radii, and to the fact that they
assume total eclipses. Evidence for the latter is not compelling, however, and
awaits further photometric measurements for confirmation. A spectroscopic $B$
luminosity ratio of $\sim$0.60 was found, which guided the choice of the
potential of the primary component in the WD analysis.
The \ion{Mg}{ii} 4481 line is clearly visible on the disentangled spectrum of
the primary. The spectrum of the secondary is too noisy to show any metallic
line. The best-fitting primary temperature was obtained simultaneously with the
temperature of the secondary, which appeared quite compatible (within $300$~K)
with the photometric one (i.e. given the spectroscopic primary temperature and
the photometric temperature ratio), and with the luminosity ratio. The
photometric temperature ratio was adopted.

On the mass-$\log g$ diagrams, both stars fall on the 30~Myr isochrone within
the errors. On the HR diagram, the primary has a position compatible with its
evolutionary track within errors, though it appears slightly too luminous. The
secondary is slightly hotter than the primary, and is more overluminous; still,
it remains compatible with its evolutionary track if the errors on both
luminosity and mass are considered.

\subsection{5 026631}
This system presents a medium-quality light curve of the EW type with minima of
unequal depth. This is clearly a semi-detached system with strongly distorted
components of unequal brightness. It presents the second lowest
inclination of the sample with $i \approx 61\degr$, implied by the rather small
amplitude of the light curve. No blend is apparent on the finding chart, except
for two or three very faint neighbours at about $2\arcsec$.

There are only 8 out-of-eclipse spectra, but these ones are sufficiently
constraining to get reliable RV curves.

A spectroscopic $B$ luminosity ratio of 0.74 was found, while the photometric
value is 0.50. The disentangled spectra show no useful metallic lines. This
is due notably to the high
$V_\mathrm{rot}\,\sin i$ values ($\sim$190 km s$^{-1}$). The temperature of the
primary was obtained by fixing the temperature and luminosity ratios to the
photometric values.

The mass-$\log g$ and HR diagrams show the typical positions for the components
of a semi-detached system, with the primary near its expected evolutionary track
and an overluminous secondary. The primary is slightly overluminous relative to
its track, as is often the case in this work, while the secondary is slightly
below its track, a rare occurrence.

This binary was studied by HHH05. Their primary temperature ($25\,500$ K) and mass ratio ($\sim$1)
differ noticeably from our values.

\subsection{5 032412}
This wide, detached system presents medium-to-high quality light curves with
minima of unequal depth, betraying components of unequal brightness. The target
appears well isolated on the finding chart.

There are 13 out-of-eclipse spectra. The RV curves are very well constrained
and the RMS scatters are low. Interestingly, both the light and velocity curves
indicate a negligible eccentricity, in spite of the small relative radii of the
components, as if circularization had taken place during the protostellar phase.
Note that this is the most massive system of our whole sample: its total mass
reaches $30~\mathcal{M}_\odot$.

A spectroscopic $B$ luminosity ratio of 0.55 was found. The disentangled spectra
are of high quality (Fig. \ref{5__32412}), even for the \ion{H}{i} lines.
Beside the \ion{H}{i} and \ion{He}{i} lines, the following lines are visible in
the spectrum of the primary: \ion{He}{ii} 4200 and 4542 (strong),
\ion{Si}{iv} 4089, \ion{Si}{iv} 4116, \ion{O}{ii} 4185, \ion{Si}{iv} 4212,
\ion{O}{ii} 4276-7 and \ion{Si}{iii} 4553. An effective temperature of about
35\,000 K was inferred from the best-fitting synthetic spectrum. 

The following metallic lines are visible in the disentangled spectrum of the
secondary:  \ion{Si}{iv} 4089, \ion{Si}{iv} 4116, \ion{O}{ii} 4185,
\ion{O}{ii} 4190, \ion{C}{ii} 4267, \ion{O}{ii} 4276-7, \ion{O}{ii} 4415-7,
\ion{Mg}{ii} 4481 and \ion{Si}{iii} 4553. \ion{He}{ii} 4542 is clearly visible
too. Comparing the relative depths of \ion{Mg}{ii} 4481 with \ion{He}{i} 4472,
\ion{He}{ii} 4542 with \ion{Si}{iii} 4553, \ion{C}{ii} 4267 with
\ion{O}{ii} 4276-7, \ion{Si}{iv} 4089 and \ion{Si}{iv} 4116 with
\ion{He}{i} 4121 allows to estimate an effective temperature close to 31\,000~K.
Thanks to the good SNR of the spectra, fitting simultaneously the temperatures
of the components and the luminosity ratio resulted in a temperature ratio very
close to the photometric one. Nevertheless, the adopted temperatures are those
obtained by imposing the photometric ratio.

The mass-$\log g$ diagram shows a very young binary with both components on the
ZAMS. On the HR diagram, the positions of both components agree fairly well with
the stellar evolutionary tracks of 17.1 and 13.1 $\mathcal{M}_{\odot}$ stars.
However, the primary appears slightly overluminous relative to its track.

\begin{figure*}[htb]
\begin{center}
\includegraphics[trim = 0mm 5mm 0mm 3mm, clip, width=6.5cm]{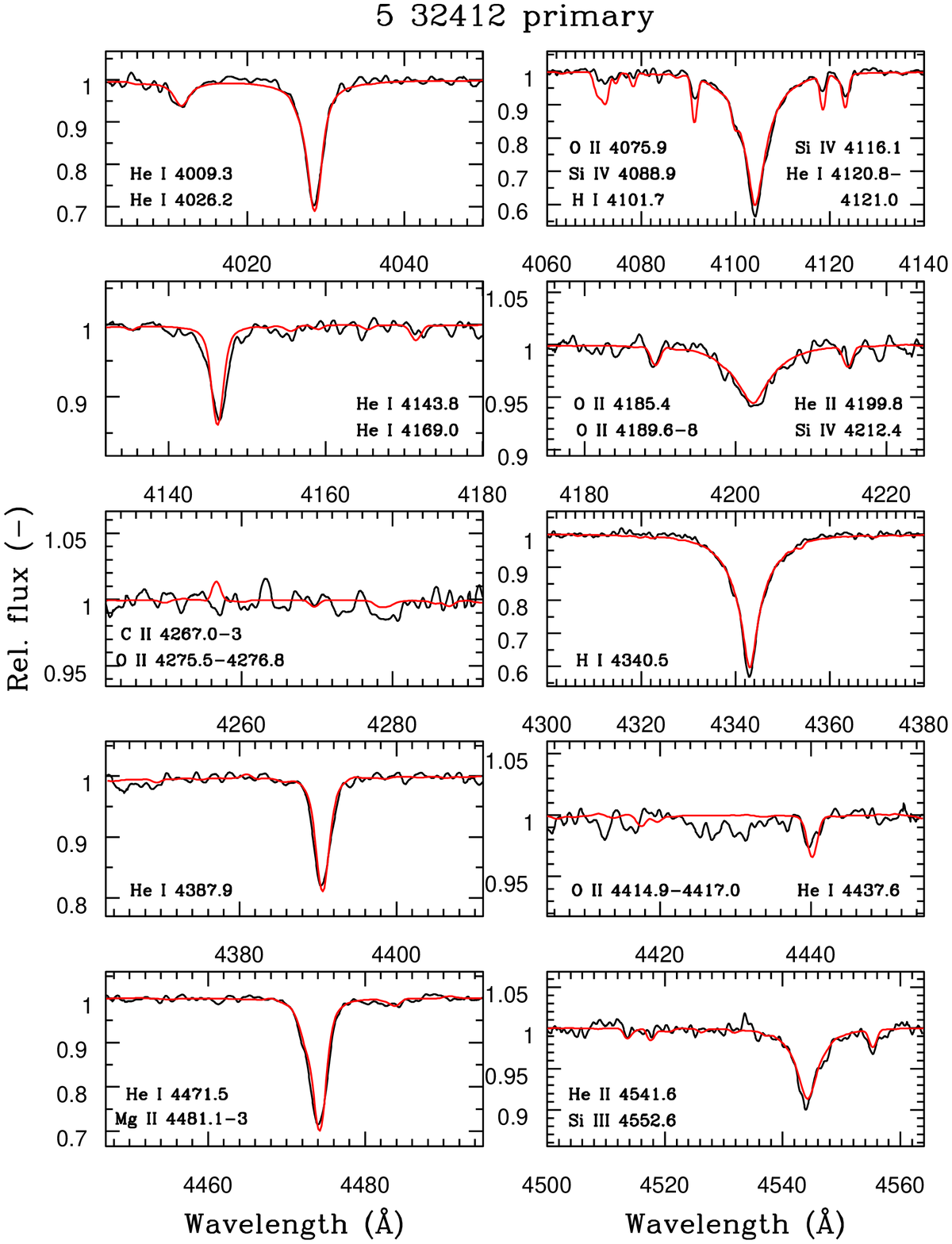}
\includegraphics[trim = 0mm 5mm 0mm 3mm, clip, width=6.5cm]{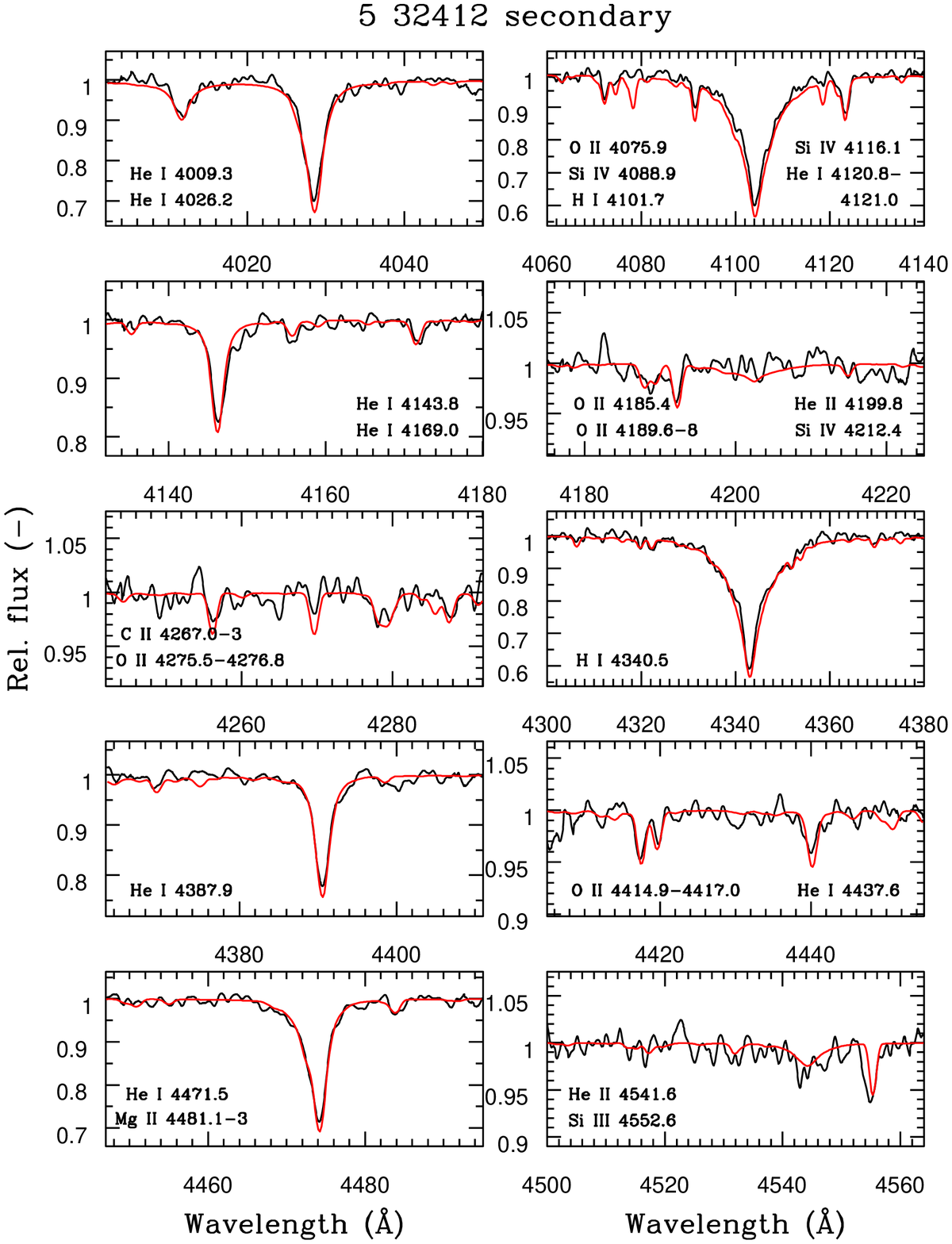}
\caption{Sections of the disentangled spectra of the primary and secondary
components of the binary 5 032412. The scale on the vertical axis is adapted to
the size of the features in each section. The red line represents a
velocity-shifted synthetic spectrum of the star. Beside prominent \ion{H}{i} and
\ion{He}{i} lines, \ion{He}{ii}, \ion{O}{ii}, \ion{Mg}{ii}, \ion{Si}{iii} and
\ion{Si}{iv} lines are visible in both spectra. The strong \ion{He}{ii} 4542,
next to a smaller \ion{Si}{iii} 4553 line, and the \ion{He}{ii} 4200 line
confirm the high temperature of the primary ($\sim$34\,900 K). The lower
temperature of the secondary is confirmed by the stronger \ion{Mg}{ii} 4481
line and the smaller \ion{He}{ii} lines.  
}
\label{5__32412}%
\end{center}
\end{figure*}

\subsection{5 038089}
This bright detached system presents medium-quality light curves with eclipses
of very similar depth. Therefore, the components are very similar.

This system has abnormal colour indices (see Table \ref{basicparam} and
Fig. \ref{colour_colour.2}), which suggests the presence of a third light. An
examination of the OGLE finding chart reveals that this binary is clearly
blended, though with much fainter stars.

There are 11 out-of-eclipse spectra. The RV curves are very well constrained
and the RMS scatter is remarkably low.

A spectroscopic $B$ luminosity ratio of 0.68 was found. The eclipses are not
very deep, so we fixed the potential of the primary to reproduce this luminosity
ratio. The disentangled spectra, shown in Fig. \ref{5__38089}, are very similar.
Strong \ion{He}{ii} 4200 and 4552
lines are visible in both spectra. The following metallic lines are equally
identifiable (Fig. \ref{5__38089}): \ion{O}{ii} 4076, \ion{Si}{iv} 4089,
\ion{Si}{iv} 4116, \ion{O}{ii} 4185, \ion{O}{ii} 4190, \ion{C}{ii} 4267,
\ion{O}{ii} 4276,\ion{O}{ii} 4415-4417, \ion{Si}{iii} 4553. This wealth of lines
allows to determine the temperatures of both components with a great accuracy.
From the best-fitting synthetic spectra, we found 30\,400 K and 30\,800 K for
the effective temperature of the primary and secondary, respectively. This is
very close (i.e. within $200$~K) to the temperatures estimated from the
composite spectra by imposing the photometric temperature and luminosity ratios;
thus, one can safely conclude from this example that the two methods are
equivalent.

The high quality of the spectroscopic observations allowed to estimate the
astrophysical parameters with a greater accuracy than most systems in our
sample. On the mass-$\log g$ diagram, both components lie just above the
10 Myr isochrone; their respective positions suggest that the ratio of radii may
be slightly underestimated. The HR diagram shows that both stars are
significantly overluminous with respect to the evolutionary tracks of 13.0 and
11.7 $\mathcal{M}_{\odot}$ stars. Invoking the blending of the binary with a
third-light contributor does not seem to help much. No clear sign of a third
light can be seen in the lightcurve; this is admittedly a weak argument, since
the light curve of a detached system with weak proximity effects cannot
constrain well a third light. But, in addition, a third light will not change
much the relative radii of the components, and the effective temperatures seem
well constrained by the relative intensities of several lines, so that the
luminosities should remain unaffected. Furthermore, the radial velocity curves
are of such quality that it is difficult to imagine how the masses could be
biased otherwise than through the inclination angle $i$. A third light would
make the photometric minima less deep and so, it would mimick a lower $i$.
Since the RV amplitudes give the product $\mathcal{M}\sin^3i$, underestimating
$i$ is equivalent to {\sl overestimating} $\mathcal{M}$, while we would need
the reverse to explain the HR diagram we see. A test with the EBOP code confirms
this qualitative argument: assuming a third light $\ell_{3,I}=0.2$ changes the
radius of the primary by $-0.26$\% only (keeping the ratio of radii constant),
so that the luminosity decreases by half a percent, while the inclination is
increased by more than three degrees, lowering the masses by $3.6$\%. That would
imply evolutionary tracks with a luminosity about 0.05 dex lower (or 11\%) in
the HR diagram. Finally, the distance modulus of this system is very close to
the one expected for the SMC. Therefore, for a third light to be the cause of
the inconsistency, it can only be via the RV curves, the amplitude of which
should be biased to small values by a stationary third spectrum.

This binary was part of the first release of 10 bright SMC systems by HHH03.
Comparing our results with theirs, we see that we have similar
estimates for the mass ratio and temperatures of the stars, but markedly
different values for the masses and radius of the secondary. A photometric
solution was proposed by dG03, who do not recommend it for distance
determination because its components do not fall on their mass-luminosity
relation.

\begin{figure*}[htb]
\begin{center}
\includegraphics[trim = 0mm 5mm 0mm 3mm, clip, width=6.5cm]{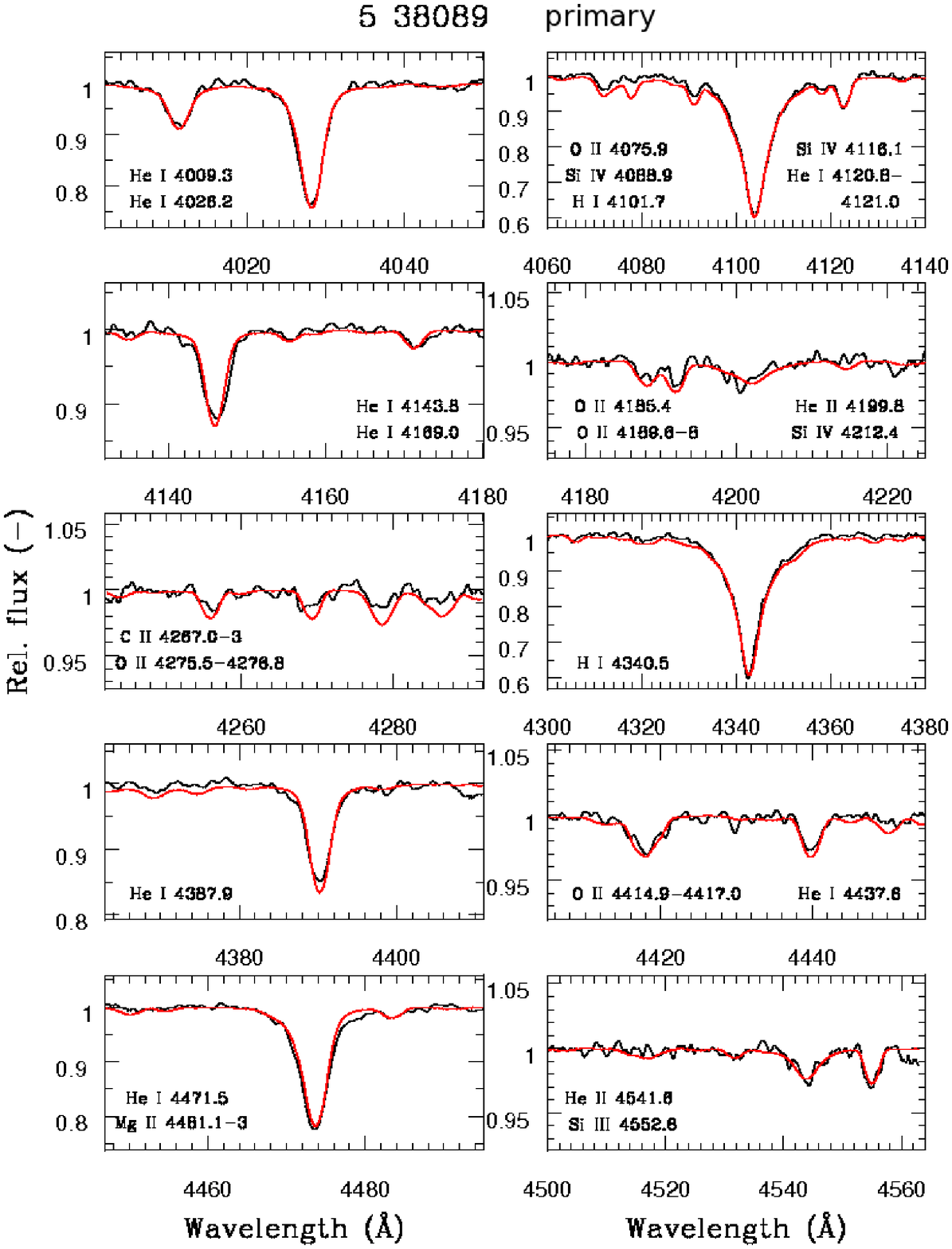}
\includegraphics[trim = 0mm 5mm 0mm 3mm, clip, width=6.5cm]{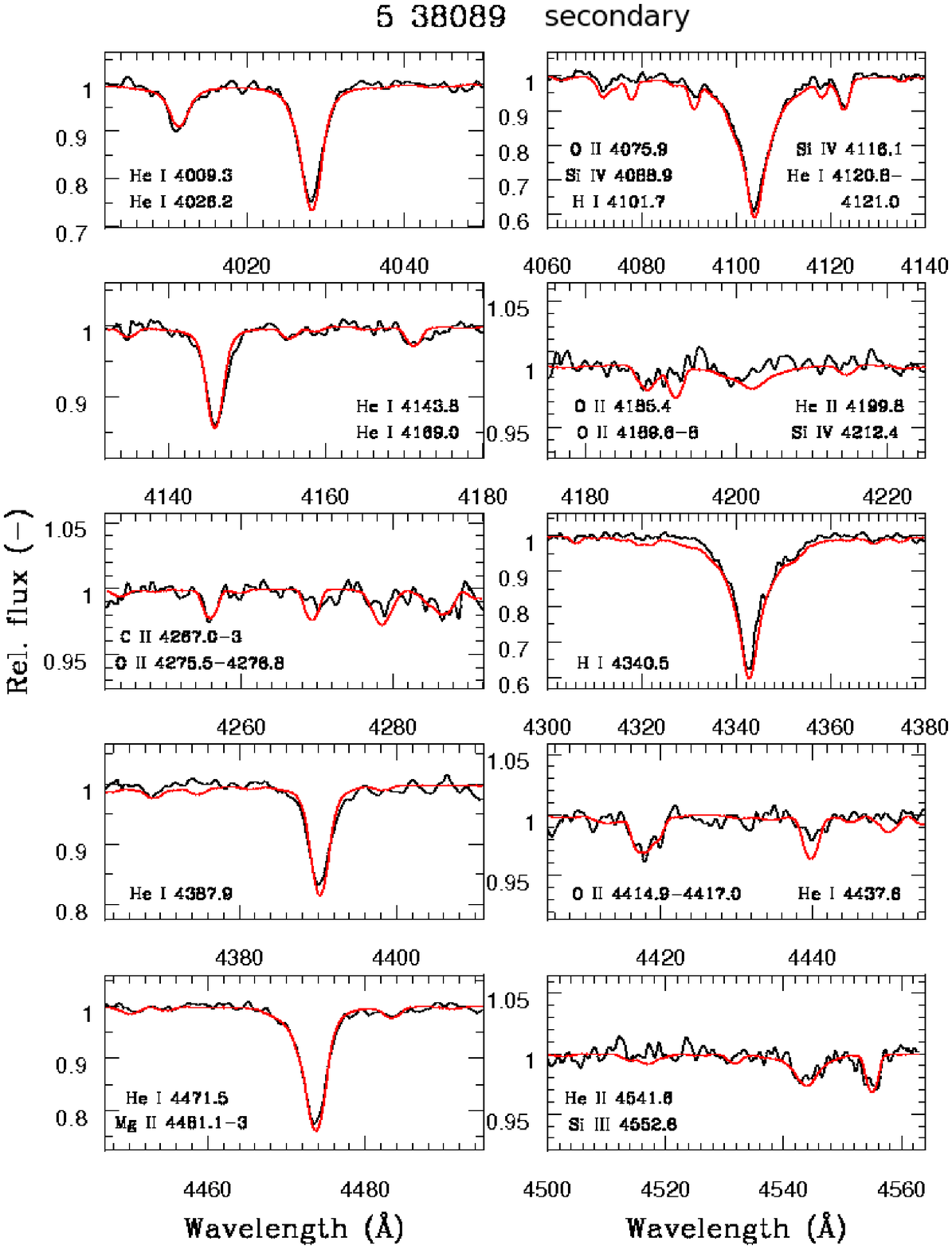}
\caption{  Sections of the disentangled spectra of the primary and secondary
components of the binary 5 038089. The scale on the vertical axis is adapted to
the size of the features in each section. The red line represents a
velocity-shifted synthetic spectrum of the star. Beside prominent \ion{H}{i} and
\ion{He}{i} lines, \ion{He}{ii} and a number of fainter metallic lines
(\ion{C}{ii}, \ion{O}{ii}, \ion{Si}{iii}, \ion{Si}{iv}) are visible. The
relative intensities of the \ion{He}{ii} 4542 and \ion{Si}{iii} 4553 are
very useful to constrain the temperature of the two stars.}
\label{5__38089}%
\end{center}
\end{figure*}

\subsection{5 095337}
This is a close but detached system, with tidally distorted components
of unequal brightness. The $O-C$ curve suggests that the primary eclipse is not
perfectly reproduced by the synthetic light curve, but there is no obvious
third-light contribution. The finding chart shows some blends with two or three
fainter stars at $1\farcs 5$ or so.

There are 10 out-of-eclipse spectra. The RV curves are well constrained with
observations close to both quadratures.

A spectroscopic $B$ luminosity ratio of 0.66 was found. The disentangled spectra
are of low quality. Because of the low $S/N$ and the high
$V_\mathrm{rot} \, \sin i$ ($\sim$200 km s$^{-1}$), no metallic line can be
positively identified. There is some nebular emission in the Balmer lines.
The best-fitting synthetic spectra allowed to estimate simultaneously the
temperatures of the primary and of the secondary, the ratio of which agrees
quite well with the photometric one. In addition, the spectroscopic $B$
luminosity ratio perfectly matches the photometric one, so that the potential of
both the primary and the secondary were left free to converge.

On the mass-$\log g$ diagram, both components define a segment which is
perfectly parallel and very close to the 10 Myr isochrone. On the HR diagram,
however, both components appear strongly overluminous compared to the
evolutionary tracks of 8.7 and 7.6 $\mathcal{M}_{\odot}$ stars. 
Decreasing the effective temperatures by about $1300$~K would reconcile the
luminosities with the tracks. However, this looks difficult. The emission in
both Balmer lines was suppressed on a 4~\AA~ range centered on each emission
line, and we verified that increasing that range to 8~\AA~ does not change the
estimated temperature in a significant way. Thus, either this system suffers
from some bias on the RV curves, or its metallicity is closer to $Z=0.001$ than
to $Z=0.004$.

\subsection{5 095557}
This is the system with the highest eccentricity, displaying a medium quality
light curve with minima of unequal depths. The target is perfectly isolated on
the finding chart.

There are 11 out-of-eclipse spectra. The RV curves are well constrained with
observations close to both quadratures, but the fit is not very good and,
unfortunately, most spectra are grouped in the phase interval with the smaller
amplitude.

A spectroscopic $B$ luminosity ratio of $\sim$0.5 was found when limiting the
fit to the seven spectra for which the radial velocity difference $\Delta RV >
250~\mathrm{km\,s^{-1}}$. This ratio increases to 0.63 if all eleven spectra are
taken into account, so that this quantity is rather poorly constrained. The
disentangled spectra are of very low quality. This is probably
due partly to the low $S/N$ of the observed spectra, and partly to some
inaccuracies in the orbital parameters. The temperature of the secondary
given by the fits to the composite spectra is about $1000$~K higher than the
photometric estimate, which was adopted. The pseudosynchronized values of
$V_\mathrm{rot} \, \sin i$ were adopted.

An apsidal motion is detected at the $5\,\sigma$ significance level, and the
WD result is confirmed by the variation of the $e\,\cos\omega$ quantity as given
by the EBOP code. The photometric data was divided into four sets and the fits
were obtained by fixing the inclination, the ratio of radii and the relative
radius of the primary. The result is displayed in Fig.~\ref{apsid_95557} and
suggests the apsidal motion to be real. The adopted apsidal motion seems
underestimated on that figure, but the constraint imposed by the RV curves has
to be kept in mind.
\begin{figure}[htb]
\begin{center}
\includegraphics[width=9cm]{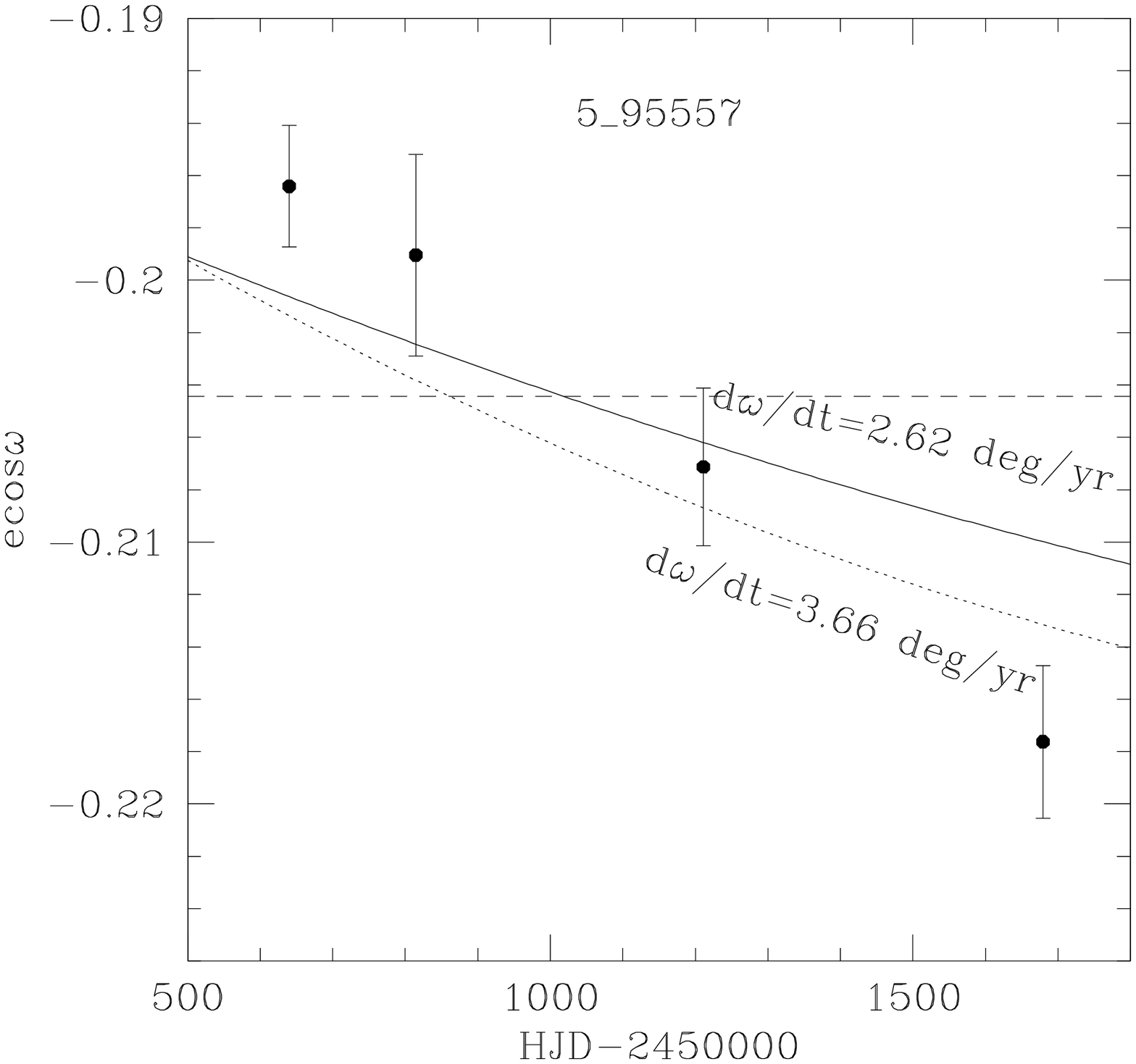}
\caption{Same as Fig.~\ref{4_160_ecosw}, but for the system 5 95557. The
filled black dots represent independent $e\cos\omega$ determinations made on
four successive times series containing about the same number of photometric
observations in the $I$ band. The solid line is based on the parameters
listed in Table~\ref{orbparam}, while the dotted line corresponds to the fitted
$\dot{\omega}+2\sigma$. The horizontal dashed line shows the result obtained
from the whole time series under the assumption of a constant $\omega$.}
\label{apsid_95557}
\end{center}
\end{figure}
On the mass-$\log g$ diagram, the two components lie right on the 30 Myr
isochrone, in spite of the rather large uncertainty on the masses. On the HR
diagram, both components appear clearly overluminous relative to their
respective evolutionary tracks, as in many other systems.

\subsection{5 100485}
This is a detached system with `twin' components and a circular orbit. The
finding chart reveals no crowding problem, the closest neighbour being at about
$2\farcs 5$.

There are 13 out-of-eclipse spectra. The RV curves are well constrained, with
observations close to both quadratures and small residuals.

A spectroscopic $B$ luminosity ratio of 0.93 was found, which matches the
photometric value (0.97) rather well. The disentangled spectra are very similar.
The following metallic lines can positively be identified: \ion{C}{ii} 4267,
\ion{Mg}{ii} 4481 and \ion{Si}{iii} 4553. These lines allowed to estimate a
temperature close to $\sim$22\,000 K for each component. The least-squares fit,
performed with both temperatures and the luminosity ratio free to converge,
provided a temperature ratio very close to the photometric one.

The two components lie right on the 20 Myr isochrone on the mass-$\log g$
diagram, and they are compatible, within the errors, with their respective
evolutionary tracks in the HR diagram. Here again, however, they are slightly
overluminous, unless their metallicity is low.

\subsection{5 100731}
This system presents low-to-medium quality light curves of the EW type, with the
smallest amplitudes among those in our sample. This is another case of binary
seen under an unfavourable inclination ($\sim$60$\degr$).
The finding chart shows a perfectly isolated target.

The most satisfying fit of the light curves was obtained with an overcontact
model.

There are only 8 out-of-eclipse spectra. Nevertheless, these observations are
sufficiently constraining to get reliable RV curves.

Spectroscopy gives a $B$ luminosity ratio of $\sim 0.44$, while the photometric
ratio amounts to $\sim 0.51$. As usual, the latter was preferred, especially
because of the strong constraints provided by a Roche-lobe-filling pair. The
disentangled spectra have a low $S/N$, and a least-squares fit to the composite
spectra with both temperatures free provided two temperatures close to
$23\,000$ K. However, the scatter of the secondary temperature was high, so the
photometric temperature ratio was adopted and fixed.

The mass-$\log g$ diagram shows a secondary component more evolved than the
primary, as expected. On the HR diagram, both components appear to be
underluminous with respect to their evolutionary tracks, though by a little
amount. This is an exceptional occurrence.

\subsection{5 106039}
This system is a typical semi-detached one. There is a small depression
occurring just before the eclipse of the primary, which is strong evidence for
a gas stream. We did not attempt to model that stream with a cool spot on the
primary, because the distorsion of the light curve remains relatively mild. The
target appears perfectly isolated on the finding chart.

There are 9 out-of-eclipse spectra. The RV curves are well constrained with
observations close to both quadratures.

A spectroscopic $B$ luminosity ratio of 1.03 was found, in excellent agreement
with the photometric value (1.01). The following metallic lines are visible in
the disentangled spectrum of the primary: \ion{C}{ii} 4267,
\ion{Mg}{ii} 4481 (faint) and \ion{Si}{iii} 4553. A stronger
\ion{Mg}{ii} 4481 line is equally visible in the spectrum of the secondary.
From these lines, the primary temperature was estimated to be close to
$25\,500$ K. The least-squares fit, performed with temperature and luminosity
ratios fixed to the photometric values, provided a primary temperature
$\sim 800$~K higher.

The position of the primary component in the HR diagram is in fair agreement
with the theoretical evolutionary track of a 8.6 $\mathcal{M}_{\odot}$ star,
though it is slightly overluminous. The evolved secondary component is
overluminous with respect to the track of a single star of the same mass.
  
\subsection{5 111649}
This is a detached system with very slightly distorted twin components.
There is a group of bright stars close to the target on the finding chart, but
they are remote enough ($3.5-4\arcsec$) that no 3rd light should be expected from them.

There are 10 out-of-eclipse spectra. The RV curves are not well sampled because
the period of 2.95955 days is very close to an integer number of days, but the
scatter is small, thanks to the slow projected rotational velocities induced by
the relatively long period.

A spectroscopic $B$ luminosity ratio of 0.89 was found, which was imposed (via
the potential of the primary) to define the ratio of radii. Indeed, the small
amplitude of the light curve prevents a purely photometric ratio to be well
constrained. Noisy \ion{C}{ii} 4267 and \ion{Mg}{ii} 4481 lines are visible in
the disentangled spectrum of the secondary. The temperature of the primary was
determined from a fit with fixed temperature and luminosity ratios. According to
both spectroscopy and photometry, the secondary appears marginally hotter than
the primary, the temperature difference being about $2\,\sigma$.

Both components have a very similar mass ($\sim$5.4 $\mathcal{M}_{\odot}$), and
according to the mass-$\log g$ diagram, the empirical mass contrast appears a
bit too high to match the $\sim 80$~Myr isochrone. Interestingly, both
components lie right on the Terminal Age Main Sequence (TAMS), where evolution
is so fast that a theoretical lower limit to the mass ratio can be settled.
Starting from the purely empirical surface gravities, and increasing their
difference by two sigma (so that $\log g_\mathrm{P}-1\,\sigma=3.72$ and
$\log g_\mathrm{S}+1\,\sigma=3.80$), one can read the corresponding masses along
the $\log t=7.9$ isochrone. Assuming that both stars have not yet passed the
``hook'' that marks the end of the mains sequence, the resulting mass ratio
is $q=0.98$. But it is quite possible that the primary has just passed the hook
while the secondary has not, in which case $q=0.966$. In both cases, the mass
ratio is closer to one than the value directly obtained from the RV curves by
$0.03$ to $0.04$. Therefore, this system can be considered as hosting real
twins.

The position of the primary in the HR diagram agrees almost perfectly with the
theoretical evolutionary track, especially if the star has just evolved beyond
the ``hook''. That of the secondary, however, is a bit too high, as if its mass
was underestimated.

\subsection{5 123390}
This slightly eccentric system ($e = 0.042$) presents low-to-medium quality
light curves with a small amplitude. This is clearly a detached system with components of
unequal brightness. The finding chart shows a neighbour at about $1\farcs 5$ to the SW
of the binary, which might have polluted the spectra slightly.

There are 14 out-of-eclipse spectra. The RV curves have a small RMS
scatter and are very well constrained by the observations.

A spectroscopic $B$ luminosity ratio of 0.58 was found. The disentangled spectra
are of fairly good quality. The following metallic lines are visible in both
spectra: \ion{O}{ii} 4076, \ion{C}{ii} 4267, \ion{O}{ii} 4276,
\ion{O}{ii} 4415-4417, \ion{Mg}{ii} 4481 and \ion{Si}{iii} 4553. For the
primary, the best-fitting temperature for these lines is $\sim$26\,000 K. The
least-squares fit of the composite spectra gave temperatures of $28400$ and
$26260$~K for the primary and secondary respectively, but with a large scatter.
A plot of the fitted temperatures versus the unnormalized chi-square shows that
in some cases, the fit switched components, i.e. attributed the high temperature
to the secondary and vice-versa, which partly explains the large scatter (see
Fig. \ref{Teff2_chi2}).
\begin{figure}[htb]
\begin{center}
\includegraphics[width=9cm]{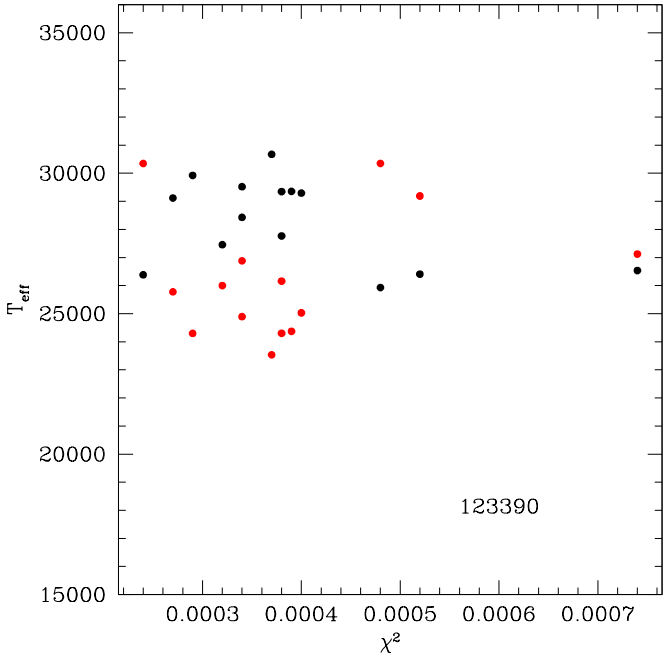}
\caption{Temperature of the primary (black dots) and of the secondary (red dots)
component of the system 5 123390,
obtained by fitting composite synthetic spectra to the observed ones, versus the
quality of the fit. The trends are roughly horizontal, at least for good fits,
which inspires confidence. In a few cases, the components are exchanged (see
text).}
\label{Teff2_chi2}
\end{center}
\end{figure}
Strangely enough, the photometric ratio of temperatures is close to one, so that
the temperatures become, when fitted while keeping this ratio fixed, $27840$ and
$28320$~K for the primary and secondary component respectively. Taken at face
value, however, Fig. \ref {Teff2_chi2} rather suggests $29000$ and $25000$~K.

According to the mass-$\log g$ diagram, this binary is $\sim$12 Myr old. The
position of the primary component in the HR diagram is a bit too high relative
to its evolutionary track. The secondary component is much too luminous, falling
on the track of the primary! If the purely spectroscopic temperatures were
adopted, the primary would be even more overluminous, but the secondary would
fall right on its track. 

There is a marginally significant apsidal motion of $4.75\pm 1.63$ $\degr$
yr$^{-1}$. We show in Fig. \ref{5_123_ecosw} the $e\cos\omega$ quantity found
with the EBOP code on the basis of the $I$ magnitudes. There is indeed a
slight trend corresponding to an increase of $\omega$ with time.
\begin{figure}[htb]
\begin{center}
\includegraphics[width=9cm]{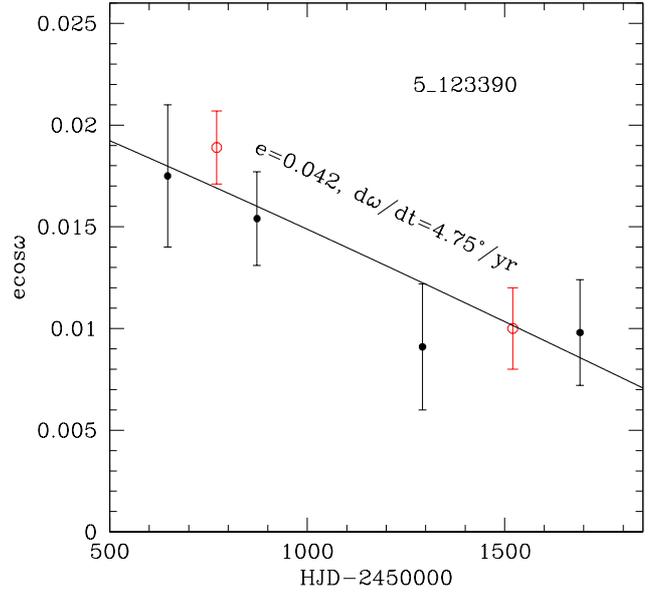}
\caption{Same as Fig.~\ref{4_160_ecosw}, but for the system 5 123390. The
filled black dots represent independent $e\cos\omega$ determinations made on
four successive times series containing about the same number of photometric
observations in the $I$ band. The open red dots refer to $e\cos\omega$
determinations based on only two time series, but which are twice longer.
The solid line is based on the parameters listed in Table~\ref{orbparam}.}
\label{5_123_ecosw}
\end{center}
\end{figure}

\subsection{5 180185}
The EA-type light curves of this system are irregularly sampled, because the
period is close to $5.5$ days, and there are only very few data in the primary
minimum. This is a typical well-detached system with twin components and a small
eccentricity.
The finding chart reveals a slightly fainter neighbour about $2\arcsec$ away from the
target, i.e. far enough that its influence on the spectra may be considered
negligible. 

This system has unreliable colour indices (see Table \ref{basicparam} and
Fig. \ref{colour_colour.2}).

There are 12 out-of-eclipse spectra. The RV curves are excellent and well
constrained by the observations, with the smallest residuals seen in our sample.
The formal errors on the resulting amplitudes are smaller than one percent,
allowing mass estimates to within $2-3$\%.

A spectroscopic $B$ luminosity ratio of 0.63 was found, which is close to the
value given by the WD analysis and finally adopted (0.71).
This is the system with the longest period ($P\sim5.5$ days) in our sample. The
long period is responsible for the low $V_\mathrm{rot} \, \sin i$ values
($\sim$40 km s$^{-1}$) compatible with synchronous rotation. 
 As a consequence, the disentangled spectra of this binary show rather sharp
lines. The following metallic lines are visible in both spectra:
\ion{C}{ii} 4267, \ion{Mg}{ii} 4481 and \ion{Si}{iii} 4553.
The temperature of the primary was obtained as usual by fixing the temperature
and luminosity ratios to the photometric values, in spite of the small number of
points in the minima. The secondary has a slightly hotter temperature than the
primary.

Both components are fairly well aligned on the 50 Myr isochrone in the
mass-$\log g$ diagram, although the primary should be slightly more evolved.
On the HR diagram, the secondary falls right on its track, while the luminosity
of the primary appears too low. This might be due to an unreliable temperature
ratio, because of the small number of photometric points in the minima.

\subsection{5 180576}
This system presents low quality light curves of the EB type. The depths of both
minima are rather low and the RMS scatter is high. This is a detached system
with components of unequal brightness and a circular orbit. The finding chart
shows a close neighbour, at about $1\farcs 5$ to the NNW, which might have distorted
the temperature estimate of the binary components.

There are 12 out-of-eclipse spectra. The RVs curves are rather good and well
constrained by the observations.

A spectroscopic $B$ luminosity ratio of 0.42 was found. This is close to the
value reached by the final WD analysis.
The observed composite spectra are very noisy ($25 \leq S/N \leq 64$) and are
contaminated by nebular emission. The disentangled spectra are of rather poor
quality. This is partly due to the low reliability of the continuum placement.
The \ion{C}{ii} 4267 and \ion{Mg}{ii} 4481 lines are visible in the spectrum of
the primary. The temperature of the primary, determined from a least-squares fit
where the temperature and luminosity ratios were fixed, depends on the quality
of the fit, as shown by Fig. \ref{180576Teff_chi2}. We adopted the temperature
of the primary corresponding to the best $\chi^2$, in view of the roughly linear
correlation between $T_{\mathrm{eff}}$ and $\chi^2$, but without attempting to
extrapolate the relation to $\chi^2=0$.
\begin{figure}[htb]
\begin{center}
\includegraphics[width=9cm]{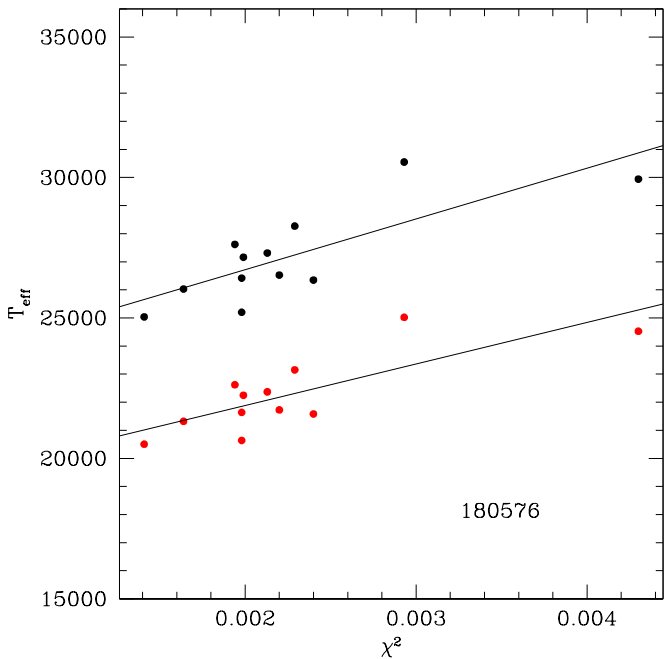}
\caption{Effective temperatures of the system 5 180576 obtained from a fit to
the composite spectra, keeping the temperature and luminosity ratios fixed.
Black dots are for the primary, red dots for the secondary. Note that the two
curves should be considered as one and the same, since their ratio is constant.}
\label{180576Teff_chi2}
\end{center}
\end{figure}

On the mass-$\log g$ diagram, the components are perfectly aligned along the
15 Myr isochrone. The position of the primary component in the HR diagram
appears slightly too high relative to the corresponding evolutionary track,
while the secondary falls right on its track.

\subsection{5 185408}
This system presents medium quality light curves of the EA type. The minima are
of similar depth and there is some ellipsoidal variation.
This is a typical detached system with closely similar components, with medium
quality light curves and low amplitude minima of similar depth. The orbit is
circular. The $I-$band light curve was cleaned from five outliers before the
PHOEBE/WD analysis. The finding chart shows a well isolated target, the closest
neighbour being at about $3\arcsec$ to the WNW.

There are 12 out-of-eclipse spectra. The RV curves are excellent, especially in
view of the faintness of the object, and well constrained by the observations. 

A spectroscopic $B$ luminosity ratio of $\sim$0.64 was found. The WD analysis
tended to raise this value and the potential of the primary had to be fixed to
keep it close to the spectroscopic one. The \ion{C}{ii} 4267, \ion{Mg}{ii} 4481
and \ion{Si}{iii} 4553 lines are visible in both disentangled spectra. The
temperatures were obtained by fixing the temperature and luminosity ratios to
the photometric values.

On the mass-$\log g$ diagram, the components are fairly well aligned on the 10
Myr isochrone. On the HR diagram, both components are significantly overluminous
with respect to their respective evolutionary tracks, as is often the case.

 \subsection{5 196565}
This detached eccentric system presents medium quality light curves with
eclipses of similar depth. The $V$ and $B$ data are missing in the OGLE
database. The finding chart shows a neighbour of similar brightness as the
target about $2\arcsec$ to the S, raising some concern regarding possible spectral
pollution. The results do not confirm these fears, however.

There are 13 out-of-eclipse spectra. The RV curves are very good and well
constrained by the observations, in spite of a lack of observations where the
expected amplitude reaches its maximum.

A spectroscopic $B$ luminosity ratio of $\sim$0.41 was found, and the potential
of the primary was adjusted and fixed in the WD analysis to maintain that value.
The \ion{C}{ii} 4267 and \ion{Mg}{ii} 4481 lines are visible in both disentangled
spectra. The least-squares fit of composite synthetic spectra to the observed
ones provided a temperature ratio larger than that provided by the photometry,
and a primary temperature of $20960\pm 600$~K. Fixing the temperature ratio to
the photometric value resulted in the slightly hotter primary temperature that
has been adopted.

The mass-$\log g$ diagram suggests that the secondary has a too large radius
relative to the primary, as if the luminosity ratio was overestimated: while the
primary lies close to the 40 Myr isochrone, the secondary lies on the 50 Myr
one. On the HR diagram, however, each component has a position compatible with
its respective evolutionary track within the error bars, though the primary is
slightly overluminous while the secondary is slightly underluminous.

This system was studied by WW01. Their relative radius of the primary ($0.204$)
is very close to ours ($0.200$) but their radius of the secondary ($0.116$) is
smaller than ours ($0.150$). The difference is due to the fact that they
consider the eclipses as total, while we consider them as partial. Additional
accurate photometry in both minima would be welcome to settle the question.

\subsection{5 261267}
This is a typical semi-detached system with high quality light curves and
eclipses of unequal depths.  The $O-C$ curve shows no detectable depression
before the primary eclipse. Although the finding chart shows a few neighbours,
they all lie beyond $2\arcsec$ of the target.

This system has peculiar colour indices (see Table \ref{basicparam} and
Fig. \ref{colour_colour.2}). Since it lies near the edge of the CCD in the
OGLE-II survey, it is also listed under the name 6 11806 in the corresponding
database. Thus, there are more than 600 data points in the $I$-band light curve,
instead of about 300.

There are 10 out-of-eclipse spectra, which constrain the RV curves relatively
well.

Both photometric and spectroscopic ratios are very similar ($\sim$0.4). Because
of the relatively high $V_\mathrm{rot} \, \sin i$ ($> $150 km s$^{-1}$)
and moderate $S/N$ of the observations, there are no exploitable metallic lines
in the disentangled spectra. The temperature of the primary was obtained, as
usual, by least-squares fit to the composite spectra, after fixing the
temperature and luminosity ratios to the values given by a preliminary WD
analysis.

The mass-$\log g$ and HR diagrams are typical of a massive Algol-type binary.
The primary is overluminous relative to the evolutionary track of an isolated
star of the same mass, as are other semi-detached systems like 4 113853 and
5 277080.

\subsection{5 265970}
This slightly eccentric detached system has medium-to-high quality light
curves of the EA type. The sampling of the light-curve is incomplete, due to an
orbital period close to $3.5$ days. In particular, the depth of the secondary
minimum is ill-defined. Therefore, the photometric temperature ratio and
inclination are not very reliable. Actually, there is a correlation between
these two parameters, in the sense that an increase of inclination implies a
decrease of the temperature ratio. The finding chart shows a well defined target, but with
an only slightly fainter neighbour about $2\farcs 6$ to the NW.

This system lies near the edge of the OGLE-II CCD, and so was measured also on
the adjacent chip under the name 6 17345, so that there are as many as 586 data
points in the $I$ band. The two data sets were merged after applying a small
magnitude offset to each. Using the EBOP code, the fitted magnitude at
quadrature and its error was defined for each set, then the mean magnitude at
quadrature weighted by the inverse of the variance was computed. Finally, the
appropriate offset was applied to each of the two sets to adjust it to this mean
magnitude.

There are 10 out-of-eclipse spectra. The RV curves are quite good and well
constrained by the observations.

Because of the loose constraints on the light curves, the proposed solution
for this system heavily relies on the spectroscopic observations.
Nevertheless, the proposed solution meets very well the spectroscopic
and photometric constraints, so we consider it as close to reality.

A spectroscopic $B$ luminosity ratio of $\sim$0.23 was found from the usual
least-squares fit, which provides the temperature of the primary with an
excellent internal precision. The temperature of the secondary is much less
certain, because of the small luminosity ratio. In order to maintain the
luminosity ratio to the spectroscopic value in the WD analysis, one has to fix
the potential of the primary to an appropriate value, and the temperature ratio
converges to a smaller value than the spectroscopic one, but still compatible
with it given the errors. Thus the photometric temperature ratio was adopted,
and the temperature of the primary determined in the usual way.

A number of metallic lines are identifiable in both disentangled spectra:
\ion{C}{ii} 4267, \ion{O}{ii} 4276, \ion{O}{ii} 4415-17, \ion{Mg}{ii} 4481 and
\ion{Si}{iii} 4553. There is no emission in the Balmer lines.  

The positions of the stars in the mass-$\log g$ diagram are not quite mutually
consistent: the ratio of radii should be decreased in order to bring the two
components on the same isochrone, which would correspond to about 26 Myr.
On the HR diagram, the primary component matches its theoretical evolutionary
track surprisingly well, while the secondary has a position consistent with its
track within the error bar. The spectroscopic constraints, which are strong, are
well fulfilled, but additional photometric data would be useful to improve our
solution. This system is especially interesting, because the primary is very
close to the TAMS while the secondary is much less evolved.

We applied the EBOP code on four subset of the total time series, after fixing
all parameters to their average value, except inclination, $e\cos\omega$,
$e\sin\omega$, magnitude at quadrature and phase shift. No significant trend can
be seen on Fig. \ref{5_265_ecosw}, which does not prove, however, that $\omega$
remains constant with time. The less reliable $e\sin\omega$ quantity does not
differ significantly from zero according to the EBOP code. In the WD solution,
we have arbitrarily imposed a small value $\dot{\omega}=1.8\times
10^{-4}~\mathrm{rad\,day^{-1}}$ which roughly corresponds to the theoretical
prediction. 
\begin{figure}[htb]
\begin{center}
\includegraphics[width=9cm]{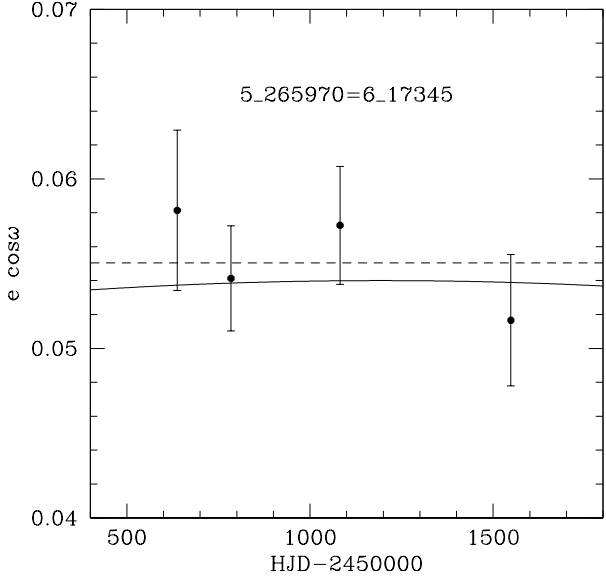}
\caption{Same as Fig.~\ref{4_160_ecosw}, but for the system 5 265970. The
horizontal dashed line shows the weighted average of the $e\cos\omega$ values.
The solid line corresponds to $\dot{\omega}\simeq
3.8^\circ\,\mathrm{yr^{-1}}$, the expected theoretical value.}
\label{5_265_ecosw}
\end{center}
\end{figure}

\subsection{5 266015}
This system presents high quality light curves of the EB type with minima of
unequal depth. Its bright $I-$band light curve is perfectly well sampled. The
small depression occurring just before the primary eclipse is indicative of a
semi-detached system with a secondary component filling its Roche lobe and
pouring matter onto the primary component. However, the amplitude of this effect
was judged too small to justify an attempt to model it through a spot.

There are 10 out-of-eclipse spectra. The RV curves are well constrained by the
observations.

A spectroscopic $B$ luminosity ratio of 0.61 was found, rather remote from the
photometric value ($0.49$), as is often the case in semi-detached systems.
The disentangled spectra of both components have a decent $S/N$
(Fig. \ref{5_266015}). Nevertheless, the $V_\mathrm{rot} \, \sin i$ being high
($> $150 km s$^{-1}$), the small metallic lines are not very conspicuous. The
\ion{He}{ii} 4120 and 4542 lines are visible in the spectrum of the primary.
The \ion{Si}{iii} line appears in both spectra. This is one of the few systems
with apparently no significant nebular emission lines, therefore the
disentangled Balmer lines can be used to find the temperature of the primary.
The spectral features of the primary point to a $32\,000$ K star, which is
confirmed by the usual least-squares fit. 

The positions of the stars in the mass-$\log g$ and HR diagrams are coherent
with an evolved system having undergone mass exchange. The primary component is
close to the track corresponding to a 15.6 $\mathcal{M}_{\odot}$ star.

This system was studied by WW02. As with most semi-detached systems they
studied, their photometric mass ratio proved to be unreliable.

\begin{figure*}[htb]
\begin{center}
\includegraphics[trim = 0mm 5mm 0mm 3mm, clip, width=6.5cm]{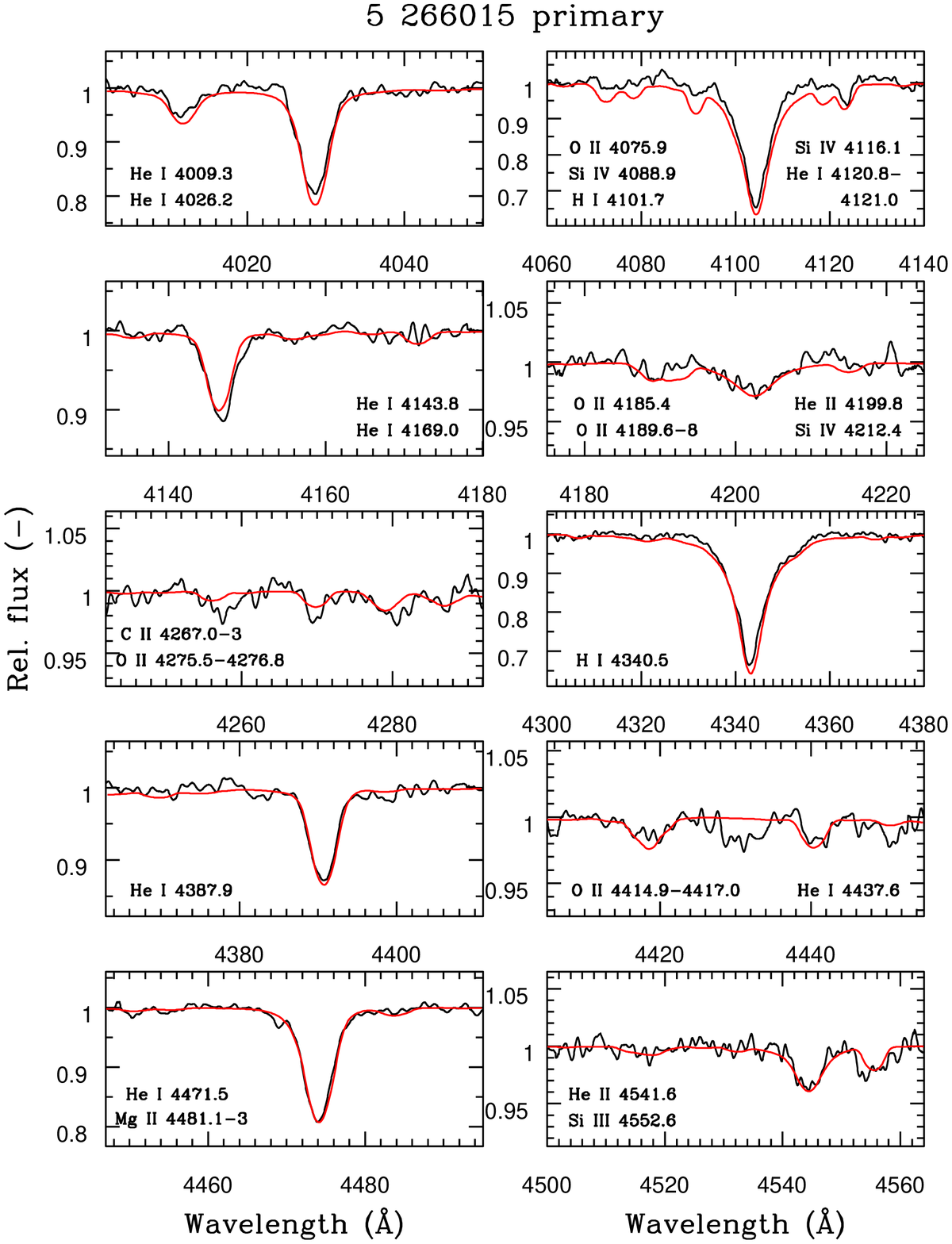}
\includegraphics[trim = 0mm 5mm 0mm 3mm, clip, width=6.5cm]{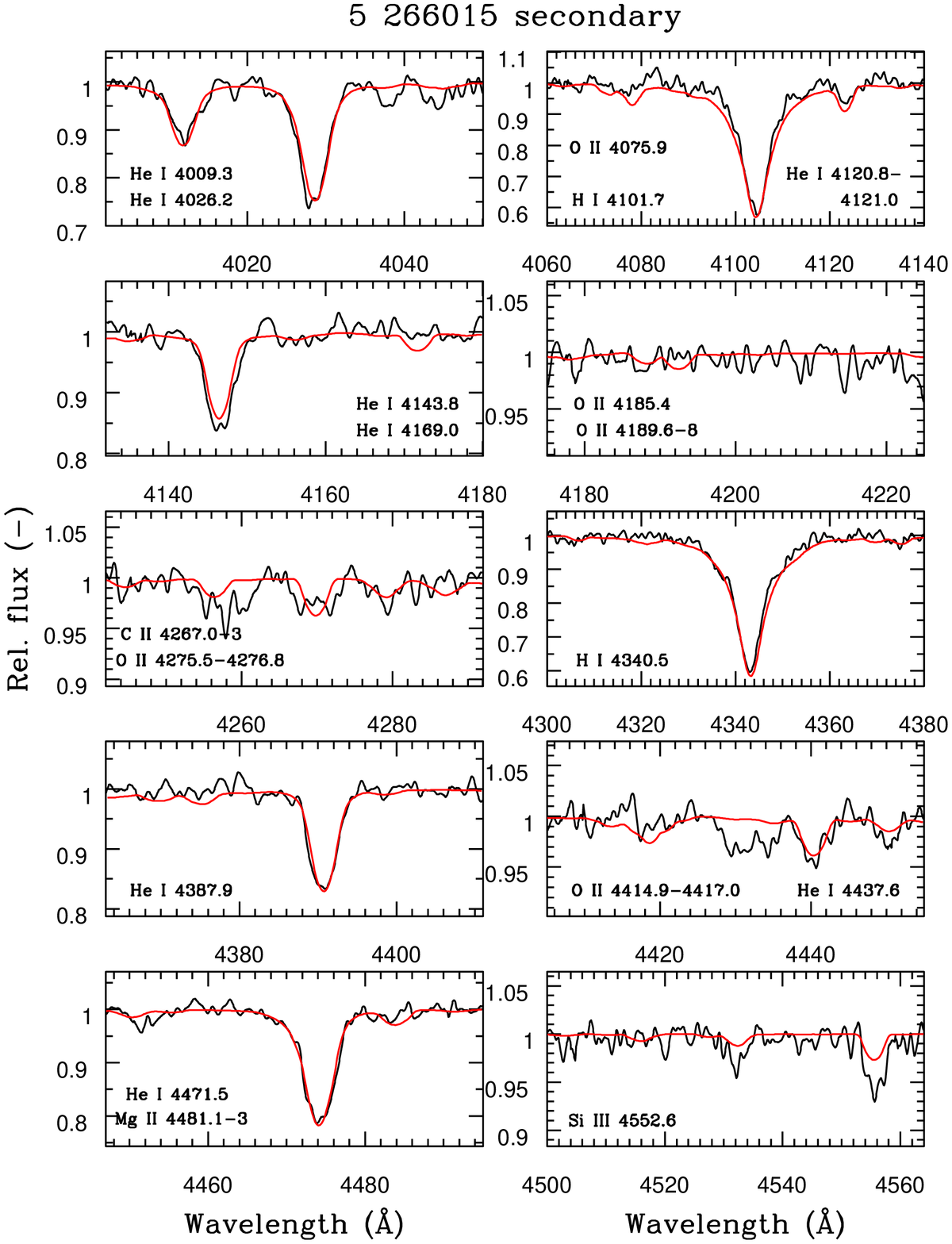}
\caption{Sections of the disentangled spectra of the primary and secondary
components of the binary 5 266015. The scale on the vertical axis is adapted to
the size of the features in each section. The red line represents a velocity-shifted 
synthetic spectrum of the star.
Beside prominent \ion{H}{i} and
\ion{He}{i} lines, \ion{He}{ii}, \ion{O}{ii}, \ion{Mg}{ii}, \ion{Si}{iii} and
\ion{Si}{iv} lines are visible in both spectra.
The strong \ion{He}{ii} 4542, next to a smaller \ion{Si}{iii} 4553 line, and
the \ion{He}{ii} 4200 line confirm the high temperature of the primary ($\sim$32\,000 K).
Poor continuum placement is responsible for the apparent $O-C$ mismatch of the H$\delta$ line.  
}
\label{5_266015}%
\end{center}
\end{figure*} 

\subsection{5 266131}
The most striking feature of this slightly eccentric detached system ($e\simeq 0.04$)
is its huge apsidal motion, which explains the apparently very bad $I-$band
$O-C$ curve.  An examination of the OGLE finding chart revealed
that this binary is slightly blended, but {\sl a posteriori}, this does not
seem to have distorted the results.

This system lying close to the edge of the CCD in the OGLE-II survey, it also
exists under the name 6 22883, which doubles the number of data points in the $I$
band. The magnitude at quadrature was determined using the EBOP code for each of
the two data sets; the two resulting values agreeing within one thousandth of a
magnitude, the two data sets were merged without applying any magnitude offset.

There are 10 out-of-eclipse spectra, which constrain quite well the RV curves.

A spectroscopic $B$ luminosity ratio of $\sim$0.65 was found. As usual with the
well detached systems, the potential of the primary was fixed at a value such
that this luminosity ratio is preserved through the WD analysis.
The $S/N$ of the disentangled spectra are too low to allow any useful metallic
line to be seen. Since the temperature ratio is well constrained by the
photometric data, the temperature of the primary was determined by fixing this
ratio at its photometric value. There is no nebular emission in the Balmer
lines.

The mass-$\log g$ diagram suggests that the ratio of radii has been slightly
underestimated, since the secondary has a too large surface gravity compared to
the primary. Still, the positions of both components are compatible, within the
errors, with an isochrone at $7-8$ Myr. The HR diagram shows a good match
between the positions of both components and the evolutionary tracks of single
9.0 and 7.7 $\mathcal{M}_{\odot}$ stars. However, both components are slightly
overluminous relative to their respective tracks.

The apsidal motion amounts to $\dot{\omega} \simeq 50$ $\degr$ yr$^{-1}$,
as evidenced by the $e\cos\omega$ values obtained with the EBOP code for four
successive subsets of the whole time serie in the $I$ band. That value is
confirmed by the WD analysis. The run of $e\cos\omega$ versus time is shown in
Fig. \ref{5_266_ecosw}, together with the best fit curve provided by the WD
code. The reality of a very fast apsidal motion is beyond any doubt. It is
further discussed in Section 5.
\begin{figure}[htb]
\begin{center}
\includegraphics[width=9cm]{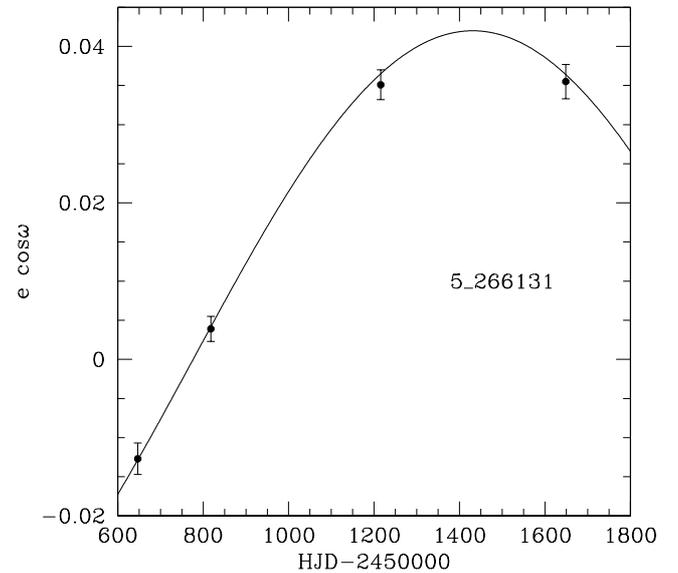}
\caption{Same as Fig.~\ref{4_160_ecosw}, but for the system 5 266131. The
solid line is the best fit provided by the WD code.}
\label{5_266_ecosw}
\end{center}
\end{figure}

\subsection{5 266513}
This a close detached system with similar components, according to its medium
quality light curves with minima of similar depth. The finding charts shows a
slightly fainter star at $1.2-1.3\arcsec$ to the W of the target.

There are 10 out-of-eclipse spectra. The RV curves are well constrained by the
observations, though the noise is rather large because this system is the
faintest in our sample.

A spectroscopic $B$ luminosity ratio of 0.73 was found, but with a large scatter
of $0.09$. The potential of the primary was fixed so that the WD analysis
preserves a ratio close to that value.
The disentangled spectra have a low $S/N$ and no metallic lines are exploitable.
Moreover, the Balmer lines are polluted by nebular emission. The photometric
temperature ratio is rather well defined, thanks to the large depth of the
minima, and was fixed for the determination of the temperature of the primary,
as usual for most detached systems.

The mass-$\log g$ diagram shows the two components fairly well aligned along the
$\sim 15$ Myr isochrone. On the HR diagram, the primary lies almost exactly on
its evolutionary track, while the secondary is slightly overluminous, though its
position is quite compatible with the evolutionary tracks within the errors. The
luminosity ratio (hence the ratio of radii) seems to have been slightly
overestimated.

\subsection{5 277080}
This system is a typical semi-detached binary with high quality light curves of
the EB type. A small depression before the primary eclipse signals the possible
presence of a mass-transfer stream. The effect is small enough, however, that
we did not deem it worth the effort to model it through a cool spot on the
primary.

This system has peculiar colour indices (see Table \ref{basicparam} and
Fig. \ref{colour_colour.2}). Indeed, an examination of the OGLE finding chart
revealed that this binary is strongly blended with a brighter star lying $1\farcs 2$
to the SW of the target.

There are 11 out-of-eclipse spectra. Because of an orbital period close to 2
days, there are no observations before the quadratures. In spite of that, the RV
curves are rather well constrained. This is one of the few systems showing no
nebular emission in the Balmer lines.

A spectroscopic $B$ luminosity ratio of 0.7 was found. This is higher than the
photometric ratio (0.57), as seems to be often the case of semi-detached
systems. The disentangled spectra are fairly good. A number of
metallic lines are visible in the spectrum of the primary: \ion{O}{ii} 4185,
\ion{C}{ii} 4267, \ion{O}{ii} 4276-7, \ion{Mg}{ii} 4481 and \ion{Si}{iii} 4553.
The \ion{Mg}{ii} 4481 line is the only metallic line detectable in the spectrum
of the secondary. The temperature of the primary was determined after fixing the
temperature and luminosity ratios to their photometric values.

The positions of the stars in the mass-$\log g$ and HR diagrams are typical of
Algol-type system with a secondary more evolved than the primary. The primary
component, however, is much more luminous than expected from the evolutionary
track of a single star of the same mass, and we cannot guarantee that its
effective temperature has not been overestimated. The peculiar HR diagram might
be the result of the blend mentioned above with a nearby bright star.
   
\subsection{5 283079}
This is a typical detached system with twin components and a circular orbit.
The target appears just isolated on the finding chart, with a companion of
similar brightness about $2\farcs 5$ away from it to the SE.

There are 10 out-of-eclipse spectra. The RV curves are well defined and
constrained by the observations. There is probably a faint nebular emission in
both H$\gamma$ and H$\delta$ lines.

With a spectroscopic $B$ luminosity ratio close to 1.0, a mass ratio of 1.003
and a temperature ratio of 0.997, this is the system with the most similar
components in our sample. Thus it is probably a real pair of ``twins", i.e.
it has a mass ratio larger than $0.95$. This is why the choice of the primary is
undecided in this system: which component is the primary was chosen on the basis
of an early iteration, and it is only in the last iteration that we obtained
$q=1.003 > 1$. The potential of the primary was fixed in the WD analysis, in
order to keep the $B$ luminosity ratio close to 0.99. The disentangled spectra
are noisy and the \ion{C}{ii} 4267 and  \ion{Mg}{ii} 4481 are barely visible.
The temperature of the primary was determined by fixing the temperature ratio to
0.997 and the luminosity ratio to 0.99.

On the mass-$\log g$ diagram, both components lie on the 3 Myr isochrone, so
this system is very young. On the HR diagram, they are slightly overluminous
with respect to the evolutionary tracks, but still within the error bars. On the
other hand, the components lie exactly on the metal-poor tracks ($Z=0.001$).

\section{Discussion}
\label{label_discussion}

\subsection{Common systems with HHH03/05}

\begin{table*}
\caption{ Comparison of our solutions (NGBR) for the five binaries in common
with HHH03/05.}            
\label{compHHH}
\tiny      
\centering          
\begin{tabular}{l l l l l l l l l l l l l}  
\hline\hline
Object & Ref. & Model & $\mathcal{M}_\mathrm{P}$  & $\mathcal{M}_\mathrm{S}$  & $R_\mathrm{P}$  & $R_\mathrm{S}$  &
 $T_\mathrm{eff}^\mathrm{P}$  &  $T_\mathrm{eff}^\mathrm{S}$ & $\log g_\mathrm{P}$  & $\log g_\mathrm{S}$ & $i$ \\
  & & & ($\mathcal{M}_{\odot}$) & ($\mathcal{M}_{\odot}$) & (R$_{\odot}$) & (R$_{\odot}$) &  (K) & (K) & (dex cgs) & (dex cgs) & ($\degr$) \\
\hline
4 110409 & NGBR  & sd & $13.93\pm0.59$ & $7.30\pm0.36 $ & $4.64\pm0.14$ & $7.81\pm0.18$ & $32370\pm816$ & $21170\pm583$ & $4.25\pm0.03$ & $3.52\pm0.03$ & $77.3\pm0.1$ \\
         & HHH05 & sd & $13.7\pm0.8  $ & $8.9\pm1.1   $ & $4.3\pm0.2  $ & $8.4\pm0.3  $ & 25500 (fixed) & $15390\pm160$ & $4.30\pm0.05$ & $3.54\pm0.06$ & $76.8\pm0.3$ \\
4 163552 & NGBR  & d  & $ 9.56\pm0.61$ & $9.10\pm0.57 $ & $5.06\pm0.16$ & $4.54\pm0.15$ & $24400\pm990$ & $24330\pm993$ & $4.01\pm0.04$ & $4.08\pm0.04$ & $80.9\pm0.2$ \\  
         & HHH05 & d  & $13.3\pm1.0  $ & $12.4\pm1.1  $ & $5.3\pm0.1  $ & $5.2\pm0.2  $ & 25500 (fixed) & $25390\pm190$ & $4.11\pm0.04$ & $4.10\pm0.05$ & $78.3\pm0.1$ \\
5 026631 & NGBR  & sd & $11.40\pm0.34$ & $7.86\pm0.25 $ & $5.19\pm0.08$ & $4.93\pm0.08$ & $28670\pm394$ & $21050\pm356$ & $4.07\pm0.02$ & $3.95\pm0.02$ & $61.4\pm0.1$ \\  
         & HHH05 & sd & $11.5\pm0.6  $ & $11.3\pm0.8  $ & $5.1\pm0.1  $ & $5.7\pm0.1  $ & 25500 (fixed) & $17130\pm300$ & $4.08\pm0.03$ & $3.99\pm0.04$ & $61.5\pm0.1$ \\
5 038089 & NGBR  & d  & $13.01\pm0.20$ & $11.70\pm0.18$ & $5.98\pm0.08$ & $4.94\pm0.07$ & $30660\pm130$ & $31020\pm301$ & $4.00\pm0.01$ & $4.12\pm0.01$ & $77.1\pm0.2$ \\  
         & HHH03 & d  & $17.1\pm1.5  $ & $19.1\pm1.6  $ & $6.1\pm0.2  $ & $6.1\pm0.3  $ & 30100 (fixed) & 29180 (fixed) & $4.10\pm0.05$ & $4.15\pm0.05$ & $76.9\pm0.3$ \\
5 277080 & NGBR  & sd & $ 9.68\pm0.55$ & $4.88\pm0.29 $ & $4.72\pm0.09$ & $5.13\pm0.09$ & $29780\pm480$ & $20410\pm372$ & $4.08\pm0.03$ & $3.71\pm0.03$ & $76.6\pm0.1$ \\  
         & HHH05 & sd & $17.4\pm0.9  $ & $11.3\pm1.0  $ & $5.0\pm0.1  $ & $6.8\pm0.2  $ & 25500 (fixed) & $15890\pm150$ & $4.27\pm0.03$ & $3.82\pm0.05$ & $76.8\pm0.3$ \\
\hline                  
\end{tabular}
\end{table*}

\begin{table*}
\caption{ Comparison of our results (NGBR) with the light-curves solutions of
Wyithe \& Wilson (\cite{WW01, WW02}) (WW01, WW02) and Graczyk (2003) (dG03).
The following parameters are listed: the model of the system, the mass ratio, the
relative radii, the inclination, the
eccentricity, the $I$ luminosity ratio, the eclipse parameter (see Eq. 2 in WW01)
and the third light (normalized to total flux at phase 0.25). 
}            
\label{compWW}
\tiny      
\centering          
\begin{tabular}{c c c c c c c c c c c}  
\hline\hline
Object & Ref. & Model & $q$ & $R_\mathrm{P}/a$ & $R_\mathrm{S}/a$  & 
$i (\degr)$ & $e$ & $(L_\mathrm{S}/L_\mathrm{P})_{I}$ & $F_{e}$ & $l_{3,I}$\\  
\hline
 
4 117831 & NGBR & d  & $0.981\pm0.013$      & $0.261\pm0.008$ & $0.287\pm0.008$ & $78.2\pm0.4$ & $0	       $ & $1.175$ & $0.599$ & $0$\\  
	 & WW02 & sd & $0.157\pm0.054$      & $0.357\pm0.032$ & $0.209\pm0.021$ & $77.9\pm2.1$ & $0.049\pm0.019$ & $0.372$ & $0.853$ & $0$\\
4 163552 & NGBR & d  & $0.952\pm0.015$      & $0.339\pm0.011$ & $0.305\pm0.010$ & $80.9\pm0.2$ & $0	       $ & $0.798$ & $0.797$ & $0.119\pm0.006$ \\ 
	 & dG03 & d  & $0.98\pm0.09  $      & $0.338\pm0.006$ & $0.326\pm0.008$ & $85.4\pm0.9$ & $0	       $ & $0.951$ & $0.895$ & $0.263\pm0.007$\\ 
	 & WW02 & sd & $0.207\pm0.010$      & $0.383\pm0.004$ & $0.234\pm0.003$ & $82.8\pm0.7$ & $0.004\pm0.002$ & $0.397$ & $1.051$ & $0$\\  
4 175333 & NGBR & d  & $0.781\pm0.014$      & $0.237\pm0.008$ & $0.203\pm0.007$ & $80.4\pm0.5$ & $0.022\pm0.007$ & $0.582$ & $0.675$ & $0$\\ 
	 & WW01 & d  & 1.0		    & $0.253\pm0.009$ & $0.143\pm0.007$ & $85.0\pm1.5$ & $0.012\pm0.003$ & $	 $ & $1.080$ & $0$\\
5 016658 & NGBR & d  & $0.889\pm0.011$      & $0.320\pm0.011$ & $0.255\pm0.010$ & $80.3\pm0.2$ & $0	       $ & $0.609$ & $0.798$ & $0$\\ 
	 & WW01 & d  & 1.0		    & $0.345\pm0.007$ & $0.200\pm0.004$ & $85.7\pm1.3$ & $0.002\pm0.002$ & $	 $ & $1.175$ & $0$\\
5 038089 & NGBR & d  & $0.899\pm0.011$      & $0.273\pm0.004$ & $0.226\pm0.003$ & $77.1\pm0.2$ & $0	       $ & $0.693$ & $0.609$ & $0$\\  
	 & dG03 & d  & $1.12 \ ^\mathrm{a}$ & $0.246\pm0.009$ & $0.255\pm0.007$ & $76.8\pm1.9$ & $0	       $ & $1.053$ & $0.554$ & $0$\\
5 196565 & NGBR & d  & $0.803\pm0.007$      & $0.200\pm0.005$ & $0.150\pm0.005$ & $83.5\pm0.2$ & $0.138\pm0.006$ & $0.43 $ & $0.789$ & $0$\\
	 & WW01 & d  & 1.0		    & $0.204\pm0.006$ & $0.116\pm0.003$ & $88.7\pm2.5$ & $0.085\pm0.002$ & $	 $ & $1.282$ & $0$\\
5 266015 & NGBR & sd & $0.441\pm0.005$      & $0.332\pm0.005$ & $0.310\pm0.005$ & $78.5\pm0.1$ & $0	       $ & $0.494$ & $0.714$ & $0$\\ 
	 & WW02 & sd & $1.014\pm0.051$      & $0.183\pm0.003$ & $0.357\pm0.004$ & $80.8\pm0.4$ & $0.000\pm0.002$ & $2.564$ & $1.039$ & $0$\\
5 266131 & NGBR & d  & $0.864\pm0.013$      & $0.281\pm0.010$ & $0.242\pm0.009$ & $83.5\pm0.2$ & $0.042\pm0.001$ & $0.647$ & $0.847$ & $0$\\  
	 & WW01 & d  & 1.0		    & $0.332\pm0.008$ & $0.216\pm0.009$ & $85.1\pm1.5$ & $0.003\pm0.003$ & $	 $ & $1.071$ & $0$\\
5 283079 & NGBR & d  & $1.003\pm0.013$      & $0.243\pm0.007$ & $0.242\pm0.007$ & $87.7\pm0.3$ & $0	       $ & $0.987$ & $0.918$ & $0$\\
         & WW01 & d  & 1.0		    & $0.272\pm0.006$ & $0.233\pm0.007$ & $88.5\pm1.8$ & $0.001\pm0.001$ & $	 $ & $1.028$ & $0$\\

\hline		   
\end{tabular}
\begin{list}{}{}
\item[$^{\mathrm{a}}$] From HHH03.
\end{list}
\end{table*}

Five systems are common to the HHH03/05's sample and ours: two detached systems
(4 163552 and 5 038089) and three semi-detached ones (4 110409, 5 026631 and
5 277080). Comparison of a number of parameters is shown in Table \ref{compHHH}.
The good news is that we agree on the model, i.e. the configuration of the
system in relation with the Roche lobes. The bad news is that, in several cases,
we do not agree on most parameters, except the inclination, a purely photometric
parameter. There is nothing surprising about that, as we have the same light
curves.

The first system, 4 110409, shows a fair agreement for the masses (excellent for
the primary, within $1.6\,\sigma$ for the secondary), the radii (within
$2\,\sigma$)
and the surface gravities ($1\,\sigma$ or better). By contrast, the effective
temperatures differ by as much as 7000 (primary) and 6000~K (secondary), those
given by HHH05 being clearly underestimated.

The second system, 5 163552, shows a good agreement of the temperatures, the
estimates by HHH05 being only $\sim 1000$~K higher than ours. On the other hand,
the masses given by HHH05 are $3\,\sigma$ larger than ours, and their radii are
larger too. Part of the problem might arise from HHH05 neglecting the third light.
In addition, HHH05 may have been more sensitive to crowding effects on their RV
curves, since the diameter of the 2dF fibers they used is $2\arcsec$, while the diameter
of the FLAMES/GIRAFFE fibers is only $1\farcs 2$. But nebular emission may be the main
reason for the discrepancy: HHH05 overestimate the amplitude of both RV curves
(relative to us), and indeed nebular emission can mimick a wider separation of
Balmer lines near quadratures.

For the semi-detached system 5 26631, there is a perfect agreement about the mass
of the primary, while the mass of the secondary given by HHH05 is $4\,\sigma$
larger than ours. Their solution appears less realistic than ours, because they
obtain a mass ratio close to unity, which would be exceptional for a semi-detached
system. Similarly, we agree on the radius of the primary, while HHH05 overestimate
the radius of the secondary. HHH05 estimate temperatures lower by $\sim 3000$~K
(primary) and by $\sim 4000$~K (secondary) than us. The amplitude of their RV
curve is overestimated for the primary, but slightly lower than ours for the 
secondary. Again, nebular emission might be responsible for the discrepancy.

The fourth system, 5 38089, shows a very large discrepancy in the mass estimates:
for the secondary, it amounts to 63\% if we take our value as the reference.
Both masses are overestimated by HHH03, while the radius of the primary is in
perfect agreement, and that of the secondary is significantly overestimated.
On the other hand, the temperatures are in nearly perfect agreement (within 2\%)
for the primary, and slightly disagree (by 1840~K) for the secondary. It appears
that HHH03 have strongly overestimated the amplitude of the RV curve of the
primary, while that of the secondary is much closer to ours. Here, one cannot
invoke any systematic effect linked with nebular emission, since the latter cannot
be seen on our spectra. On the other hand, this system has peculiar colour
indices (Fig. \ref{colour_colour.2}) and its position is unsatisfactory on
our own HR diagram. Crowding problems causing what could be called
``spectroscopic third light'' are probably the main source of problems in both
studies, and may have been enhanced in that of HHH03 by the wider 2dF fiber.

The last system, 5 277080, holds the record of mass discrepancy: relative to our
values, HHH05 overestimate the mass by factors as large as 1.8 and 2.3 for the
primary and the secondary respectively. The radii are overestimated as well,
but by a much smaller amount. The temperatures differ by more than $4000$~K.
Nebular emission cannot be the culprit, since it remains undetectable in our
spectra.

The problem lies in the RV curves and the disentangled spectra. There are three
basic spectroscopic parameters on which all the subsequent analysis is built:
the two velocity semi-amplitudes and one of the temperatures (the other one
being constrained by the photometric mean surface brightness ratio). If the velocity
semi-amplitudes are wrong, the masses, radii, $q-$ratio, $\log g-$values and
luminosities are wrong. The luminosities are obviously highly dependent on the
temperatures, despite a partial cancelling of a possible error by the bolometric
correction (see Section \ref{uncertainties}). We are confident on our values for
the velocity semi-amplitudes and the primary temperature. The RV curves of the
five systems shared with HHH03/05 are shown in Figs \ref{allRV1},
\ref{allRV2} and \ref{allRV3}. They are well sampled, there are good constraints
at the quadratures and the $O-C$ curves are flat (except for 5 277080 where some
systematics appears). Consequently, there is no
reason to doubt about the quality of our velocity semi-amplitudes. The RV curves
are not shown in HHH03/05, nevertheless there are certainly two reasons to
suspect that they have a lower quality than ours. The first one is
instrument-related. As already mentioned, our instrumental setup has better
resolution ($0.67$ \AA\ instead of 2 \AA). Moreover, for the systems in common,
our data have a higher $S/N$ (typically a factor of about two, from the
$S/N-$calculators of the respective instruments), in spite of a lower
$S/N$ on average for the whole sample, as we observed mostly fainter systems.
Thus our raw data are intrinsically better. The second reason is related to the
measurement of the radial velocities. Our radial velocities were determined
exclusively from the \ion{He}{i} lines mentioned in Section \ref{RV}. From the
procedure described in HHH03/05, it seems clear that in most cases (i.e. with
no conspicuous emission lines) both Balmer and \ion{He}{i} and/or the whole
spectrum were used to determine the radial velocities. One can suspect that
unrecognized nebular emission in the \ion{H}{i} lines is partly
responsible for bad RV measurements. Actually, in a number of cases, nebular
emission in the Balmer lines can easily be confused with the typical ``SB2
cleavage" of blended lines. The disentangling of a composite spectrum being
tightly linked to the radial velocities, the quality of the component spectra
and their subsequent use for fixing the temperature of one component will be
badly influenced by poorly determined radial velocities.                

\begin{figure}[!ht]
\centering
\includegraphics[width=9.0cm]{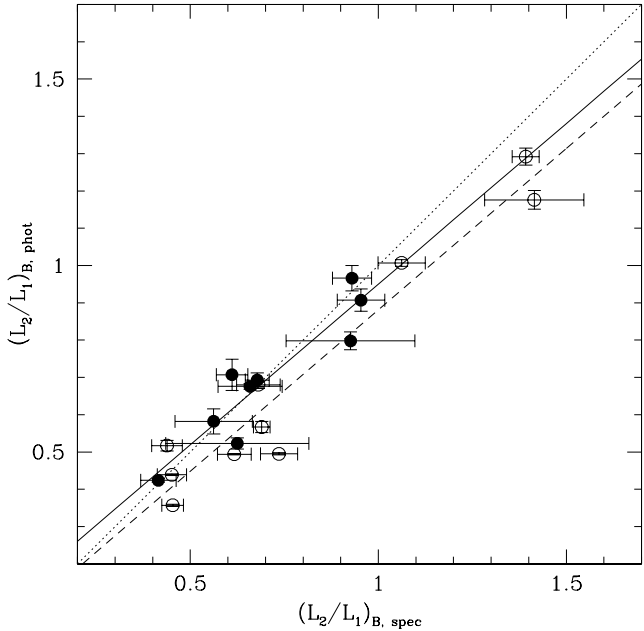}
\caption{Photometric vs. Spectroscopic values of the $B$ luminosity ratio
(full symbol: detached; open symbol: semi-detached/contact). The dotted line
corresponds to $\mathcal{L}_{B}^{\mathrm{spec}} = \mathcal{L}_{B}^{\mathrm{phot}}$.
The best linear fits of the two samples are shown (solid: detached;
dashed: semi-detached/contact ). Note that the detached systems shown are
only those for which the photometric ratio agrees well with the spectroscopic
one.}
\label{lumratio}%
\end{figure}  

An important point of divergence in the methodology used to ``solve" a binary is
the use (or non-use) of the spectral $B$ luminosity ratio. Comparing the
spectral and photometric values of the $B$ luminosity ratio actually reveals
that they often disagree for detached systems. The reason is the near-degeneracy
of the ratio of radii with the inclination in the case of partial eclipses
(Wyithe \& Wilson \cite{WW01}). In principle, the photometric luminosity ratio
of semi-detached systems should be better constrained, since they have less
free photometric parameters than the detached ones (see Fig. 3 in Wyithe
\& Wilson \cite{WW01} and \cite{WW02}). The regression shown in Fig.
\ref{lumratio} clearly shows that the spectroscopic and photometric ratios of
the population of semi-detached systems agree fairly well, except that the slope
is slightly lower than one. The latter fact may be caused by the secondary tidal
distorsion being larger than that of the primary: the total luminosity, which is
the parameter provided by the WD code, may be smaller than suggested by the
apparent brightness at the quadrature. Among the detached systems, Fig.
\ref{lumratio} shows only those for which the spectroscopic and photometric
luminosity ratios were found to agree, so the good match is artificial. 

Thus, the spectroscopic luminosity ratios were adopted systematically for the
detached systems, except in the cases where they agree with the photometric one
within the uncertainties (full dots in Fig. \ref{lumratio}). For the
semi-detached and contact ones, the photometric ratio was preferred.
In HHH05, the spectroscopic ratio was only used to settle
the case between two conflicting photometric solutions. 
  
Finally, it must be emphasized that the quality of our spectral analysis owes
greatly to the excellent library of synthetic NLTE O- and B-spectra.

\subsection{Nebular emission and kinematics}
\label{nebularemission}
 H$\gamma$ and H$\delta$ emission lines are visible in most binary spectra of
 our sample. FWHM values and heliocentric radial velocities are presented
 in Table \ref{binaemission}. These are mean values inferred from spectra close
 to conjunction in order to separate the real emission from the ``SB2 cleavage". 
 The strengths are not indicated, as in normalized spectra the nebular lines are
diluted by the stellar continua. Therefore, it would not be possible to compare
line strengths measured on two spectra of two different systems. Even in the
case of two spectra of the same binary at two different epochs, the strength
measurement is not reliable because of the dependency of the
 diluting stellar continuum on the seeing\footnote{Note that the sky
 subtracted from the stellar spectra was limited to the continuum component.
 Therefore, sky subtraction did not alter the nebular emission lines.}.
 
\begin{table*}
\caption{ Nebular emission: mean radial velocities and mean FWHM of the
H$\gamma$ and H$\delta$ lines. The quoted uncertainties are standard deviations. }  
\label{binaemission}
\centering          
\begin{tabular}{c c c c c | c c c c c}  
\hline\hline
Object & \multicolumn{2}{c}{$\mathrm{H}\gamma$} & \multicolumn{2}{c}{$\mathrm{H}\delta$} \vline &
Object & \multicolumn{2}{c}{$\mathrm{H}\gamma$} & \multicolumn{2}{c}{$\mathrm{H}\delta$}\\
\cline{2-3}
\cline{4-5}
\cline{7-8}
\cline{9-10}
 & $V_{\mathrm{rad}}$ & $FWHM$ & $V_{\mathrm{rad}}$ & $FWHM$ & 
 & $V_{\mathrm{rad}}$ & $FWHM$ & $V_{\mathrm{rad}}$ & $FWHM$ \\
 & (km s$^{-1}$) & (\AA) & (km s$^{-1}$) & (\AA) & 
 & (km s$^{-1}$) & (\AA) & (km s$^{-1}$) & (\AA) \\
\hline 
4 110409 & $-$	          & $-$	           & $-$ 	              &  $-$	        & 5 100485 & $159.1\pm   4.7$  & $1.01\pm 0.10$ & $166.9\pm 17.1$ & $0.83\pm 0.19$  \\
4 113853 & $142.6\pm1.3$  & $0.93\pm 0.05$ & $147.4\pm   6.7$ &  $0.77\pm 0.08$ & 5 100731 & $163.7\pm   3.6$  & $1.02\pm 0.12$ & $161.6\pm  7.3$ & $0.95\pm 0.11$  \\
4 117831 & $145.8\pm1.8$  & $0.99\pm 0.08$ & $152.2\pm  14.2$ &  $0.92\pm 0.13$ & 5 106039 & $159.1\pm   2.5$  & $0.91\pm 0.03$ & $159.3\pm  6.3$ & $0.83\pm 0.09$  \\
4 121084 & $147.9\pm2.0$  & $1.22\pm 0.04$ & $152.9\pm   2.7$ &  $1.05\pm 0.02$ & 5 111649 & $149.4\pm   3.0$  & $0.77\pm 0.09$ & $152.0\pm 12.9$ & $0.64\pm 0.08$  \\
4 121110 & $136.8\pm1.4$  & $1.44\pm 0.07$ & $141.4\pm   6.0$ &  $1.20\pm 0.07$ & 5 123390 & $-$	       &  $-$           &  $-$	          & $-$	    \\
4 121461 & $147.6\pm2.0$  & $1.11\pm 0.07$ & $150.5\pm   0.6$ &  $1.00\pm 0.06$ & 5 180185 & $164.5\pm   5.7$  & $0.75\pm 0.08$ &  $-$            & $-$	    \\
4 159928 & $146.7\pm1.1$  & $0.99\pm 0.03$ & $148.9\pm   2.5$ &  $0.94\pm 0.06$ & 5 180576 & $157.4\pm   2.9$  & $0.99\pm 0.15$ & $163.8\pm 3.7$  & $0.96\pm 0.13$  \\
4 160094 & $143.6\pm1.2$  & $1.03\pm 0.03$ & $144.2\pm   3.7$ &  $0.95\pm 0.03$ & 5 185408 & $-$               &  $-$           &  $-$	          & $-$	    \\
4 163552 & $143.6\pm0.3$  & $0.96\pm 0.02$ & $147.1\pm   2.4$ &  $0.95\pm 0.04$ & 5 196565 & $-$	       &  $-$           &  $-$	          & $-$	    \\
4 175149 & $-$	          &  $-$       	   &$-$	      	      & $-$ 	      & 5 261267   & $160.7\pm   3.9$  & $0.88\pm 0.14$ & $170.3\pm7.6$   & $0.78\pm 0.22$  \\
4 175333 & $146.9\pm3.3$  & $1.11\pm 0.07$ & $151.1\pm   1.5$ &  $0.99\pm 0.02$ & 5 265970 & $-$               &  $-$           &  $-$            & $-$ \\
5 016658 & $168.3\pm9.4$  & $1.11\pm 0.06$ & $    	    $ &  $-$	        & 5 266015 & $-$               &  $-$           &  $-$            & $-$ \\
5 026631 & $148.8\pm5.7$  & $0.93\pm 0.21$ & $151.8\pm   4.0$ &  $0.66\pm 0.04$ & 5 266131 & $-$               &  $-$           &  $-$            & $-$ \\
5 032412 & $-$	          & $-$	    	   &$-$	      	      &  $-$	        & 5 266513 & $-$               &  $-$           &  $-$            & $-$ \\
5 038089 & $-$	          & $-$	    	   &$-$	      	      &  $-$	        & 5 277080 & $-$               &  $-$           &  $-$            & $-$  \\
5 095337 & $171.6\pm3.1$  & $1.07\pm 0.11$ & $172.7\pm   6.5$ &  $0.87\pm 0.12$ & 5 283079 & $-$               &  $-$           &  $-$            & $-$  \\
5 095557 & $164.2\pm2.5$  & $1.01\pm 0.07$ &  $-$                &  $-$ \\
\hline                  
\end{tabular}
\end{table*}

\begin{figure}[!ht]
\includegraphics[trim = 5mm 180mm 80mm 10mm, clip, width=8.5cm]{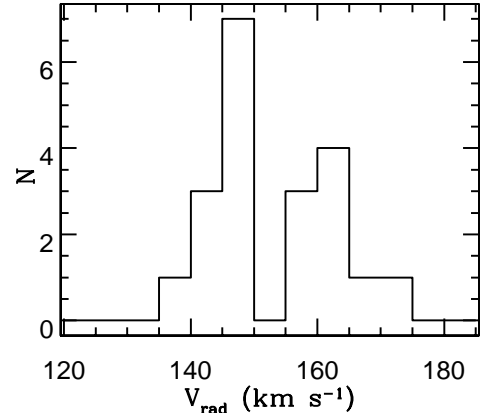}
\caption{Histogram of the mean heliocentric radial velocities of the nebular H$\gamma$
lines observed in the spectra of 20 binaries. The sample is arranged in 5 km s$^{-1}$ bins.}
\label{histo_velocityfield_bina}%
\end{figure}

 It is beyond the scope of this paper to make a detailed kinematic study of the
SMC. Nevertheless, it is interesting to examine the distribution of the 20 RVs
derived from the nebular H$\gamma$ lines listed in Table \ref{binaemission}. The
distribution is illustrated in Fig. \ref{histo_velocityfield_bina}.
The 5 km s$^{-1}$-binned histogram shows a bimodal distribution with two modes at
145-150 and 160-165 km s$^{-1}$. Since the size of the sample is modest, one can
wonder about the significance of this observation. Therefore, the RVs of the
nebular lines in the sky spectra were also investigated (21 spectra per epoch).
The H$\gamma$ data are presented in Table \ref{skyemission}. It is readily
apparent that the RV-distribution shows two peaks as well, at 140-145 and
160-165 km s$^{-1}$ (Fig.\ref{histo_velocityfield}). Therefore, these
observations provide compelling evidence that the nebular \ion{H}{ii} of the SMC
displays a bimodal velocity distribution (at least in our small
25$\arcmin$-diameter field).
  
\begin{figure}[!ht]
\includegraphics[trim = 5mm 180mm 80mm 10mm, clip, width=8.5cm]{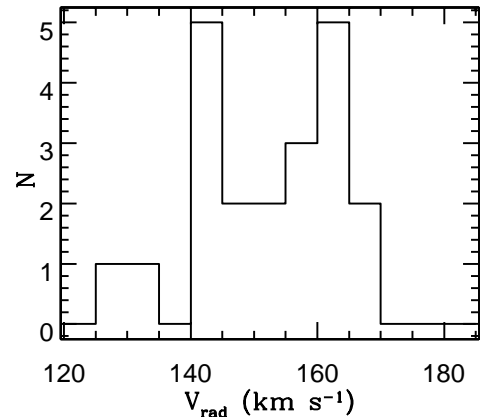}
\caption{ Histogram of the mean heliocentric radial velocities of the nebular H$\gamma$
lines observed in the 21 sky spectra. The sample is arranged in 5 km s$^{-1}$ bins.}
\label{histo_velocityfield}%
\end{figure}
  
Comparing the systemic velocities of our 33 systems with the combined nebular \ion{H}{ii}
velocities (41 values) shows that the two samples have very different
distributions. The nebular velocities were fitted by two Gaussian functions of equal
amplitudes with modes at 145.2 and 161.4 km s$^{-1}$ and standard deviations of
3.44 and 4.14 km s$^{-1}$, respectively. The corresponding FWHMs are about one
fifth of a resolution element ($\sim$47 km s$^{-1}$). The systemic velocities
are best fitted with a Gaussian function centred on 162 km s$^{-1}$ with a
standard deviation of 12.9 km s$^{-1}$. That last fit is less convincing than
for the \ion{H}{i}, as there is less data and the velocity dispersion
is 3-4 $\times$ higher. Further nebular and systemic velocity data from the
analysis of the remaining SB1 systems contained in the GIRAFFE field is
expected to substantially improve the statistics.
  
\begin{figure}[!ht]
\includegraphics[trim = 1mm 180mm 80mm 10mm, clip, width=9.5cm]{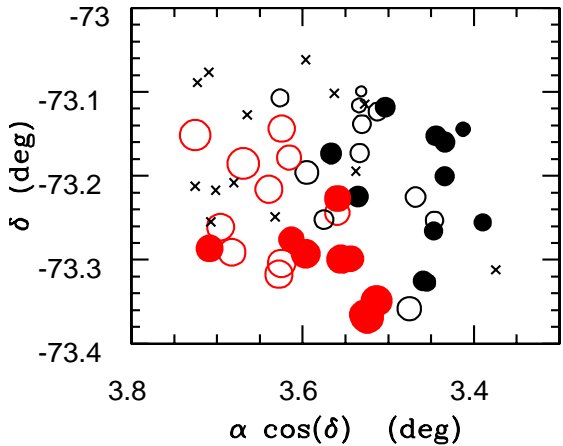}
\caption{Velocity field of the nebular \ion{H}{ii}. The size of symbol is proportional to
the value of the velocity. Values greater than 155 km s$^{-1}$ are in red (electronic
version only). Filled symbols are for binary systems while open symbols are for skies.
Crosses are for binary systems showing no emission.}
\label{sky}%
\end{figure} 
  
It is worth mentioning that the observed bimodal distribution of \ion{H}{ii}
is compatible with the kinematic study of \ion{H}{i} in the SMC by
Stanimirovi\'{c} et al. (\cite{SSJ04}), even though the gas we see in our work
is certainly hotter than the gas observed at 21 cm wavelength. Indeed,
the authors mention the existence of a bimodal velocity field with central
velocities of 137 and 174 km s$^{-1}$. Interestingly, Fitzpatrick (\cite{elF85})
already mentioned two \ion{H}{i} complexes with velocities 134 and 167
km s$^{-1}$, which our emitting gas is probably related with. 
Figure \ref{sky} shows the mean (heliocentric) velocity field obtained from
nebular \ion{H}{ii} data. Filled and open symbols correspond to binary and sky
(nebular) data, respectively. The size of the symbol is drawn according to the
value of the radial velocity. A cross indicates the position of a binary without
significant emission. The map has the same orientation as in Fig.\ref{photodss}.
We can tentatively draw an oblique line from the NE to the SW corner demarcating
high-velocity points from low-velocity points. This line has a $\sim 60\degr$
slope in the EN (clockwise) direction. On the east (i.e. left) side of this
line, nebular RVs are greater than $\sim 155~\mathrm{kms^{-1}}$. This velocity
gradient is consistent with Fig. 3 in Stanimirovi\'{c} et al., i.e. the velocity
of \ion{H}{i} increases from west to east in this part of the SMC. Therefore,
the hot and cool phases of the gas seem to share very similar spatial and
kinematic distributions.
 
\begin{figure}[!ht]
\centering
\includegraphics[width=8.0cm]{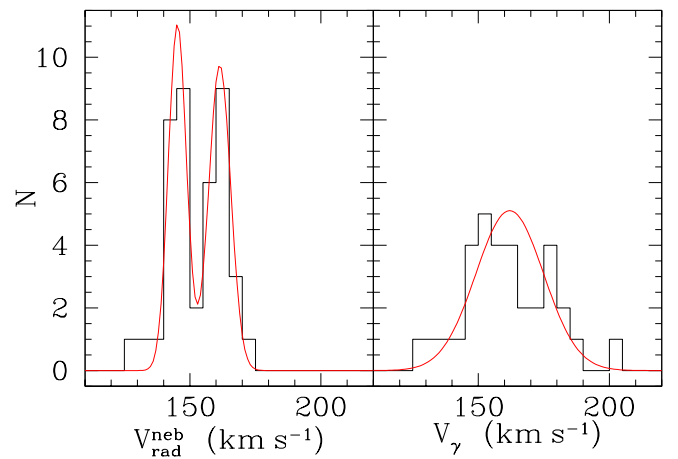}
\caption{Comparison between two kinematic tracers: radial velocities of nebular
\ion{H}{ii} (left) and systemic velocities of binary systems (right). The data
are arranged in 5 km s$^{-1}$ bins. The best-fitting double-Gaussian and
Gaussian models are over-plotted.}
\label{histo_systemicvelocities}
\end{figure} 

\begin{table}[!ht]
\caption{
Sky positions: astrometry, sky background excesses (see Section
\ref{nebularemission}) and mean radial velocities derived from the H$\gamma$
emission line. 
}              
\label{skyemission}      
\centering          
\begin{tabular}{c c c r c} 
\hline\hline        
id & $\alpha$ (J2000) & $\delta$ (J2000) & $\Delta_\mathrm{sky}$  &
$V_\mathrm{rad}$ \\ 
 & ($\degr$) & ($\degr$) & (\%) & (km s$^{-1}$)\\  
\hline                    
 S1  & 12.20608 & $-73.17275$ &  $ +3.8\pm 5.4 $  & $143.9\pm4.2$ \\
 S2  & 12.68487 & $-73.18533$ &  $ +13.4\pm 4.1 $ & $169.4\pm5.8$ \\ 
 S3  & 12.43554 & $-73.19611$ &  $ -5.8\pm 4.7 $  & $152.8\pm3.7$ \\ 
 S4  & 12.49946 & $-73.14372$ &  $ -4.8\pm 3.8 $  & $160.0\pm4.4$ \\ 
 S5  & 12.49446 & $-73.17817$ &  $  +0.5\pm 7.0 $ & $155.8\pm6.2$ \\ 
 S6  & 12.34712 & $-73.24400$ &  $ -11.4\pm 2.0$  & $156.0\pm3.7$ \\ 
 S7  & 12.40693 & $-73.25211$ &  $ -3.5\pm 1.5 $  & $145.2\pm5.5$ \\ 
 S8  & 12.63725 & $-73.31728$ &  $ +19.8\pm 5.4 $ & $161.9\pm2.4$ \\ 
 S9  & 12.61700 & $-73.30436$ &  $  +7.1\pm 5.5 $ & $162.2\pm6.5$ \\
 S10 & 12.60371 & $-73.21592$ &  $ -6.8\pm 3.2 $  & $161.4\pm7.3$ \\
 S11 & 12.83250 & $-73.26103$ &  $ +15.9\pm 3.8 $ & $160.9\pm6.1$ \\
 S12 & 12.48021 & $-73.10711$ &  $  +4.2\pm 3.1 $ & $141.1\pm2.7$ \\
 S13 & 12.85400 & $-73.15156$ &  $ +34.9\pm 7.3 $ & $168.2\pm6.6$ \\ 
 S14 & 12.80883 & $-73.29111$ &  $ -1.5\pm 2.7 $  & $161.0\pm3.9$ \\ 
 S15 & 12.17254 & $-73.13839$ &  $ -6.0\pm 3.4 $  & $142.8\pm2.9$ \\ 
 S16 & 12.09937 & $-73.12347$ &  $ -5.7\pm 3.0 $  & $143.2\pm1.1$ \\ 
 S17 & 12.16858 & $-73.11614$ &  $  +7.3\pm 5.5 $ & $134.9\pm2.2$ \\ 
 S18 & 12.14775 & $-73.09939$ &  $ +10.2\pm 6.1 $ & $127.1\pm1.5$ \\ 
 S19 & 11.95912 & $-73.25294$ &  $ +20.2\pm 5.2 $ & $142.1\pm4.0$ \\
 S20 & 12.01675 & $-73.22519$ &  $ +25.4\pm 11.8$ & $146.3\pm3.2$ \\ 
 S21 & 12.13529 & $-73.35828$ &  $ -2.3\pm 6.8 $  & $154.3\pm3.6$ \\ 
\hline                  
\end{tabular}
\end{table}
  
\subsection{Comparison with evolutionary models and the mass-luminosity relation}
\label{compar_evol}
Comments on the individual systems were given in Section \ref{individual}. Here, the
collective properties of the sample are examined more thoroughly. For each of the 33
systems, Table \ref{compevol} lists the differences $\Delta \mathcal{M}$ between the evolutionary
mass interpolated, in the HR diagram, in the evolutionary tracks from Charbonnel et al.
(\cite{CMMSS93}), and the observed dynamical mass determined from the simultaneous
RV-curves and light-curves analysis. Figure \ref{delta_Mobs_Thispaper} shows these
differences versus the observed masses for the 23 binaries that have a detached
configuration. The agreement between the observed and evolutionary masses is acceptable,
though some systematics does appear. The most discrepant objects are the secondary
components of 5 38089 and 5 123390, and the primary of 5 38089. These systems have a
peculiar position in the HR diagram, perhaps because of third-light contamination.

For comparison, the sample of 18 detached systems of HHH05 is shown 
on Fig. \ref{delta_Mobs_HHH}. Figures \ref{delta_Mobs_Thispaper} and
\ref{delta_Mobs_HHH} are drawn at the same scale. Beside the very large
scatter of HHH05's $\Delta \mathcal{M}$ values (they span a range of almost $10~\mathcal{M}_\odot$),
the most striking feature is the correlation between $\Delta \mathcal{M}$ and
the observed mass for a number of primary and secondary components, especially
between $7$ and $15~\mathcal{M}_\odot$. Explaining this trend is not straightforward, but
it might be related to the temperature determination. Indeed, HHH05 mention in
their Section 3.1.2: ``We found that the O---B1 stars were quite easy to
classify, but that spectral classifications were more difficult for later B
types due to the absence of He\,\textsc{ii} lines and, in these low metallicity
stars, the lack of detectable Si\,\textsc{ii} and other metal lines''. We faced
the same difficulty, but with much less severe consequences. The larger
telescope and better resolution clearly have brought a dramatic improvement in
the quality of the results.

There is a slight trend in our mass residuals: first of all, most
$\Delta \mathcal{M}$ values are positive. Second,
while the mass difference is negligible for low masses
(around $4-5~\mathcal{M}_\odot$), it seems to gradually increase to almost $2~\mathcal{M}_\odot$ at about
$13~\mathcal{M}_\odot$. The single point at $17~\mathcal{M}_\odot$ corresponds to $\Delta \mathcal{M}\sim 1~\mathcal{M}_\odot$
only, but it is not sufficient to define the possible relation beyond $13~\mathcal{M}_\odot$. It
might be part of a parallel sequence defined by 4 other stars. One may speculate that, if
real, the latter sequence might correspond to e.g. systems for which the sky background was
overestimated, which would have caused an underestimate of $T_{\mathrm{eff}}$. The main
trend is surprising, because the more massive systems are also more luminous on average and
so appear brighter since they all lie at about the same distance. Hence the
signal-to-noise ratio of their spectra is larger, and the determination of the
$T_{\mathrm{eff}}$ should be more reliable. On the other hand, $T_{\mathrm{eff}}$ is better
constrained in stars cooler than $\sim 20\,000$~K, because the He\,\textsc{i} lines increase
in strength with increasing temperature, while the H\,\textsc{i} Balmer lines decrease. In
hotter stars, both the H\,\textsc{i} and He\,\textsc{i} lines decrease with increasing
temperature, so that even a slight error on the continuum placement or (equivalently) on the
sky subtraction may cause a significant error on $T_{\mathrm{eff}}$. At the hot end of the
temperature range of our sample (i.e. for $T_{\mathrm{eff}} > 30\,000$~K), the effective
temperature is again better constrained, thanks to the fast raise of He\,\textsc{ii} lines.
Therefore, one may expect that the least reliable temperatures -- hence the least reliable
luminosities -- will occur for stars with intermediate $T_{\mathrm{eff}}$ or masses. This
might explain why the scatter of the $\Delta \mathcal{M}$ values in Fig. \ref{delta_Mobs_Thispaper} is
large between 7 and $14~\mathcal{M}_\odot$.

The fact that $\Delta \mathcal{M} > 0$ in most cases is not easy to explain. Possible explanations can
be grouped in three categories:
\begin{itemize}
\item {\sl Systematic error in the $T_{\mathrm{eff}}$ determination:} As mentioned above,
the temperature determination is very sensitive to continuum placement or sky subtraction,
especially when all lines vary in the same way with $T_{\mathrm{eff}}$. The continuum has
been determined both automatically and manually for each star, and we have verified that the
least-squares fit used to determine the temperature always gives the same result, unless the
continuum is clearly wrong. All fits were done under visual monitoring, so the quality of
the continuum placement could be judged with confidence. In addition, plotting the estimated
temperature against the sum of the squared residuals generally results in a horizontal line,
as shown in Section \ref{individual} (the system 5 180576 is a notable exception). This is
a proof that the $T_{\mathrm{eff}}$ determination does not depend on the SNR of the
spectra or, in other words, that the continuum was defined in the same way, whatever the SNR
of the spectrum. The normalisation, as well as the temperature determination, were done
(in the so-called second step) on the original unsmoothed spectra. In order to further
explore the possible role of the SNR, we smoothed the spectra of the system 4 121110 with a
3-points boxcar (0.6~\AA~ wide) and performed the $T_{\mathrm{eff}}$ determination again.
The result was the same within $\sim 100$~K, which confirms that it is insensitive to the
SNR of the spectra, since the latter, in this test, was improved by a factor of $\sqrt{3}$.
Therefore, it is doubtful that the explanation resides in the way the continuum was defined.

The sky subtraction may be more problematic. Indeed, we have seen that the sky does vary
slightly in the FLAMES field, and all we could do was to subtract an average sky spectrum
from the stellar spectra. Underestimating the sky would have left some continuum flux that
would have slightly diluted the stellar absorption lines, leading to an overestimate of
$T_{\mathrm{eff}}$. The only cure to this problem would have been to use the IFUs instead of
the MEDUSA fibres, in order to be able, at least in principle, to measure the sky
surrounding each object. But this would have been at the expense of a drastic reduction of
the number of targets. Although sky subtraction can explain some of the scatter we see on
Fig. \ref{delta_Mobs_Thispaper}, it remains to understand why it could lead to a systematic
overestimate of $T_{\mathrm{eff}}$ (causing the overestimate of $L$ and, ultimately, of the
interpolated $\mathcal{M}$). Since an average sky was subtracted, one would expect as many negative
values of $\Delta \mathcal{M}$ as positive ones. However, the sky positions had been chosen, on the
DSS chart, on the basis of their ``darkness'', so that the selected skies may be darker than
the ones that typically surround the targets.

\item {\sl Inadequacy of the stellar evolution models:} In Figs. \ref{allHR1}--\ref{allHR3},
are plotted not only the evolutionary tracks for the metallicity $Z=0.004$ of Charbonnel et
al. (\cite{CMMSS93}), but also those at $Z=0.001$ (Schaller et al. \cite{SSMM92}). It is
interesting to see that the more metal-poor models are slightly more luminous and hotter,
and some systems fit them better. It is possible that some stars have indeed a metallicity
approching $Z=0.001$ (especially for the light elements, which are the main contributors to
$Z$), as shown e.g. by Peters \& Adelman (\cite{PA06}). However, it is doubtful that all our
systems have such a low metallicity, and a look at the HR diagrams shows that this
explanation cannot hold for all of them.

Also plotted on Figs.  \ref{allHR1}--\ref{allHR3} are the evolutionary tracks of
Claret \& Gimenez (\cite{CG98}) for a helium content $Y=0.28$ and a metallicity
$Z=0.004$. These models have a helium enhancement $\Delta Y=0.028$ relative to those of
Charbonnel et al. (\cite{CMMSS93}), since the latter have $Y=0.252$. One can see that
increasing the helium content increases the luminosity, withouth changing the
temperature. Again, some stars would fit better such models than the standard ones,
while others (like the components of 4 175333) would need so large a helium
enhancement as to make this solution untenable.

\item {\sl Inadequacy of the synthetic spectra:} If the true metallicity of most
system is close to $Z=0.004$ and the helium content is normal, and if the standard
evolutionary models can be trusted, then the synthetic spectra used
to estimate the effective temperatures may be suspected.
An overestimate of $T_\mathrm{eff}$ would certainly have resulted from the
use of LTE atmosphere models: as shown by Hunter et al. (\cite{HDS07}) in their
Table~13, the temperatures of stars analyzed by Kilian-Montenbruck et al.
(\cite{KM94}), on the basis of LTE atmosphere models and non-LTE line formation
calculations, are about $29\,000$~K, instead of $26\,000-27\,000$~K using NLTE
models. Nevertheless, the synthetic spectra we used are precisely based on the
latter, so that no obvious bias on temperatures is expected. Using the FASTWIND
code, Massey et al. (\cite{MZM09}) found temperatures of O stars $\sim 1000$~K hotter
than those obtained by Bouret et al. (\cite{BLH03}), who used the TLUSTY and
CMFGEN codes. Part of the discrepancy might be explained, however, by moonlight
continuum contamination in the spectra of Bouret et al. (\cite{BLH03}).
On the other hand, Mokiem et al. (\cite{MKE06}, \cite{MKE07}) found temperatures
$\sim 1000$~K hotter than those of Massey et al. (\cite{MZM09}), while they also
used FASTWIND. Finally, Massey et al. (\cite{MZM09}) still find some amount of
mass discrepancy for their LMC stars. Therefore, some debate still exists
regarding the temperature scales, though the above references all deal with O
stars, rather than with the B stars we are more concerned with. Unfortunately,
similar studies for metal poor B stars in the range we are interested in
($20\,000 < T_\mathrm{eff}< 30\,000$~K) are lacking, so it is difficult to assess
the reliability of the temperature scale. One may expect, however, that it is
more reliable than for O stars, since the NLTE effects are less important.

The OSTAR2002/BSTAR2006 libraries we have used are recent and are
probably the best available. The only
reservation one could possibly raise is the choice of the microturbulence
(2~\kms for $T_{\mathrm{eff}}\leq 30\,000$~K and 10~\kms for
$T_{\mathrm{eff}}> 30\,000$~K), which may be a rough approximation of
reality.

Conversely, even perfectly realistic models may prove inappropriate, if
superficial abundance anomalies exist in the components of our binary systems.
It is well known that early B-type stars may show a large overabundance of
helium -- the so-called helium-rich stars, of which HD 37776 is the prototype
-- but no example is known among components of close binaries. In any case,
stronger He lines would mimick a lower effective temperature, while it seems
that the temperatures are overestimated. Thus, only helium-weak stars could
explain an overestimate of temperatures, but none are known at such high masses.
\end{itemize}

In conclusion, we consider the first explanation (sky subtraction problem)
as the most probable one, until proven otherwise by other studies.

\begin{table}
\caption{Comparison with theoretical evolutionary models: difference between the
evolutionary mass and the observed mass.}             
\label{compevol}      
\centering          
\begin{tabular}{c c r r} 
\hline\hline  
Object & Model & $\Delta \mathcal{M}_\mathrm{P}$ & $\Delta \mathcal{M}_\mathrm{S}$  \\
	&  & ($\mathcal{M}_{\odot}$) & ($\mathcal{M}_{\odot}$)  \\
\hline                    
 4 110409 & sd & 1.076  & 1.923 \\
 4 113853 & sd & 1.500  & 0.855  \\	    
 4 117831 & d  & 0.392  & 0.366 \\
 4 121084 & d  & 1.503  & 1.404 \\
 4 121110 & d  & 1.482  & 0.923  \\
 4 121461 & d  & 0.731  & 0.429  \\
 4 159928 & sd & 1.089  & 0.566  \\
 4 160094 & d  & 0.855  & 0.261  \\
 4 163552 & d  & 0.222  & 0.197  \\
 4 175149 & sd & 4.143  & 5.091  \\
 4 175333 & d  & 1.012  & 0.797  \\
 5 016658 & d  & 0.296  & 0.646  \\
 5 026631 & sd & 0.924  & 0.459  \\
 5 032412 & d  & 0.922  & 0.208  \\
 5 038089 & d  & 1.764  & 2.349  \\  
 5 095337 & d  & 1.239  & 0.934 \\  
 5 095557 & d  & 1.215  & 0.639 \\   
 5 100485 & d  & 0.406  & 0.628   \\
 5 100731 & c  & 0.248  & 0.686   \\ 
 5 106039 & sd & 0.807  & 1.119    \\
 5 111649 & d  & 0.158  & 0.479   \\
 5 123390 & d  & 1.236  & 2.574    \\
 5 180185 & d  & 0.387  & 0.045    \\
 5 180576 & d  & 1.119  & 0.082    \\
 5 185408 & d  & 0.767  & 0.939    \\
 5 196565 & d  & 0.446  & 0.241  \\
 5 261267 & sd & 1.921  & 1.569   \\
 5 265970 & d  & 0.002  & 0.257  \\
 5 266015 & sd & 0.387  & 2.744   \\
 5 266131 & d  & 0.637  & 0.447    \\
 5 266513 & d  & 0.193  & 0.458   \\
 5 277080 & sd & 3.138  & 2.331   \\		  
 5 283079 & d  & 0.704  & 0.674  \\ 
\hline
\end{tabular}
\end{table}

\begin{figure}[!ht]
\centering
\includegraphics[width=7.5cm]{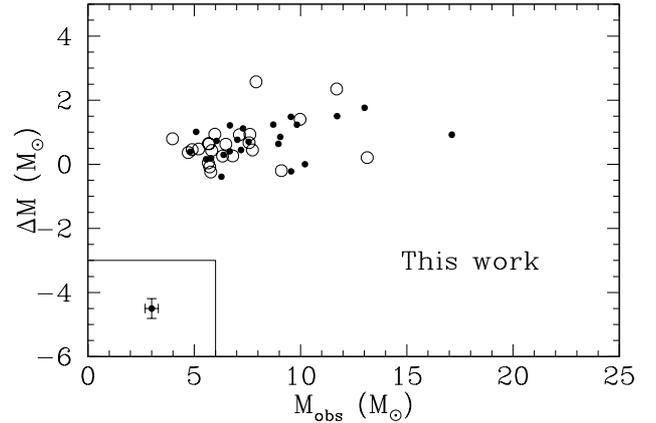}
\caption{Detached systems: difference between evolutionary mass and observed
mass vs. observed mass. Filled symbols: primary; open symbols: secondary. The
point in the inset, to the lower left corner, shows the median error bars for
the mass of the primary. The vertical bar does not include the errors on
luminosity and $T_{\mathrm{eff}}$ (hence on the interpolated mass), so it must
be considered as a lower limit.}
\label{delta_Mobs_Thispaper}%
\end{figure}

 \begin{figure}[!ht]
\centering
\includegraphics[width=7.5cm]{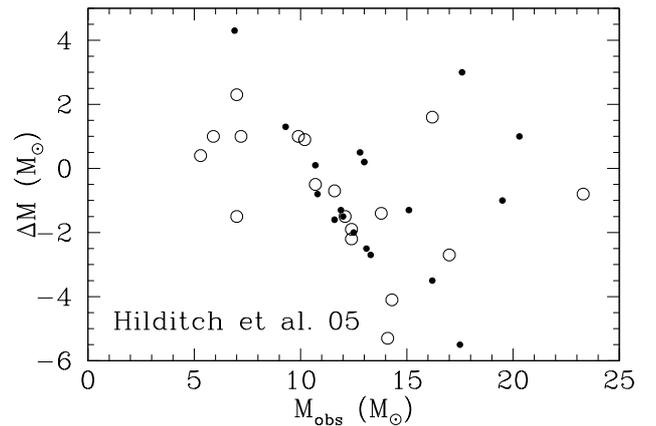}
\caption{ Same as Fig. \ref{delta_Mobs_Thispaper}, for the detached systems of
Hilditch et al. (\cite{HHH05}). }
\label{delta_Mobs_HHH}%
\end{figure} 

The astrophysical parameters of our sample of detached binaries allow us to give
a mass-luminosity relation for the SMC valid for masses between $\sim$4 and
$\sim$18 $\mathcal{M}_{\odot}$. A linear regression fit gives, on all 46 points:
\begin{equation}
\log L / L_\odot = (3.04\pm0.11) \log \mathcal{M} / \mathcal{M}_\odot +
(0.90\pm0.09)
\label{eq:mass_lum}
\end{equation}
with a RMS of $0.095$ (see the red line on Fig. \ref{mass_lum}). This is
almost exactly the same relation as that obtained by Gonz\'{a}les et al.
\cite{GOMM05} for the LMC.
The mass-luminosity diagram is plotted in
Fig. \ref{mass_lum} together with the isochrones computed from the Geneva
models for $Z = 0.004$ and $Y = 0.252$ (Charbonnel et al. \cite{CMMSS93}). 
All objects are, within the $1\,\sigma$ uncertainty, on or above the
theoretical zero-age main sequence. 

Most stars lie below the 50 Myr isochrone. The exceptions are some low-mass
objects ($\lesssim 6 \mathcal{M}_\odot$), i.e. both components of 5 111649 and
of 4 175333. These two systems appear to be $\gtrsim$70 Myr old. They clearly
bias the regression line to a low value of the slope. Graczyk (\cite{dG03})
adopts the relation
\[\log L/L_\odot = (3.664\pm 0.047)\log\mathcal{M}/\mathcal{M}_\odot+
(0.380\pm0.027)\]
for the LMC and the SMC, which is shown as the full blue line in
Fig. \ref{mass_lum}. The latter relation appears slightly too steep: it clearly
runs above unevolved stars with $\mathcal{M}\gtrsim 8-10~\mathcal{M}_\odot$.
 \begin{figure}[!ht]
\centering
\includegraphics[width=9cm]{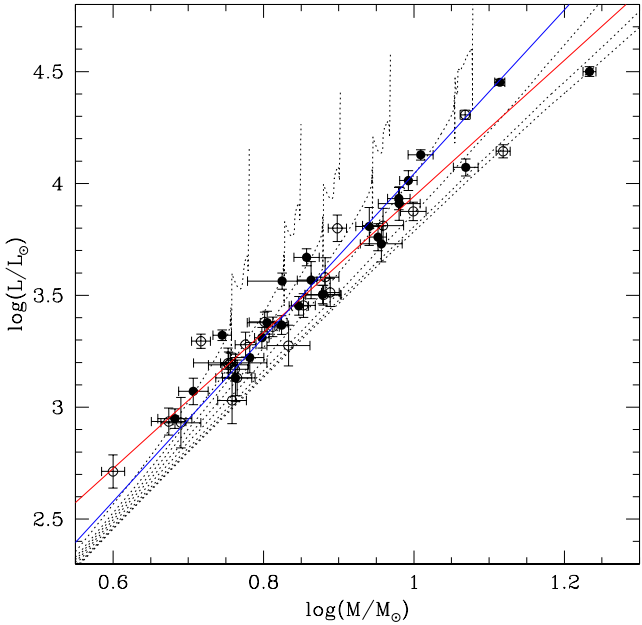}
\caption{The mass-luminosity relation for our 23 detached systems. Full dots
represent the primary components, while open dots represent the secondary ones.
The dotted lines are the isochrones with age of 0, 5, 10, 20, 30, 40, 50 and
70 Myr for $Z = 0.004$ (Charbonnel et al. \cite{CMMSS93}). The red line is a
simple least-squares fit of all 46 points, without weighing. The steeper blue
line is the relation adopted by Graczyk (\cite{dG03}) for the LMC and SMC.}
\label{mass_lum}%
\end{figure} 

\subsection{Apsidal motion}
Four of the nine eccentric systems have a negligible or marginal
(i.e $\lesssim 2\,\sigma$) apsidal motion, while four others show a
formally significant one (see Table~\ref{orbparam}); the apsidal motion of the
ninth system was arbitrarily fixed to a value close to the theoretical
expectation. Interestingly, the system
5 266131 has a very fast apsidal motion, and probably holds the record among
those hosting non-degenerate components (Petrova \& Orlov \cite{PO99}, Bulut \&
Demircan \cite{BD07}). It is also worth noticing that in
two of these systems (5 123390 and 5 266131), the relative radius
of the primary component is very close to, or even slightly exceeds, the
limiting radius above which orbital circularization occurs quickly
(North \& Zahn \cite{NZ03}, \cite{NZ04}): they have $r_p=0.238$ and
$0.281$ respectively, while the limiting radius is $\sim 0.25$. Thus,
those systems are being caught in the act of circularizing their orbit, and
indeed their eccentricities are among the smallest of all our eccentric
binaries.

In order to compare our observed apsidal rotation rates with those predicted by
theory, we used the equations (1) to (8) of Claret \& Gim\'enez (\cite{CG93}),
which allow to obtain the average tidal--evolution constant
$\bar{k_2}_\mathrm{obs}$
of the system from the orbital and stellar parameters, including the measured
apsidal period $P_\mathrm{apsid}= 2\pi/\dot{\omega}$. On the other hand, we
used the grid of stellar models for the metallicity $Z=0.004$ computed by Claret
(\cite{C05}). After interpolation in surface gravity and mass of the theoretical
$\log(k_2)$ values, the same formulae were used to obtain the predicted average
constant $\bar{k_2}_\mathrm{theo}$. We corrected for the relativistic precession term,
even though it is small for short periods (Mazeh \cite{tM08}), since it
contributes for no more than 2\% to the total precession period. The result is
shown in Fig.~\ref{apsidal}, where we plot $\bar{k_2}_\mathrm{obs}$ as a function of
$\bar{k_2}_\mathrm{theo}$. The one--to--one relation is the straight diagonal line,
and 3 systems fall on or very close to it. Two other systems are clearly
discrepant. One of the latter, 5 95557, has a formally significant, but small
apsidal motion; the masses are not very precisely determined, which may explain
part of the difference. It is quite possible that the apsidal motion of 5 95557
is affected by the gravitational perturbation of a small, unseen third
companion. The other discrepant system, 5 123390. is only marginally so, since
it is only $2\,\sigma$ away from the equality line. In any case, one has to keep
in mind that the error bars shown represent lower limits to the real
uncertainties, because they include only the error on the apsidal period, while
errors on the masses and radii of the components also contribute.
The latter errors certainly dominate in the case of 5 266131, which lies below
the equality line. The open dot represents the system 5 265970, for which an
apsidal motion close to the theoretical value was assumed.

\begin{figure}[!ht]
\centering
\includegraphics[width=8.5cm]{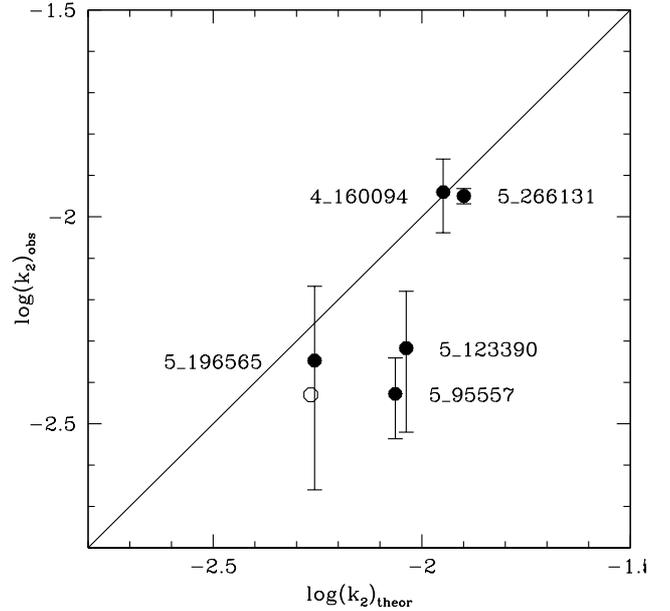}
\caption{Observed versus theoretical apsidal constant $\bar{k}_2$, averaged
over the two companions of each binary system according to the prescriptions of
Claret \& Gimenez (\cite{CG93}). The internal structure models used are those of
Claret (\cite{C05}) for the metallicity $Z=0.004$. The oblique line is the
equality. The error bars reflect the uncertainty on $\dot{\omega}$ alone, so
they represent lower limits to the real error. The open dot without error bar
represents the system $5\;265970$, for which an arbitrary apsidal motion was
assumed.}
\label{apsidal}
\end{figure}

It would be easy, by reobserving with photometry the few systems showing
significant apsidal motion, to strongly reduce the error bar on the average
$\bar{k_2}_\mathrm{obs}$. Further spectroscopic data would be needed as well to obtain
more precise masses and radii, in order to determine a meaningful value of
$\bar{k_2}_\mathrm{obs}$. That would bring an additional constraint on the metallicity
and on the extent of overshooting, for instance.

\subsection{Distribution of the mass ratios of the detached systems: do binaries
 really like to be twins ?}
From the 21 detached systems of the HHH03/05 sample, Pinsonneault \& Stanek
(\cite{PS06}) suggest that the proportion of massive detached systems with a
mass ratio close to unity is far larger than what would be expected from a classic
Salpeter-like ($p(q) \propto q^{-2.35}$) or a flat ($p(q) = \mathrm{const}$)
$q-$distribution. This statement is based on two arguments. The first one is
the striking difference between the median mass ratio of the detached sample
($q = 0.87$) and of the semi-detached/contact sample ($q = 0.65$).
The surprising point is not the low $q-$value for the semi-detached sample, as
this is expected for a post mass-transfer system, but rather the high proportion
of systems with $q > 0.85$ in the detached sample. Moreover, they argue that this
difference is real, i.e. not due to an observational bias such as the easier
detection of systems with components of similar size and brightness. Their argument
is that, beside the detached binaries with $q > 0.85$, there are two detached
systems with a $q-$value as low as $\sim$0.55. Indeed, it seems to be
a very strong argument in favour of the reality of a population of twins,
because the observation of detached systems with such a low $q$ value would
mean that the cut-off value  of the $q$ distribution of the detached systems is
small enough ($\lesssim 0.55$) to exclude an observational bias. Let us remind
that the cut-off value is the $q$ value which marks the transition between a
double-lined and a single-lined binary (SB2-SB1 transition), i.e. a function of
the resolution and of the $S/N$, beside stellar parameters. Therefore, as our
sample of 23 detached systems is comparable and even slightly larger than
HHH03/05's, this is an excellent opportunity to compare both statistics
and thus to shed some light on the controversial topic of the $q$ distribution
of detached and semi-detached/contact systems.

\begin{figure}[!ht]
\centering
\includegraphics[width=8.5cm]{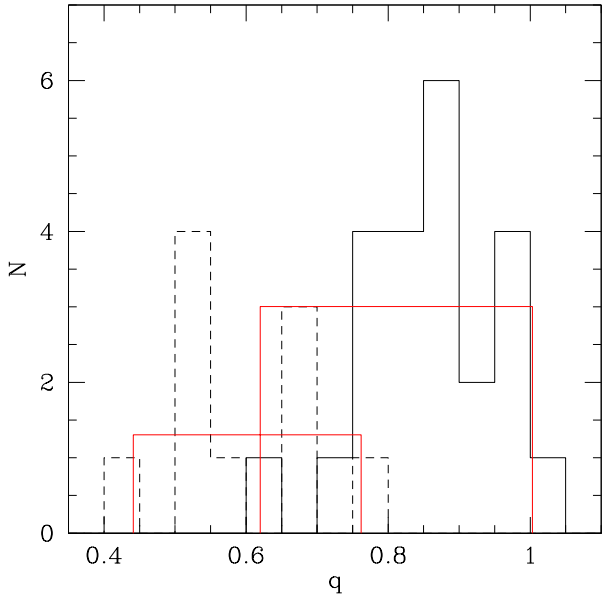}
\caption{ Distribution of the 33 observed mass ratios (solid: detached,
dashed: semi-detached/contact) with 0.05 bins. The best flat distribution is
over-plotted for each category (detached and semi-detached or contact systems).
Both distributions are truncated at a cut-off value of $q\sim0.7$ and $q\sim0.4$,
respectively. }
\label{histo_q}
\end{figure}

The $q-$distributions for our detached and semi-detached/contact binaries are
given in Fig. \ref{histo_q}. One sees that the two samples have quite
different distributions. The semi-detached/contact distribution is patchy
because of the small size of the sample (10 objects), but it seems to be
compatible with a constant law. The cumulative probability distribution of the
mass ratios (Fig. \ref{qmul}, top) confirms that it is consistent with a flat
distribution, limited to the interval $0.44 < q < 0.76$. The statistics of
semi-detached system is not discussed by Pinsonneault \& Stanek (\cite{PS06}).
The subject was computationally investigated by van Rensbergen et al.
(\cite{vRdLV}). Interestingly, they mention that 80\% of the observed Algols
have a mass ratio in the 0.4-1 range. The low-value for this range coincides
with the observed cut-off of our small sample. The problem is how to interpret
this value. Through our experience
in processing spectral data, we assume in this paper that this is the detection
threshold for SB2 systems, but van Rensbergen et al. seem to consider this value
not as an observational bias but rather as a parameter of real astrophysical
relevance. Indeed, they have
reproduced the observed Algolid $q-$distribution (from Budding et al.'s catalog
\cite{eB04}) assuming a ``liberal" model of evolution, i.e. a lot of mass loss
and a little loss of angular momentum during mass transfer. Their simulations
clearly depict a distribution with a broad peak in the 0.4-0.6 range and decreasing
for higher $q-$values (see Fig. 5 in van Rensbergen et al. \cite{vRdLV}). That is
in excellent agreement with our results too. The only problem is one of their
initial assumptions: they consider that the $q-$distribution of the detached
Algol-progenitor systems obeys a Salpeter-like power law! That is not what we
observe with our sample of 23 detached binaries. Figures
\ref{histo_q} and \ref{qmul} (top) show that the $q-$distribution of these
systems is hardly compatible with a decreasing power law. Indeed, the data are
very well modeled by a flat distribution truncated at $q = 0.72$. The higher
value for the SB1-SB2 cut-off compared to semi-detached systems is related to
the almost $q^3-$dependence of the luminosity ratio for main-sequence stars.
Let us mention that a flat $q-$distribution is assumed by most population
synthesis studies (Pinsonneault \& Stanek \cite{PS06}).

\begin{figure}[!ht]
\centering
\includegraphics[width=8.5cm]{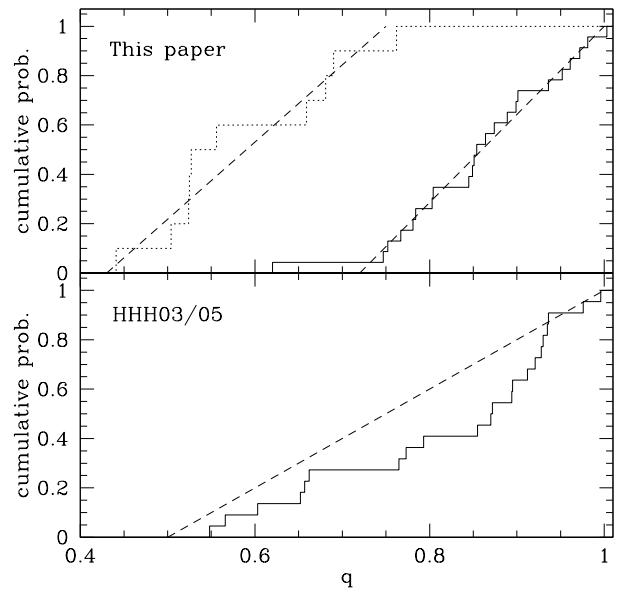}
\caption{Top: cumulative probability distribution as a function of the mass ratio
for our 23 detached (solid line) and 10 semi-detached/contact (dotted line)
binaries. The dashed oblique line plotted over the detached distribution represents
a flat distribution with a cutoff at $q = 0.72$. In the case of the
semi-detached/contact systems, a flat distribution with a cutoff at $q=0.43$
provides a reasonable fit as well.     
Bottom: cumulative probability distribution for HHH03/05's 21 detached binaries.
The over-plotted flat distribution with a cutoff at $q = 0.5$ shows the
incompatibility of a flat distribution with the HHH03/05 data. There is a similar
figure in Pinsonneault \& Stanek (\cite{PS06}). They show that the data are
best-fitted with a 55\% flat-45\% twin distribution (not shown in this figure).}
\label{qmul}
\end{figure}   

The cumulative probability distribution of the mass ratios for the HHH03/05 data
is given in Fig. \ref{qmul} (bottom). A similar figure is shown in
Pinsonneault \& Stanek. This distribution differs significantly from ours on
two points: the low cut-off value of $\sim$0.5 and the relatively strong
deviation from a flat distribution. The low cut-off is rather surprising. The mean
apparent magnitude of the HHH03/05 sample ($\overline{I^\mathrm{q}} = 15.47$ mag)
is 1.38 mag lower than ours ($\overline{I^\mathrm{q}} = 16.85$ mag). A rough
estimation of the ratio of our mean $S/N$ to their mean $S/N$, taking into account
our 2-times bigger telescope and longer exposure time via Eq. 3.2 of Hilditch
(\cite{rH01}), gives a ratio of $\sim$0.8. That means that our mean spectral $S/N$
is 20\% smaller than theirs despite the bigger optics because of significantly
fainter targets. Nevertheless, their 20\%-better $S/N$ hardly explains a luminosity
ratio cut-off of $\sim$1/8 to be compared with our $\sim$1/3-value (for detached
main-sequence binaries, $\mathcal{L} \sim q^3$). Moreover, the instrumental
resolution of the LR2 setup in MEDUSA mode is 3-times the value of the 2dF
spectrograph with the 1200B grating. Therefore, either our sample is deficient
in low-$q$ systems or there are some badly determined $q-$values in one of the
samples. The first explanation cannot be entirely ruled out because of the
small-sample statistics. Nevertheless, the comparison of the RV-related parameters
for the five systems common to both HHH03/05's program and ours can certainly help
to settle the issue.

Can the ``twin hypothesis'' be excluded by our data? The thorough study by
Lucy (\cite{L06}) points to a negative answer. He discusses both the ``weak''
hypothesis ($\mathcal{H}_w$) of an excess of binary systems with $q > 0.8$
relative to a constant distribution, and the ``strong'' hypothesis
($\mathcal{H}_s$) of an excess of systems with $q > 0.95$. He shows that
$\mathcal{H}_s$ is confirmed for binary systems in our Galaxy with an accurate
enough mass ratio ($\sigma_q < 0.01$). Therefore, it would be of great interest
to test the same hypothesis in another galaxy, and our study might, at first
sight, be considered a second step towards that goal, after HHH03/05. But a
realistic typical error on our $q$ values is $\sim 0.05$, and our sample of 15
systems with $q > 0.84$ is so small, relative to the 102 systems used by Lucy,
as to make impossible any confirmation or rejection of $\mathcal{H}_s$. Our data
might be more useful to constrain $\mathcal{H}_w$, but extensive simulations
would be necessary to estimate the detection biases, and such an effort does not
appear justified by our small sample. Therefore, the twin hypothesis remains an
open question as far as the SMC (and any galaxy other than ours) is concerned.

\subsection{ Extinction and distances}
\label{distsection}
\subsubsection{Colour excess and extinction} The histogram of the 28 colour
excesses is given in Fig. \ref{histo_Ebv}. Note that one system (5 196565)
does not have $B$ and $V$ light curves and four systems have an unreliable
observed $B-V$ index (see Section \ref{qualitycheck}). The mean value is
$\overline{E}_{B-V} = 0.134\pm0.051$
mag with individual values in the range 0.052-0.252 mag. By comparison, for four
stars in the bar, Gordon et al. (\cite{GCMLW03}) found values ranging from 0.147
to 0.218 mag. From their extinction map across the SMC, Zaritsky et al.
(\cite{ZHTGM02}) give $E_{B-V}\sim$0.05-0.25 mag. Sasselov et al. (\cite{dS97})
found a mean value of $\overline{E}_{B-V} = 0.125\pm0.009$ mag. Thus, our
results are in very good agreement with these previous determinations. In order to
investigate the spatial variation of the extinction across the studied field,
$E_{B-V}$ was plotted  against $\theta$, the distance to the optical center along
the projected axis (not shown). No correlation was found. The GIRAFFE field
is likely too small to detect a trend, if any.
 
\begin{figure}[!ht]
\centering
\includegraphics[width=8.5cm]{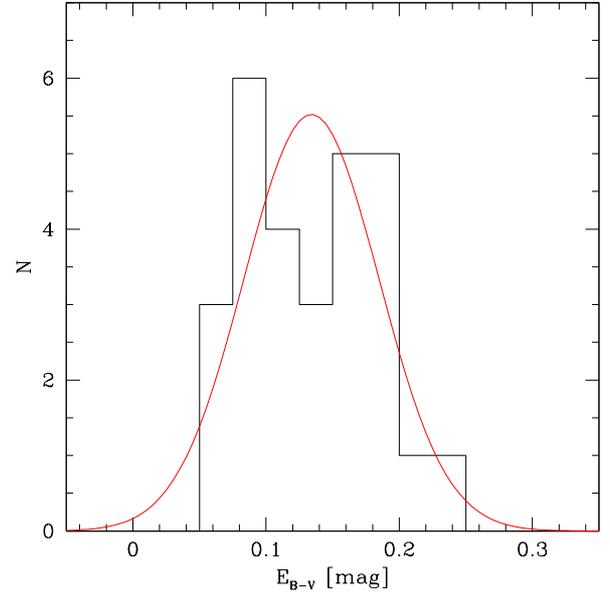}
\caption{Histogram of colour excesses of 32 binaries in 0.025 bins.
The best Gaussian fit is over-plotted. }
\label{histo_Ebv}%
\end{figure}
\subsubsection{Distance} The parameters relevant to the determination of the
distance moduli are shared between Tables \ref{synthphot} (synthetic photometry)
and \ref{dmi}. The DM was computed in both $V-$ and $I-$band, as explained in
Section \ref{distancemodulus}. One system (5 196565) has no value because of only one
light curve available. Fortunately, for most systems the difference between the two
computations of the individual DM is low and well within the uncertainties. As
expected in Section \ref{distancemodulus}, the $I-$values have a slightly lower
absolute uncertainty. The discrepancy between the $V$ and $I$ moduli is
negligible: $\overline{\Delta_{V-I}} \simeq -0.01$~mag with a scatter
$\sigma=0.02$~mag for the 28 reliable systems.

The DMi were checked for a possible dependence upon a number of parameters: apparent
visual magnitude, absolute visual magnitude, systemic velocity, colour excess and
distance to the optical center along the projected axis. Statistical tests were
performed (calculus of the Spearman rank correlation coefficient, followed by a
Student T test) and no significant correlation was found. The $DM_V$ vs.
$M_V^\mathrm{q}$ and $DM_V$ vs. $V^\mathrm{q}$ diagrams are shown in Fig.
\ref{histo_dmi} (\textit{bottom}). In order to calculate an unbiased mean DM, the HR
diagrams (Fig. \ref{allHR1}-\ref{allHR3}) of the individual systems were used
to select the systems suitable for this purpose. This is actually a more objective
criterion than relying solely on the quality of the light curves and spectra.
Therefore, 8 objects showing strong discrepancies with theoretical models were
conservatively discarded from the sample of 27 binaries (which remain from the
initial sample of 33 binaries, after suppression of stars with unreliable or no
colour index, or with a 3rd light, see below). Some of the primaries of these 8
systems may have
luminosities really higher or lower than expected, but we can lucidly
assume that most discrepancies are the direct consequences of low $S/N$ data
and/or third light contamination. These suspicious systems are indicated by
open symbols in Fig. \ref{histo_dmi}. 

From the sample of 33 objects, six at least must be discarded for the
computation of the mean distance modulus: 5 196565 because of the total lack
of $B$ and $V$ data, 4 163552 because of a third light contribution,
and 4 systems which do not have reliable observed colour indices (5 038089,
5 180185, 5 261267 and 5 277080). At that stage, there remain 27 systems with
presumably good distance moduli, without consideration of their HR diagrams.
Removing further the 8 systems which have a dubious position in the HR diagram
(i.e. 4 113853, 4 121084, 4 121110, 4 175149, 4 175333, 5 95557, 5 123390 and
5 185408) leaves us with only 19 systems, a bit more than half of the initial
sample. By ``dubious position", we mean systems for which the error bar of the
mass does not overlap that of the luminosity/temperature on the HR diagram.
A third subsample was defined, which contains the nine ``best" systems, defined
as those detached systems for which, either the luminosity/temperature error bar
overlaps the $Z=0.004$ evolutionary track, or the mass error bar overlaps the
representative point of the corresponding component, for at least one of the
components. These systems are 4 121461, 4 160094, 5 16658, 5 32412, 5 100731,
5 111649, 5 265970, 5 266131 and 5 266513.

Another important point is to identify possible biases acting upon the
DM distribution. Actually, there is a small magnitude cut-off.
This is the apparent magnitude cut-off, close to $V^\mathrm{q}\sim18.1$ mag.
It almost coincides with the apparent magnitude of 5 266513, the faintest binary
of our sample, whose spectra have (smoothed) $S/N$ in the 18-50 range. This
$V$ cut-off sets a higher limit for the absolute magnitude a binary should have
in order to be seen across the whole depth of the SMC field. If we take
$\sim$19.5 as the upper limit for the DM, then we find that the 
$M_V^\mathrm{q} $cut-off is $-1.82\pm0.16$ mag with $\overline{E}_{B-V} =
0.134\pm0.051$ mag. The cut-off strip for $V^\mathrm{q} = 18.1$ mag is indicated
in Fig. \ref{histo_dmi} (\textit{bottom-left}). Systems to the right of the
oblique lines are too faint to be observed as SB2 with good enough spectra.      

For example, in spite of a reliable solution, the binary 4 175333 should be
discarded from the calculus of the mean DM in order to remove the observational
bias just discussed. Indeed, with $M_V^\mathrm{q} = -1.217$ and
$(V-M_{\mathrm{V}})_0=18.61$, this system would not be perceived as an SB2 if it
were more distant, i.e. if it had  $(V-M_{\mathrm{V}})_0\gtrsim 19.0$ (see
Fig. \ref{histo_dmi}, \textit{bottom-left}). In fact this object had already
been discarded on the basis of the peculiar position of its components in the HR
diagram.

At the other end of the $M_V^\mathrm{q}-$ spectrum, there is a second cut-off,
but this one does not bias the DM-distribution. This is the cut-off of the
luminosity function, responsible for the depleted left-part of  
Fig. \ref{histo_dmi} (\textit{bottom-right}). Close to $M_{V}^\mathrm{q}\sim-4.2$
mag, this is roughly the low-limit for late O stars: intrinsically brighter
stars should be earlier O stars, but they are very scarce, so the probability
of finding one in a sample of 33 objects is very low.  
  
The histograms of the individual DMi in $V$ and $I$ are shown in Fig.
\ref{histo_dmi} (\textit{top-left}). For a given band, both the whole sample
(i.e. the 27 systems remaining after the exclusion of the obviously problematic
ones) and the ``good systems" (i.e. the 19 best ones) distributions are plotted. 
Although 8 systems are considered as less reliable, one can see that all but
three (4 175149, 4 175333 and 5 123390) discarded values are within the
``good systems" range of DMi. This range spans 0.45 mag, from 18.87 to 19.32.
The modes of the $V$ and $I$ distributions of the ``good systems" are 19.15 and
19.20 respectively. In addition the distribution of the 9 ``best" systems is
plotted in red.
 
One can wonder how far the average value of the DMs is the most pertinent
indicator of the mean distance of our stars. There is a $\sim 0.05$ to $0.1$~mag
deviation between the modal and mean values of the distributions (for both $V$
and $I$), but we verified that these distributions, in spite of their apparent
asymmetry, do not depart from gaussians in a significant way. Since the modal
value is not a robust one for such small samples, we rather considered the
weighted average and the median values.

The average and median distance moduli for the three subsamples defined above
are given in Table \ref{final_distmod} for the $V$ and $I$ bands. The average
values are weighted by the reciprocal variance of the individual moduli. The
RMS scatter is given beside each average value; beside each median value is
given the width of the distribution, defined as the half difference between the
upper and lower quartiles.

\begin{table}
\caption{Summary of the average and median distance moduli, for the three
subsamples defined in the text. The given uncertainties represent the RMS
scatters, not the errors on the respective average distance moduli.}             
\label{final_distmod}      
\centering          
\begin{tabular}{l l l l}     
\hline\hline       
Subsample & Stat. type & $\overline{(V-M_{\mathrm{V}})_0}$
                       & $\overline{(I-M_{\mathrm{I}})_0}$ \\
\hline
all 27    & w. average & $19.07\pm 0.19$ & $19.07\pm 0.19$ \\
          &   median   & $19.11\pm 0.12$ & $19.11\pm 0.12$ \\ \hline
19 good   & w. average & $19.10\pm 0.14$ & $19.11\pm 0.14$ \\
          &   median   & $19.13\pm 0.10$ & $19.12\pm 0.13$ \\ \hline
 9 best   & w. average & $19.10\pm 0.15$ & $19.12\pm 0.16$ \\
          &   median   & $19.13\pm 0.07$ & $19.12\pm 0.07$ \\ \hline
\end{tabular}
\end{table}
We see that the $V$ and $I$ distance moduli of any given subsample are perfectly
consistent. More surprisingly, the sample of all 27 systems has a slightly
shorter distance modulus than the two other subsamples, while it contains many
systems with large positive $\Delta\mathcal{M}$ values (see Section
\ref{compar_evol}). Likewise, the sample of the 9 best systems (i.e. those which
match the standard $Z=0.004$ evolutionary tracks best) have the same distance
modulus as the sample of the 19 good systems. These results suggest that the
components which are overluminous with respect to their $Z=0.004$ track are not
necessarily overluminous in reality. Otherwise, if their luminosity were
overestimated, they should also lie at a larger apparent distance.
\begin{figure}[!ht]
\centering
\includegraphics[width=9.0cm]{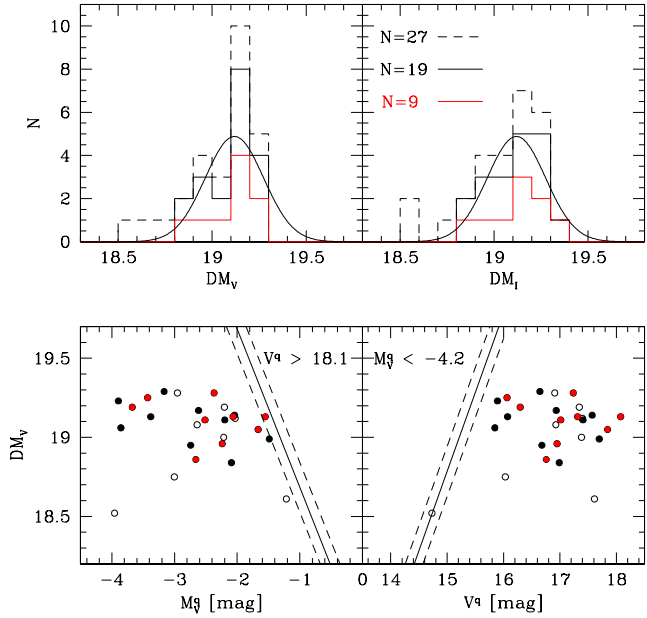}
\caption{\textit{Top}: distribution of the $V$- (\textit{left}) and $I$-band (\textit{right})
distance moduli. The distribution of the whole sample of 27 systems with good colours and
presumably without 3rd light (dashed line) is shown. Only binaries with good HR diagrams
are selected for the smaller statistics.
The best fitting Gaussian curves for the 19-binary distributions are over-plotted.
\textit{Bottom}: distance modulus vs. absolute visual magnitude (\textit{left}) and  
distance modulus vs. apparent visual magnitude (\textit{right}). The estimated magnitude
cut-off for a mean extinction is indicated by the diagonal solid line; the dashed
lines correspond to the cut-off lines for $\overline{E}_{B-V} = 0.134\pm0.051$ mag.
 Ten binaries with dubious luminosities are shown with open symbols.}
\label{histo_dmi}%
\end{figure}

Finally, combining both $V$ and $I$ results, we adopt

\begin{equation}
\langle(m-M)_0\rangle =19.11\pm0.03\;\;\;\; (66.4\pm0.9\; \mathrm{kpc})
\label{adopted_distance}
\end{equation}
as the most probable average distance modulus for our sample. That value is
$0.20\pm0.03$ mag ($5.8\pm0.9$ kpc) higher than the mean DM derived by HHH05,
that is to say $18.912\pm0.035$ (s.e.) ($60.59\pm0.98$ kpc) from two samples
totalizing 39 binaries. Our DM value does not agree HHH05's, but contrary to
theirs, it is not representative of the whole SMC. Our DM refers to the
25\arcmin-diameter GIRAFFE field at
$\sim$0.45\degr \, of the SMC optical center in a SW direction. Is this result
coherent with the current knowledge of the 3-D structure of the SMC? The SMC
bar is currently thought to be NE to SW-oriented, with the NE part closer to us
than the SW part and the optical center occupying a mid position (Groenewegen
\cite{mG00}, HHH05). According to HHH05, there is a $\sim$2-3 kpc variation of
the mean DM over their 2\degr \, field. Given that gradient, our sample can be
expected to be $\sim$0.5 kpc further than the optical center, since our field is
$\sim$0.45\degr to the SW from it; but this difference is actually smaller than the
uncertainty on our mean distance. If we took at face value
the average distances moduli obtained by HHH05 and by us respectively, and assumed
that HHH05 give a distance modulus representative of the SMC optical centre, one
could in principle compute the real distance between the optical centre and the
region corresponding to our field of view. Thus, taking into account the angular
separation between the optical center and the center of the GIRAFFE field of
view, the separation would correspond to $5.8\pm1.3$ kpc along the SMC bar (see
Fig. \ref{photodss}). That would mean that the bar is almost directed towards
us, with a mean distance varying by as much as $25.8$~kpc over the $2^\circ$
covered by HHH05's sample! Such a huge distance gradient, if real, would have
been detected by Groenewegen \cite{mG00}. Thus our larger distance cannot be
reconciled with that obtained by
HHH05. It is most probably due to the better observational material used here,
as well as to the excellent synthetic spectra of Lanz \& Hubeny (\cite{LH03},
\cite{LH07}). On the other hand, our DM perfectly agrees with that obtained by
Groenewegen (\cite{mG00}) using cepheids ($19.11\pm 0.11$ from the W index, and
$19.04\pm 0.17$ from the K band). Taken at face value, our distance modulus
clearly favors the so-called long distance scale.

\subsubsection{Distance according to the detached systems versus distance
according to the others}
In the context of a search for systematic effects, one may wonder whether
detached systems give a different distance modulus than the semi-detached or
contact ones. Separating our 19 ``good'' systems into a group of 13 detached
ones and a group of 6 semi-detached ones, we have computed the weighted average
distance modulus (in the $I$ band) for each group. For the detached systems, we
obtain $\overline{(I-M_I)_0(d)}=19.07\pm0.04$~mag, the error being the RMS
scatter divided by $\sqrt{13}$. For the semi-detached and contact ones, we
obtain $\overline{(I-M_I)_0(sd,c)}=19.15\pm0.05$~mag. Thus, our 6 semi-detached
and contact systems are, on average, $0.08$~mag more remote than our 13 detached
systems. Given the small statistics and the error bars, the difference is not
significant since it amounts to only about $1.6\,\sigma$.

Nevertheless, we have looked at the distance moduli determined by HHH03/05, in
order to see whether a similar difference exists in their data. Interestingly,
almost the same difference appears: for 21 detached systems, one has
$\overline{DM(d)}=18.85\pm 0.06$, while for the 29 others one has
$\overline{DM(sd,c)}=18.91\pm 0.05$. So, here too, the semi-detached and contact
systems are more remote than the detached ones on average, and by a similar
amount, of $0.06$~mag. The effect is only of the order of $1\,\sigma$, so it is
formally not significant. The coincidence between our results and those of
HHH03/05 remains intriguing, however.
 
\subsubsection{Depth}
The observed dispersion of the DMi is the convolution of the true depth of our
population of binaries with the average uncertainty on the DM determined for one
binary system. Hence, under the assumption of a Gaussian distribution of both
the cosmic and the error standard deviations, the true $1\,\sigma$ depth is given
by the quadratic difference between the observed standard deviation of the
moduli and the typical error on the modulus of an individual system. The
observed standard deviation being 0.144~mag, and taking the average of the
squares of the estimated errors as representative of the typical error, the
$1\,\sigma$ depth becomes $\sqrt{0.144^{2} - 0.00479} = 0.126$ mag RMS
($\sim$3.9 kpc), since the average variance of an individual $I$ distance
modulus is $0.00479$ mag$^2$. Groenewegen gives
0.11 mag (3.4 kpc) as the intrinsic $1\,\sigma$ spread for his sample of SMC
Cepheids, so our result is in perfect agreement with his.
The HHH05 sample (29 binaries) displays a higher dispersion, $\sim$0.28 mag
($\sim$7.8 kpc), but the authors consider that value as an overestimate. Our
depth estimate is only 15\% larger than that of Groenewegen (\cite{mG00}).

\subsubsection{Temperature scale and distance modulus}
We have seen that a suspicion remains, that the effective temperatures may have
been overestimated in a number of systems. What is the effect of such an
overestimate on the resulting distance modulus? For the sake of simplicity, let
us consider a single star rather than a binary system. Its distance modulus can
be written (under the assumption that we use the $V$ magnitude):
\begin{eqnarray*}
m-M&=&V_q+2.5\,\log\left(10^{-0.4\,M_V}\right)-3.1\,E_{B-V} \\
   &=&\mathrm{const}+2.5\,\,\log\left(10^{-0.4\,(M_{bol}-BC)}\right) \\
   &&  -3.1\,\left[(B-V)_q-(B-V)_0\right]
\end{eqnarray*}
since $V_q=\mathrm{const}$. Invoking Eqns. \ref{eqCI} and \ref{eqnBC}, which allow
to express $(B-V)_0$ and $BC$ as a function of $T_{\mathrm{eff}}$, expressing
the bolometric magnitude as a function of radius $R$ and $T_{\mathrm{eff}}$, and
remembering that $(B-V)_q$ is constant for a given object, one obtains
\begin{eqnarray*}
m-M&=&\mathrm{const}\,+2.5\,
\log\left(10^{-0.4\,\left[-5\log R-10\log T_{\mathrm{eff}}+\mathrm{cte}
+5.51\log T_{\mathrm{eff}}\right]}\right) \\
&&-1.395\,\log T_{\mathrm{eff}} \\
&=&\mathrm{const}\,+3.095\,\log T_{\mathrm{eff}}
\end{eqnarray*}
where we have taken into account that the radius is constant. Introducing a second star
will essentially not change the result, because the ratio of the temperatures
will appear (as well as the ratio of radii), which can be considered constant.
Therefore, by differentiating the last equation above, one finally gets a very
simple relation between a change $\Delta T_{\mathrm{eff}}$ of effective
temperature, and the resulting change of the distance modulus:
\begin{equation}
\Delta (m-M)_0 \simeq 1.344 \,\frac{\Delta T_{\mathrm{eff}}}{T_{\mathrm{eff}}}
\end{equation}
In the case of 4 175333, for instance, the temperature excess is especially
large since it reaches $\sim 2000$~K. If the temperature is indeed
overestimated, then the system is $\Delta (m-M)_0\simeq 1.344 \times 0.092
=0.123$~mag too far. Strangely enough, its modulus is only 18.60, so that it
would be reduced to 18.48 under this assumption. The system 4 121110 has
$\Delta T_{\mathrm{eff}}\sim 1580$~K and $\Delta (m-M)_0\sim 0.08$, so its
modulus should be $19.00$ instead of $19.08$. Likewise, 4 121461 and 4 160094
have $\Delta (m-M)_0\sim 0.08$ and $0.06$ respecively, while they are listed as
``good'' systems.

A quick estimate of the average temperature excess of the 9 ``best'' systems
leads to an average modulus excess of about $0.04$~mag, so that the corrected
distance modulus would be reduced to 19.07, which still remains rather long
compared to the frequently quoted value of $18.9$. In any case, it appears
difficult to reduce the distance modulus to less than $19.0$. Thus, our data do
favour the ``long'' rather than the ``short'' distance scale to the Magellanic
Clouds.
                  
\section{Conclusions}
The first goal of the present paper was to extract the fundamental stellar
parameters characterizing 33 SB2 systems and then to test the reliability of
stellar evolution models at low metallicity. We have shown that most components
of detached systems have properties in fair agreement with the predictions
of the Geneva single-star models at $Z =0.004$, although several systems appear
slightly overluminous for their mass. From the subsample of detached
systems, we give a mass-luminosity relation valid for masses between $\sim 4$
and $\sim 18 \mathcal{M}_{\odot}$. Despite the lack of evolutionary cross
checks, the semi-detached binaries appear to satisfy the typical morphology of
Algol-like systems: a main-sequence primary with a large, less massive and more
evolved secondary companion.

Our binaries were then used as primary standard candles.
Average distance moduli obtained from the $I-$band data show no systematic
difference compared to values obtained from the $V$ data. 
For each band, the frequency distribution of the DMi is not significantly different
from a Gaussian. Combining $V$ and $I$ results and considering both the mean and
the median values, we adopt $\langle(m-M)_0\rangle =19.11\pm 0.03$
as the average DM for the 19 most reliable systems. We have shown that, even if
the effective temperatures are overestimated, so as to make the components of
our systems overluminous relative to their mass, the average DM would still
remain above $19.00$, so that our results support rather the long distance scale
to the Magellanic Clouds than the short one. The true depth of our sample is
estimated to be $2\times 0.126=0.25~\mathrm{mag}$ ($\sim$7.6 kpc), if we
consider the $2\;\sigma$ interval as representative.

We have found a significant apsidal motion for at least three systems. For
one system, the apsidal motion might be affected by an unseen third
companion, but at least two other systems most probably owe it to the stellar
structure of their components alone. The observed and predicted precession rates
agree perfectly for the latter two systems, which are now circularizing their
orbit, and the agreement extends to one or two other systems showing much less
precise apsidal motion. The system 5 266131 shows probably the largest
precession motion of all systems hosting main sequence stars, which is due
exclusively to the non-point mass nature of the components.
 
Beside the astrophysical parameters and the distance moduli, other interesting
results have been obtained.
We have shown that the twin hypothesis for massive binaries is not
supported by our results, although the latter are not strong enough to reject it
either. Our sample of 23 detached systems agree quite well with a standard flat
mass ratio spectrum. There is a clear cut-off of the $q-$distribution at
$\sim0.72$. Therefore, the high proportion of detached binaries having a
$q$ value close to unity is more probably the result
of an observational bias. The discrepancy observed by Pinnsonneault \& Stanek
(\cite{PS06}) may well be due to a number of underestimated mass ratios in
the HHH03/05 sample. The distribution of the mass ratios of the
semi-detached/contact binaries is consistent with a flat distribution restricted
to the interval $0.4 < q < 0.8$. The sample is too small (10 objects) to tell
more.

Another important result is the bimodal distribution of the radial velocity of
the \ion{H}{ii} gas, with peak values at $145.2\pm 0.5$ and $161.4\pm 0.6$ km s$^{-1}$. 

Finally, we must emphasize that the accomplishment of this project was possible
thanks to the public availability of the OGLE-II catalog of EB light curves, the
OSTAR2002/BSTAR2006 libraries of synthetic spectra, the KOREL code and the
PHOEBE interface to the WD codes.

\acknowledgements This work was supported by the Swiss National Fund for
Scientific Research. RG warmly thanks K. Pavlovski and S. Iliji\'{c} for
fruitful discussions and skilled advice during his one-week stay at the
Department of Physics of the University of Zagreb (Croatia) in September 2005.
P. Hadrava and A. Pr\v{s}a are thanked for their help with, respectively, the
program KOREL  and the PHOEBE interface to the WD program. We are indebted to
Prof. André Blecha (now retired) for having led the creation of the Data
Reduction Software of the FLAMES-GIRAFFE instrument, which allowed PN and FR to
obtain Guaranteed Time Observations, part of which are analyzed here. PN thanks
Dr. Paul Bartholdi, who had initiated him to the SM macro {\sl amoeba}.
We also thank the anonymous referee of a previous version of this paper;
one of his remarks led to a re-analysis of our spectra, which resulted in
more robust masses and effective temperatures.

\begin{table*}
\caption{Orbital parameters: number of spectra used in the solution,
minimum and maximum signal-to-noise ratio of the smoothed spectra,
eccentricity, argument of the periastron for the epoch $t_0$ given in
Table~\ref{basicparam} and first derivative of it,
inclination, semi-major axis, mass ratio, phase shift and systemic velocity.}             
\label{orbparam}      
\centering          
\begin{tabular}{c c c c c c c c c c}     
\hline\hline       
  Object&$n_\mathrm{sp}$&S/N& $e$  &$\omega_{0}$ ($\degr$)&$i$     &   $a$         &$q\equiv \mathcal{M}_\mathrm{S}/\mathcal{M}_\mathrm{P}$& $\Delta\phi$      &$V_\mathrm{\gamma}$\\ 
        &               &   &      &$\dot{\omega}$ ($\degr$ yr$^{-1}$)& ($\degr$)  & ($R_{\odot}$)   & &                  & (km s$^{-1}$)     \\
\hline                    
  4 110409 & 11 & $65-140$ & $0      $ &  90	       & $77.3\pm0.1$ &$24.07\pm0.20$ &$0.524\pm0.009$& $ -0.00048 \pm 0.00045 $ & $176.1\pm1.3$\\
  4 113853 & 7  & $25- 65$ & $0      $ &  90	       & $60.1\pm0.3$ &$10.95\pm0.10$ &$0.681\pm0.011$& $ -0.00447 \pm 0.00125 $ & $134.1\pm1.4$\\
  4 117831 & 12 & $28- 71$ & $0      $ &  90	       & $78.2\pm0.4$ &$ 9.87\pm0.06$ &$0.981\pm0.013$& $ -0.00149 \pm 0.00072 $ & $129.6\pm1.1$\\
  4 121084 & 9  & $43- 86$ & $0      $ &  90	       & $83.0\pm0.1$ &$10.30\pm0.06$ &$0.851\pm0.009$& $ -0.00246 \pm 0.00035 $ & $147.2\pm1.6$\\
  4 121110 & 11 & $52- 95$ & $0      $ &  90	       & $81.9\pm0.3$ &$11.53\pm0.06$ &$0.747\pm0.007$& $ -0.00404 \pm 0.00033 $ & $163.0\pm1.1$\\
  4 121461 & 15 & $30- 54$ & $0.185$   & $222.8\pm1.4$ & $81.1\pm0.2$ &$14.95\pm0.08$ &$0.961\pm0.009$& $ -0.04792 \pm 0.00071 $ & $148.5\pm0.8$\\
           &    & $      $ & $\pm0.004$& $0.0$~(fixed) &	      &	              &	              &	                         &      \\
  4 159928 & 8  & $50-108$ & $0      $ &  90	       & $61.1\pm0.2$ &$11.99\pm0.12$ &$0.556\pm0.010$& $ -0.00283 \pm 0.00062 $ & $151.3\pm2.0$\\
  4 160094 & 11 & $29- 97$ & $0.089$   & $36.5\pm6.7$  & $77.0\pm0.2$ &$15.04\pm0.24$ &$0.752\pm0.015$& $  0.0170  \pm 0.0006  $ & $149.0\pm2.4$\\
           &    & $   $ & $\pm0.007$ & $9.8\pm1.9$     &              &		      &               &	                         &      \\
  4 163552 & 9  & $73-132$ & $0      $ &  90	       & $80.9\pm0.2$ &$14.91\pm0.19$ &$0.952\pm0.015$& $ -0.00037 \pm 0.00020 $ & $154.4\pm2.4$\\
  4 175149 & 15 &1$17-168$ & $0      $ &  90	       & $78.8\pm0.1$ &$17.98\pm0.22$ &$0.659\pm0.015$& $ -0.00308 \pm 0.00054 $ & $163.2\pm2.2$\\
  4 175333 & 14 & $26- 74$ & $0.022$   & $232.9\pm15.8$& $80.4\pm0.5$ &$10.18\pm0.09$ &$0.781\pm0.014$& $ -0.00087 \pm 0.00075 $ & $139.9\pm1.4$\\
           &    & $   $ & $\pm0.006$ & $0.0$~(fixed)   &              &               &               &                          &      \\
  5 016658 & 11 & $30- 62$ & $0      $ &  90	       & $80.3\pm0.2$ &$11.16\pm0.09$ &$0.889\pm0.011$& $  0.00002 \pm 0.00035 $ & $158.6\pm1.4$\\
  5 026631 & 8  & $77-120$ & $0      $ &  90	       & $61.4\pm0.1$ &$14.18\pm0.11$ &$0.690\pm0.006$& $  0.00264 \pm 0.00051 $ & $150.8\pm1.5$\\
  5 032412 & 13 & $58- 92$ & $0      $ &  90	       & $83.4\pm0.1$ &$30.82\pm0.14$ &$0.767\pm0.006$& $  0.00014 \pm 0.00019 $ & $177.4\pm0.7$\\
  5 038089 & 11 &1$28-216$ & $0      $ &  90	       & $77.1\pm0.2$ &$21.88\pm0.16$ &$0.899\pm0.011$& $ -0.00047 \pm 0.00039 $ & $152.9\pm1.2$\\
  5 095337 & 10 & $48- 89$ & $0      $ &  90	       & $88.8\pm0.4$ &$ 9.98\pm0.08$ &$0.874\pm0.013$& $  0.00278 \pm 0.00033 $ & $183.8\pm1.8$\\
  5 095557 & 11 & $33- 75$ & $0.218$   & $155.7\pm2.5$ & $84.7\pm0.3$ &$17.54\pm0.32$ &$0.854\pm0.026$& $  0.00837 \pm 0.00062 $ & $203.3\pm2.6$\\
           &    & $   $ & $\pm0.003$   & $2.6\pm0.5$   &              &               &               &                          &      \\
  5 100485 & 13 & $46- 86$ & $0      $ &  90           & $84.3\pm0.2$ &$13.12\pm0.04$ &$0.972\pm0.006$& $ -0.00109 \pm 0.00026 $ & $159.2\pm0.6$\\
  5 100731 &  8 & $35- 74$ & $0      $ &  90           & $59.7\pm1.1$ &$11.36\pm0.16$ &$0.762\pm0.012$& $ -0.01465 \pm 0.00173 $ & $173.0\pm1.6$\\
  5 106039 &  9 & $52-102$ & $0      $ &  90           & $75.7\pm0.1$ &$16.81\pm0.21$ &$0.527\pm0.012$& $ -0.00182 \pm 0.00041 $ & $153.3\pm2.1$\\
  5 111649 & 10 & $52- 89$ & $0      $ &  90           & $74.4\pm0.2$ &$19.14\pm0.11$ &$0.936\pm0.010$& $ -0.00162 \pm 0.00040 $ & $149.4\pm0.8$\\
  5 123390 & 14 & $91-179$ & $0.042$   & $62.3\pm5.2$  & $74.3\pm0.2$ &$18.39\pm0.19$ &$0.804\pm0.010$& $  0.00470 \pm 0.00046 $ & $141.7\pm1.8$\\
           &    & $   $ & $\pm0.006$   & $4.8\pm1.6$   &              &               &               &                          &      \\
  5 180185 & 12 & $37- 64$ & $0.036$   & $88.6\pm2.8$  & $84.9\pm0.2$ &$29.91\pm0.14$ &$0.901\pm0.009$& $ -0.00182 \pm 0.00051 $ & $165.2\pm0.6$\\
           &    & $   $ & $\pm0.005$ & $0.0$~(fixed)   &              &               &               &                          &      \\
  5 180576 & 12 & $25- 64$ & $0      $ &  90           & $75.0\pm0.4$ &$13.31\pm0.08$ &$0.784\pm0.009$& $  0.00350 \pm 0.00079 $ & $168.1\pm1.0$\\
  5 185408 & 12 & $38- 66$ & $0      $ &  90           & $76.2\pm0.3$ &$12.69\pm0.06$ &$0.849\pm0.008$& $ -0.00141 \pm 0.00057 $ & $174.9\pm0.8$\\
  5 196565 & 13 & $44- 86$ & $0.138$   & $231.9\pm2.7$ & $83.5\pm0.2$ &$24.65\pm0.12$ &$0.803\pm0.007$& $ -0.03068 \pm 0.00038 $ & $162.3\pm0.5$\\
           &    & $   $ & $\pm0.006$   & $1.2\pm0.6$   &              &               &               &                          &      \\
  5 261267 & 10 & $41- 88$ & $0      $ &  90           & $89.7\pm0.2$ &$11.92\pm0.08$ &$0.525\pm0.005$& $  0.00289 \pm 0.00018 $ & $176.0\pm1.3$\\
  5 265970 & 10 & $67-111$ & $0.054$   & $362.5\pm4.7$ & $82.5\pm0.1$ &$24.67\pm0.24$ &$0.620\pm0.008$& $  0.01600 \pm 0.00030 $ & $159.5\pm1.3$\\
           &    & $   $ & $\pm0.001$ & $3.8$~(fixed)   &              &               &               &                          &      \\
  5 266015 & 10 & $69-118$ & $0      $ &  90           & $78.5\pm0.1$ &$17.60\pm0.09$ &$0.441\pm0.005$& $ -0.00215 \pm 0.00028 $ & $188.0\pm1.1$\\
  5 266131 & 10 & $37- 71$ & $0.042$   & $227.2 \pm2.3$& $83.5\pm0.2$ &$12.82\pm0.15$ &$0.864\pm0.013$& $ -0.00962 \pm 0.00028 $ & $183.1\pm2.6$\\
           &    & $   $ & $\pm0.001$   & $50.2\pm2.0$  &              &               &               &                          &      \\
  5 266513 & 10 & $18- 50$ & $0      $ &  90           & $84.0\pm0.4$ &$ 9.91\pm0.06$ &$0.845\pm0.010$& $ -0.00073 \pm 0.00056 $ & $164.4\pm1.1$\\
  5 277080 & 11 & $74-130$ & $0      $ &  90           & $76.6\pm0.1$ &$15.96\pm0.13$ &$0.504\pm0.009$& $ -0.00020 \pm 0.00034 $ & $155.7\pm1.3$\\
  5 283079 & 10 & $28- 68$ & $0      $ &  90           & $87.7\pm0.3$ &$12.28\pm0.08$ &$1.003\pm0.013$& $  0.00232 \pm 0.00030 $ & $179.4\pm1.3$\\
\hline                  
\end{tabular}
\end{table*}

\begin{table*}
\caption{Wilson-Devinney analysis, as obtained on the basis of both photometry and
spectroscopy: system type (detached, semi-detached or contact), temperature ratio, normalized
surface potentials and passband luminosity ratios. Third light contribution is given for
the system 4 163552 (normalized to total flux at phase 0.25).
The given potentials and the luminosity ratios are those obtained \textit{including} the
spectroscopic constraints (final WD/PHOEBE analysis).}             
\label{WDana}      
\centering          
\begin{tabular}{l l l l l l l l l}     
\hline\hline       
Object & Model & $T_\mathrm{eff}^\mathrm{S} / T_\mathrm{eff}^\mathrm{P}$ & $\Omega_\mathrm{P}$ & $\Omega_\mathrm{S}$ &
$(L_\mathrm{S}/L_\mathrm{P})_{B}$ & $(L_\mathrm{S}/L_\mathrm{P})_{V}$ & $(L_\mathrm{S}/L_\mathrm{P})_{I}$ & $(L_\mathrm{S}/L_\mathrm{P})_B^{\mathrm{sp}}$ \\ 

  &  & & & & $\ell_{3,B}$ & $\ell_{3,V}$ & $\ell_{3,I}$ &  \\
\hline                    
  4 110409 & sd &  $0.654 \pm0.007$  & $5.732\pm0.092$  & $2.922$ (fixed)  & $1.292\pm0.023$ & $1.353\pm0.024$ &  $1.367\pm0.024$ & $1.392\pm0.036$ \\
  4 113853 & sd &  $0.737 \pm0.015$  & $3.868\pm0.090$  & $3.209$ (fixed)  & $0.680\pm0.010$ & $0.713\pm0.009$ &  $0.753\pm0.009$ & $0.681\pm0.058$ \\
  4 117831 & d  &  $0.981 \pm0.023$  & $4.855$ (fixed)  & $4.479\pm0.108$  & $1.175\pm0.052$ & $1.177\pm0.050$ &  $1.180\pm0.049$ & $1.009\pm0.103$ \\
  4 121084 & d  &  $0.940 \pm0.007$  & $3.770$ (fixed)  & $3.817\pm0.043$  & $0.707\pm0.061$ & $0.713\pm0.010$ &  $0.720\pm0.010$ & $0.704\pm0.061$ \\
  4 121110 & d  &  $0.894 \pm0.007$  & $3.820\pm0.035$  & $4.184\pm0.059$  & $0.424\pm0.005$ & $0.433\pm0.005$ &  $0.442\pm0.005$ & $0.415\pm0.047$ \\
  4 121461 & d  &  $0.949 \pm0.023$  & $6.980\pm0.133$  & $6.822\pm0.099$  & $0.907\pm0.030$ & $0.914\pm0.029$ &  $0.923\pm0.029$ & $0.954\pm0.063$ \\
  4 159928 & sd &  $0.723 \pm0.010$  & $3.138\pm0.024$  & $2.983$ (fixed)  & $0.357\pm0.003$ & $0.376\pm0.003$ &  $0.397\pm0.004$ & $0.453\pm0.029$ \\
  4 160094 & d  &  $0.854 \pm0.018$  & $5.517$ (fixed)  & $5.545\pm0.100$  & $0.476\pm0.050$ & $0.489\pm0.007$ &  $0.502\pm0.007$ & $0.428\pm0.101$ \\
  4 163552 & d  &  $0.997 \pm0.004$  & $3.983\pm0.026$  & $4.214\pm0.044$  & $0.798\pm0.024$ & $0.799\pm0.016$ &  $0.800\pm0.014$ & $0.926\pm0.171$ \\
  	   &    &   		  &		     &		           & $0.142\pm0.014$ & $0.112\pm0.012$ &  $0.119\pm0.006$ &                 \\
  4 175149 & sd &  $0.849 \pm0.006$  & $4.411\pm0.056$  & $3.171$ (fixed)  & $1.176\pm0.025$ & $1.200\pm0.023$ &  $1.229\pm0.023$ & $1.415\pm0.132$ \\
  4 175333 & d  &  $0.879 \pm0.022$  & $5.054\pm0.151$  & $5.010\pm0.196$  & $0.582\pm0.034$ & $0.595\pm0.035$ &  $0.609\pm0.036$ & $0.562\pm0.103$ \\
  5 016658 & d  &  $1.007 \pm0.009$  & $4.086$ (fixed)  & $4.590\pm0.068$  & $0.609\pm0.074$ & $0.612\pm0.008$ &  $0.614\pm0.008$ & $0.604\pm0.074$ \\
  5 026631 & sd &  $0.734 \pm0.007$  & $3.507\pm0.030$  & $3.226$ (fixed)  & $0.495\pm0.003$ & $0.519\pm0.003$ &  $0.547\pm0.003$ & $0.736\pm0.049$ \\
  5 032412 & d  &  $0.880 \pm0.008$  & $7.284$ (fixed)  & $6.949\pm0.073$  & $0.578\pm0.050$ & $0.584\pm0.007$ &  $0.583\pm0.007$ & $0.546\pm0.005$ \\
  5 038089 & d  &  $1.012 \pm0.009$  & $4.606\pm0.047$  & $5.065\pm0.067$  & $0.693\pm0.019$ & $0.692\pm0.018$ &  $0.691\pm0.018$ & $0.678\pm0.032$ \\
  5 095337 & d &   $0.936 \pm0.007$  & $3.685\pm0.022$  & $3.835\pm0.040$  & $0.676\pm0.008$ & $0.684\pm0.008$ &  $0.691\pm0.008$ & $0.659\pm0.085$ \\
  5 095557 & d  &  $0.908 \pm0.013$  & $5.991\pm0.107$  & $6.602\pm0.165$  & $0.523\pm0.015$ & $0.532\pm0.015$ &  $0.541\pm0.015$ & $0.625\pm0.190$ \\
  5 100485 & d  &  $1.008 \pm0.007$  & $5.193\pm0.076$  & $5.216\pm0.080$  & $0.966\pm0.034$ & $0.965\pm0.034$ &  $0.963\pm0.034$ & $0.930\pm0.052$ \\
  5 100731 & c  &  $0.828 \pm0.035$  & $3.851\pm0.058$  & $3.351$ (fixed)  & $0.517\pm0.014$ & $0.533\pm0.013$ &  $0.552\pm0.012$ & $0.438\pm0.041$ \\
  5 106039 & sd &  $0.638 \pm0.006$  & $5.178\pm0.057$  & $2.928$ (fixed)  & $1.007\pm0.009$ & $1.076\pm0.009$ &  $1.168\pm0.009$ & $1.062\pm0.063$ \\
  5 111649 & d  &  $1.025 \pm0.011$  & $4.600$ (fixed)  & $4.736\pm0.067$  & $0.889\pm0.056$ & $0.886\pm0.015$ &  $0.882\pm0.014$ & $0.880\pm0.052$ \\
  5 123390 & d  &  $1.017 \pm0.020$  & $5.077$ (fixed)  & $5.631\pm0.106$  & $0.587\pm0.060$ & $0.587\pm0.011$ &  $0.585\pm0.010$ & $0.584\pm0.061$ \\
  5 180185 & d  &  $1.036 \pm0.029$  & $7.724\pm0.169$  & $8.571\pm0.249$  & $0.707\pm0.042$ & $0.705\pm0.025$ &  $0.700\pm0.024$ & $0.611\pm0.042$ \\
  5 180576 & d  &  $0.813 \pm0.027$  & $4.918\pm0.170$  & $5.071\pm0.207$  & $0.450\pm0.027$ & $0.465\pm0.028$ &  $0.483\pm0.029$ & $0.45 \pm0.07 $ \\
  5 185408 & d  &  $0.944 \pm0.018$  & $5.052$ (fixed)  & $4.957\pm0.094$  & $0.757\pm0.047$ & $0.764\pm0.017$ &  $0.772\pm0.016$ & $0.619\pm0.045$ \\
  5 196565 & d  &  $0.867 \pm0.016$  &$5.956$ (fixed)&$6.639\pm0.120$&$0.43 \pm0.04^{\mathrm{a}}$&$^{\mathrm{b}}$&$0.453\pm0.009$ & $0.413\pm0.040$ \\
  5 261267 & sd &  $0.728 \pm0.004$  & $3.353\pm0.013$  & $2.923$ (fixed)  & $0.439\pm0.002$ & $0.461\pm0.003$ &  $0.487\pm0.001$ & $0.451\pm0.039$ \\
  5 265970 & d  &  $0.901 \pm0.012$  & $4.157$ (fixed)  & $5.370\pm0.056$  & $0.219\pm0.024$ & $0.264\pm0.001$ &  $0.227\pm0.001$ & $0.233\pm0.024$ \\
  5 266015 & sd &  $0.757 \pm0.005$  & $3.512\pm0.026$  & $2.759$ (fixed)  & $0.494\pm0.002$ & $0.515\pm0.002$ &  $0.539\pm0.002$ & $0.617\pm0.045$ \\
  5 266131 & d  &  $0.936 \pm0.007$  & $4.510$ (fixed)  & $4.727\pm0.065$  & $0.647\pm0.060$ & $0.655\pm0.008$ &  $0.661\pm0.007$ & $0.644\pm0.060$ \\
  5 266513 & d  &  $0.959 \pm0.018$  & $4.390$ (fixed)  & $4.530\pm0.084$  & $0.692\pm0.070$ & $0.697\pm0.016$ &  $0.703\pm0.015$ & $0.733\pm0.076$ \\
  5 277080 & sd &  $0.685 \pm0.004$  & $3.932\pm0.047$  & $2.884$ (fixed)  & $0.567\pm0.015$ & $0.600\pm0.008$ &  $0.641\pm0.008$ & $0.690\pm0.022$ \\
  5 283079 & d  &  $0.997 \pm0.010$  & $5.162$ (fixed)  & $5.184\pm0.074$  & $0.987\pm0.086$ & $0.987\pm0.015$ &  $0.988\pm0.015$ & $0.936\pm0.088$ \\
\hline                  
\end{tabular}
\begin{list}{}{}
\item[$^{\mathrm{a}}$]$B$ data missing, ratio given by the WD code on the basis
of the $I$ data, via the atmosphere models
\item[$^{\mathrm{b}}$]$V$ data missing
\end{list}
\end{table*}

\begin{table*}
\caption{Astrophysical parameters for the primary components: masses, radii, effective temperatures, surface gravities, luminosities and 
equilibrium equatorial velocity (i.e. synchronous velocity for circular systems and pseudo-synchronous velocity for eccentric systems).}             
\label{astroparam1}      
\centering          
\begin{tabular}{l r l l l l r}     
\hline\hline       
Object & $\mathcal{M}_\mathrm{P}$   & $R_\mathrm{P}$   & $T_\mathrm{eff}^\mathrm{P}$  &  $\log g_\mathrm{P}^{\ \mathrm{a}}$   &
$\log L_\mathrm{P}/L_{\odot}^{\ \mathrm{b}}$  & $V_\mathrm{rot}^{P}$  \\ 
  & ($\mathcal{M}_{\odot}$) &  (R$_{\odot}$) &  (K) & (dex cgs) &  & (km s$^{-1}$) \\     

\hline                    
  4 110409 &  $13.93 \pm 0.59$ &  $4.64 \pm 0.14$ &  $32370 \pm  816$  & $4.25 \pm 0.03$ &  $4.326 \pm 0.051$ & $ 79 \pm 2$  \\
  4 113853 &  $ 6.03 \pm 0.38$ &  $3.50 \pm 0.12$ &  $23150 \pm  761$  & $4.13 \pm 0.04$ &  $3.499 \pm 0.065$ & $134 \pm 5$   \\
  4 117831 &  $ 4.81 \pm 0.25$ &  $2.71 \pm 0.08$ &  $19190 \pm  389$  & $4.26 \pm 0.04$ &  $2.949 \pm 0.044$ & $118 \pm 4$	     \\
  4 121084 &  $11.72 \pm 0.44$ &  $3.64 \pm 0.09$ &  $31580 \pm  552$  & $4.39 \pm 0.03$ &  $4.072 \pm 0.038$ & $223 \pm 6$  \\
  4 121110 &  $ 9.55 \pm 0.33$ &  $3.84 \pm 0.07$ &  $28390 \pm  767$  & $4.25 \pm 0.02$ &  $3.933 \pm 0.050$ & $175 \pm 3$  \\
  4 121461 &  $ 6.05 \pm 0.26$ &  $2.59 \pm 0.10$ &  $22970 \pm  783$  & $4.40 \pm 0.04$ &  $3.222 \pm 0.068$ & $ 67 \pm 3$   \\
  4 159928 &  $11.27 \pm 0.46$ &  $4.85 \pm 0.12$ &  $29080 \pm  870$  & $4.12 \pm 0.03$ &  $4.178 \pm 0.056$ & $213 \pm 5$  \\
  4 160094 &  $ 9.05 \pm 0.58$ &  $3.23 \pm 0.15$ &  $27550 \pm 1106$  & $4.38 \pm 0.05$ &  $3.730 \pm 0.081$ & $ 96 \pm 4$   \\
  4 163552 &  $ 9.56 \pm 0.61$ &  $5.06 \pm 0.16$ &  $24400 \pm  990$  & $4.01 \pm 0.04$ &  $3.910 \pm 0.075$ & $166 \pm 5$  \\
  4 175149 &  $11.78 \pm 0.47$ &  $4.85 \pm 0.16$ &  $33100 \pm  272$  & $4.14 \pm 0.03$ &  $4.402 \pm 0.032$ & $123 \pm 4$  \\
  4 175333 &  $ 5.09 \pm 0.23$ &  $2.41 \pm 0.08$ &  $21800 \pm  644$  & $4.38 \pm 0.03$ &  $3.071 \pm 0.059$ & $ 98 \pm 3$   \\
  5 016658 &  $ 6.38 \pm 0.39$ &  $3.57 \pm 0.12$ &  $21390 \pm  519$  & $4.14 \pm 0.04$ &  $3.378 \pm 0.051$ & $145 \pm 5$   \\
  5 026631 &  $11.40 \pm 0.34$ &  $5.19 \pm 0.08$ &  $28670 \pm  394$  & $4.07 \pm 0.02$ &  $4.212 \pm 0.028$ & $186 \pm 3$   \\
  5 032412 &  $17.12 \pm 0.34$ &  $4.74 \pm 0.11$ &  $35410 \pm  174$  & $4.32 \pm 0.02$ &  $4.500 \pm 0.022$ & $ 66 \pm 2$   \\
  5 038089 &  $13.01 \pm 0.20$ &  $5.98 \pm 0.08$ &  $30660 \pm  130$  & $4.00 \pm 0.01$ &  $4.452 \pm 0.014$ & $127 \pm 2$  \\
  5 095337 &  $ 8.72 \pm 0.36$ &  $3.68 \pm 0.12$ &  $26990 \pm 1250$  & $4.25 \pm 0.03$ &  $3.808 \pm 0.085$ & $206 \pm 7$  \\
  5 095557 &  $ 6.68 \pm 0.71$ &  $3.60 \pm 0.10$ &  $23710 \pm  363$  & $4.15 \pm 0.05$ &  $3.564 \pm 0.036$ & $ 75 \pm 2$  \\
  5 100485 &  $ 6.67 \pm 0.14$ &  $3.14 \pm 0.06$ &  $22670 \pm  498$  & $4.27 \pm 0.02$ &  $3.367 \pm 0.042$ & $104 \pm 2$  \\
  5 100731 &  $ 8.73 \pm 0.51$ &  $3.76 \pm 0.18$ &  $24550 \pm  473$  & $4.23 \pm 0.05$ &  $3.664 \pm 0.054$ & $168 \pm 8$		      \\
  5 106039 &  $ 8.69 \pm 0.37$ &  $3.63 \pm 0.08$ &  $26320 \pm  335$  & $4.26 \pm 0.03$ &  $3.754 \pm 0.029$ & $ 84 \pm 2$  \\
  5 111649 &  $ 5.56 \pm 0.16$ &  $5.30 \pm 0.11$ &  $17000 \pm  168$  & $3.74 \pm 0.02$ &  $3.322 \pm 0.024$ & $ 91 \pm 2$   \\
  5 123390 &  $ 9.83 \pm 0.26$ &  $4.38 \pm 0.14$ &  $27840 \pm  549$  & $4.15 \pm 0.03$ &  $4.013 \pm 0.044$ & $102 \pm 3$  \\
  5 180185 &  $ 6.28 \pm 0.14$ &  $4.41 \pm 0.14$ &  $18500 \pm  400$  & $3.95 \pm 0.03$ &  $3.311 \pm 0.046$ & $ 41 \pm 1$   \\
  5 180576 &  $ 7.30 \pm 0.31$ &  $3.25 \pm 0.14$ &  $25000 \pm 1057$  & $4.28 \pm 0.04$ &  $3.568 \pm 0.083$ & $105 \pm 5$   \\
  5 185408 &  $ 7.03 \pm 0.21$ &  $3.05 \pm 0.09$ &  $24170 \pm  475$  & $4.32 \pm 0.03$ &  $3.453 \pm 0.042$ & $106 \pm 3$   \\
  5 196565 &  $ 7.20 \pm 0.28$ &  $4.93 \pm 0.13$ &  $21530 \pm  368$  & $3.91 \pm 0.03$ &  $3.670 \pm 0.038$ & $ 63 \pm 2$   \\
  5 261267 &  $ 9.17 \pm 0.39$ &  $4.32 \pm 0.10$ &  $27930 \pm  904$  & $4.13 \pm 0.03$ &  $4.008 \pm 0.060$ & $171 \pm 4$  \\
  5 265970 &  $10.21 \pm 0.39$ &  $7.14 \pm 0.16$ &  $23280 \pm  153$  & $3.74 \pm 0.03$ &  $4.128 \pm 0.023$ & $103 \pm 2$   \\
  5 266015 &  $15.58 \pm 0.31$ &  $5.84 \pm 0.08$ &  $32120 \pm  545$  & $4.10 \pm 0.02$ &  $4.512 \pm 0.032$ & $163 \pm 2$  \\
  5 266131 &  $ 8.96 \pm 0.23$ &  $3.60 \pm 0.13$ &  $26510 \pm  774$  & $4.28 \pm 0.03$ &  $3.760 \pm 0.060$ & $140 \pm 5$  \\
  5 266513 &  $ 5.79 \pm 0.35$ &  $2.84 \pm 0.13$ &  $20800 \pm 1167$  & $4.30 \pm 0.05$ &  $3.131 \pm 0.106$ & $130 \pm 6$   \\
  5 277080 &  $ 9.68 \pm 0.55$ &  $4.72 \pm 0.09$ &  $29780 \pm  480$  & $4.08 \pm 0.03$ &  $4.196 \pm 0.032$ & $123 \pm 2$  \\
  5 283079 &  $ 7.56 \pm 0.41$ &  $2.98 \pm 0.08$ &  $25160 \pm  475$  & $4.37 \pm 0.03$ &  $3.504 \pm 0.041$ & $118 \pm 3$  \\
\hline                  
\end{tabular}
\begin{list}{}{}
\item[$^{\mathrm{a}}$] Computed from $g = \frac{G\mathcal{M}}{R^{2}}$
\item[$^{\mathrm{b}}$] Computed from $\log \frac{L}{L_{\odot}} = 2 \log \frac{R}{R_{\odot}} + 4 \log \frac{T_\mathrm{eff}}{T_\mathrm{eff,\odot}}$
\end{list}
\end{table*}

\begin{table*}
\caption{Astrophysical parameters for the secondary components: masses, radii, effective temperatures, surface gravities, luminosities and 
equilibrium equatorial velocity (i.e. synchronous velocity for circular systems and pseudo-synchronous velocity for eccentric systems).}             
\label{astroparam2}      
\centering          
\begin{tabular}{l r l l l l r}     
\hline\hline       
Object &  $\mathcal{M}_\mathrm{S}$  &  $R_\mathrm{S}$  & $T_\mathrm{eff}^\mathrm{S \ a}$  &  $\log g_\mathrm{S}^{\ \mathrm{b}}$  & 
$\log L_\mathrm{S}/L_{\odot}^{\ \mathrm{c}}$ & $V_\mathrm{rot}^{S}$ \\ 
  &  ($\mathcal{M}_{\odot}$) &  (R$_{\odot}$) &   (K) &  (dex cgs) &  & (km s$^{-1}$) \\     

\hline                    
  4 110409 & $ 7.30 \pm 0.36$ &  $7.81 \pm 0.18$ &  $21170 \pm 582$  & $3.52 \pm 0.03$ & $4.041 \pm 0.052$ & $133 \pm 3$\\
  4 113853 & $ 4.10 \pm 0.28$ &  $3.79 \pm 0.13$ &  $17060 \pm 664$  & $3.89 \pm 0.04$ & $3.038 \pm 0.074$ & $145 \pm 5$\\
  4 117831 & $ 4.72 \pm 0.25$ &  $2.77 \pm 0.08$ &  $18830 \pm 588$  & $4.23 \pm 0.04$ & $2.936 \pm 0.060$ & $120 \pm 4$\\
  4 121084 & $ 9.98 \pm 0.39$ &  $3.28 \pm 0.09$ &  $29680 \pm 560$  & $4.41 \pm 0.03$ & $3.875 \pm 0.041$ & $202 \pm 6$\\
  4 121110 & $ 7.13 \pm 0.26$ &  $2.85 \pm 0.07$ &  $25010 \pm 703$  & $4.38 \pm 0.03$ & $3.455 \pm 0.053$ & $130 \pm 3$\\
  4 121461 & $ 5.82 \pm 0.26$ &  $2.59 \pm 0.10$ &  $21790 \pm 909$  & $4.38 \pm 0.04$ & $3.131 \pm 0.080$ & $ 67 \pm 3$\\
  4 159928 & $ 6.27 \pm 0.30$ &  $4.07 \pm 0.12$ &  $21030 \pm 686$  & $4.02 \pm 0.03$ & $3.462 \pm 0.062$ & $179 \pm 5$\\
  4 160094 & $ 6.81 \pm 0.45$ &  $2.63 \pm 0.14$ &  $23500 \pm1059$  & $4.43 \pm 0.05$ & $3.276 \pm 0.091$ & $ 78 \pm 4$\\
  4 163552 & $ 9.10 \pm 0.57$ &  $4.54 \pm 0.15$ &  $24330 \pm 992$  & $4.08 \pm 0.04$ & $3.812 \pm 0.077$ & $149 \pm 5$\\
  4 175149 & $ 7.76 \pm 0.31$ &  $6.18 \pm 0.20$ &  $28110 \pm 306$  & $3.75 \pm 0.03$ & $4.330 \pm 0.034$ & $156 \pm 5$\\
  4 175333 & $ 3.98 \pm 0.14$ &  $2.06 \pm 0.07$ &  $19170 \pm 741$  & $4.41 \pm 0.03$ & $2.713 \pm 0.074$ & $ 84 \pm 3$\\
  5 016658 & $ 5.67 \pm 0.35$ &  $2.84 \pm 0.11$ &  $21510 \pm 553$  & $4.29 \pm 0.04$ & $3.190 \pm 0.057$ & $115 \pm 5$\\
  5 026631 & $ 7.86 \pm 0.25$ &  $4.93 \pm 0.08$ &  $21050 \pm 355$  & $3.95 \pm 0.02$ & $3.631 \pm 0.033$ & $177 \pm 3$\\
  5 032412 & $13.14 \pm 0.28$ &  $4.06 \pm 0.11$ &  $31160 \pm 315$  & $4.34 \pm 0.02$ & $4.144 \pm 0.029$ & $ 57 \pm 2$\\
  5 038089 & $11.70 \pm 0.18$ &  $4.94 \pm 0.07$ &  $31020 \pm 301$  & $4.12 \pm 0.01$ & $4.307 \pm 0.021$ & $105 \pm 2$\\
  5 095337 & $ 7.62 \pm 0.32$ &  $3.23 \pm 0.12$ &  $25270 \pm1185$  & $4.30 \pm 0.04$ & $3.581 \pm 0.087$ & $181 \pm 6$\\
  5 095557 & $ 5.70 \pm 0.64$ &  $2.86 \pm 0.08$ &  $21530 \pm 446$  & $4.28 \pm 0.05$ & $3.198 \pm 0.044$ & $ 60 \pm 2$\\
  5 100485 & $ 6.49 \pm 0.13$ &  $3.06 \pm 0.06$ &  $22840 \pm 527$  & $4.28 \pm 0.02$ & $3.359 \pm 0.044$ & $102 \pm 2$\\
  5 100731 & $ 6.65 \pm 0.40$ &  $3.23 \pm 0.16$ &  $20300 \pm 931$  & $4.24 \pm 0.05$ & $3.201 \pm 0.091$ & $144 \pm 7$\\
  5 106039 & $ 4.58 \pm 0.22$ &  $5.46 \pm 0.11$ &  $16800 \pm 265$  & $3.63 \pm 0.03$ & $3.328 \pm 0.032$ & $126 \pm 2$\\
  5 111649 & $ 5.21 \pm 0.15$ &  $4.89 \pm 0.10$ &  $17420 \pm 260$  & $3.78 \pm 0.02$ & $3.295 \pm 0.032$ & $ 84 \pm 2$\\
  5 123390 & $ 7.91 \pm 0.22$ &  $3.31 \pm 0.13$ &  $28320 \pm 784$  & $4.30 \pm 0.04$ & $3.800 \pm 0.059$ & $ 77 \pm 3$\\
  5 180185 & $ 5.66 \pm 0.13$ &  $3.60 \pm 0.13$ &  $19170 \pm 657$  & $4.08 \pm 0.03$ & $3.197 \pm 0.067$ & $ 33 \pm 1$\\
  5 180576 & $ 5.73 \pm 0.25$ &  $2.65 \pm 0.13$ &  $20310 \pm1090$  & $4.35 \pm 0.05$ & $3.030 \pm 0.103$ & $ 86 \pm 4$\\
  5 185408 & $ 5.97 \pm 0.19$ &  $2.80 \pm 0.08$ &  $22830 \pm 640$  & $4.32 \pm 0.03$ & $3.280 \pm 0.055$ & $ 97 \pm 3$\\
  5 196565 & $ 5.78 \pm 0.23$ &  $3.69 \pm 0.13$ &  $18670 \pm 461$  & $4.07 \pm 0.03$ & $3.170 \pm 0.052$ & $ 47 \pm 2$\\
  5 261267 & $ 4.81 \pm 0.23$ &  $3.86 \pm 0.10$ &  $20330 \pm 666$  & $3.95 \pm 0.03$ & $3.359 \pm 0.061$ & $153 \pm 4$\\
  5 265970 & $ 6.33 \pm 0.30$ &  $3.72 \pm 0.15$ &  $20980 \pm 340$  & $4.10 \pm 0.04$ & $3.381 \pm 0.044$ & $ 54 \pm 2$\\
  5 266015 & $ 6.86 \pm 0.16$ &  $5.46 \pm 0.08$ &  $24320 \pm 438$  & $3.80 \pm 0.02$ & $3.971 \pm 0.034$ & $153 \pm 2$\\
  5 266131 & $ 7.74 \pm 0.21$ &  $3.10 \pm 0.12$ &  $24820 \pm 750$  & $4.35 \pm 0.04$ & $3.513 \pm 0.063$ & $120 \pm 5$\\
  5 266513 & $ 4.90 \pm 0.30$ &  $2.45 \pm 0.13$ &  $19940 \pm1184$  & $4.35 \pm 0.05$ & $2.931 \pm 0.113$ & $112 \pm 6$\\
  5 277080 & $ 4.88 \pm 0.29$ &  $5.13 \pm 0.09$ &  $20410 \pm 372$  & $3.71 \pm 0.03$ & $3.612 \pm 0.035$ & $134 \pm 2$\\
  5 283079 & $ 7.58 \pm 0.41$ &  $2.97 \pm 0.08$ &  $25160 \pm 534$  & $4.37 \pm 0.03$ & $3.501 \pm 0.044$ & $117 \pm 3$\\
\hline                  
\end{tabular}
\begin{list}{}{}
\item[$^{\mathrm{a}}$] Computed from $T_{\mathrm{eff}}^{\mathrm{S}} = (T_{\mathrm{eff}}^{\mathrm{S}} / T_{\mathrm{eff}}^{\mathrm{P}})_{\mathrm{phot}} T_{\mathrm{eff}}^{\mathrm{P}}$
\item[$^{\mathrm{b}}$] Computed from $g = \frac{G\mathcal{M}}{R^{2}}$
\item[$^{\mathrm{c}}$] Computed from $\log \frac{L}{L_{\odot}} = 2 \log \frac{R}{R_{\odot}} + 4 \log \frac{T_\mathrm{eff}}{T_\mathrm{eff,\odot}}$
\end{list}
\end{table*}

\begin{table*} 
\caption{Distance determination: bolometric absolute magnitudes, bolometric corrections,
visual absolute magnitudes, visual extinction and reddening-free distance
modulus in the $V-$ and $I-$band. The $I-$band distance modulus is available
only for systems for which the $B$ and $V$ magnitudes are
available. The ``q" superscript denotes values at quadrature.}    
\label{dmi}
 \tiny     
\centering           
\begin{tabular}{l c c c c c c c c c c}     
\hline\hline       
Object & $M_\mathrm{bol}^\mathrm{P\ a}$ & $M_\mathrm{bol}^\mathrm{S\ a}$ & $BC_{V}^\mathrm{P\ b}$ & $BC_{V}^\mathrm{S\ b}$ & 
  $M_{V}^\mathrm{P\ c}$ & $M_{V}^\mathrm{S\ c}$ & $E_{B-V}^\mathrm{q\ d}$ & $A_{V}^\mathrm{q\ e}$ & $5 \log[$d$] -
  5_V^\mathrm{\ f}$ & $5 \log[$d$] - 5_I^\mathrm{\ g}$ \\
	& (mag) & (mag) & (mag) & (mag) & (mag) & (mag) & (mag) & (mag) & (mag) & (mag)  \\
\hline                    
 4 110409 & $-6.06\pm0.13$  & $-5.35\pm0.13$ & $-3.10\pm0.06$ & $-2.12\pm0.07$ & $-2.96\pm0.08$ & $-3.23\pm0.07$ & $0.207\pm0.009$ & $0.642\pm0.029$ & $19.06\pm0.061$ & $19.06\pm0.057$\\
 4 113853 & $-4.00\pm0.16$  & $-2.85\pm0.18$ & $-2.36\pm0.08$ & $-1.60\pm0.09$ & $-1.64\pm0.10$ & $-1.25\pm0.10$ & $0.194\pm0.009$ & $0.601\pm0.029$ & $19.00\pm0.078$ & $18.99\pm0.075$\\
 4 117831 & $-2.62\pm0.11$  & $-2.59\pm0.15$ & $-1.90\pm0.05$ & $-1.85\pm0.07$ & $-0.73\pm0.08$ & $-0.74\pm0.09$ & $0.064\pm0.006$ & $0.198\pm0.018$ & $18.99\pm0.062$ & $18.97\pm0.061$\\
 4 121084 & $-5.43\pm0.09$  & $-4.94\pm0.10$ & $-3.06\pm0.04$ & $-2.93\pm0.05$ & $-2.37\pm0.07$ & $-2.00\pm0.07$ & $0.187\pm0.005$ & $0.580\pm0.017$ & $19.28\pm0.051$ & $19.27\pm0.049$\\
 4 121110 & $-5.08\pm0.12$  & $-3.89\pm0.13$ & $-2.83\pm0.06$ & $-2.55\pm0.07$ & $-2.25\pm0.07$ & $-1.34\pm0.08$ & $0.158\pm0.008$ & $0.490\pm0.023$ & $19.08\pm0.057$ & $19.08\pm0.054$\\
 4 121461 & $-3.30\pm0.17$  & $-3.08\pm0.20$ & $-2.34\pm0.08$ & $-2.22\pm0.10$ & $-0.96\pm0.11$ & $-0.86\pm0.12$ & $0.147\pm0.009$ & $0.456\pm0.027$ & $19.05\pm0.083$ & $19.09\pm0.080$\\
 4 159928 & $-5.69\pm0.14$  & $-3.90\pm0.15$ & $-2.88\pm0.07$ & $-2.12\pm0.08$ & $-2.81\pm0.08$ & $-1.78\pm0.09$ & $0.168\pm0.007$ & $0.521\pm0.021$ & $19.29\pm0.066$ & $19.30\pm0.064$\\
 4 160094 & $-4.57\pm0.20$  & $-3.44\pm0.23$ & $-2.77\pm0.10$ & $-2.40\pm0.11$ & $-1.80\pm0.13$ & $-1.04\pm0.14$ & $0.075\pm0.009$ & $0.233\pm0.027$ & $18.96\pm0.102$ & $18.97\pm0.100$\\
 4 163552 & $-5.02\pm0.19$  & $-4.78\pm0.19$ & $-2.48\pm0.10$ & $-2.47\pm0.10$ & $-2.55\pm0.10$ & $-2.30\pm0.11$ & $0.187\pm0.008$ & $0.580\pm0.026$ & $18.35\pm0.079$ & $18.36\pm0.077$\\
 4 175149 & $-6.26\pm0.08$  & $-6.07\pm0.09$ & $-3.15\pm0.02$ & $-2.79\pm0.03$ & $-3.11\pm0.07$ & $-3.29\pm0.07$ & $0.053\pm0.007$ & $0.164\pm0.022$ & $18.52\pm0.057$ & $18.54\pm0.054$\\
 4 175333 & $-2.93\pm0.15$  & $-2.03\pm0.19$ & $-2.22\pm0.07$ & $-1.89\pm0.09$ & $-0.71\pm0.09$ & $-0.14\pm0.11$ & $0.070\pm0.007$ & $0.217\pm0.022$ & $18.61\pm0.074$ & $18.60\pm0.072$\\
 5 016658 & $-3.70\pm0.13$  & $-3.23\pm0.14$ & $-2.17\pm0.06$ & $-2.19\pm0.06$ & $-1.53\pm0.08$ & $-1.04\pm0.10$ & $0.080\pm0.006$ & $0.248\pm0.019$ & $19.13\pm0.068$ & $19.15\pm0.066$\\
 5 026631 & $-5.78\pm0.07$  & $-4.33\pm0.08$ & $-2.85\pm0.03$ & $-2.12\pm0.04$ & $-2.93\pm0.04$ & $-2.20\pm0.05$ & $0.105\pm0.004$ & $0.326\pm0.013$ & $19.13\pm0.036$ & $19.17\pm0.034$\\
 5 032412 & $-6.50\pm0.05$  & $-5.61\pm0.07$ & $-3.32\pm0.01$ & $-3.03\pm0.02$ & $-3.18\pm0.05$ & $-2.58\pm0.06$ & $0.250\pm0.006$ & $0.775\pm0.019$ & $19.19\pm0.044$ & $19.22\pm0.041$\\
 5 038089 & $-6.38\pm0.04$  & $-6.02\pm0.05$ & $-2.98\pm0.01$ & $-3.01\pm0.02$ & $-3.40\pm0.03$ & $-3.01\pm0.04$ & $0.057\pm0.011$ & $0.177\pm0.034$ & $18.89\pm0.042$ & $18.83\pm0.032$\\
 5 095337 & $-4.77\pm0.21$  & $-4.20\pm0.22$ & $-2.72\pm0.11$ & $-2.57\pm0.11$ & $-2.05\pm0.11$ & $-1.64\pm0.12$ & $0.124\pm0.016$ & $0.384\pm0.050$ & $19.17\pm0.097$ & $19.21\pm0.089$\\
 5 095557 & $-4.16\pm0.09$  & $-3.24\pm0.11$ & $-2.42\pm0.04$ & $-2.19\pm0.05$ & $-1.74\pm0.07$ & $-1.06\pm0.07$ & $0.116\pm0.004$ & $0.360\pm0.013$ & $19.19\pm0.052$ & $19.20\pm0.051$\\
 5 100485 & $-3.67\pm0.10$  & $-3.65\pm0.11$ & $-2.31\pm0.05$ & $-2.32\pm0.06$ & $-1.36\pm0.06$ & $-1.32\pm0.06$ & $0.077\pm0.009$ & $0.239\pm0.028$ & $18.84\pm0.052$ & $18.87\pm0.047$\\
 5 100731 & $-4.41\pm0.13$  & $-3.25\pm0.23$ & $-2.50\pm0.05$ & $-2.04\pm0.11$ & $-1.91\pm0.11$ & $-1.21\pm0.14$ & $0.106\pm0.005$ & $0.329\pm0.016$ & $19.28\pm0.090$ & $19.32\pm0.089$\\
 5 106039 & $-4.63\pm0.07$  & $-3.57\pm0.08$ & $-2.66\pm0.03$ & $-1.56\pm0.04$ & $-1.97\pm0.05$ & $-2.01\pm0.05$ & $0.154\pm0.010$ & $0.477\pm0.032$ & $18.95\pm0.050$ & $18.95\pm0.042$\\
 5 111649 & $-3.56\pm0.06$  & $-3.49\pm0.08$ & $-1.59\pm0.02$ & $-1.65\pm0.04$ & $-1.97\pm0.05$ & $-1.84\pm0.05$ & $0.177\pm0.009$ & $0.549\pm0.028$ & $18.86\pm0.046$ & $18.88\pm0.040$\\
 5 123390 & $-5.28\pm0.11$  & $-4.75\pm0.15$ & $-2.78\pm0.05$ & $-2.83\pm0.07$ & $-2.50\pm0.08$ & $-1.92\pm0.10$ & $0.092\pm0.015$ & $0.285\pm0.048$ & $18.75\pm0.078$ & $18.76\pm0.068$\\
 5 180185 & $-3.53\pm0.12$  & $-3.24\pm0.17$ & $-1.80\pm0.05$ & $-1.89\pm0.08$ & $-1.73\pm0.08$ & $-1.35\pm0.10$ & $0.067\pm0.012$ & $0.208\pm0.036$ & $19.45\pm0.073$ & $19.31\pm0.067$\\
 5 180576 & $-4.17\pm0.21$  & $-2.82\pm0.26$ & $-2.54\pm0.10$ & $-2.04\pm0.13$ & $-1.63\pm0.13$ & $-0.79\pm0.15$ & $0.150\pm0.013$ & $0.465\pm0.039$ & $19.14\pm0.106$ & $19.13\pm0.101$\\
 5 185408 & $-3.88\pm0.11$  & $-3.45\pm0.14$ & $-2.47\pm0.05$ & $-2.32\pm0.07$ & $-1.41\pm0.07$ & $-1.13\pm0.09$ & $0.100\pm0.005$ & $0.310\pm0.015$ & $19.12\pm0.057$ & $19.13\pm0.056$\\
$5 196565\ ^\mathrm{g}$&$-4.42\pm0.09$&$-3.17\pm0.13$&$-2.18\pm0.04$&$-1.82\pm0.06$&$-2.25\pm0.07$&$-1.35\pm0.09$ &               &	             &  	       &	        \\   
 5 261267 & $-5.27\pm0.15$  & $-3.65\pm0.15$ & $-2.79\pm0.08$ & $-2.04\pm0.08$ & $-2.48\pm0.08$ & $-1.61\pm0.08$ & $0.085\pm0.009$ & $0.263\pm0.027$ & $19.35\pm0.068$ & $19.29\pm0.064$\\
 5 265970 & $-5.57\pm0.06$  & $-3.70\pm0.11$ & $-2.36\pm0.02$ & $-2.12\pm0.04$ & $-3.21\pm0.05$ & $-1.58\pm0.09$ & $0.078\pm0.005$ & $0.242\pm0.016$ & $19.25\pm0.048$ & $19.28\pm0.046$\\
 5 266015 & $-6.53\pm0.08$  & $-5.18\pm0.08$ & $-3.08\pm0.04$ & $-2.46\pm0.04$ & $-3.45\pm0.05$ & $-2.71\pm0.05$ & $0.179\pm0.005$ & $0.555\pm0.017$ & $19.23\pm0.038$ & $19.25\pm0.036$\\
 5 266131 & $-4.65\pm0.15$  & $-4.03\pm0.16$ & $-2.68\pm0.07$ & $-2.53\pm0.07$ & $-1.97\pm0.10$ & $-1.50\pm0.11$ & $0.136\pm0.012$ & $0.422\pm0.036$ & $19.11\pm0.081$ & $19.12\pm0.075$\\
 5 266513 & $-3.08\pm0.26$  & $-2.58\pm0.28$ & $-2.10\pm0.13$ & $-1.99\pm0.14$ & $-0.98\pm0.15$ & $-0.58\pm0.16$ & $0.159\pm0.012$ & $0.493\pm0.038$ & $19.13\pm0.118$ & $19.11\pm0.114$\\
 5 277080 & $-5.74\pm0.08$  & $-4.28\pm0.09$ & $-2.93\pm0.04$ & $-2.04\pm0.04$ & $-2.81\pm0.05$ & $-2.24\pm0.05$ & $0.225\pm0.013$ & $0.698\pm0.042$ & $18.52\pm0.056$ & $18.71\pm0.046$\\
 5 283079 & $-4.01\pm0.10$  & $-4.00\pm0.11$ & $-2.56\pm0.05$ & $-2.56\pm0.05$ & $-1.45\pm0.07$ & $-1.44\pm0.07$ & $0.161\pm0.005$ & $0.499\pm0.017$ & $19.11\pm0.054$ & $19.07\pm0.052$\\
\hline                  
\end{tabular}
\begin{list}{}{}
\item[$^{\mathrm{a}}$] Computed from $M_\mathrm{bol} = -5 \log \frac{R}{R_{\odot}} - 10 \log \frac{T_\mathrm{eff}}{T_\mathrm{eff,\odot}} + M_\mathrm{bol,\odot}$
\item[$^{\mathrm{b}}$] Interpolated from Lanz \& Hubeny (\cite{LH03}, \cite{LH07})
\item[$^{\mathrm{c}}$] Computed from $M_{V} = M_\mathrm{bol} - BC_{V}$
\item[$^{\mathrm{d}}$] Computed from $E_{B-V} = (B-V)-(B-V)_{0}$
\item[$^{\mathrm{e}}$] Computed from $A_{V} = \mathcal{R}_{V} E_{B-V}$ with $\mathcal{R}_{V} = 3.1\pm0.3$
\item[$^{\mathrm{f}}$] Computed from $5 \log[d] - 5 = V^\mathrm{q} - M_V^\mathrm{q} - A_V^\mathrm{q}$
where $M_{V}^\mathrm{q} = -2.5 \log \left( 10^{-0.4 M_{V}^\mathrm{P}} + 10^{-0.4 M_{V}^\mathrm{S}} \right)$
\item[$^{\mathrm{g}}$] Computed from $5 \log[d] - 5 = I^\mathrm{q} - M_V^\mathrm{q} + (V-I)_0^\mathrm{q} - 0.600\,A_V^\mathrm{q}$
\item[$^{\mathrm{h}}$] $B$ and $V$ light curves missing
\end{list}
\end{table*}

\begin{table*}
\caption{Light curves: ratio of the primary minimum to the RMS ($I-$band), RMS scatters and minimum chi-squared values from WD/PHOEBE code.}             
\label{LCscatter}
\tiny      
\centering          
\begin{tabular}{l c c c c | l c c c c} 
\hline\hline  
Object & $\Delta I_{\mathrm{min I}} / \sigma_I$ & $\sigma_I$ & $\sigma_V$ & $\sigma_B$
   & Object & $\Delta I_{\mathrm{min I}} / \sigma_I$ & $\sigma_I$ & $\sigma_V$ & $\sigma_B$\\
	& & (mag) & (mag) & (mag)   &  & & (mag) & (mag) & (mag)   \\
\hline                    
  4 110409 &$\sim$65&0.016 & 0.028& 0.042 &  5 100485 & 30.1 &  0.017 & 0.018 & 0.020 \\
  4 113853 &  9.3 & 0.017 & 0.019 & 0.025 &  5 100731 & 19.2 &  0.021 & 0.013 & 0.022 \\
  4 117831 & 10.8 & 0.028 & 0.018 & 0.025 &  5 106039 & 23.7 &  0.016 & 0.107 & 0.114 \\
  4 121084 & 34.6 & 0.019 & 0.013 & 0.016 &  5 111649 & 19.7 &  0.011 & 0.017 & 0.029 \\
  4 121110 & 31.7 & 0.015 & 0.016 & 0.024 &  5 123390 & 13.8 &  0.010 & 0.050 & 0.063 \\
  4 121461 & 10.5 & 0.031 & 0.025 & 0.017 &  5 180185 & 18.6 &  0.019 & 0.040 & 0.038 \\
  4 159928 & 18.7 & 0.013 & 0.008 & 0.009 &  5 180576 &  9.7 &  0.021 & 0.021 & 0.030 \\
  4 160094 & 12.5 & 0.019 & 0.008 & 0.011 &  5 185408 & 10.0 &  0.023 & 0.012 & 0.013 \\
  4 163552 & 46.1 & 0.009 & 0.011 & 0.017 &$5 196565\ ^\mathrm{a}$&21.0&0.017& &       \\
  4 175149 & 36.1 & 0.024 & 0.021 & 0.028 &  5 261267 & 49.7 &  0.017 & 0.052 & 0.024 \\
  4 175333 & 12.1 & 0.027 & 0.020 & 0.022 &  5 265970 & 33.4 &  0.012 & 0.015 & 0.016 \\
  5 016658 & 22.4 & 0.020 & 0.009 & 0.020 &  5 266015 & 49.2 &  0.012 & 0.018 & 0.015 \\
  5 026631 & 24.5 & 0.010 & 0.011 & 0.013 &  5 266131 &  7.8 &  0.019 & 0.026 & 0.101 \\
  5 032412 & 33.4 & 0.009 & 0.012 & 0.028 &  5 266513 & 15.5 &  0.034 & 0.020 & 0.025 \\
  5 038089 & 22.4 & 0.014 & 0.042 & 0.049 &  5 277080 & 43.1 &  0.013 & 0.027 & 0.100 \\
  5 095337 & 17.9 & 0.018 & 0.186 & 0.206 &  5 283079 & 26.5 &  0.020 & 0.017 & 0.024 \\
  5 095557 & 21.1 & 0.026 & 0.022 & 0.019  \\
\hline                  
\end{tabular}
\begin{list}{}{}
\item[$^{\mathrm{a}}$] $B$ and $V$ light curves missing
\end{list}
\end{table*}

\begin{table*}
\caption{Radial velocity curves: RMS scatters.}             
\label{RVscatter}
\tiny      
\centering          
\begin{tabular}{l c c  | l c c } 
\hline\hline  
Object & $\sigma_\mathrm{P}$ & $\sigma_\mathrm{S}$  & 
  Object & $\sigma_\mathrm{P}$ & $\sigma_\mathrm{S}$ \\
	& (km s$^{-1}$) & (km s$^{-1}$)   & & (km s$^{-1}$) & (km s$^{-1}$)  \\
\hline                    
  4 110409 & 11.5 &  4.2   &   5 100485 &  6.1 &  4.4    \\
  4 113853 &  7.8 & 12.4   &   5 100731 &  7.0 & 17.4    \\
  4 117831 & 11.0 & 12.6   &   5 106039 &  8.1 & 12.2    \\
  4 121084 &  7.6 & 13.8   &   5 111649 &  4.3 &  5.5    \\
  4 121110 &  7.6 &  9.1   &   5 123390 &  8.4 & 14.1    \\
  4 121461 &  7.6 & 10.1   &   5 180185 &  4.0 &  2.9    \\
  4 159928 &  8.3 & 11.9   &   5 180576 &  8.0 & 12.8    \\
  4 160094 & 13.7 & 29.4   &   5 185408 &  7.1 & 12.7    \\
  4 163552 & 10.5 &  8.8   &   5 196565 &  5.7 &  9.5    \\
  4 175149 &  8.6 & 10.5   &   5 261267 &  7.3 & 13.5   \\
  4 175333 & 19.0 & 23.4   &   5 265970 & 12.8 & 11.8    \\
  5 016658 &  7.7 & 17.2   &   5 266015 &  5.2 &  9.8   \\
  5 026631 &  7.8 &  7.4   &   5 266131 &  9.9 &  8.4   \\
  5 032412 &  3.3 &  4.9   &   5 266513 & 14.6 &  9.8   \\
  5 038089 &  3.1 &  2.9   &   5 277080 &  8.3 &  8.0   \\
  5 095337 &  8.3 & 12.0   &   5 283079 & 12.4 & 11.7   \\
  5 095557 & 31.0 & 20.6   \\
\hline                  
\end{tabular}
\end{table*}

\begin{figure*}[htb]
\centering
\includegraphics[trim = 5mm 10mm 5mm 5mm, clip, width=17cm]{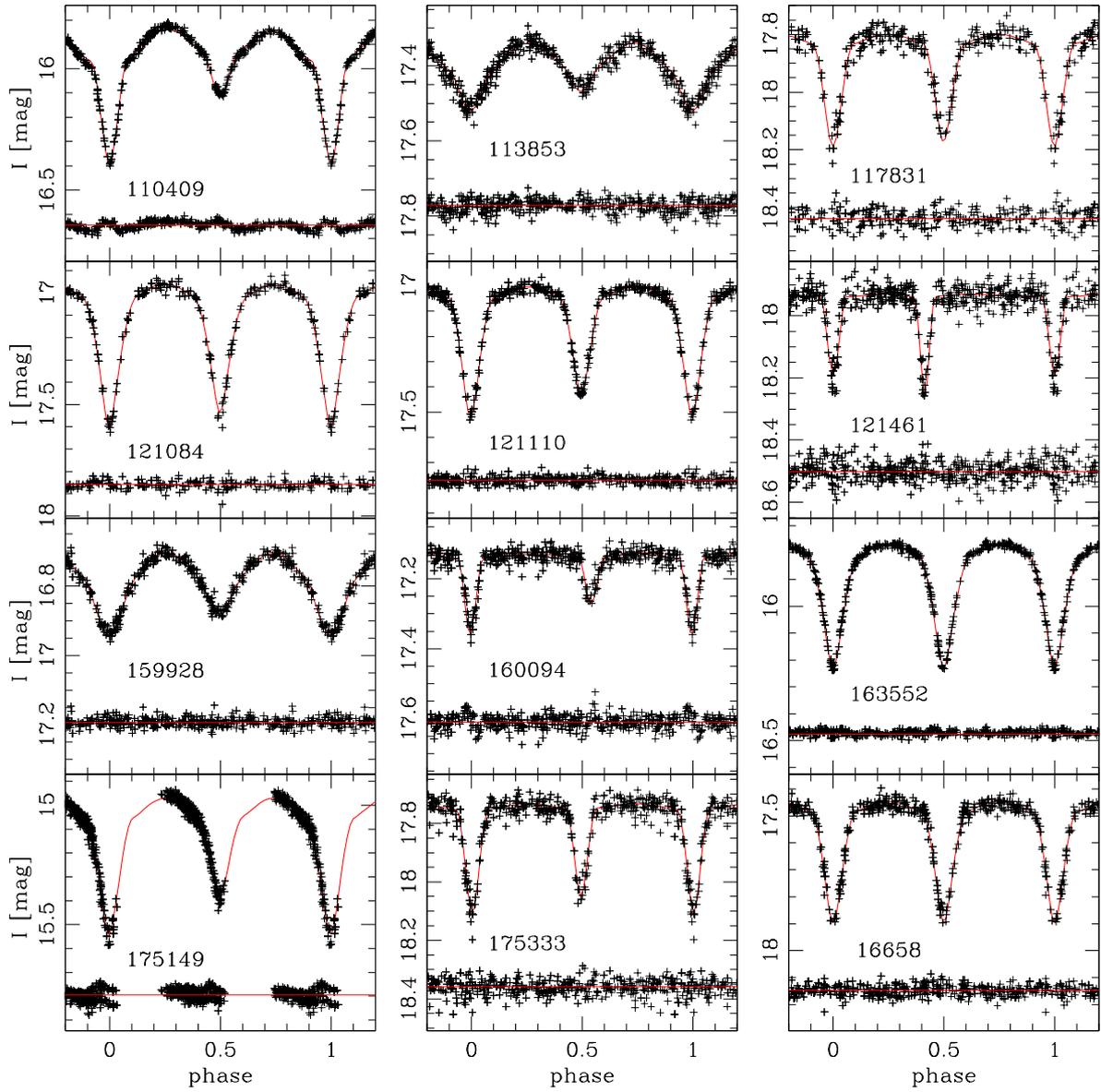}
\caption{$I-$band light curves with $O-C$ residuals.
See text for comments on individual stars. }
\label{allLC1}%
\end{figure*}

\begin{figure*}[htb]
\centering
\includegraphics[trim = 5mm 10mm 5mm 5mm, clip, width=17cm]{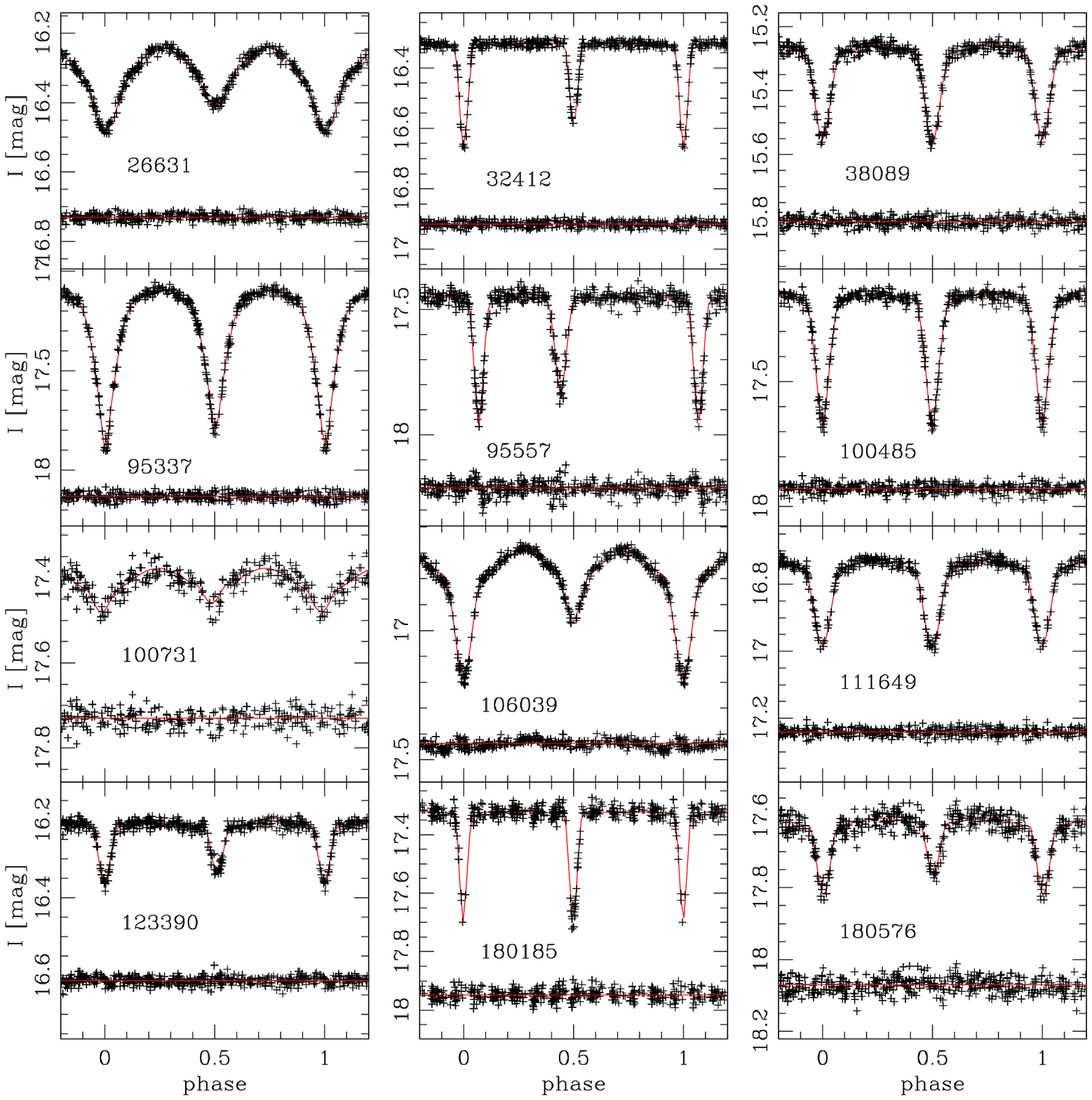}
\caption{Same as Fig. \ref{allLC1}, for 12 more stars. }
\label{allLC2}%
\end{figure*}

\begin{figure*}[htb]
\centering
\includegraphics[trim = 5mm 10mm 5mm 5mm, clip, width=17cm]{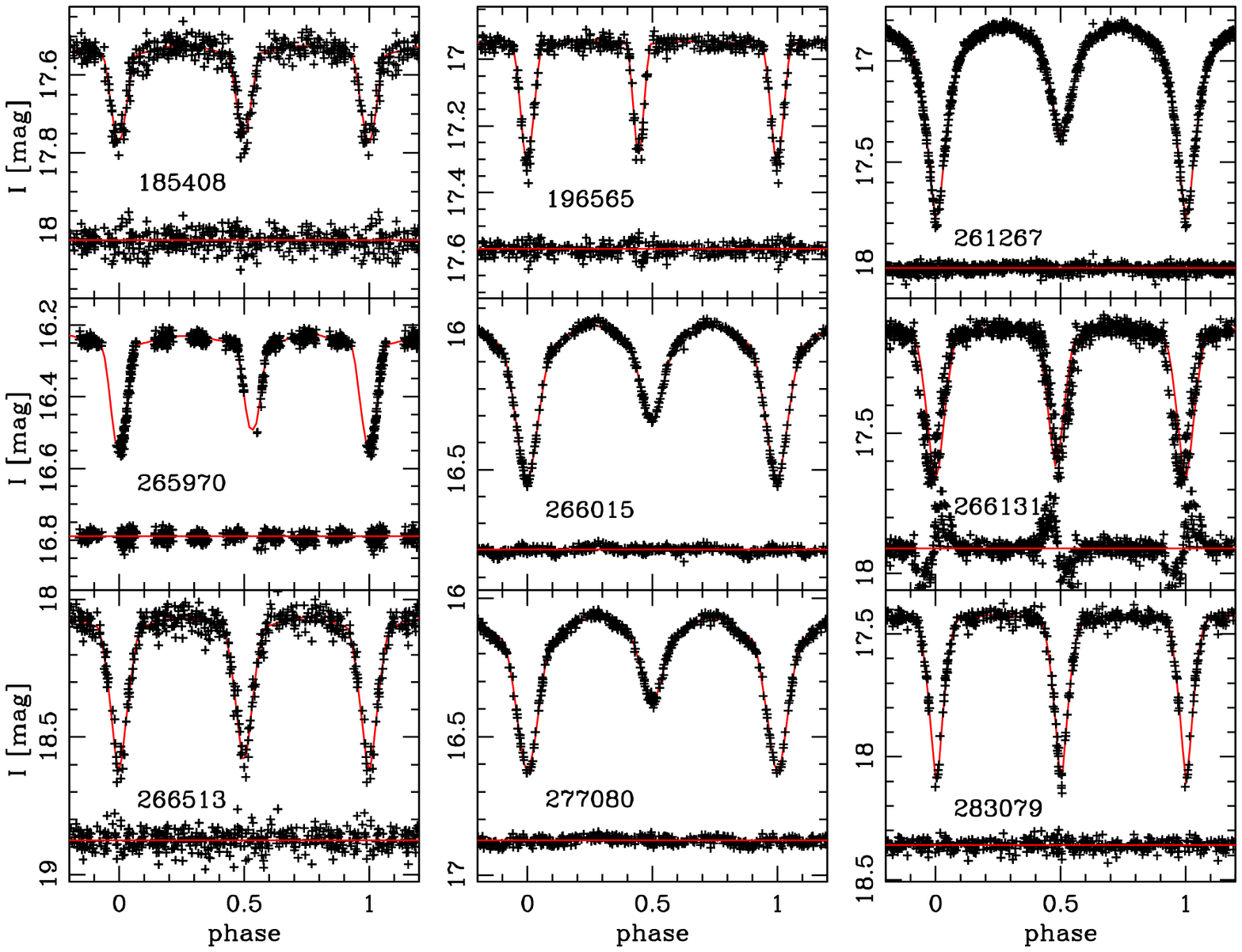}
\caption{Same as Fig. \ref{allLC1}, for nine more stars. }
\label{allLC3_small}%
\end{figure*}

\clearpage

\begin{figure*}[htb]
\centering
\includegraphics[trim = 5mm 10mm 5mm 5mm, clip, width=17cm]{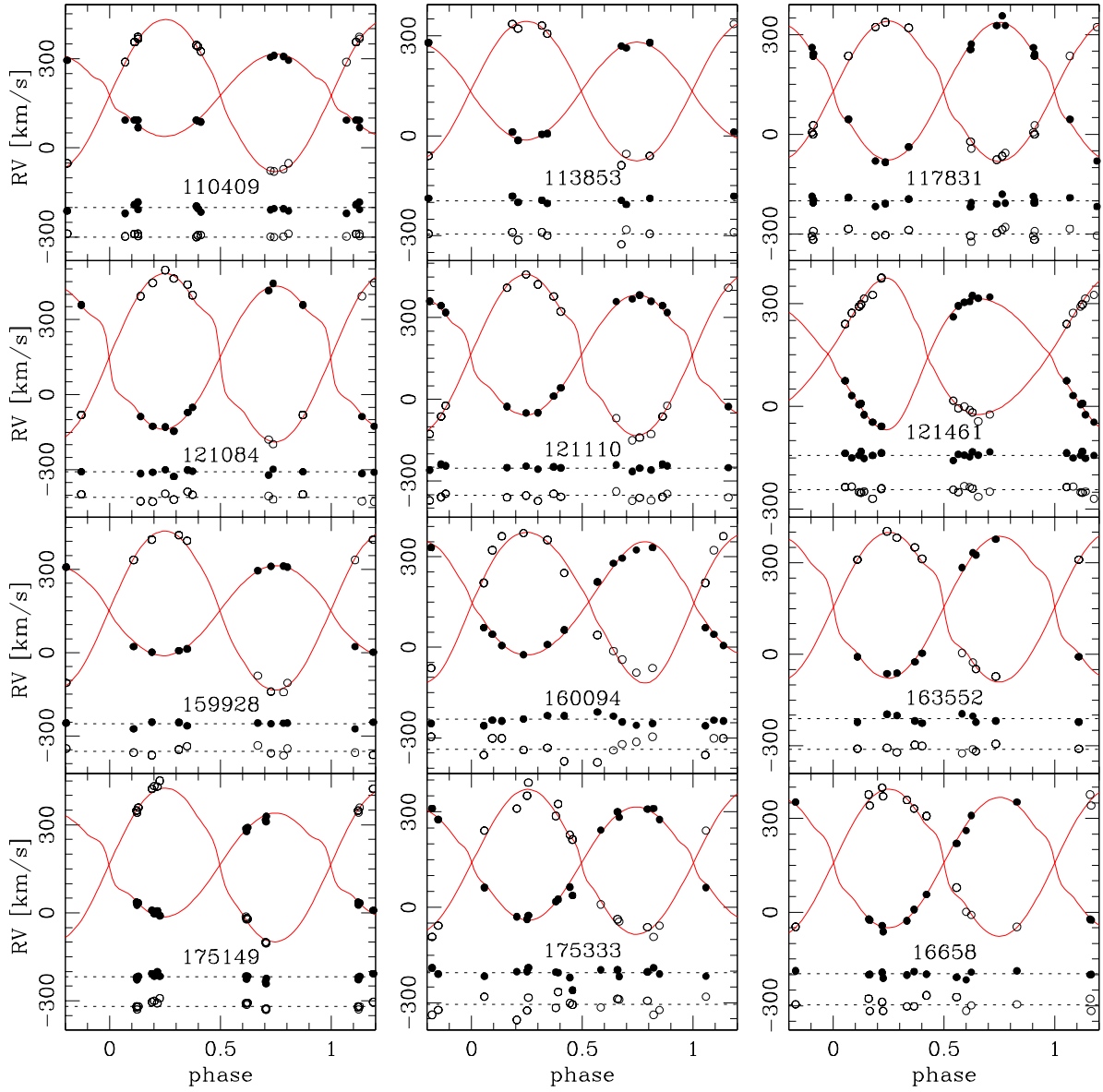}
\caption{Radial velocity data and best-fitting RV curves.The $O-C$ residuals
are shown with a different arbitrary offset for each component. See text for
comments on individual stars.}
\label{allRV1}%
\end{figure*}

\begin{figure*}[htb]
\centering
\includegraphics[trim = 5mm 10mm 5mm 5mm, clip, width=17cm]{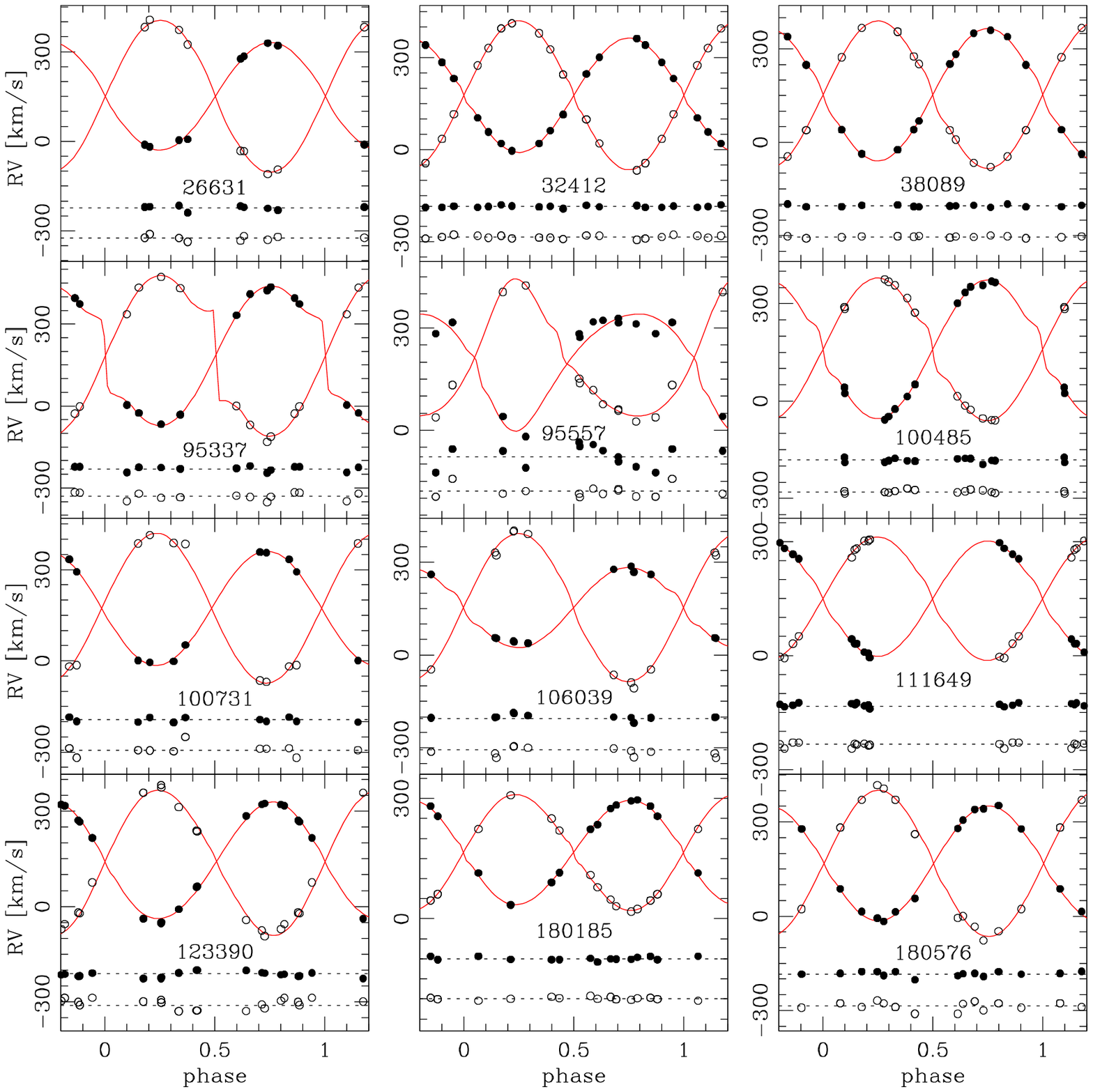}
\caption{ Same as Fig. \ref{allRV1}, for 12 more stars. }
\label{allRV2}%
\end{figure*}

\begin{figure*}[htb]
\centering
\includegraphics[trim = 5mm 10mm 5mm 5mm, clip, width=17cm]{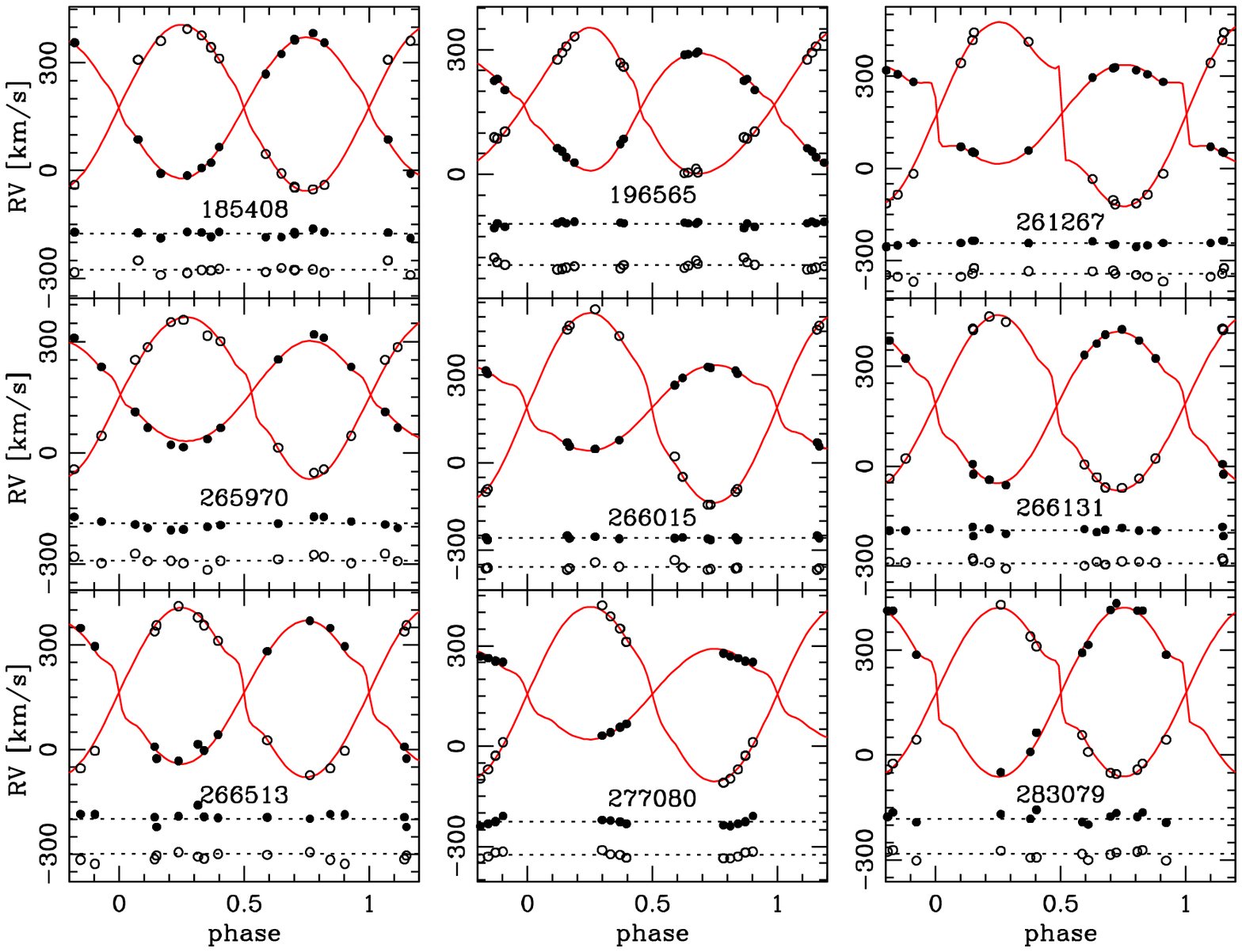}
\caption{Same as Fig. \ref{allRV1}, for nine more stars. }
\label{allRV3}%
\end{figure*}

\begin{figure*}[htb]
\centering
\includegraphics[trim = 5mm 10mm 5mm 5mm, clip, width=17cm]{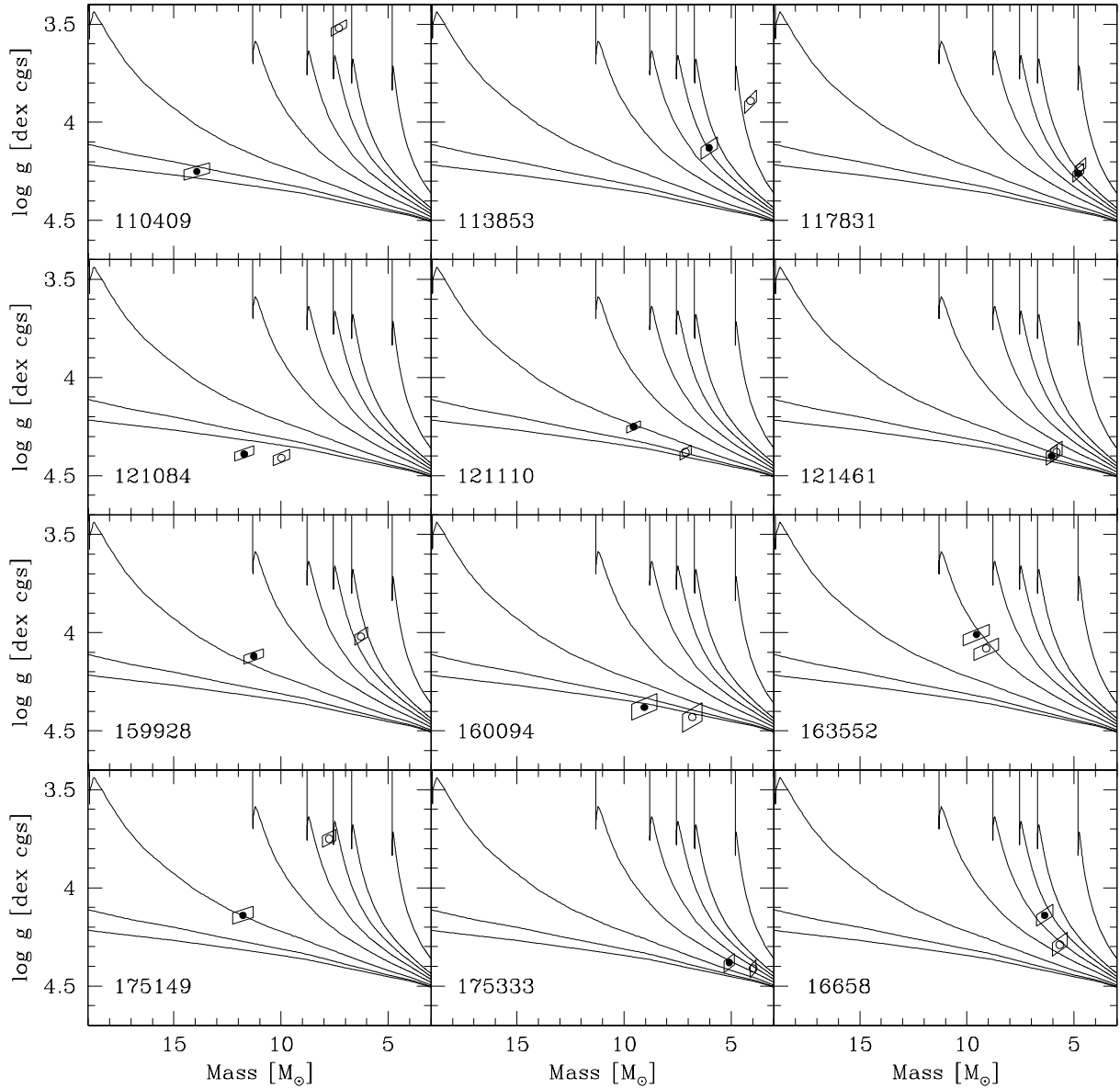}
\caption{Mass--surface gravity diagrams: the positions of the primary and the
secondary components are indicated by filled and open symboles,
respectively. The dotted lines ares isochrones from Charbonnel et al. (\cite{CMMSS93})
at $Z = 0.004$, with ages of 3, 5, 10, 20, 30, 40, 50 and 100 Myr. See text for
comments on individual stars.  }
\label{allGM1}%
\end{figure*}

\begin{figure*}[htb]
\centering
\includegraphics[trim = 5mm 10mm 5mm 5mm, clip, width=17cm]{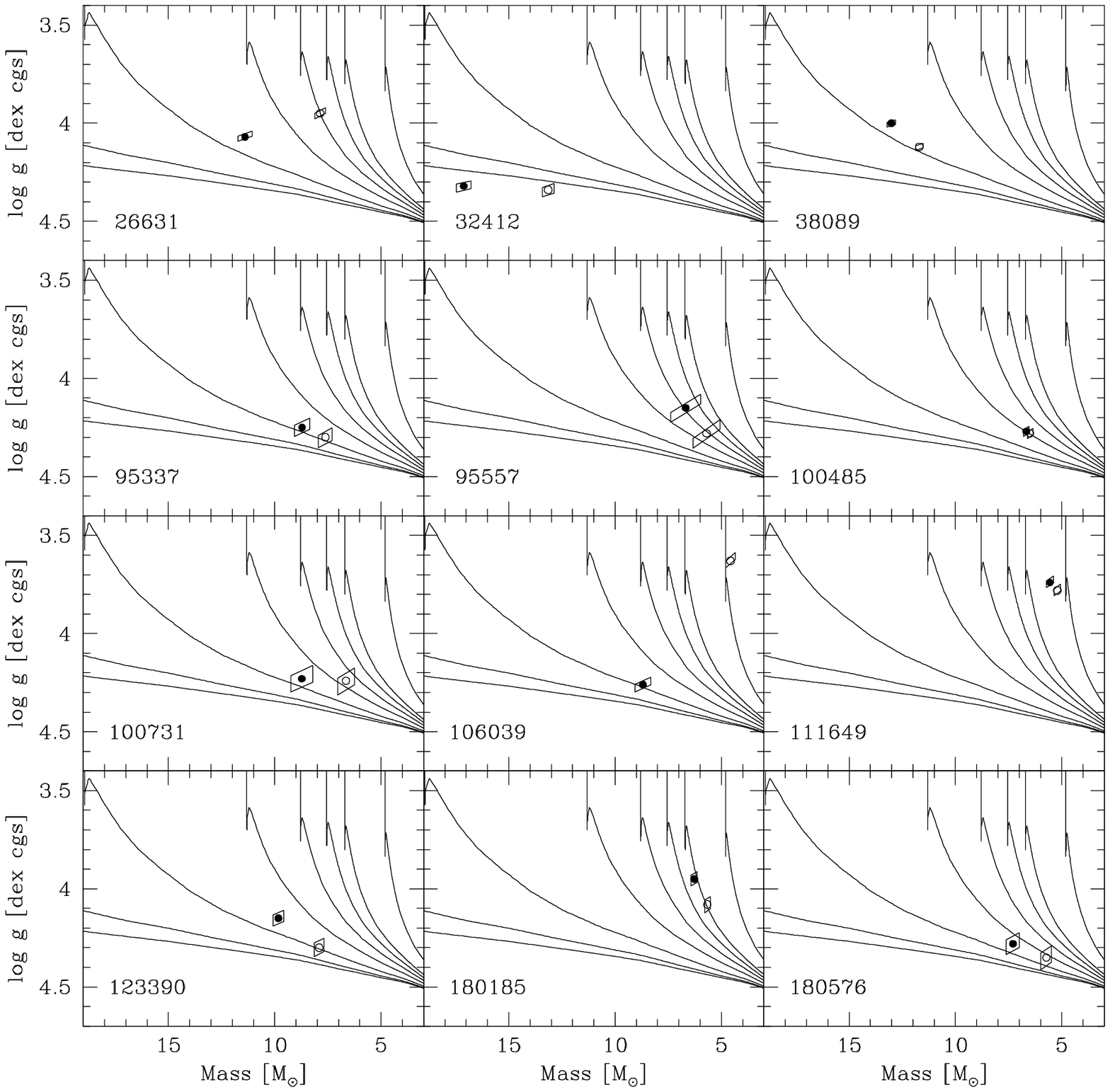}
\caption{Same as Fig. \ref{allGM1}, for 12 more stars.}
\label{allGM2}%
\end{figure*}

\begin{figure*}[htb]
\centering
\includegraphics[trim = 5mm 10mm 5mm 5mm, clip, width=17cm]{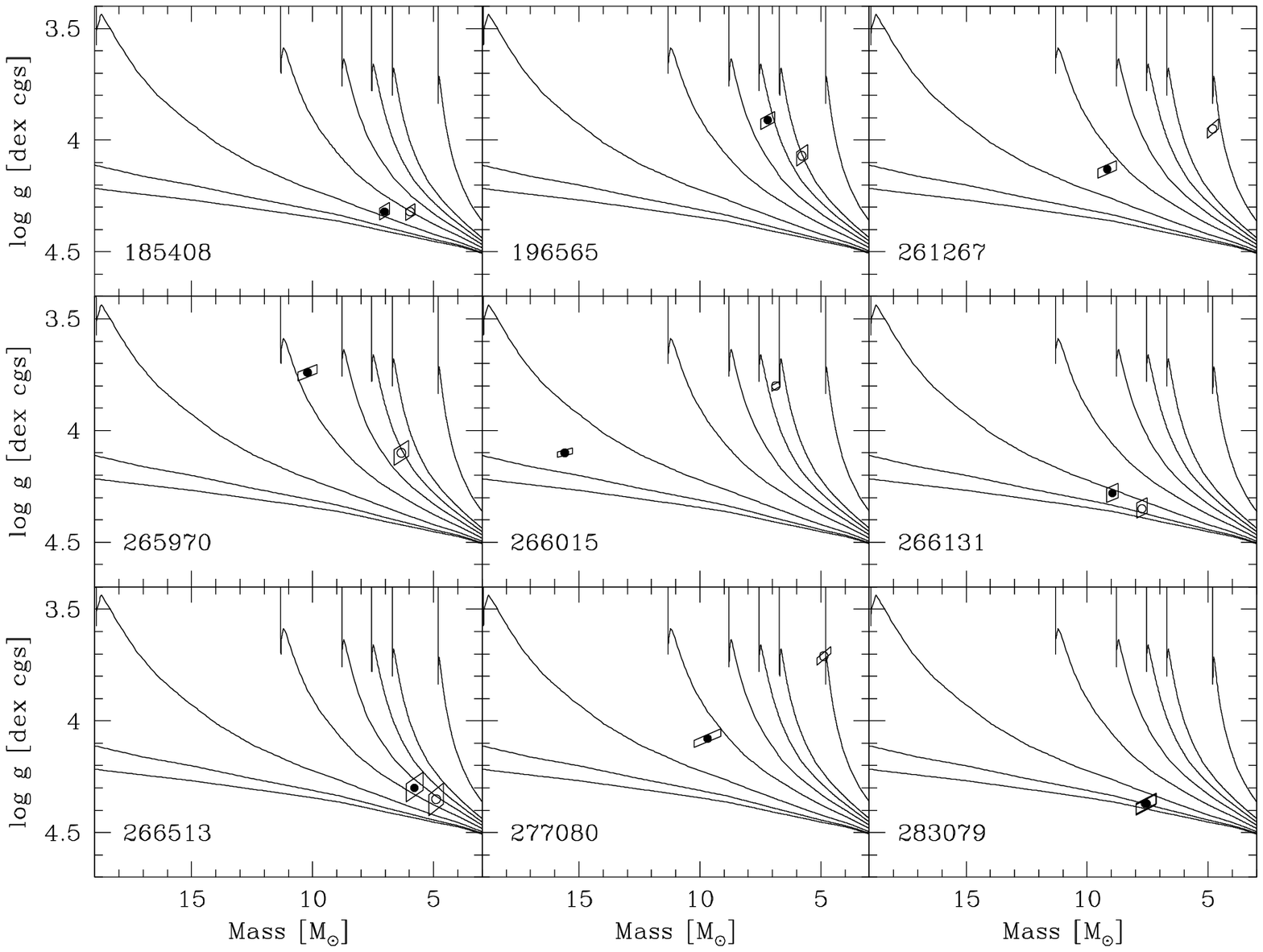}
\caption{Same as Fig. \ref{allGM1}, for nine more stars.}
\label{allGM3_small}%
\end{figure*}

\begin{figure*}[htb]
\centering
\includegraphics[trim = 5mm 10mm 5mm 5mm, clip, width=17cm]{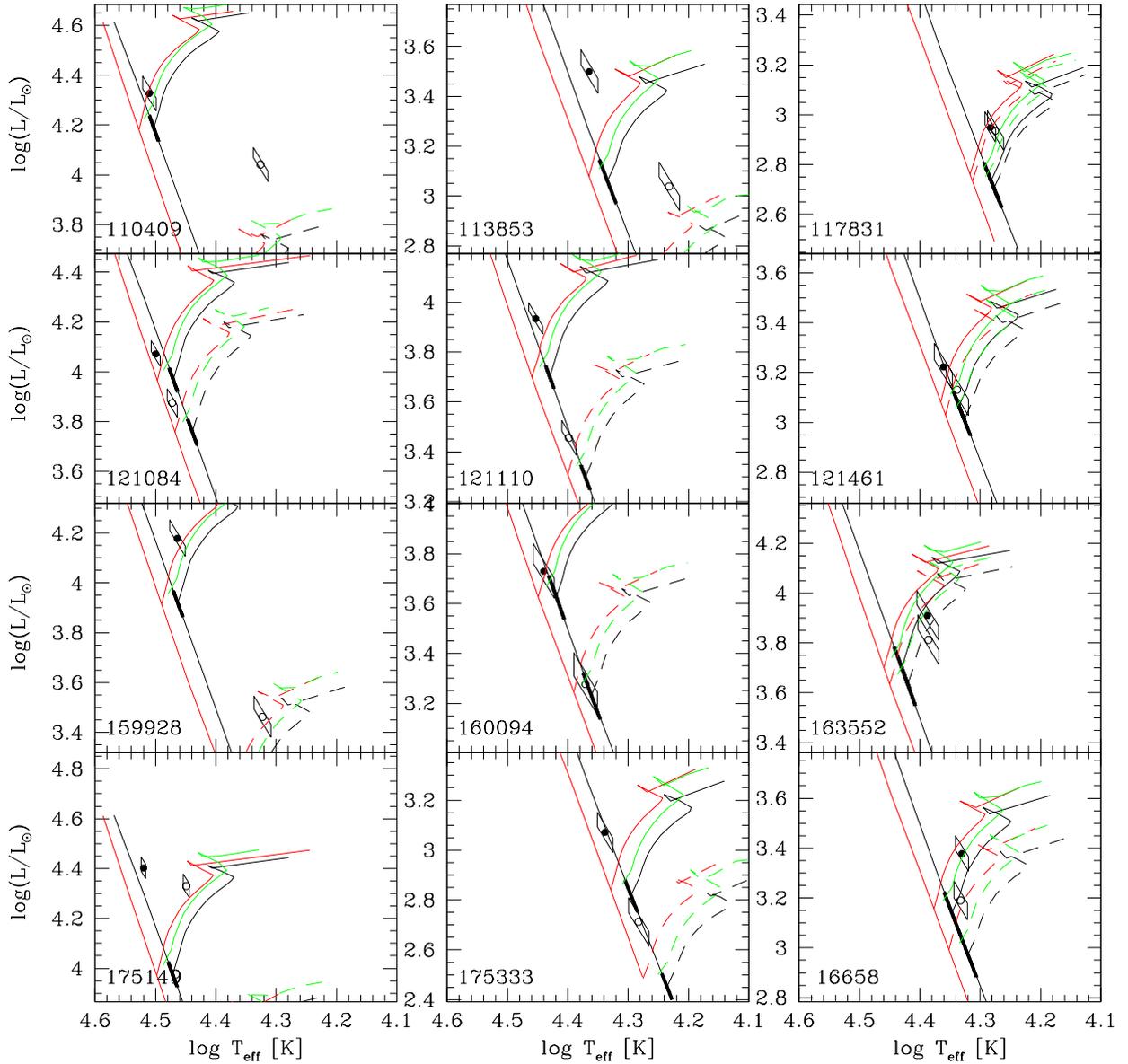}
\caption{HR diagrams: the positions of the primary and the secondary components
are indicated by filled and open symbols, respectively. The primary is the
component with the larger mass. The evolutionary tracks from Charbonnel et al.
(\cite{CMMSS93}) at $Z = 0.004$, corresponding to the observed masses, are
indicated by solid (primary) and dashed black lines (secondary). Since
these authors adopt a helium content $Y=0.24+3\times \Delta Z$, the helium
content of these models is $Y=0.252$. The
oblique line corresponds to the ZAMS. The bold segments on the ZAMS, at the
departure point of the evolutionary tracks, indicate the $\pm 1\,\sigma$ error
on the mass. The red tracks and ZAMS correspond to a poorer metallicity
$Z=0.001$ (Schaller et al. \cite{SSMM92}) and a helium content $Y=0.243$. The
green tracks are interpolated from the $Z=0.004$ models of Claret \& Gimenez
(\cite{CG98}) for a helium content $Y=0.28$; thus, they show the effect of a
helium enhancement $\Delta Y=0.028$ relative to the tracks of Charbonnel et al.
(\cite{CMMSS93}). See text for comments on individual stars.}
\label{allHR1}%
\end{figure*}

\begin{figure*}[htb]
\centering
\includegraphics[trim = 5mm 10mm 5mm 5mm, clip, width=17cm]{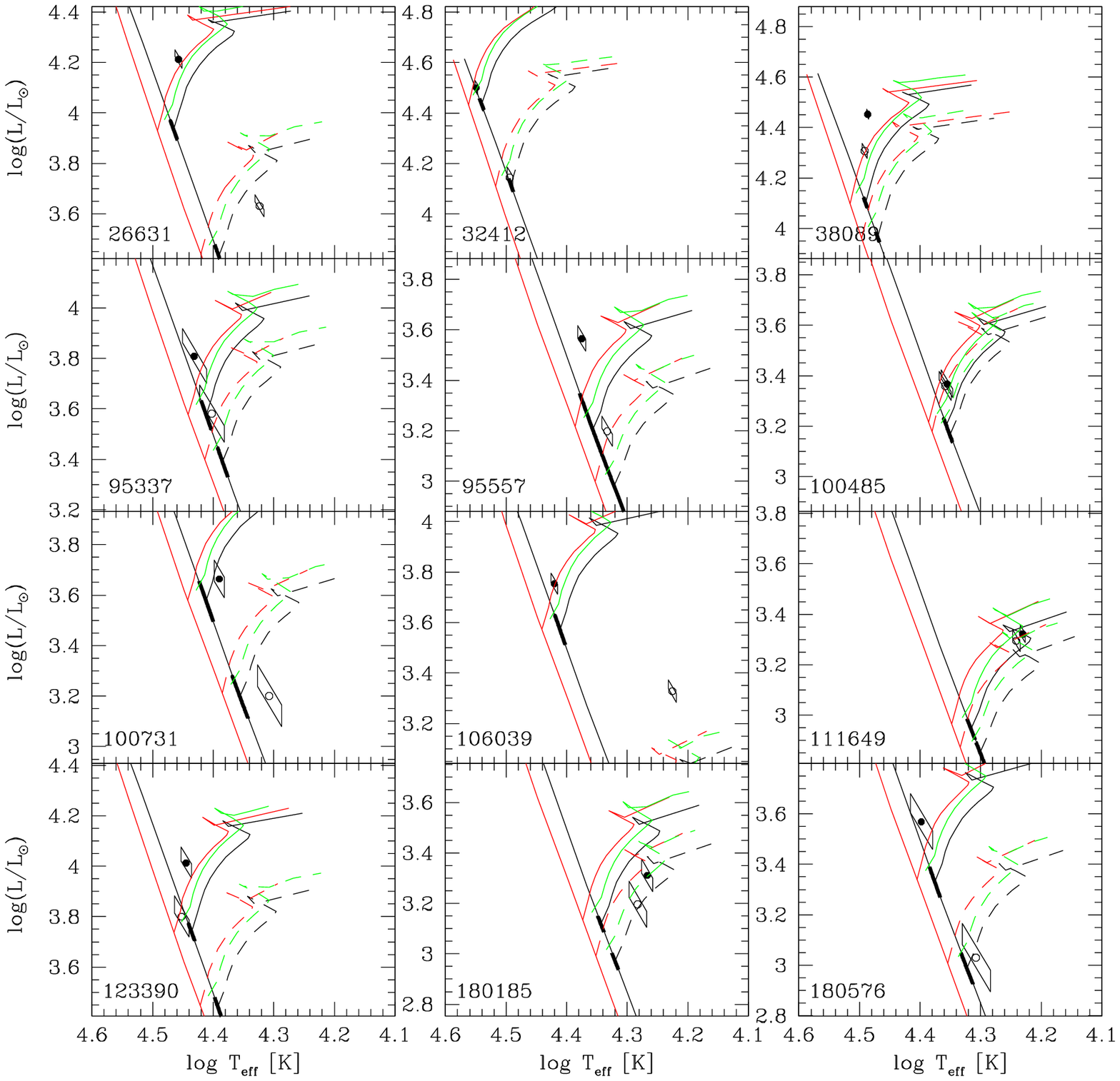}
\caption{Same as Fig. \ref{allHR1}, for 12 more stars.}
\label{allHR2}%
\end{figure*}

\begin{figure*}[htb]
\centering
\includegraphics[trim = 5mm 10mm 5mm 5mm, clip, width=17cm]{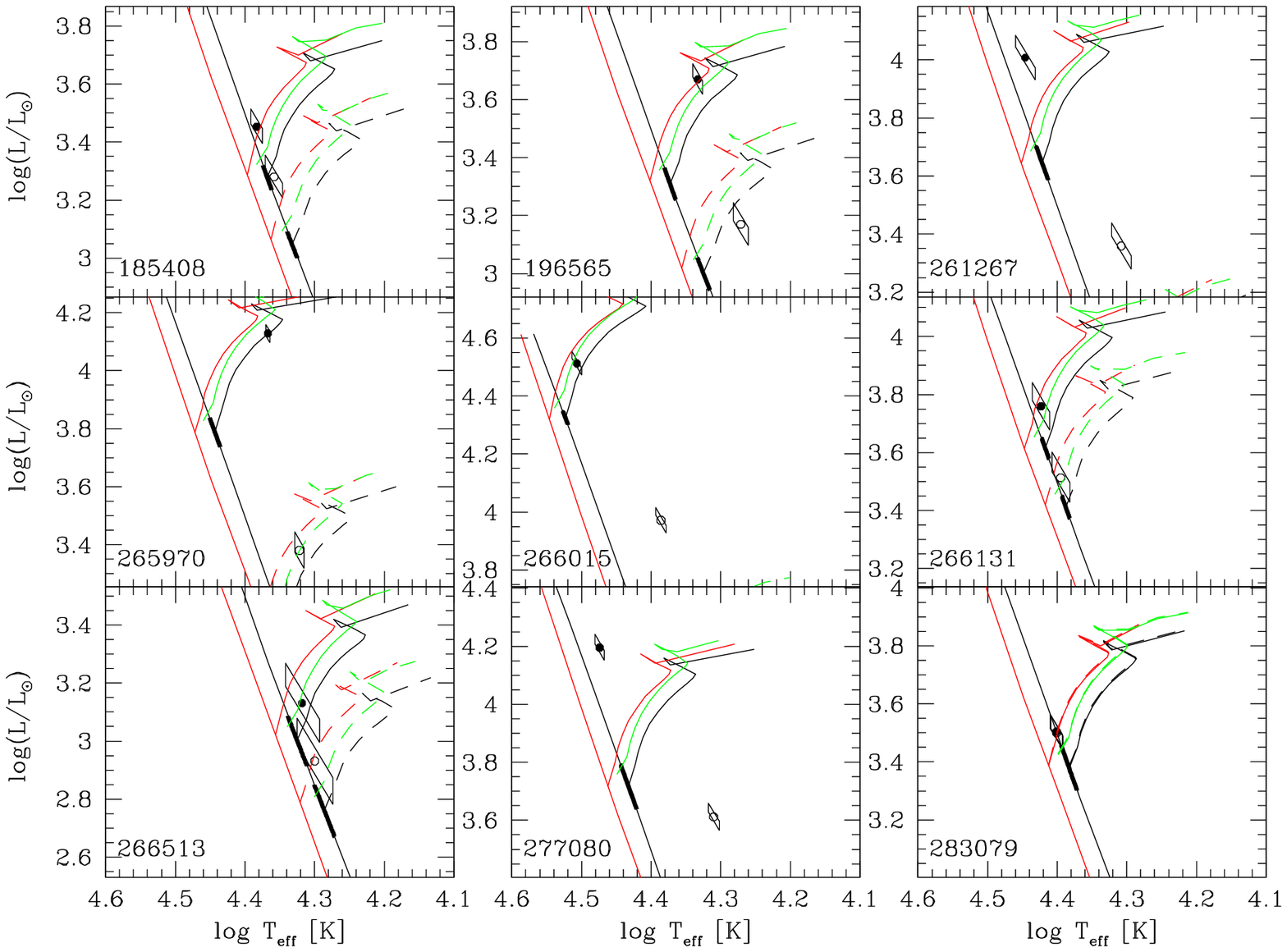}
\caption{Same as Fig. \ref{allHR1}, for nine more stars.}
\label{allHR3}%
\end{figure*}
\end{document}